%% file: axion_at_capp.tex
\documentclass{WileySev}

\usepackage{w-bookps}
\usepackage{amsmath}
\usepackage{graphicx}
\usepackage{caption}
\usepackage{floatrow}
\usepackage{comment}
\usepackage{amssymb}
\usepackage[linktocpage=true]{hyperref}
\usepackage{subcaption}
\usepackage{adjustbox}
\usepackage{multirow}
\usepackage{rotating}
\usepackage{makecell}
\graphicspath{{./}{./figures/}}
\usepackage{enumitem}
\usepackage{tabularx}
\usepackage[usenames, dvipsnames]{color}
\usepackage{gensymb}
\usepackage[bottom]{footmisc}
\usepackage{lineno}
\usepackage{chapterbib}

\begin{document}

\booktitle{Axion Dark Matter Research \\ \centerline{with IBS/CAPP}}
\author{Center for Axion and Precision Physics Research\\\\ \affil{Institute for Basic Science\\Korea Advanced Institute of Science and Technology}}

\titlepage

\offprintinfo{Axion Dark Matter Search with IBS/CAPP}{CAPP}

\begin{contributors} 
\name{Yannis K. Semertzidis}\footnote{Principal investigator ({\it yannis@ibs.re.kr})}  \\{\it Center for Axion and Precision Physics Research, Institute for Basic Science, Daejeon, Republic of Korea\\Korea Advanced Institute of Science and Technology, Daejeon, Republic of Korea}\\
\name{Jihn E. Kim} \\{\it Center for Axion and Precision Physics Research, Institute for Basic Science, Daejeon, Republic of Korea\\Department of Physics, Kyung Hee University, Seoul, Republic of Korea}\\

\name{Jihoon Choi, Woohyun Chung, Selcuk Haciomeroglu, Dongmin Kim, Jingeun Kim, ByeongRok Ko, Ohjoon Kwon, Andrei Matlashov, Lino Miceli, Hiroaki Natori, Seongtae Park, MyeongJae Lee, Soohyung Lee, Elena Sala, Yunchang Shin, Taehyeon Seong, Sergey Uchaykin, SungWoo Youn}\footnote{Corresponding author ({\it swyoun@ibs.re.kr})} \\{\it Center for Axion and Precision Physics Research, Institute for Basic Science, Daejeon, Republic of Korea}\\

\name{Danho Ahn, Saebyeok Ahn, Seung Pyo Chang, Wheeyeon Cheong, Hoyong Jeong, Junu Joeng, Dong Ok Kim, Jinsu Kim, On Kim, Younggeun Kim, Caglar Kutlu, Doyu Lee, Zhanibek Omarov, Chang-Kyu Sung, Beomki Yeo, Andrew Kunwoo Yi, Merve Yildiz} \\{\it Korea Advanced Institute of Science and Technology, Daejeon, Republic of Korea}

\end{contributors}

\tableofcontents

\begin{introduction}
\input{exec_sum/main.tex}

\end{introduction}

\chapter[Theoretical Motivation]{Theoretical Motivation}
\input{1.1/newcommand.tex}
\input{1.1/main.tex}

\chapter[Axion Dark Matter Experiments at CAPP]{Axion Dark Matter Experiments}
\section{Overview}
\input{1.3.1/main.tex}

\section{CAPP-PACE}
\input{1.3.2/main.tex}

\section{CAPP-8TB}
\input{1.3.3/main.tex}
\section{CAPP-9T MC}
\label{sec:capp-9T_MC}
\input{1.3.4/main.tex}
\section{CAPP-12TB}
\input{1.3.6/main.tex}
\section{CAPP-25T}
\input{1.3.7/main.tex}
\clearpage

\chapter[Experimental Research Activities]{Experimental Research Activities}
\section{Low noise amplifiers in GHz range}
\input{1.4.1/main.tex}

\section{Phase-matching for efficient high frequency axion search}
\label{sec:pizza_cavity}
\input{1.4.2/main.tex}

\section{Exploiting higher-order resonant modes}
\label{sec:higher_mode}
\input{1.4.3/main.tex}

\section{Study of magnetoresistance}
\label{sec:magnetoresistance}
\input{1.4.5/main.tex}
\section{R\&D for high Q cavity} 
\input{1.4.6/main.tex}
\section{Data acquisition} 
\input{1.4.7/main.tex}

\section{Experimental control system}
\label{sec:control_system}
\input{1.4.8/main.tex}
\clearpage

\chapter[Global Axion Research]{Global Axion Research}
\section{Global Network of Optical Magnetometers for Exotic Physics (GNOME)}
\input{1.6.1/main.tex}
\section{Axion Resonant Interaction Detection Experiment (ARIADNE)}
\label{sec:ARIADNE}
\input{1.6.2/main.tex}
\section{Dark Matter Axion Haloscope Search with the CAST Dipole Magnet at CERN (CAST-CAPP)}
\input{1.6.3/main.tex}

\end{document}

%% file: exec_sum/main.tex
The Center for Axion and Precision Physics Research was established in October 16, 2013, a little over five-years ago.  What brought us here was the love of Physics and to be able to contribute to the advancement of knowledge.  Axions are the result of the solution to the so-called strong-CP problem, i.e., why the electric dipole moment (EDM) of the neutron is too small; at least some 9 to 10 orders of magnitude below theory expectations!  A beautiful and extremely successful theory of strong interactions, confirmed at the highest level with measurements at DESY/Germany and elsewhere, seemed to fail miserably when it came to CP-violation!  A new, dynamic mechanism was proposed by Peccei and Quinn, based on induction from observations, resulting in a new particle first suggested by Weinberg and Wilczek independently.  The original name suggested by Weinberg was Higgslet as there are similarities to the Higgs mechanism in vacuum, with the obvious differences being that the Higgs is a scalar particle, while the axion is pseudoscalar and possibly their masses are very different.  However, Wilczek won the argument on this, with the name axion, after a detergent, since the mechanism ``...cleaned-up the mess..." and the word axion sounds like axial as in the axial current.

Our highest priority was to operate axion dark matter experiments with the highest possible sensitivity.  Since it is extremely difficult to make progress in this field, as witnessed by the very long history of the main players already in it, we had to make bold decisions in our research and development priorities.  Using microwave cavities, a method already used for a couple of decades, originally proposed by Pierre Sikivie, it was clear that we had to push in all fronts in order to be able to make a significant and meaningful contribution.  So, we chose to build a state of the art facility, capable of supporting several experiments in parallel.  The first experiments were meant to become learning benches, since the axion dark matter field is only deceivingly simple, and the difference between the design goals and reality could be way too large to be useful.  Later on, they were meant to host a number of experiments using high power magnets, but also several axion dark matter experiments using conventional superconductors phased locked and combined together offline.  So, the plan was to

\begin{enumerate}
\item Acquire several dilution refrigerator systems (dry) and install the infrastructure for it.  We benefited from advice of the experts in this field.  The goal was to achieve as low temperature as possible for our cavities, with a target below 50~mK.  Our cavities are routinely below that level even with the magnet on.
\item Acquire several high-tech instruments and equipment, which are expensive and easier to afford before the labor cost became large.
\item Prioritize the major high field, high volume magnets needed and start R\&D where it's needed to see what can be built.  The R\&D program was very successful and we came up with a path to success.   We decided to commission an HTS-25T/100mm magnet from Brookhaven National Lab and an LTS-12T/320mm magnet from Oxford company, based on $\rm Nb_3 Sn$ cable.  The probability of both magnets to function properly is very high, but the HTS magnet is currently funding limited due to issues at the review committee at IBS/HQ.
\item Hire a Young Scientist through the corresponding program to work on reaching high efficiency for high frequency axions using conventional microwave cavities, which he's proven to work at room temperature and currently he's testing at cryogenic temperatures.  In addition, he collaborated with scientists from a different team at CAPP and came up with an additional method using dielectric to raise the frequency while maintaining high efficiency.  The combination of the two methods promises to bring us well into the 20~GHz range using the same HTS and LTS magnets.
\item Acquire the helium re-liquefiers for the wet systems, needed for the high field magnets.
\item Attract the best scientists and personnel to build the required teams.  This involved several talks in many institutions around the world to attract the best talent.
\item Collaborate with professors and scientists from KAIST and KRISS, who were intimately involved in the build-up of the Center.  We targeted to build near quantum-noise limited RF-SQUID amplifiers in the 2$-$5~GHz range, and we succeeded to have the first ones in the world built right here.  Interference from outside CAPP forces derailed this effort before its completion.  We then went forward to hire our own world-class experts who are collaborating with external suppliers for our needs with quantum-noise limited amplifiers in the 1$-$10~GHz range.  Understanding and using efficiently this technology is a must in order to hope a chance of obtaining the best sensitivity.  We expect to fully integrate this technology with our axion dark matter experiments within the first half of 2019 reaching within a factor of two of the quantum-noise limit for the lower frequencies of our axion frequency target range and very near it, at the higher frequencies. 
\item Start R\&D on high quality cavity resonators that can tolerate large magnetic fields.  Even though this project suffered from interference from outside CAPP forces, we currently have a small, but still promising program going on.
\item Host several RF schools at CAPP to bring our people's expertise into the state of the art.
\item Collaborate in a small number of international collaborations as a means to train and challenge our students, but also to keep up with the latest developments in RF techniques and microwave-cavity related issues.
\end{enumerate}

\begin{figure}[b]
\centering
\includegraphics[width=0.8\textwidth]{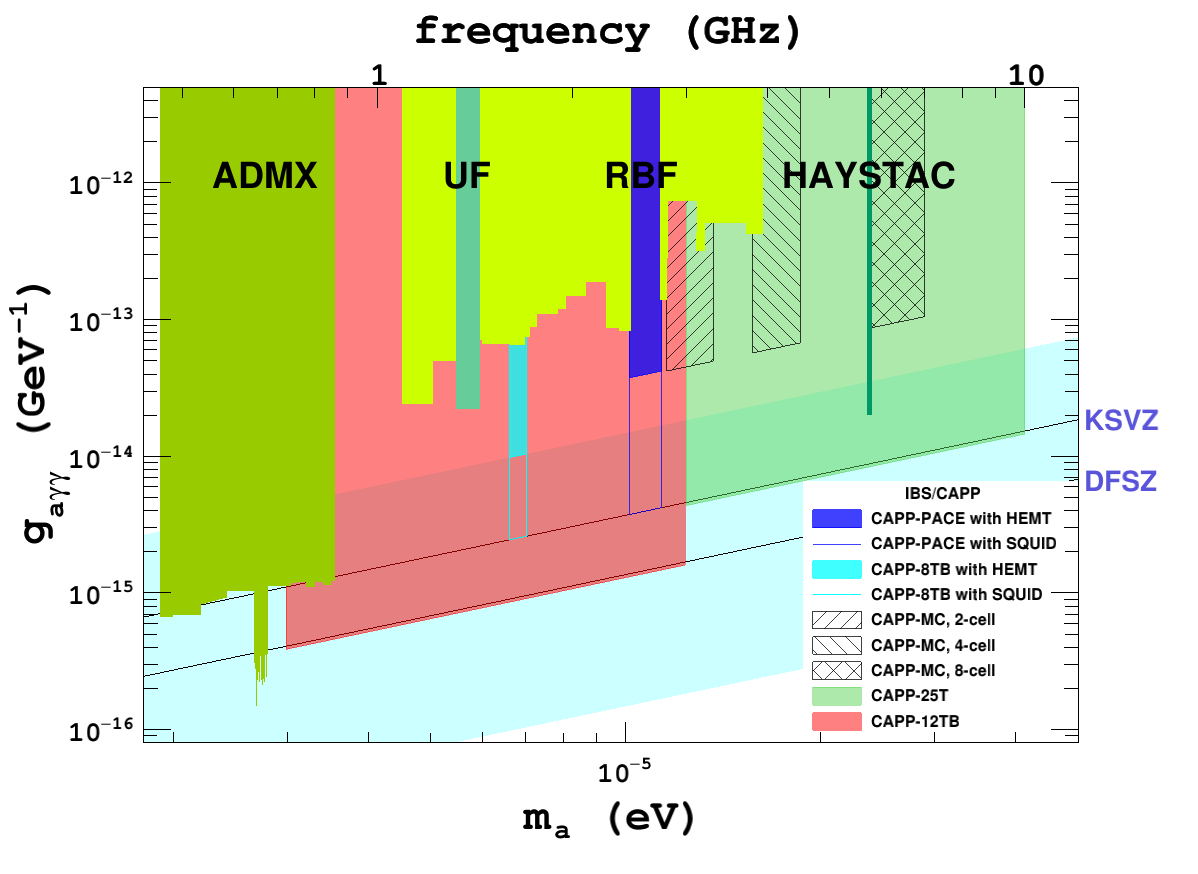}
\caption{\label{CAPP_sensitivity}Sensitivity to the axion dark matter of the various efforts at CAPP with the relevant axion mass range.  The limits refer to currently data taking efforts, CAPP-PACE (using an 8T magnet), CAPP-8TB (with a larger aperture 8T magnet), and CAPP-MC (for multi-cavity detectors with phase matched using a 9T magnet), and future projected limits with the HTS-25T/100mm and the LTS-12T/320mm magnets.
}
\end{figure}

\begin{figure}[t]
\centering
\includegraphics[width=0.8\textwidth]{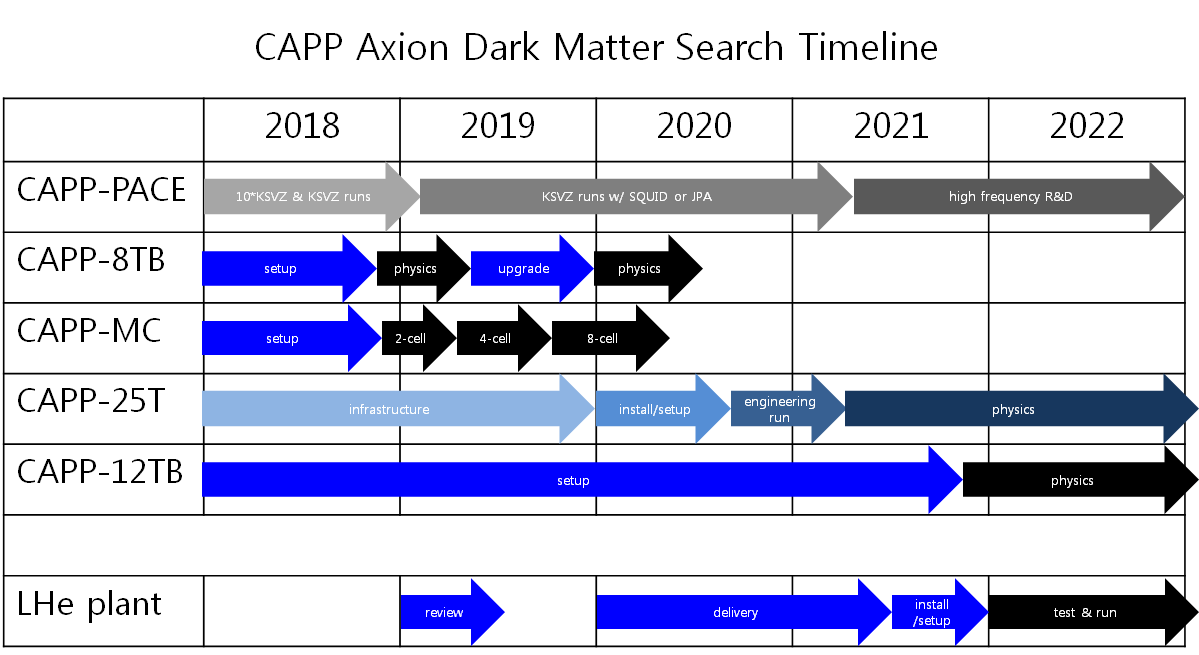}
\caption{\label{CAPP_timeline}Technically driven timeline of the axion dark matter experiments at IBS/CAPP.  The recent IBS budget reductions had a major effect in the high-field magnet delivery timelines.
}
\end{figure}

The CAPP projected sensitivity of the first experimental phase, to the axion dark matter as a function of its axion mass, is shown in Figure~\ref{CAPP_sensitivity} and the technically limited timeline of the program is summarized in Figure~\ref{CAPP_timeline}.

A small number of efforts is described next, whose hardware investments are either completed or are winding down.  
Part of our effort to probe the Strong-CP problem is to develop a new experimental concept regarding the proton EDM capable of improving the sensitivity to the relevant Physics by three to four orders of magnitude.  For this experiment we have played a leading role at CERN and have included it in the Physics Beyond Colliders (PBC) process, involving a competent group of experts from the accelerator dept. of CERN and a strong international community.  We were in charge of developing high-tech hardware, capable of detecting coherent beam motion with 1\,nm/$\sqrt{\rm Hz}$, an unprecedented sensitivity using SQUID-based magnetometers.  This concept was entirely developed at IBS/CAPP, while the SQUID gradiometers used for it were developed at KRISS.  We have also completed a number of systematic error studies demonstrating the feasibility of the experiment.  The next step is to write a full blown proposal to a suitable laboratory.

Our involvement in the muon $g-2$ experiments at Fermilab and J-PARC is small but significant in terms of systematic error studies and small scale hardware development. At Fermilab, we came up with a new method to reduce the coherent betatron oscillations of the stored beam by a factor of ten and scrape the beam in an improved way.  We have already implemented the needed hardware and we are planning to commission it within 2019.  Our Faraday magnetometer hardware, with a large dynamic range needed to measure the fast transient field from the kicker and its eddy currents, is the standard used at the experiment. Even smaller efforts include the COMET experiment at J-PARC (studying muon to electron conversion in the presence of a nucleus), and we are part of GNOME, a network distributed around the world of axion detectors based on optical magnetometry.  Our system is operational and it's reporting.  ARIADNE, is looking for axion mediated monopole-dipole interactions without the requirement that they are the dark matter.  For ARIADNE we are responsible for the development of the SQUID-based gradiometers, which is at an advanced stage. Finally, a new concept in the proposal stage, under development at CAPP, is to look for axion conversion to RF-photon transients, caused by the passage of axion stars in the vicinity of neutron stars, using large aperture RF-dish antennas.  They can reveal the axion frequency, which can then be confirmed by laboratory experiments.

Finally, a few words regarding the character of the Center.  We tried to bring in the culture similar to the one in the Physics department of BNL, which is based in doing only what you are best at in the world, while keeping high integrity and safety standards.  The scientists need to compete internationally at all times, need to learn to function in a diverse environment and they need to feel safe to come up with innovating, high-risk with high physics potential ideas.  Currently, the Center is mature, confident and it can compete in any aspect of our research subjects with any institution around the world.  Our people are competent, confident, respectful of each other and the regulations and as long as the IBS budgets do not deteriorate any further, we are going to keep a prominent position internationally in the axion dark matter and high precision physics.  The culture of institutions is mostly set at the very beginning, so starting this Center from nothing was a blessing in disguise, despite the immense amount of work involved.  Looking at our scientists, our students and the support personnel, we are confident that our goals are well within reach.

%% file: 1.1/newcommand.tex
\newcommand{\dis}[1]{\begin{equation}\begin{split}#1\end{split}\end{equation}}
 
\newcommand{\cred}[1]{{\color{red}{#1}}}
\newcommand{\cblue}[1]{{\color{blue}{#1}}}

\newcommand{\axem}{{\it axiem}}
   
 \newcommand{\cagg}{c_{a\gamma\gamma}}
\newcommand{\caggb}
{\bar{c}_{a\gamma\gamma}}
  
\newcommand\etal{{\it et al.}}
 \newcommand\ie{{\it i.e.}~}
   
\newcommand\fa{f_{a}}
\newcommand\vew{v_{\rm ew}}
\newcommand\Mp{M_{\rm P}}
\newcommand\MG{M_{\rm GUT}}
 \newcommand{\Qem}{Q_{\rm em}}
\newcommand\UPQ{U(1)$_{\rm PQ}$}
\newcommand\UG{U(1)$_{\rm global}$}
\newcommand\Uanom{U(1)$_{\rm anom}$}
\newcommand\NDW{N_{\rm DW}}
   
 \newcommand{\dell}{\delta_{\rm PMNS}}
\newcommand{\delq}{\delta_{\rm CKM}}

\newcommand\tev{\,{\rm TeV}}
\newcommand\gev{\,{\rm GeV}}
\newcommand\mev{\,{\rm MeV}}
\newcommand\kev{\,{\rm keV}}
\newcommand\ev{\,{\rm eV}}

\newcommand\axino{{\tilde{a}}}
\newcommand\maxino{{m_{\axino}}}
 
\newcommand{\Z}{{\bf Z}}

%% file: 1.1/main.tex
\section{Introduction on CP symmetry and related ideas}\label{sect:Intro}
The theory background of CAPP is centered around the discrete symmetry CP, which involves the search of the ``invisible'' axion and the detection of proton electric dipole moment.  Parity  P  is the most well-known example for the definition and violation of a discrete symmetry. Parity violation in weak interactions has been accepted in physics after T.D. Lee and C.N. Yang (who got Nobel Prize in 1957) proposed it and the subsequent discoveries of parity violation in Co$^{60}$  and the strange $\Lambda$ particle decays.  CP is a product of P and another discrete operation C. C stands for ``charge conjugation''.  In particle physics, it means that C operation is changing a particle to its anti-particle. In condensed matter physics, it is not so because all material considered in condensed matter physics is composed of particles in our Universe. Anyway, C operation changes a particle to its anti-particle. For a charged particle, its anti-particle has the opposite charge. So, it is obvious in case of charged particles. Even for  neutral particles, there are examples where anti-particles can be defined. The well-known example is the neutral K-meson. So, a better way to define a nontrivial particle-antiparticle is for a particle having a nonzero charge of some continuous symmetry. Continuous symmetries are classified into ``gauge'' and ``global'' symmetries. If this continuous symmetry is a gauged U(1) as QED, then the non-vanishing electromagnetic charge can define particle--antiparticle system as in the case of proton--antiproton. For the neutral K-meson, the global quantum number in consideration is ``strangeness''. CP is the product of these two discrete operations, C and P.

The first experimental observation of CP violation was in the neutral K meson system. There are two neutral K mesons, $K$ and its anti-particle $\bar K$. Out of these two neutral K mesons, one can linear-combine to have CP eigenstates, $K_1$ and $K_2$ which have CP even and odd, respectively.  The dominant decay channels of these K mesons are   two pions and three pions. Two pions have CP even.  $K_1$ decays predominently to two pions, and $K_2$ decays predominently to three pions.  So, if  $K_2$ decays to two pions, then CP is violated. The exact mass eigenstate close to  $K_2$ is called   ``long-lived'' K meson, $K_L$. The other orthogonal state is called   ``short-lived'' K meson, $K_S$. Because CP is violated, $K_L$ is not exactly $K_1$ but very close to it. The $K_L$ decay to two pions was discovered in 1964 by Christenson, Cronin, Fitch, and Turlay. Among these Cronin and Fitch got  Nobel Prize in 1980. It was known in the neutral K meson system  that ``CP violation is an interference phenomenon''.

{\it CP violation is an interference phenomenon.  }
 Theory of electromagnetic interaction, quantum electrodynamics (QED), introduces the photon coupling to charged particles through the vector current, known as the ``electromagnetic current''. Current is the quantity appearing in the RHS of Maxwell's equations. This four component vector current behaves like a 4-vector under the Lorentz transformation. Furthermore, the ``electromagnetic current'' written in terms of fields corresponding to charged particles is diagonal, which means that photon does not change flavors. For example, photon cannot change electron to muon, etc. Therefore, the interference phenomenon for CP violation cannot be introduced through QED alone. Modern theory of strong interaction, quantum chromodynamics (QCD), introduces the gluon coupling to colored particles through the vector current, known as the ``colored quark current'', which does not change flavors. Because it is a flavor-conserving vector current, QCD by current alone cannot introduce CP violation.
 
\section{CP violation in weak interactions}\label{sect:Weak}
 In the above discussion, we restriced to the currents in QED and QCD. To have interference phenomenon for CP violation through currents, we must consider flavor changing gauge bosons, i.e. charged weak gauge bosons $W^\pm_\mu$. The currents coupled to W bosons are called ``charged currents'', which describe weak interactions by charged currents. In weak interactions, there is also the neutral heavy gauge boson $Z_\mu$ coupling to the neutral currents, but the weak neutral currents conserve flavors and do not help in introducing the interference phenomenon for CP violation. Modern theory of weak interactions is the so-called Standard Model (SM) of particle physics in which the weak interaction part was proposed by Glashow, Salam and Weinberg who shared Nobel Prize in 1979. Discoverers of W and Z bosons, Rubbia and van der Meer, got Nobel Prize in 1984.

In the SM, it was an issue to introduce the weak CP violation.  Some ideas of weak CP violation were
\begin{eqnarray}
&1.& \textrm{by light colored scalars  by Kobayashi and Maskawa \cite{KM73}}, \nonumber \\
&2.& \textrm{by right-handed current(s) by Mohapatra \cite{Mohapatra72}, and Kobayashi and Maskawa \cite{KM73},} \nonumber  \\ 
&3.& \textrm{by three left-handed families   by Kobayashi and Maskawa \cite{KM73}} \\ 
&4.& \textrm{by propagators of light color-singlet scalars by Weinberg \cite{Weinberg76},}\nonumber  \\ 
&5.& \textrm{by an extra-U(1) gauge interaction.}  \nonumber 
\end{eqnarray}
The old Wolfenstein's superweak interaction \cite{Wolfenstein64} can be reproduced by item 5  in modern gauge theories. Item 3, known as the Cabibbo-Kobayashi-Maskawa (CKM) model, got Nobel Prize in 2008.

The CKM model defines the charged currents by the $3\times 3$ matrix known as the CKM matrix. It describes the charged currents from $Q=-\frac13$ quarks ($d,s,b$) to $Q=+\frac23$ quarks ($u,c,t$), written with the parametrization given in \cite{KimSeo11},
 \dis{
 V_{\rm CKM}&=\left(
\begin{array}{ccc}
c_1 & s_1c_3 & s_1s_3  \\
-c_2s_1 & e^{-i\delq}s_2s_3+c_1c_2c_3 & -e^{-i\delq}s_2c_3+c_1c_2s_3  \\
-e^{i\delq}s_1s_2 & -c_2s_3+c_1s_2c_3 e^{i\delq} & c_2c_3+c_1s_2s_3e^{i\delq} \\
\end{array}\right) \label{eq:CKM}
}
where $c_1=\cos\theta_1, s_1=\sin\theta_1,$ etc. $ V_{\rm CKM}$ is a unitary matrix with 4 parameters, $\theta_1,\theta_2,\theta_3,$ and $\delq$. The interference pheonomenon is given by the invariant quantity known as the Jarlskog determinant $J$ \cite{Jarlskog85}. $J$ is given with the form (\ref{eq:CKM}) instead of the widely used one \cite{PDG15},  as \cite{KimNam14} 
\dis{
J=|{\rm Im}V_{31}V_{22}V_{13}|=|c_1c_2c_3 s_1^2s_2 s_3 \sin\delq|.
}
Obviously, if the CP phase $\delq$ is zero, then $J=0$ and there is no CP violation. In addition, for a nonzero $J$ the product of $c_1s_1$, $c_2s_2$, and $c_3s_3$ must be nonzero, meaning the participation of all three families. It is an interference phenomenon.
 
The origin of $V_{\rm CKM}$ is   the Yukawa couplings. In the mass generating process via the vacuum expectation value of the Higgs field, the CKM matrix is given with respect to the mass eigenstate quarks, $u,d,s,c,b,t$. This is the theory of the weak CP violation via the charged weak currents in the modern SM in particle physics.

In the SM, we considered currents and only one scalar, the Higgs boson $h$ discovered in 2012. Higgs and Englert got Nobel Prize for the prediction of $h$ in 2013. So, any other scalar particle can introduce CP violation as item 1, since the Yukawa couplings of these extra scalars may not be flavor diagonal.

\section{Supersymmetry and CP violation}\label{sect:QCDcp}

Supersymmetry, in particular $N=1$ supersymmetry for physical applications, introduces superpartners of the known particles of spin differing by $\frac12$. The so-called MSSM (minimal supersymmeric SM) introduces a superpartner to every SM particle. For example,  spin-$\frac12$ electron introduces spin-0 selectron,  spin-$\frac12$ quark introduces spin-0 squark, etc. Therefore, so many spin-0 particles beyond the SM are introduced. Then, four-fermion densities mediated by these spin-0 superpartners can violate CP.  Static properties of proton and neutron are good properties to observe these tiny CP violation effect. The static property, magnetic moment, is conserving CP. But, the hypothetical static property, the electric dipole moment (EDM), violates CP. A naive estimate of the nucleon EDM is 
\dis{
d &\sim \textrm{(nucleon size)}\cdot e\cdot\textrm{(interaction strength)}\\
&\approx \frac{1}{M_p}\cdot e\cdot   \frac{M_p^2}{(100\,\gev)^2}\sim e\cdot 10^{-4}\cdot(10^{-13\,}\textrm{cm})f^2\sim 10^{-27}e\,\textrm{cm},
}
where $M_p$ is the proton mass, $f$ is the order of the Yukawa couplings of the first family members, $O(10^{-5})$, and the scalar mass of 100 GeV is used for an illustration.

\section{QCD and CP violation}\label{sect:QCDcp}

In Sect. \ref{sect:Weak}, the weak CP violation via the currents was given. We mentioned that {\it CP violation is an interference phenomenon}. This also applies when we consider subjects beyond the currents. In QED and QCD, we mentioned that there is no CP violation via currents because the corresponding currents are flavor diagonal. But one can find physical situation beyond the currents. It may not involve particles defining the currents, but the structure of vacuum. Vacuum can have many possible non-vanishing values of integer spin fields, i.e. for bosons.

For example, the Einstein equation is the equation satisfied by the vacuum value of   spin-2 field graviton, i.e. the vacuum value of $g_{\mu\nu}$. The vacuum value of the Higgs field, an example for spin-0 boson, gives all the SM particles mass. What about the vacuum value of spin-1 field?

In most earlier studies of QCD, the vacuum value of strongly interacting gluon field was taken as 0. But, one can consider vacuum value of gluons such that, at infinity the field strength decays sufficiently fast. This is most symmetrically presented in the Euclidian space and it can be said, ``It is localized''. This configuration of gluon field is called ``instanton'', implying that it is localized in the space-time coordinate. Field strength of a non-Abelian gauge field $A_\mu^a$ is denoted as $F_{\mu\nu}^a$, where $a$ runs for the dimension of the adjoint representation of the non-Abelian gauge group, i.e. $N^2-1$ for SU($N$). The instanton solution describes non-trivial  $F_{\mu\nu}^a$, meaning spherically symmetric hedge-hog type solution. So, one needs $a\ge 3$, and U(1) gauge group cannot have such an instanton solution. It means that QED does not have extra physical effect due to vacuum values of the photon field. But, QCD can have an effect because QCD of SU(3), being  non-Abelian, has $N^2-1=8$. Such instanton solutions are described most easily by the first order constraint, the (anti-)selfdual condition,
\dis{
F_{\mu\nu}^a=\pm \tilde{F}_{\mu\nu}^a
}
where $\tilde{F}_{\mu\nu}^a=\frac12 \epsilon_{\mu\nu\rho\sigma} F^{a\,\rho\sigma}$. It is known that for these instanton solutions one can consider the gauge invariant effective interaction, the $\theta$ term
\dis{
\frac{1}{4}F^{a\,\mu\nu} \tilde{F}_{\mu\nu}^a +\frac{g^2\theta}{32\pi^2}F^{a\,\mu\nu} \tilde{F}_{\mu\nu}^a \label{eq:thetaterm}
}
where $g$ is the QCD coupling constant and we added the original kinetic energy term of the gluon field. The first term has CP even and the second term has CP odd. Thus, QCD violates CP. Here again the CP violation is an interference because both the CP even and odd terms are considered together. The CP violation in QCD is so strong because the QCD coupling $g$ in Eq. (\ref{eq:thetaterm}) is very large. The CP violation in QCD can surface up by the non-zero neutron EDM. The CP violating pion-nucleon-nucleon coupling $\overline{g_{\pi NN}}$ is present due to the term in Eq.  (\ref{eq:thetaterm}). Then, the neutron electric dipole moment(nEDM) is calculated as  (with the CP conserving ${g_{\pi NN}}$ term),
\begin{equation}
\frac{d_n}{e}=\frac{{g_{\pi NN}}\overline{g_{\pi NN}}}{4\pi^2 m_N} \ln\left(\frac{m_N}{m_\pi}\right).\label{eq:nEDM}
\end{equation}
But the observed bound of the neutron EDM is so small, $\lesssim 10^{-26\,}e\,$cm, that we face a dilemma, the so-called the strong CP problem,  ``Why is the nEDM so small?''
 The strong CP problem was tried to be understood by three methods:  (1) calculable models, (2) massless up quark, and (3) ``invisible'' axion.
 
 
   The non-observation of nEDM put a limit on $|\bar\theta|$ as less than $10^{-10}$. For the class of calculable solutions, the so-called Nelson-Barr type weak CP violation is close to a solution \cite{Nelson,Barr}, but the limit $10^{-10}$ is difficult to be realized.
For the massless up-quark solution, it seems not favored by the measured current quark masses   \cite{Manohar14}.  At present, the remaining `natural solution' is ``invisible'' axion \cite{KimRMP10}.
    
This leads us to a   brief historical introduction, eventually leading to the ``invisible'' axion \cite{KSVZ1,KSVZ2,DFSZ}. Pre- ``invisible'' axion developments are the following.
 
 Satisfying the Glashow-Weinberg condition that up-type quarks couple to $H_u$ and down-type quarks couple to $H_d$ \cite{GW77}, he introduced many Higgs doublets.   Then, the weak CP violation introduced in the potential, with a discrete symmetry $\phi_I\to-\phi_I$,
\begin{equation}
\begin{split}
V_{\rm W}&=\frac12\sum_I m_I^2\phi_I^\dagger\phi_I\\
&+\frac14 \sum_{IJ} \left\{a_{IJ} \phi_I^\dagger\phi_I\phi_J^\dagger\phi_J+b_{IJ}\phi_I^\dagger\phi_I\phi_J^\dagger\phi_J +(c_{IJ} \phi_I^\dagger\phi_I\phi_J^\dagger\phi_J+{\rm H.c.})
\right\}\label{eq:Weinberg}
\end{split}
\end{equation}
Weinberg's necessary condition for the existence of CP violation is non-zero $c_{IJ}$ terms \cite{Weinberg76}. If one removes the $c_{IJ}$ terms, Peccei and Quinn (PQ) noticed that there emerges a global symmetry which is now called the  \UPQ~symmetry \cite{PQ77}. This is an example that keeping only a few terms among the discrete symmetry allowed terms in the potential produces a global symmetry.

With the PQ symmetry by removing the $c_{ij}$ term in Eq. (\ref{eq:Weinberg}) \cite{PQ77},  Weinberg and Wilczek at Ben Lee Memorial Conference noted the existence of a pseudoscalar, the PQWW axion 
\cite{PQWW77}, which was soon declared to be non-existent \cite{Peccei78}. This has led to calculable models during 1978 
\cite{Beg78,Georgi78,Moha78,
Segre79,Langacker79}, before the advent of ``invisible'' axions \cite{KSVZ1,KSVZ2}.

\section{The ``invisible'' axion}
 
``Invisible'' can be made ``visible'' if one invents a clever cavity detector
\cite{Sikivie83}, which is used in many axion search labs now \cite{AxionLabs}.
An SU(2)xU(1) singlet housing the ``invisible'' axion gives the effective Lagrangian of $a$ as
 \begin{eqnarray}
 {\cal L}=c_1\frac{\partial_\mu a}{f_a} \sum_q\bar{q}\,\gamma^\mu\gamma_5\, q-\sum_q(\bar{q}_L\, m\, q_R\, e^{ic_2 a/f_a}  +\textrm{h.c.})+\frac{c_3}{32\pi^2 f_a}a\, G_{\mu\nu}\tilde{G}^{\mu\nu}
 \label{eq:invAxion}\\
 +\frac{c_{aWW}}{32\pi^2 f_a}a\, W_{\mu\nu}\tilde{W}^{\mu\nu}
 +\frac{c_{aYY}}{32\pi^2 f_a}a\, Y_{\mu\nu}\tilde{Y}^{\mu\nu} + {\cal L}_{\rm\, leptons},\nonumber
 \end{eqnarray}
where $\tilde{G}^{\mu\nu}, \tilde{W}^{\mu\nu}$, and $\tilde{Y}^{\mu\nu}$ are dual  field strengths of gluon, $W$, and hypercharge fields, respectively. 

It is a key question how the PQ symmetry is defined.  These couplings arise from the following renormalizable couplings,
\begin{eqnarray}
  {\cal L}_{\rm KSVZ}&=&- f\overline{Q}_R Q_L+\textrm{h.c.}, \label{eq:KSVZ}\\
  {V}_{\rm DFSZ}  &=&  -\mu_1^2 H_u^*H_u -\mu_2^2 H_d^*H_d+\lambda_1(   H_u^*H_u)^2+\lambda_2(   H_d^*H_d)^2\\
  &&+{\sigma\rm~terms} (+M H_uH_d\sigma) +\lambda' H_uH_d\sigma^2+\textrm{h.c.}\nonumber
\label{eq:DFSZ}
\end{eqnarray}

 In the KSVZ model, the heavy quark $Q$ is introduced and the $f$-term Yukawa coupling is the definition of the PQ symmetry. In the DFSZ model, the $\lambda'$-term is the definition of the PQ symmetry.
Note, however, that  there must be a fine-tuning in the coefficient $\lambda'$ such that $\vew\ll f_a$. The axion-photon-photon couplings are listed in Table \ref{tab:Inv}.
\begin{table} 
\begin{tabular}{|r|c|} 
\hline &\\[-1em]
KSVZ:  $Q_{\rm em}$ ~&$c_{a\gamma\gamma}$\\\hline
$0$ ~~ &$-2$~ \\[0.3em]
$\pm \frac13$~~  &$-\frac{4}3$~  \\[0.3em]
$\pm \frac23$~~  &$ \frac{2}3$   \\[0.3em]
$\pm 1 $~~  &$4$  \\[0.3em]
$(m,m)$~&$-\frac{1}3$~  \\[0.2em] 
\hline
\end{tabular} 
\hskip 1cm
\begin{tabular}{|r c|c|} 
\hline &&\\[-1em]
DFSZ:  $(q^c$-$e_L)$ pair ~& Higgs &$c_{a\gamma\gamma}$\\\hline
\\[-1.35em]
&&  \\[-0.85em]
non-SUSY $(d^c,e)$ ~~ & $H_d$&$\frac23$~ \\[0.3em]
non-SUSY $(u^c,e)$ ~~  & $H_u^*$&$-\frac43$~ 
  \\[0.2em] 
GUTs  ~~ &  &$\frac23$~ \\[0.3em]
SUSY   ~~  &  &$\frac23$~ 
  \\[0.2em] 
\hline
\end{tabular} 
\caption{$c_{a\gamma\gamma}$ in the KSVZ and DFSZ models. For the $u$ and $d$ quark masses, $m_u=0.5\, m_d$  is assumed for simplicity. $(m,m)$ in the last row the KSVZ means $m$ quarks of $Q_{\rm em}=\frac23\,e$ and  $m$ quarks of $Q_{\rm em}=-\frac13\,e$.  SUSY in the DFSZ includes contributions of color partners of Higgsinos. If we do not include the color partners, \ie in the MSSM without heavy colored particles, $c_{a\gamma\gamma}\simeq 0$. }
\label{tab:Inv}
\end{table}
 
The fine-tuning problem in the DFSZ model is resolved in the supersymmetric(SUSY) extension of the model. There is no renormalizable term of the singlet superfield $\sigma$ with the SM fields. The leading term is the so-called Kim-Nilles term \cite{KN84},
 \begin{eqnarray}
W_{\rm KN}= \frac{1}{M}H_uH_d\sigma^2,
 \end{eqnarray}
 where $M$ is determined from a theory. It is shown in Table \ref{tab:Inv} as the SUSY $c_{a\gamma\gamma}$.
In Table \ref{tab:Inv}, $H_d$ and $H_u^*$ imply that they give mass to $e$.  GUTs and SUSY choose appropriate Higgs doublets and always give $c_{a\gamma\gamma}=\frac23$.

In the discussion of ``invisible'' axion, gravity effects was considered to be crucial. It started with the wormhole effects in the Euclidian quantum gravity. In fact, gravity equation with the antisymmetric tensor field $B_{\mu\nu}$ gives wormhole solutions. This triggered the discrete symmetries allowable as subgroups of gauge groups  \cite{Krauss89}, and \UPQ~global symmetry needed for ``invisible'' axion was considered to be problematic \cite{GravSpoil92}.  
 This has to be resolved.
  
\begin{figure}[h]
\centering
\includegraphics[width=0.75\linewidth]{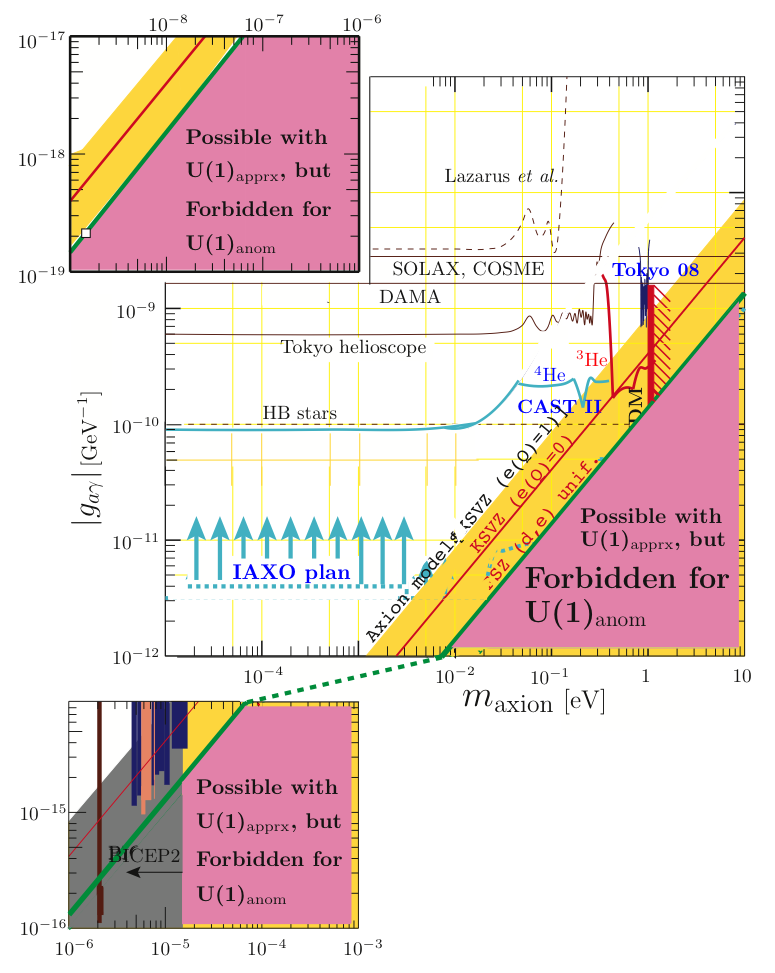}
\caption{The limits of axion searches.} \label{fig:AxData}
\end{figure}

In this road toward detecting an ``invisible'' QCD axion, there has been a few theoretical development starting from an ultra-violet completed theory. The scale must be intermediate.  The model-indepent(MI) superstring axion \cite{MIaxion} is not suitable for this because the decay constant is about $10^{16\,}\gev$ \cite{ChoiKim85} which is the white square on the upper left corner in Fig. \ref{fig:AxData}.

The question is, ``is it possible to obtain exact global symmetries?''
From string compactification, there is one way to make the ``invisible'' QCD axion to be located at the intermediate scale starting with an exact global symmetry,
\begin{equation} 
10^9\,\gev\le f_a\le 10^{11.5\,}\gev.
\end{equation}
It starts from the appearance of
an anomalous U(1) gauge symmetry in string compactification. In  compactifying the E$_8\times$E$_8'$ heterotic string, there appears an anomalous U(1)$_a$ gauge symmetry in many cases \cite{Anom86},
  \begin{equation} 
{\rm E}_8\times {\rm E}_8\to {\rm U(1)}_a\times \cdots
\end{equation}
Thus, the anomalous  U(1)$_a$ is belonging to a gauge symmetry of  E$_8\times$E$_8'$. In the original E$_8\times$E$_8'$ heterotic string, there is also the MI-axion degree $B_{\mu\nu}$. The gauge boson corresponding to this anomalous   \Uanom~ obtains mass by absorbing the MI-axion degree as its longitudinal degree. Therefore, the harmful MI-axion disappears, but not quite completely. Below the compactification scale of $10^{18\,}\gev$, there appears a global symmetry  which works as the PQ symmetry. This PQ symmetry can be broken by a SM singlet Higgs scalar(s), producing the ``invisible'' axion. So, this   ``invisible'' axion arises from an exact global symmetry  \Uanom, and is free from the gravity spoil problem because its origin is gauge symmetry.

Within this string compactification scheme, the axion-photon-photon coupling has been calculated   \cite{Kim88,KimPLBagg15,KimNamPLB16} and presented in Table \ref{tab:StAxion}.
   
\begin{table} 
\begin{tabular}{|l| r|c|} 
\hline  &&\\[-1em]
String:  &$c_{a\gamma\gamma}$  &Comments\\\hline\\[-1.35em]
&&  \\[-0.9em]
Ref. \cite{ChoiIWKim07}   &  $-\frac13$~ &Approximate\\[0.3em] 
Ref. \cite{KimPLBagg15,KimNamPLB16} &  $\frac23$~ &Anom. U(1)\\[0.3em]
\hline
\end{tabular}
\caption{String model prediction of $c_{a\gamma\gamma}$.  In the last line, $c_{a\gamma\gamma}= (1-2\sin^2\theta_W)/\sin^2\theta_W$ with $m_u=0.5\, m_d$.
}
\label{tab:StAxion}
\end{table}

\section{CP and cosmology}
 
The axion solution of the strong CP problem is a cosmological solution.    QCD
axions oscillate with the CP violating
vacuum angle $\bar\theta$, but the average value is 0.  If the axion vacuum starts from $a/f_a=\theta_1\ne 0$, then the vacuum oscillates and this collective motion behaves like cold dark matter(CDM).
 
The axion vacuum is identified by the shift of axion field by $2\pi\NDW f_a$,
\begin{equation}
a\to a+2\pi \,\NDW\,f_a.
\end{equation}
It is because the $\bar\theta$ term has the periodicity $2\pi$,
\begin{equation}
{\cal L}_{\bar\theta}=-\frac{a}{32\pi^2\,f_a}\int d^4x \,G^a_{\mu\nu}\tilde{G}^{a\,\mu\nu},~~a=a+2\pi f_a,
\end{equation}
while matter fields $\Phi$ may not have the periodicity of $2\pi f_a$, but only after $2\pi\NDW f_a$,
\begin{equation}
\Phi\to e^{i\bar\theta/\NDW}\Phi,~~
\bar\theta=\frac{a}{f_a}.
\end{equation}
Between different vacua, there are domain walls.
 
Topological defects of global \UG~produce an additional  axion energy density by the decay of string-wall system, $\rho_{st}$. Contribution of axionic string to energy density was known for a long time \cite{Davis86}. In addition, axionic domain walls carry huge energy density \cite{Vilenkin82, Sikivie82}. Because of the difficulty of removing comological scale domain walls for $\NDW\ge 2$, it was suggested that
the axionic domain wall number should be 1 \cite{Sikivie82}.   

Computer simulations use axion models with $\NDW=1$.
Three groups have calculated these which vary from O(1) to O(100),
\begin{eqnarray}
&&\textrm{Florida group:}~\textrm{O(1)} \cite{Florida01},\nonumber\\
&&\textrm{Cambridge group:}~\textrm{O(100)}   \cite{Cambridge94}, \label{eq:StContri}\\
&&\textrm{Tokyo group:}~\textrm{O(10)}    \cite{Kawasaki12}.\nonumber
\end{eqnarray}
A recent calculation for $\NDW=1$ models has been given  $\rho_{st}\sim  {\rm O(10)}\,\rho_a $ \cite{SekiguchiT}.

\begin{figure}[t]
\begin{center}
\includegraphics[width=0.55\linewidth]{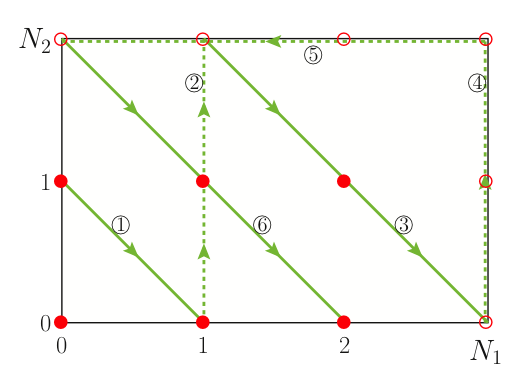}
\end{center}
\caption{The limits of axion searches.} \label{fig:Goldstone}
\end{figure}

Therefore, it is important to realize axion models with $\NDW=1$. The KSVZ axion model with one heavy quark achieves $\NDW=1$. There are two other methods. One is identifying different vacua modulo the center number of the GUT gauge group   \cite{LS82}. Another important one is obtaining $\NDW=1$ by  the Goldstone boson direction \cite{Choi85,KimPLB16}, which is shown in Fig. \ref{fig:Goldstone}.
There are two degrees for the shifts, $N_1$ and $N_2$ directions. For the torus of $N_1=3$ and $N_2=2$ models, seemingly there are 6 vacua represended by red bullets in Fig. \ref{fig:Goldstone}. The Goldstone boson directions are shown as arrow lines and torus identifications are shown as dashed arrows. So, all six vacua are connected by one way or the other, and the  $N_1=3$ and $N_2=2$ model gives $\NDW=1$. One always obtain $\NDW=1$ if $N_1$ and $N_2$ are relatively prime. The reason that the ``invisible'' axion from \Uanom~ has $\NDW=1$ is because $N_{\rm from~E_8\times E_8'}=\rm large~integer$ but $N_{\rm MI~axion}=1$ \cite{Witten85,KimPLB16}, and  $N_{\rm from~E_8\times E_8'}$ and $N_{\rm MI~axion}$ are relatively prime.

Axionic string contribution is important if strings are created after PQ
symmetry breaking. On the other hand, with a high scale inflation this string contribution to energy density is important, as shown in Eq. (\ref{eq:StContri}). There can be a more important constraint if a large $r(=\,$tensor/scalar ratio) is observed. Two groups reported this constraint \cite{Gondolo14} after the BICEP2 report \cite{BICEP2}. Probably, this is the most significant impact of BICEP2 result on
$\NDW=1$ axion physics. The region is marked around $m_a\sim 71\,\mu$eV in Fig. \ref{fig:AxData}.

%% file: 1.3.1/main.tex
The axion, postulated in the 70's as a consequence of the PQ mechanism to provide a dynamic solution to the strong-CP problem in particle physics, is a theoretically well motivated fundamental particle~\cite{bib:PRL_38_1440_1977}\cite{bib:PRD_16_1791_1977}\cite{bib:PRL_40_223_1978}\cite{bib:PRL_40_279_1978}.
In the 80's, revealing its cosmological implications, the axion emerged as an excellent candidate for cold dark matter, and is now getting much more attention, in particular under the circumstance that the popular WIMP dark matter has not been observed to scientists around the world for decades. 
As a late starter in axion research, CAPP is now establishing the state-of-the-art microwave cavity dark matter axion experiment in Korea, mainly based on the scheme that was proposed by P. Sikivie in 1983~\cite{bib:PRL_51_1415_1983}\cite{bib:PRD_32_2988_1985}. Korea never had any dark matter research until 5 years ago. 

\begin{figure}[t]
      \includegraphics[width=\textwidth]{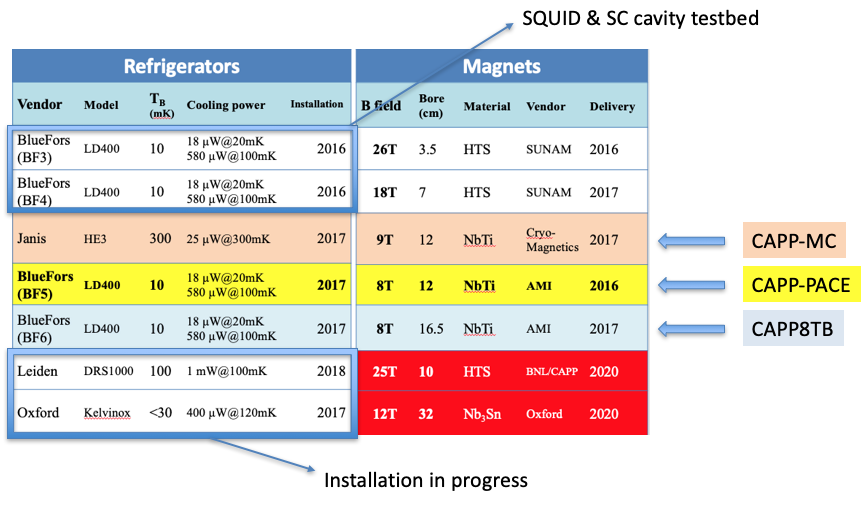}
\centering
   \caption{CULTASK refrigerators and superconducting magnets}
  \label{fig:CULTASK}
\end{figure}

CAPP has built an axion research facility at KAIST (Korea Advanced Institute of Science and Technology) Munji Campus in Daejeon, Korea in the beginning of 2016 with 7 low vibration pads (LVP). We have now 7 refrigerators and 5 superconducting magnets installed in the facility and 4 low temperature microwave axion dark matter detectors are operating on the LVP. All of our axion detectors are designed to reach very low physical temperature (mK range for resonant cavities), so we call our axion research, CULTASK (CAPP's Ultra Low Temperature Axion Search in Korea). Figure~\ref{fig:CULTASK} shows refrigerators and magnets installed and operating at CAPP. Two high power superconducting magnets (25\,T with 10\,cm bore and 12\,T 32\,cm bore) are to be delivered in 2020 and should be our workhorses in frontline axion research. 

The powerful magnets with high fields and large apertures (CAPP25T and CAPP12T) should be able to boost the axion-to-photon conversion power.
Unique designs of resonant cavities are expected to provide a capability of probing high frequency regions with maximal detection volume.
These would enables us to explore wide ranges of axion mass with enough sensitivities to detect or exclude axion models when convoluted with highly sensitive quantum amplifiers whose noise performance approaches the fundamental limits imposed by the laws of quantum mechanics.
These are the technologies that never existed or not mature enough 10 years ago. The combination of those breakthroughs will put CAPP’s flagship axion experiment, CULTASK in the front row and push the frontiers of particle astrophysics.

%% file: 1.3.2/main.tex
CAPP axion research program’s final goal is to prove or disprove axions as dark matter once and for all. With powerful (25\,T, 10\,cm bore) and large volume (12\,T, 32\,cm bore) superconducting magnets which will be available in 2020, CAPP should be able to explore the wide range of axion mass, 1\,GHz to 10\,GHz, with enough sensitivity to discover or exclude axions. 

CAPP-PACE, started as an R\&D machine to prepare for CAPP25T axion experiment, would provide the necessary experience in ultra-low temperature cryogenics, the fabrication of high Q-factor resonant cavities, a reliable frequency tuning system, highly sensitive cryo-RF electronics and a DAQ/control system including monitors ensuring the quality of data and safe environment of data taking. 
CAPP-PACE detector’s frequency range is similar to CAPP25T and all of associated RF receiver electronics will be shared with CAPP25T. CAPP-PACE detector has grown into a complete axion detector at the beginning of 2018 and we are now testing every aspect of the axion dark matter experiment in around 2.5\,GHz frequency range while taking physics quality data. 

Our main research focus in 2019 will be to prepare for CAPP25T experiment and continue to take axion data at CAPP-PACE detector with enough sensitivity to search QCD axions. Once 25\,T HTS magnet is delivered from BNL in 2020, along with the development of quantum amplifiers with near quantum-limited performance, we should be able to search QCD axions in a much wider range of frequencies with improved scanning speed. By the end of 2020 CAPP25T should be able to take the world’s best quality axion dark matter data. In order to achieve that goal, the infrastructure for CAPP25T should be complete and quantum amplifiers should be ready in 2019.

\begin{figure}[h]
      \includegraphics[width=0.8\textwidth]{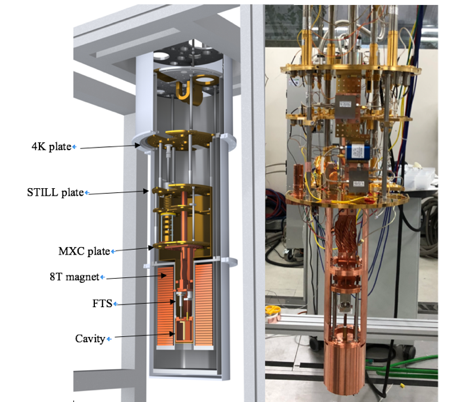}
\centering
   \caption{CAPP-PACE detector with BluFors LD400 DR}
  \label{fig:DR}
\end{figure}

The main elements of the CAPP-PACE detector are depicted schematically in Figure \ref{fig:DR} with a picture, BluFors LD400 cryogen-free dilution refrigerator (DR) system with an 8 T superconducting magnet, a resonant cavity with a frequency tuning system and a cryogenic RF receiver chain to read out the power spectrum from the cavity.

\subsection{Cryogenics and Magnet}

The DR system was installed on one of the low vibration pads constructed with one of seven 20 Ton concrete blocks each supported by four air springs to eliminate external vibrations. 
A dilution refrigerator is a refrigerator that uses \textsuperscript{4}He-\textsuperscript{3}He mixture to lower the temperature. It can reach cryogenic temperatures of about 2\,mK and is the only cooling method that allows continuous cooling in this temperature range. The operation principle of this refrigerator is as follows. Liquid \textsuperscript{4}He becomes superfluid at 2.18\,K. On the other hand, liquid \textsuperscript{3}He which is an isotope of \textsuperscript{4}He does not become superfluid at this temperature because it is Fermi fluid. When these two are mixed with each other, phase separation occurs at 0.867\,K according to the concentration of \textsuperscript{4}He / \textsuperscript{3}He. For this reason, there is a prohibited section where a mixture cannot exist at a specific mixing ratio at a certain temperature. Because there is a boundary between \textsuperscript{3}He rich phase and \textsuperscript{4}He rich phase, energy is needed to cross this boundary. We use a pump to separate the \textsuperscript{3}He from the \textsuperscript{3}He rich phase then it converts to the \textsuperscript{4}He rich phase, which can lower the temperature by absorbing the surrounding heat during cross the boundary.

The temperature of\,mK is reached using a dilution refrigerator, but first, He needs to be liquefied by making it 4\,K as a prerequisite for this to work. There are two commonly used devices for this purpose, Pulse tube cooler and Gifford-McMahon (GM) coolers are. In this experiment, a pulse tube is used. Its advantage is that the pulse tube does not have an internal moving part at the cold head. A vibration of the pulse tube cooler is much less than GM’s because in the case of GM cooler, the moving part is embedded in the cold head. However, due to the presence of He gas that continues to expand in the cold head, it is impossible to zero the vibration even in the case of a pulse tube. Since the vibration has a great influence on the sensitivity of the detector, our detector is installed on the low vibration pad (LVP) to minimize the influence of the vibration. LVP is made up of a heavy concrete block floating in the air through an air spring. The frame of the dilution refrigerator is installed on the concrete block which is separated from the floor to minimize the vibration that can be transmitted from the outside.

There are 4 dilution refrigerator units installed from $BlueFors$. All four are operating smoothly without any trouble for about 2 years of operation. By running a predefined script, it can reach the\,mK temperature without any extra action. It boasts a short (\textgreater 2 days) cool-down time using a $Cryomech$ pulse tube with a cooling power of 1.5\,W at 4\,K. It also shows very good cooling power of 580 $\mu$W at 100\,mK by using He mixture of high \textsuperscript{3}He ratio. When using amplifiers with quantum-limited noise such as Microstrip SQUID Amplifier (MSA) or Josephson Parametric Amplifier(JPA), it is important to lower the base temperature sufficiently because the amplifier's bath temperature has a large impact on noise. The temperature of each part of MXC, STILL and Cavity is measured using RuO2 thermometers. 

The microwave resonant cavity hangs in the center of the magnet bore supported by a copper structure, thermally anchored to the DR’s mixing (MXC) plate which is maintained at T\textsubscript{MXC} $\sim$ 25\,mK (cavity at T\textsubscript{CAVITY} $\sim$ 40\,mK) during the operation of the detector. The magnet maintains a temperature of about 3.6\,K and is separated thermally from the cavity by a radiation shield made of copper-brass. This allows the cavity to maintain a temperature of below 40\,mK throughout the normal operation. The frequency and antenna tuning system with piezoelectric actuators are designed to have a thermal link to the mixing plate, but wouldn't generate heat more than 40\,mK at cavities, even under a magnetic field of 8\,T. The HEMT Amplifier is mounted on the STILL plate for lower noise temperature. In particular, it is mounted on an L-shaped plate to minimize the distance from the cavity to the preamplifier, which is connected to the STILL plate but is also close to the MXC plate.  A superconducting cable is used to connect the cavity and the amplifier to minimize signal loss and heat exchange between the MXC plate and the STILL plate. The MSA is mounted on the MXC plate. In order to minimize the influence of the magnetic field, there is a cancelation coil on top of the magnet. In this way, the magnetic field of 8\,T is applied to the cavity, but only 50\,G is reached to the center of the MXC plate.

In order to reduce the system noise temperature, it is necessary to lower the physical temperature as much as possible. At\,mK range, temperature transfer by phonon-phonon scattering is not a major effect. Most of the heat is transferred through the conduction electron. Therefore, in order to smoothly cool down cavity which is farthest from the base plate, a support structure should be made by a material with high conductivity and have a large flat contact area. We used gold-plated OFHC to create a structure that best meets the above conditions. The piezoelectric actuator, which is the largest source of heat, is also fixed to this support structure using a copper rod to maintain the temperature at\,mK range.

We use a radiation shield to block the heat transmitted through the radiation. 50\,K, 4\,K, and 1\,K stages are equipped with thermal radiation shields. In particular, the 1\,K shield is made of two parts. Oxygen free copper is generally used for high thermal conductivity. However, the part that enters the magnet is made of brass to prevent heat generation during a ramp up or down. It combines with the main shield.
The goal of the cryogenic system is to provide enough cooling power to lower and maintain the physical temperature of the resonant cavity and components of the cryo-RF receiver chain as much as possible to ensure the minimal system noise (mainly from amplifiers in our frequency range).

\begin{figure}[h]
      \includegraphics[width=0.8\textwidth]{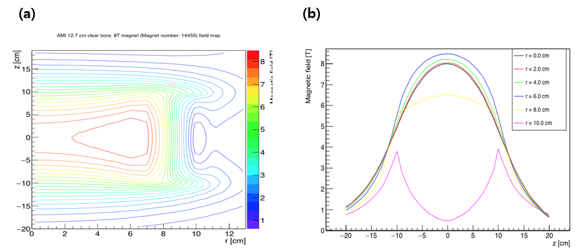}
\centering
   \caption{Magnetic field distribution of AMI’s 8\,T, 12\,cm bore magnet (AMI provided)}
  \label{fig:AMI}
\end{figure}

The 118\,mm bore of AMI’s 8\,T superconducting (NbTi) magnet sets the scale for the available axion-sensitive volume. The outer diameter of the cavity was limited to 100\,mm by placing a gap of 9\,mm between inner bore of the magnet and the outer wall of the cavity and the thickness was designed to be 5\,mm, enough to reduce the risk of breakage due to the smooth nature of the copper. The cavity height was 93\,mm to make sure the average magnetic field intensity is maintained at 7.9\,T, but has been redesigned to 180\,mm (average field intensity of 7.6\,T) to maximize the axion-sensitive volume and eventually to increase the scanning efficiency. Figure \ref{fig:AMI} shows the magnetic field distribution of AMI’s 8\,T NbTi magnet.

We keep the magnet in a superconducting state in the form of conduction cooling rather than using a cryogen, which is optimized for the BlueFors refrigerator, which does not operate as a separate cooler but delivers cooling power directly from the cold head of pulse tube through a 4\,K shield. It usually takes three and a half hours to reach 8\,T in normal operation and it is also possible to ramp down quickly in an emergency in an hour. The cavity needs a magnetic field to detect the axion, but other RF components have a bad influence on the magnetic field. Thus it is necessary to adjust the area of the magnetic field. To do this, a cancellation coil is installed at the top of the magnet. Because of this, the center of the MXC plate shows a low magnetic field of 50\,G even when the center of the magnet is 8\,T. Therefore, the MSA SQUID amplifier located on the MXC plate operates normally without the influence of the magnetic field by using only a simple lead shield, and the circulators work well with a simple shielding using a permalloy.


\subsection{Microwave cavity}

The microwave cavity is located at the center of the AMI 8\,T magnet, i.e., 400\,mm below the MXC plate of Bluefors dilution refrigerator. The microwave cavity is a very important part of the axion haloscope where the axion reacts with the magnetic field and changes directly to the microwave photon. However, since the axion to photon conversion power strongly depends on the magnitude of the magnetic field and the size of the space influenced by the magnetic field, not only the shape and size of the cavity but also the material is limited by the magnet.

\begin{figure}[h]
\begin{center}
\includegraphics[width=.75\textwidth]{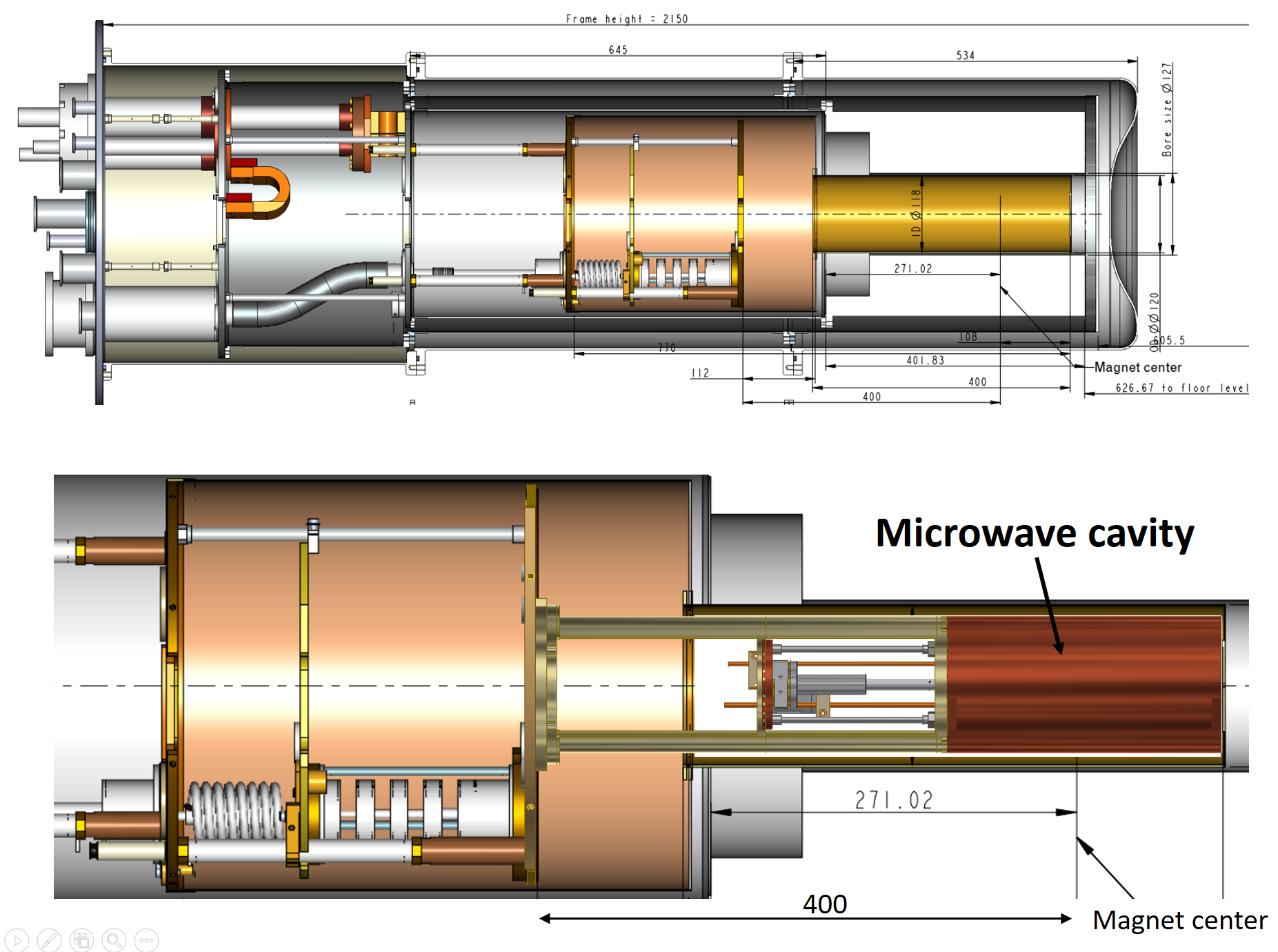}
\caption{Schematic of the geometrical relation of a microwave cavity of CAPP-PACE.  }
\label{fig:whereiscavity}
\end{center}
\end{figure}

Therefore, we selected oxygen-free high thermal conductivity copper (OFHC) as the material to be used in the CAPP-PACE cavity because of its high electric and thermal conductivity among non-magnetic materials. OFHC is a completely non-magnetic, even in extreme environments such as high-speed machining, annealing, and cryogenic temperatures. In addition, the electric conductivity rises very fast as the temperature gets lower so that the $Q$ factor below 4\,K, which directly affects the axion-to-photon conversion power, should become $4-5$ times higher than the value at the room temperature even considering the anomalous skin effect~\cite{bib:anomalous_London}\cite{bib:anomalous_pippard}\cite{bib:anomalous_nature}. Moreover, it has been reported that there is no adverse effect of magneto-resistance with copper in the environment below 9T in the previous study~\cite{bib:magneto_yannis}\cite{bib:magneto_sungwoo}. As shown in Fig.~\ref{fig:whereiscavity}, The CAPP-PACE cavity is designed so that the center of the AMI 8\,T magnet coincides with the center of the cavity, maximizing its influence on the magnetic field, and using a supporting structure of copper material we physically and thermally link the coldest part of the refrigerator to minimize the noise generated in the cavity. As mentioned earlier in the refrigerator section, the cavity temperature was stably maintained at less than 40\,mK in an experimental situation where the actual maximum magnetic field of 8\,T was turned on and frequency tuning was being operational.

\begin{figure}[h]
\begin{center}
\includegraphics[width=.55\textwidth]{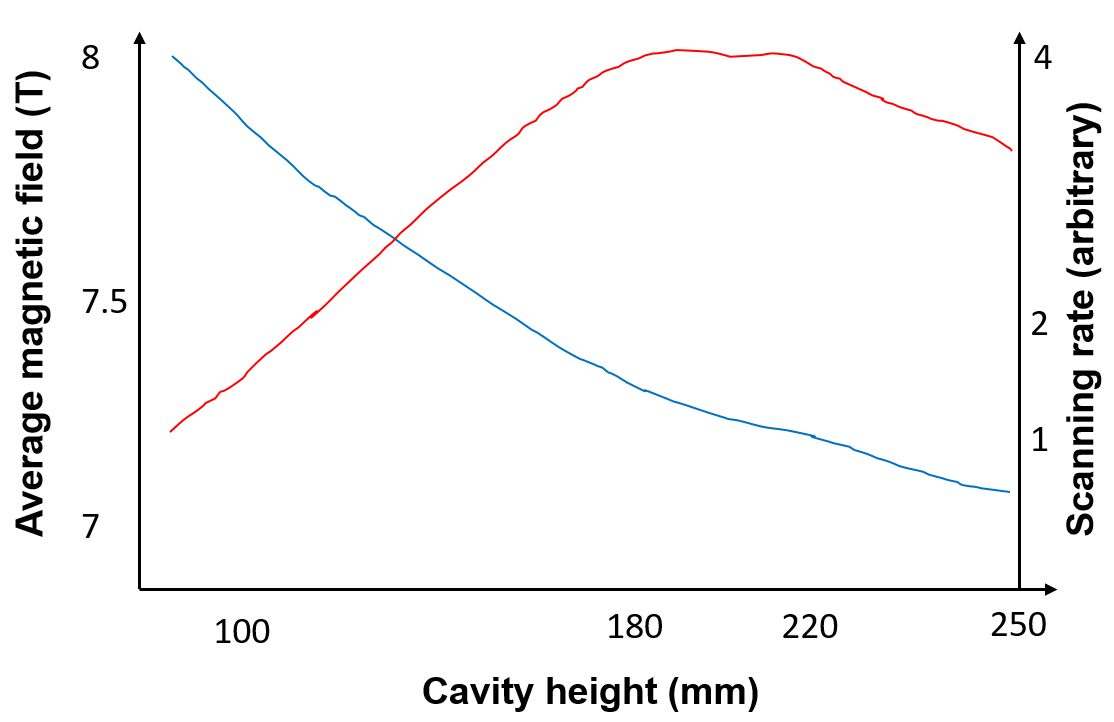}
\caption{Plot of cavity size versus average magnetic field in Tesla, scanning rate in arbitrary unit}
\label{fig:cavsize_Bfield_PACE}
\end{center}
\end{figure}

At this time, we assigned a gap of 9\,mm between the cavity and the inner wall of the magnet in order to prevent heat exchange between the cavity and the magnet and to provide a physical stability when assembling them to the refrigerator, which consequently limit the outer diameter of the cavity to 100\,mm. The side wall thickness was designed to be 5\,mm thick enough to reduce the risk of breakage due to the smooth nature of the copper. The cavity height was decided as 93\,mm for the 1st experimental run to avoid the mechanical burden of piezoelectrics and later it was optimized to be 180\,mm for fast axion searching. The average magnetic field for 93\,mm height cavity was 7.9\,T and for 180\,mm height cavity, 7.6\,T. Figure \ref{fig:cavsize_Bfield_PACE} shows that the magnitude of the average magnetic field decreases with the height of the cavity, and when the height exceeds a certain level, the axion scanning rate is rather reduced.

\begin{figure}[h]
\begin{center}
\includegraphics[width=.65\textwidth]{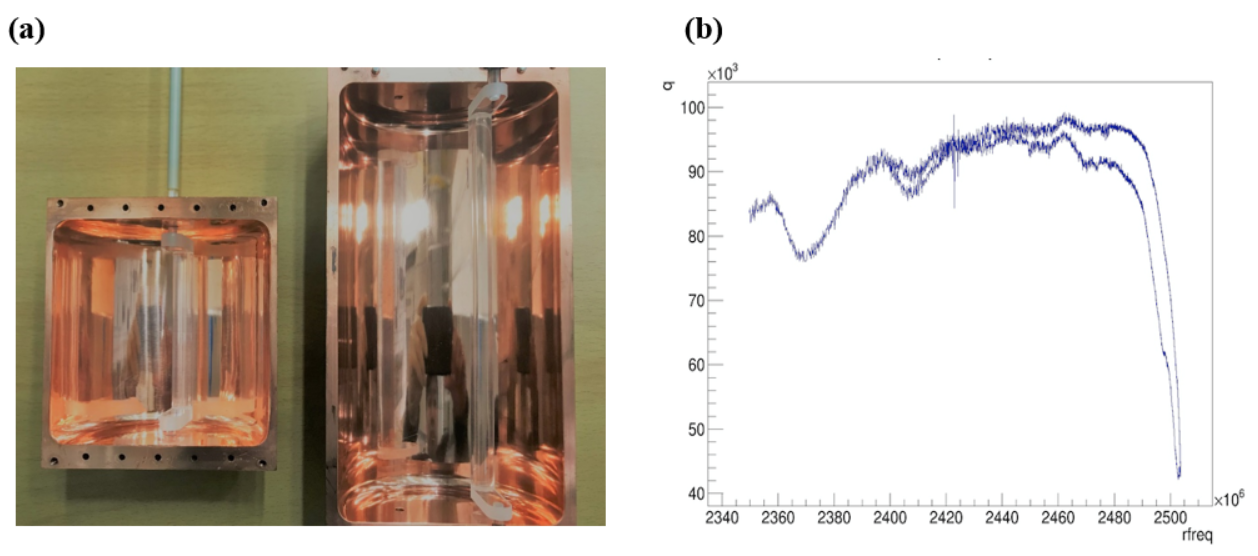}
\caption{Split cavity structure of CAPP-PACE used at (a) first experimental run, and (b) 2nd and 3rd run.}
\label{fig:splitcav_picture}
\end{center}
\end{figure}

The CAPP-PACE cavity also employs a split cavity structure, which is one of CAPP’s R\&D achievement. Typical cylindrical cavities are usually assembled in the form of a circular pipe-shaped body with discs at the top and bottom. At this time, due to the TM$_{010}$ mode characteristics, the direction of the electron at the cavity surface causing the ohmic loss becomes perpendicular to the assembly surface, which increases the resistance and consequently causes the $Q$ factor to drop, which is a so-called ``contact problem." A cavity with a high $Q$ factor is usually required to minimize contact surface separation to address this contact problem. In many cases, a large number of bolts are used to densely press the contact surface, while one of the bonding surfaces is made of a knife edge or indium is placed between the bonding surfaces to give a gasket effect. However, the higher the $Q$ value, the less effective it is and if it is repeatedly disassembled and assembled, it will not be able to play its role. In order to solve this problem, we sought a method of fabricating the cavity assembling surface parallel to the TM$_{010}$ mode current. We decided to use milling machining instead of lathe processing that is the usual manufacturing method for cylindrical structure. We prepared a cylinder, not a pipe, and cut it vertically, and dug a semi-cylindrical space in each. The two semi-cylindrical pieces thus made were aligned and assembled with bolts. Surprisingly, but as expected, it was confirmed that the measured $Q$ factor value agrees with the theoretical value~\cite{bib:split_KPSposter}. This difference is more pronounced at low temperature. The $Q$ value of the cavity produced by the conventional method is about 3 times higher at the cryogenic condition than at room temperature, whereas it  gets an increased factor of $4-5$ for the split cavity. CAPP-PACE has applied this design to all cavities during experimental runs to date and achieved a very stable $Q$-factors. Figure \ref{fig:splitcav_picture} shows the cavity used for the actual CAPP-PACE data runs and the $Q$-factor measurement without any tuning device. It shows that the $Q$-factor measurement agrees with the theoretical prediction within an error range of 1\%.

In addition, the split cavity structure solves the heat problem that occurs when turning magnets on and off. This is because the flow of eddy current is suppressed because the area of the closed loop through which the magnetic flux passes is significantly reduced. One more benefit when using a split cavity structure is to suppress TE mode excitation. The split cavity structure is not compatible with most TE modes. Contrary to the TM$_{010}$ mode, the TE mode has a contact problem in the vertically split cavity structure, resulting in a lower $Q$ value and less TE mode excitation. Consequently, the split cavity structure is strongly recommended for the axon haloscope.

\subsection{Frequency tuning system}

\begin{figure}[h]
\begin{center}
\includegraphics[width=.65\textwidth]{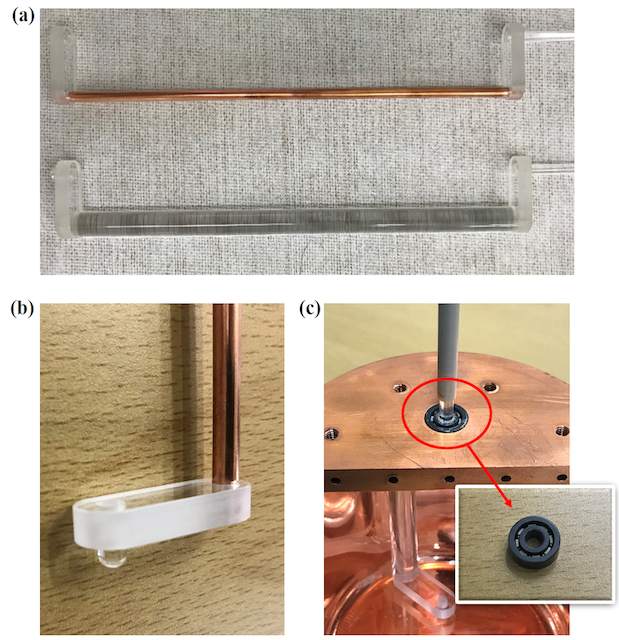}
\caption{(a) Frequency tuning rods assemblies of CAPP-PACE, the centered rods are sapphire and OFHC from left to right, respectively.  (b) Microwave cavity with sapphire tuning rod installed. (c) Magnified view of a bottom shaft. (d) Single row full ceramic bearing (4mm x 12mm x 4mm) made of silicon nitride. }
\label{fig:tuningrod_pic}
\end{center}
\end{figure}

In CAPP-PACE, the section to be scanned is about 300\,MHz, ranging from 2.45 to 2.75\,GHz, which is scattered on both sides of the resonant TM$_{010}$ mode frequency of 2.55\,GHz in a cylindrical cavity of 90mm diameter. In general, if you want to tune the frequency range higher than the resonant frequency of the hollow cavity, you can tune the cavity using a metal rod, or use a dielectric rod if you want the opposite. In CAPP-PACE, dielectric and metal rods were configured differently according to the tuning interval so that scanning over 1\,GHz frequency range is available. As shown in Figure~\ref{fig:tuningrod_pic}, the tuning device consists of 4\,mm diameter of a copper rod or 10\,mm sapphire rod for frequency perturbation and cranks at both ends of the rods for rotational movement. A sapphire rod with a diameter of 4\,mm serves as a rotary shaft for each crank. The upper rotary shaft extends out of the cavity and is connected to a rotary actuator, with a ceramic bearing between the shaft and the cavity, so that the rotary shaft is aligned in a straight line. The bottom rotary shaft was hemi-spherically closed, allowing smooth contact with the bottom of the cavity, replacing the role of bearings. This structure solves the hot rod problem, which was raised in previous studies, by bringing the sapphire and the cavity directly into contact, but also prevent the mechanical vibration that causes frequency fluctuation. These mechanical issues are covered in more detail in the next section~\cite{bib:hotrod_die}.

\begin{figure}[h]
\begin{center}
\includegraphics[width=.75\textwidth]{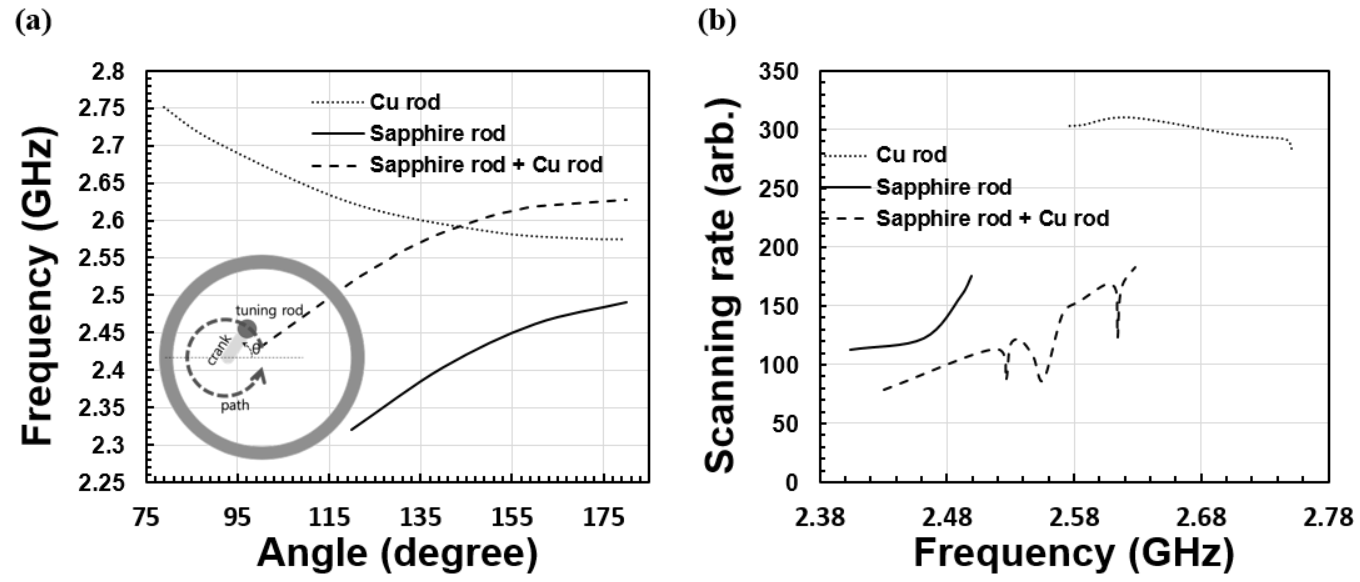}
\caption{Cavity properties with respect to frequency tuning. The solid line represents simulation results and the dotted line does measurements }
\label{fig:results_tuning}
\end{center}
\end{figure}

Since the tuning rod rotating axis is 19\,mm apart from the cavity center, the tuning rod can be positioned from the center of the cavity to 38\,mm distant from it. Figure \ref{fig:results_tuning} shows actual measured values of $Q$ factor as well as numerically calculated $Q$ factor and $C$ factor in all frequency regions scanned or to be scanned in CAPP-PACE . The calculation of the $Q$ factor and $C$ factor were obtained using CST suite, and the $Q$ factor was measured using the Keysight network analyzer~\cite{bib:CST_web}\cite{bib:keysight}. The $Q$ factor was increased by using sapphire with a very low loss tangent in the region of 2.45 $\sim$ 2.5\,GHz which is the first section. In the intermediate region of 2.5 $\sim$ 2.6\,GHz is a blind spot that cannot be scanned with the original method, but even with difficulties such as mode crossing problems and low form factor problems, a metal rod is additionally arranged appropriately and we could achieve fairly high scanning rates. In the region above 2.6\,GHz, a copper-plated thin rod is used in stainless steel to reduce the probability of occurrence of the hybrid mode and to minimize the area of the conducting area, thereby enabling high $C$ factor and $Q$ factor.

\subsection{Accurate mechanical control}

Developing a high $Q$ factor is a challenge in and of itself, but there are other difficulties when tuning devices are introduced there. The higher the $Q$, the higher the reliability of the measured resonant frequency should be. In case of quite long time data acquisition at a single frequency, essential in axion search experiment, you lose the meaning of making a high $Q$ cavity if the frequency fluctuates more than the cavity bandwidth. However, it is not easy to maintain a stable center frequency when a tuning material moves independently in the cavity at cryogenic temperature with a high mechanical limit. In addition, the frequency needs to be tuned more finely than the cavity bandwidth, thus the higher the $Q$, the more elaborate the tuning bar movement. For the CAPP-PACE cavity, the loaded $Q$ is about $>$30k, so if we want to tune the cavity frequency by 1/5 of cavity bandwidth, we must be able to move the resonant frequency within about 15\,kHz per step, that is, a super-precision rotational stage capable of moving with a resolution of 1/100 degree or less per step is required. CAPP-PACE has solved all these problems applying following advanced technologies. 
 
 Since early 2014, In CAPP, we have been using Attocube’s piezo actuators for high-resolution frequency tuning and for precise cavity coupling. Piezoelectric actuators convert electrical energy directly into mechanical energy and allow operation in the sub-nanometer range and the motion occurs instantly~\cite{bib:piezo_basic}. Theoretically, there is no resolution limit, and it does not affect the operation even in an ultra-high vacuum (UHV) and a high magnetic field. In addition, the piezo effect occurs in cryogenic condition even below 1\,K. The Attocube's piezo-electric devices, which we have used, not only has all of these advantages but also solves some of the short travel distances and weak forces that have been pointed out as weaknesses~\cite{bib:attocube_web}. They are made of titanium to be operational in high magnetic field and they can support a newton of the mechanical load which is sufficient to make the movement of tuning rod and coupling antenna. The resolution of the movement can be controllable by adjusting the amplitude and frequency of the applying signal which has a periodic saw-tooth pattern. Especially in a step mode operation, decreasing frequency increases the energy of a unit tooth inversely so that the step resolution and the force of the piezo actuator increase together. The smaller the frequency is, the piezo actuator gets stronger. Important specification of the attocube piezo actuators is listed in Table~\ref{table:piezo_spec}.

\begin{table}[h]
\centering
\begin{tabular}{c|c}
\hline
~~~~~~~~Property~~~~~~~~ & ~~~~~~~~value~~~~~~~~  \\
\hline
travel range         & $360\,^{\circ}$ endless (rotator), 12\,mm (linear positioner) \\
resolution(@4\,K)      & $0.5\,m,^{\circ}$ (rotator), 10\,nm (linear positoner)  \\
magnetic field       & 0$\sim$31\,T                                    \\
tempearture          & 10\,mK$\sim$373\,K                                \\
minimum pressure     & 5E-11 bar                                     \\
maximum load (@300\,K) & 2\,N (200\,g)   \\

\hline
\end{tabular}
\caption{\label{table:piezo_spec}Specification of attocube piezo actuators.}
\end{table}

\begin{figure}[t]
\begin{center}
\includegraphics[width=.65\textwidth]{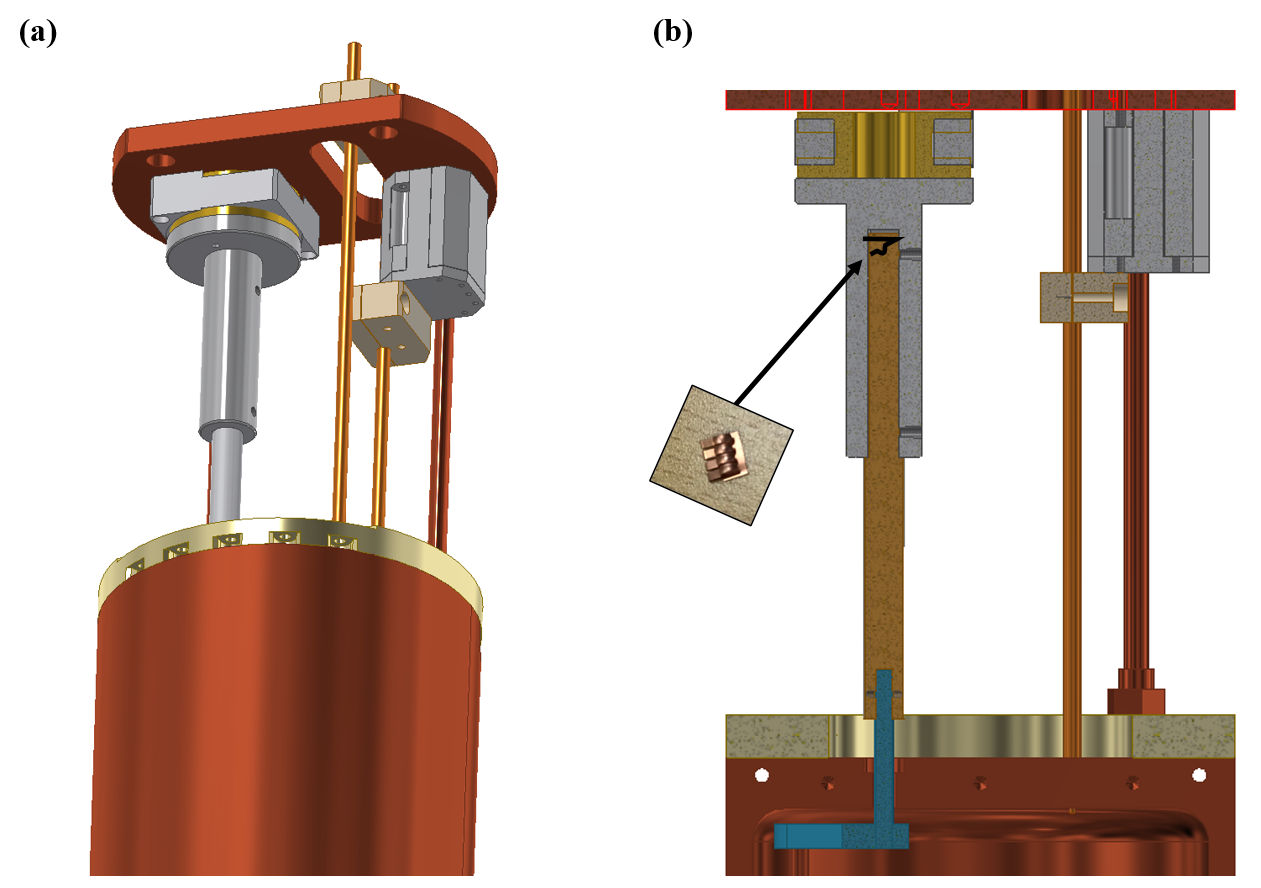}
\caption{(a) 3D rendering of attocube piezo actuators with a cavity and supporting jig. (b) Cut view of (a), Beryllium-copper springs are located between PEEK medium and rotary piezo actuator }
\label{fig:pic_piezoetc}
\end{center}
\end{figure}

\begin{figure}[h]
\begin{center}
\includegraphics[width=.75\textwidth]{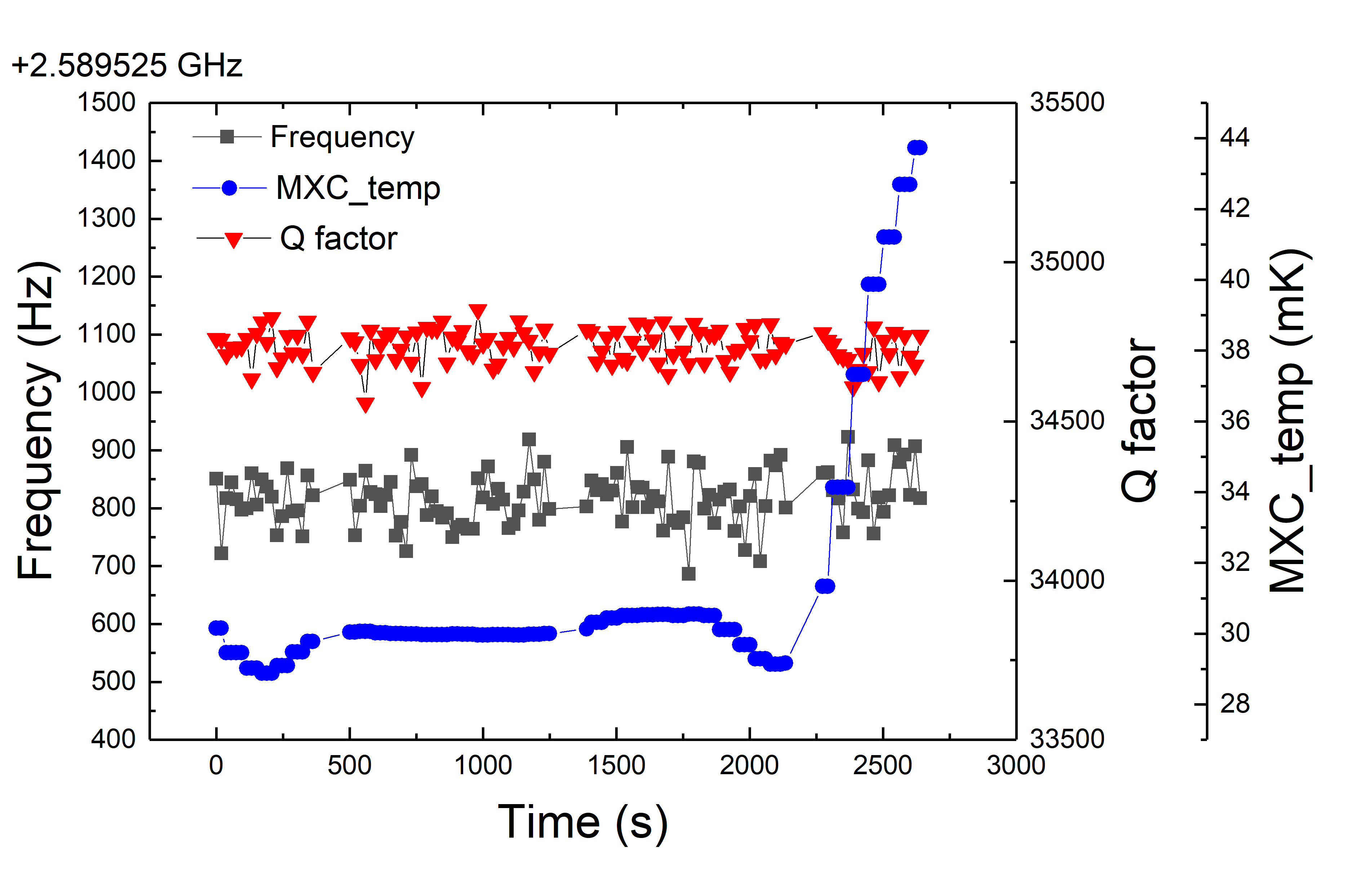}
\caption{Time versus resonant frequency, loaded Q factor where the coupling strength $\beta$ is 2, and the temperature of the cavity. For a measurement time of more than 30 minutes, the frequency and Q factor show deviations of less than 1/200,000 and 1/100, respectively, indicating that the components of the cavity system are stable without relative vibration. The temperature also showed a deviation of less than 2\,mK at around 30\,mK and a sudden increase in temperature at 2400 seconds occurred due to applying position sensing voltage.}
\label{fig:piezoread_onoff}
\end{center}
\end{figure}

\begin{figure}[h]
\begin{center}
\includegraphics[width=.85\textwidth]{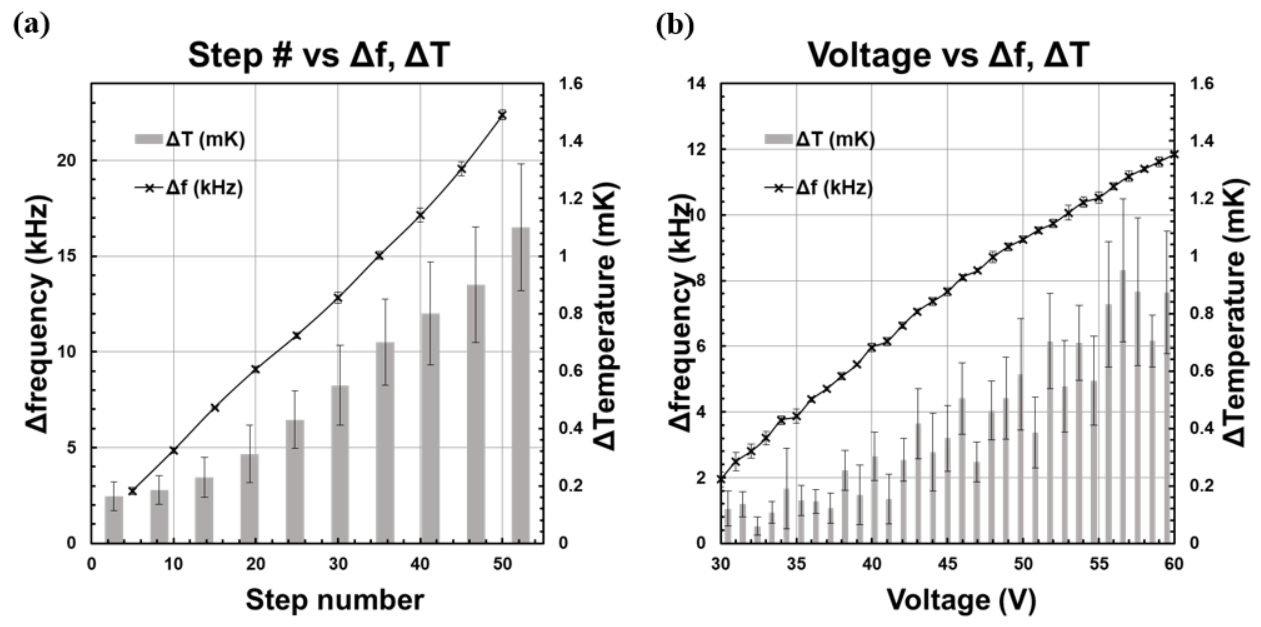}
\caption{ Cavity temperature and moved frequency as functions of piezo parameters such (a) bias voltage, (b) step numbers.    }
\label{fig:T_and_df_vspiezo}
\end{center}
\end{figure}

We fixed a rotary piezo actuator to the cavity-supporting structure like Fig.~\ref{fig:pic_piezoetc} and connected the rotary piezo with the upper shaft of the tuning device through the PEEK rod. This is to prevent the temperature of the sapphire rod from fluctuation by allowing the heat to flow through the MXC plate without flowing through the rod. We put several pieces of the beryllium copper (CuBe2) contact finger strip in the middle of the actuator and PEEK rod to assure the stable contact between cavity and sapphire shaft even with thermal contraction below 100\,mK. As a result, we could tune the cavity with great precision, as shown in the measurement results of Fig.~\ref{fig:piezoread_onoff}. Fine adjustment of the magnitude and frequency of the bias voltage allows tuning even at higher resolutions, and it is noteworthy that the tremor of the frequency is completely eliminated, resulting in a very stable resonance frequency despite tuning in\,kHz units as shown in Fig.~\ref{fig:T_and_df_vspiezo}.

Nevertheless, another consideration is that the heat generated by the piezo devices can increase the cavity temperature. We blocked the attocube piezo from always sending a bias to ask for position information so that it was about 20\,mK lower than when it was not, that is, the temperature when no piezo was installed. Additionally, by operating them in a single step mode, the temperature could only be changed to less than 5\,mK during frequency tuning even in the condition that the 8\,T magnet was on.


\subsection{RF Receiver Chain and NT measurement}

\begin{figure}[h]
      \includegraphics[width=0.6\textwidth]{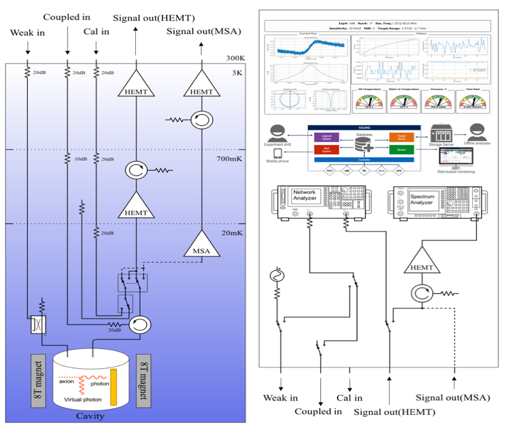}
\centering
   \caption{RF receiver chain of the PACE experiment}
  \label{fig:RFchain}
\end{figure}

One of the most important parts of the PACE experiment is the receiver chain. The simplified receiver chain is shown in Figure \ref{fig:RFchain}. The excitation signal is put in through the ``Weak-in" port of the resonant cavity to measure the cavity parameters, like Q-factor, and the coupling strength of the antenna is measured through the ``Coupled-in" port. The gain of the amplifier is measured through the ``Cal-in" port and the signal from the axion is extracted through the ``Signal-out" port. 

The first RF component which signal generated from cavity encounters is the circulator. It prevents reflected waves from the back of the preamplifier and the background noise coming from the amplifier's temperature entering the cavity. This also makes it possible to measure the coupling strength of the cavity antenna. The next component is a switch that gives us a choice of preamplifier - a HEMT (High Electron Mobility Transistor) or an MSA SQUID amplifier without a warm-up and a cool-down thermos-cycle. In addition, the DPDT switch allows one input to be connected to a noise source, allowing a more precise measurement of the noise temperature of the whole RF receiver chain during the experiment.

 In principle, it is advantageous not to place these components before the preamplifier to reduce noise as much as possible, but the impact on the overall noise temperature is negligible because the component's loss values ​​are small and because of the ultra-low temperature inside DR. In addition, each component between the cavity and preamplifier is connected by superconducting cables (connection between different temperatures) or a short, thick high purity copper RF cable (connection between same temperatures) to minimize the loss as much as possible. 

 The magnitude of the acquisition signal power is very small in the experiment. With our detector setup, the signal power corresponding to KSVZ sensitivity is about $\sim$ $2\times10\textsuperscript{-24}$ W, which is equivalent to $-202.3\,$dBm. In order to measure this tiny signal, a highly advanced RF receiver chain is required. Our setup is designed with state-of-the-art components for this purpose. HEMT or MSA SQUID amplifier is used as the preamplifier of the experiment.

The HEMT amplifier is a type of field effect transistor (FET) that is a transistor based amplifier. It works well even at low temperature and has an advantage of less fluctuation of operating current which lead less noise temperature. In general, the HEMT amplifiers used in usual experiment show noise temperatures between 2\,K and 4\,K. However, the HEMT amplifier that we use in this experiment (developed by the Low Noise Factory recently) shows about 1\,K noise temperature. This is the lowest noise level among the amplifiers that do not use the quantum devices so far. In this experiment, we successfully adapt this HEMT amplifier.

The Microstrip SQUID amplifier (MSA) are amplifier based on the Superconducting Quantum Interference Device (SQUID). Under appropriate current and flux bias, the SQUID can operate as an amplifier using the flux-to-voltage transfer characteristics. It can be used at the\,GHz frequency band through the microstrip resonator structure. It is known that the typical gain should be about 25\,dB and the noise temperature can reach about twice the Standard Quantum Limit (SQL). Near the frequency we use, the SQL is ~100\,mK, which is a much lower noise level than the 1K HEMT. We have many SQUIDs from various sources such as KRISS, IPHT, ezSQUID and the optimization study is currently in progress.\\The room temperature electronics used in this experiment is as follows. Signals coming out of the DR refrigerator are recorded via the spectrum analyzer. We also use a vector network analyzer to measure various parameters of the cavity. For evaluating the detector's performance, we generate a fake axion signal using a function generator and a signal generator. Each line is connected to the switch controlled by the computer so that you can do all this process automatically using DAQ

If the gain of the preamplifier is high enough, the preamplifier is the dominant noise source added to the ideal receiver chain. In this experiment, we use a HEMT amplifier (LNF\_LNC 1\_12) as the amplifier of the second stage, which shows noise temperature of about 6\,K. When 1\,K HEMT (LNF\_LNC\_2\_4) is used as a preamplifier, its noise temperature is about 1.1\,K and its gain is about 40 dB. Thus when this noise reaches the second stage it will be 11000\,K. Then 6\,K noise is negligibly small. When MSA is used as preamplifier even if we assume ideal case which its noise reaches Standard Quantum Limit (SQL), 1\,K HEMT(LNF\_LNC\_2\_4A) can be used as a second stage amplifier without influence. Assuming that the noise of MSA reaches to SQL and the gain of $\sim$ 20\,dB, still the noise contribution of the preamplifier is dominant because in the second stage, Noise temperature due to MSA is about 20\,K, which is much larger than 1\,K HEMT’s. In our experiments, however, there are also circulators, switches, and cables connecting them between preamplifier and cavity. First, a superconducting cable consisting of NbTi-NbTi is used to achieve a negligible loss. In the case of circulators and switches, there is insertion loss about $\sim$ 0.1\,dB, but since they are kept at $\sim 20\,$mK, they do not add a lot of noise. There may also be a Johnson noise coming from an external room temperature of 300\,K. In this experiment, we put a 20\,dB attenuator at 4\,K,  and an MXC plate on every input lines. Thus the noise contribution  corresponding to 300\,K is only about 7\,mK, which is negligible also. Therefore, in this experiment, it is concluded that the noise of the preamplifier is most important as in the ideal case.

\begin{figure}[b]
      \includegraphics[width=0.35\textwidth]{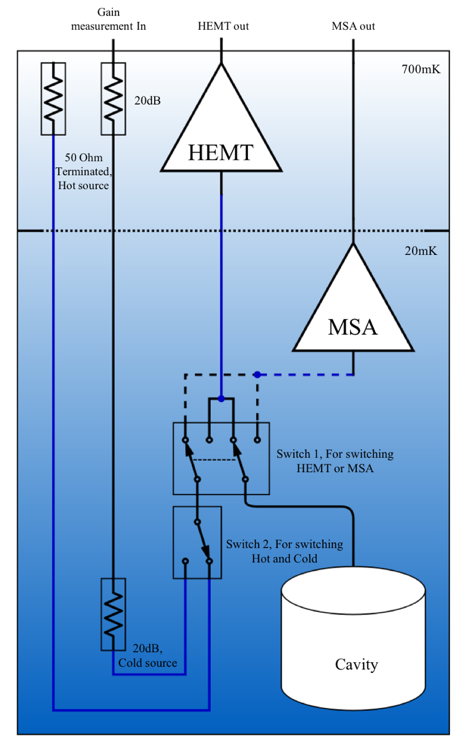}
\centering
   \caption{Noise temperature measurement setup}
  \label{fig:NT}
\end{figure}

Now let's discuss how to measure the noise of the receiver chain. Since the signal from the axion is extremely weak, the system noise temperature($T\textsubscript{sys}$) plays a critical role in scanning frequencies. Figure \ref{fig:NT} shows the RF chain setup for measuring the noise temperature of our preamplifier, 1\,K HEMT.  $T\textsubscript{sys}$ is the sum of  $T\textsubscript{receiver chain}$ and $T\textsubscript{cavity}$.

\begin{figure}[t]
      \includegraphics[width=0.6\textwidth]{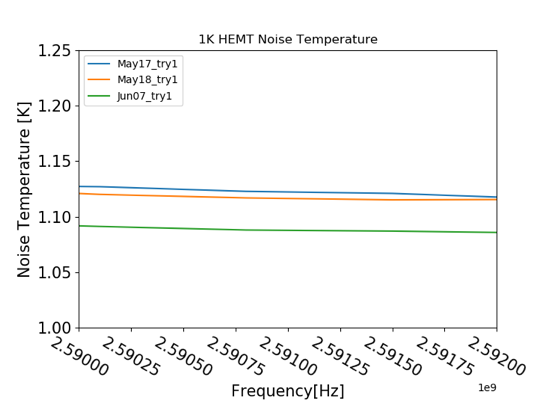}
\centering
   \caption{Noise temperature measurement result of the 1\,K HEMT amplifier}
  \label{fig:NTdata}
\end{figure}

$T\textsubscript{cavity}$ is determined by the temperature of the cavity. For the $T\textsubscript{receiver chain}$ measurement, our experiment including {\it in-situ} measurement using the Y-factor method. In usual Y-factor method, a hot and cold source is Excess Noise Ratio (ENR) a.k.a noise diode. However, in our case, It’s hard to connect this for calibration due to cryogenic temperature(below 1\,K). Thus we need different temperature noise source for measuring noise temperature. We use a terminator or attenuator installed in different temperature stages. The switch1 connects amplifiers for noise measurements, and the switch2 could choose between Hot (on) state and Cold (off) state. We can measure the noise temperature of the whole RF chain from the power difference two states. In the case of Cold state, an attenuator is used instead of a terminator. With this additional line, we can measure gain directly. Figure \ref{fig:NTdata} shows the result of our ``cold terminator method" noise temperature measurement of 1\,K HEMT.

\hypertarget{daq-and-operations}{%
\subsection{DAQ and Operations}\label{daq-and-operations}}

\begin{figure}[bp]
\begin{center}
\includegraphics[width=\textwidth]{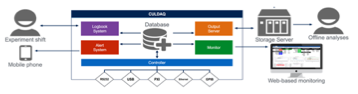}
\caption{Overview of the PACE experiments DAQ system.}
\label{fig:pace_daq_system}
\end{center}
\end{figure}

The DAQ system is composed of acquisition, controls and monitoring
subsystems. While controls and acquisition software is run on the main
computer, monitoring is decoupled from this system and can be accessed
from outside easily. The main DAQ software is responsible for running
the frequency tuning algorithm, taking calibration measurements,
performing physics data acquisition. It has an easy to use interface
that helps us smoothly start the data taking with the click of a button.
Its modular structure promotes future extensibility.  Figure
\ref{fig:pace_daq_system} depicts a simplified picture of our overall DAQ
system.

\begin{figure}[h]
\begin{center}
\includegraphics[width=.55\textwidth]{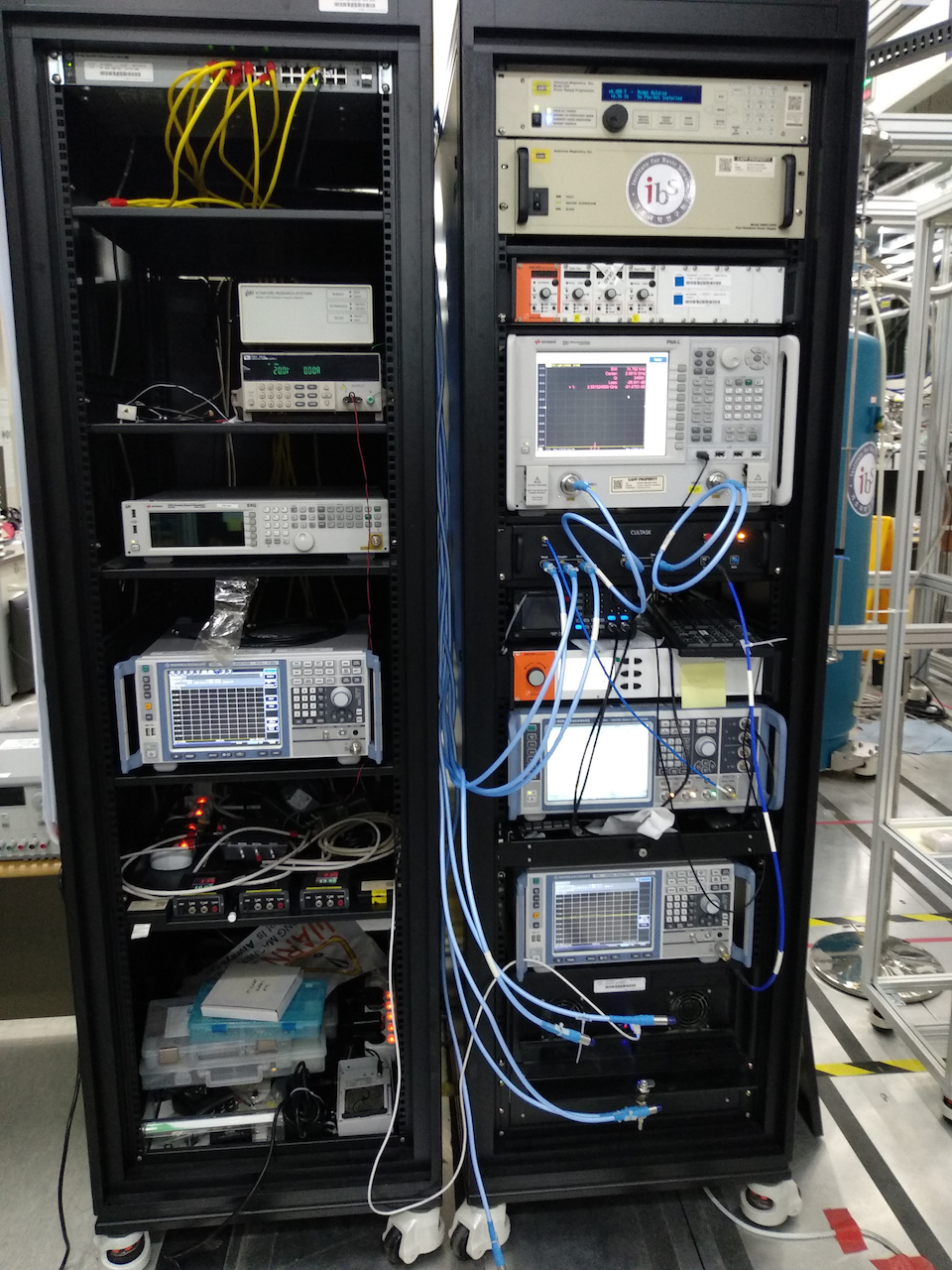}
\caption{A view of the PACE experiment's instrument rack.}
\label{fig:pace_daq_rack}
\end{center}
\end{figure}

For cavity parameter extractions we use a commercially available network
analyzer from Keysight.  The microwave signal coming
from the cavity is downconverted, digitized and recorded as averaged
power spectrum using FSV7 series spectrum analyzer from Rohde \&
Schwarz. An SMW series vector signal generator R\&S along with an
arbitrary waveform generator from Teledyne Lecroy is used to generate
various calibration signals and virialized fake axion signals. Along
with these, we have various temperature sensors.  We use commercial instruments
mainly for their convenience. Using reliable and industry proven measurement
devices allows us to focus on more important aspects of our experiment.  See
Figure \ref{fig:pace_daq_rack} for a view of our instrument rack.

The software we use is mainly divided into three categories: controls
and acquisition (dubbed simply DAQ) software and individual scripts. DAQ
software is capable of controlling all the measurement instruments and
directing the experiment flow. After starting an experimental run, we
proved that there is no need for human interaction for weeks. DAQ
program is written in Python 2.7 with a user interface utilizing a Qt4
framework library. By utilizing an industry proven object-oriented
design, a failsafe operation is ensured. Also, it's modularized
structure provides us with an easily improvable codebase. Individual
scripts running in the DAQ computer make sure that the extracted cavity
parameters including coupling constant are uploaded to the database.
Also, an alert and safety script is continuously running on the main
computer. Scripts running in the dilution refrigerator's control machine
is responsible for uploading temperature and pressure information to the
database server. Overall programming language for our project is Python.
Given that it's an interpreted language, it ought to be slower than
bare metal languages such as $C$ and FORTRAN. We mitigate this slow
running problem by using numerical python libraries heavily which are
written in $C$ and optimized. Since we record averaged spectra, the data
storage and data bandwidth are not the main problems for us. Regardless we
use a binary packaged data structure, namely ROOT binary files, to save
our physics and some auxiliary data. ROOT is a C++ data analysis library
mainly built and used by the HEP community.



The measurement efficiency is one of the most important metrics of our
experimental setup. The total scanning time for a given frequency range
is directly proportional to efficiency. We can simply define efficiency
as the ratio between time ideally required to run the experiment and the
time actually took to run the experiment.

\begin{align}
    \mathrm{SCAN}\,\mathrm{RATE} = \frac{d\nu}{dt} = \eta \mathrm{f}(\mathrm{SNR}, ... )
\end{align}

where \(\eta\) is efficiency and f is a function of the parameters of
our experimental setup.

Efficiency is directly affected by the performing times of individual
components of DAQ. Without going into details, the three main time-consuming operations are frequency tuning, instrument communication,
actual data acquisition time. Tuning usually takes maximum 30 seconds and 
communications takes less than 10 seconds whereas our data acquisition
is usually hours for a single tuning step. Thus, we may safely say that
efficiency bottleneck is the spectrum analyzer efficiency \(\eta_{acq}\).

\begin{figure}[h]
\begin{center}
\includegraphics[width=.85\textwidth]{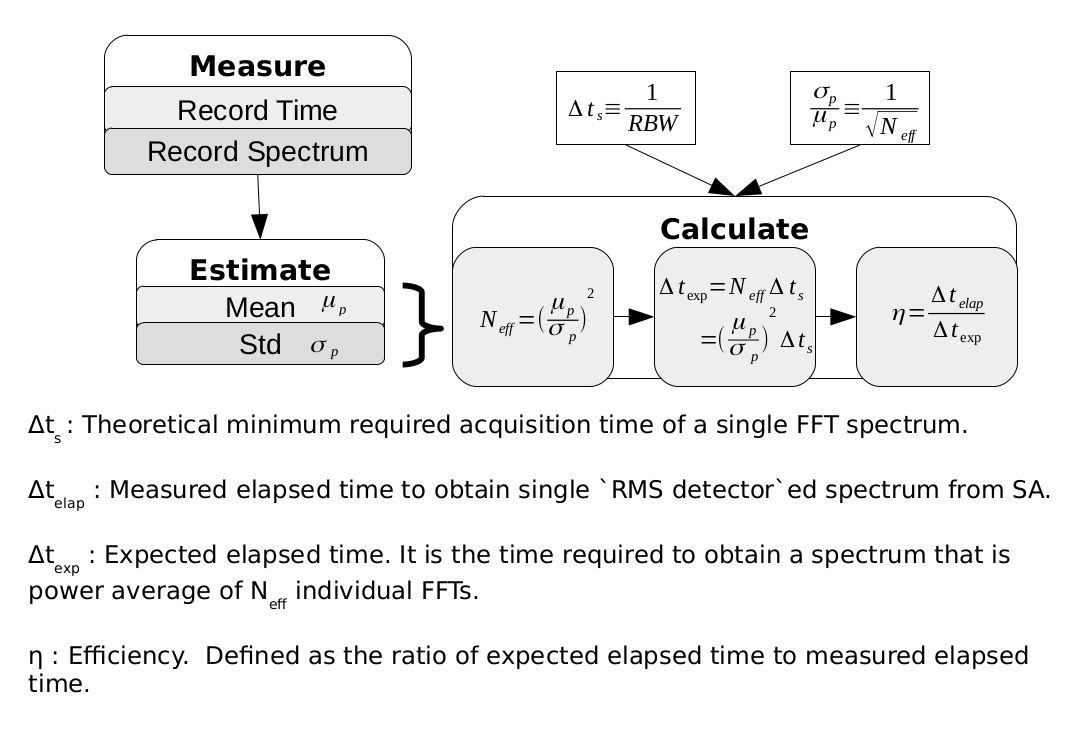}
\caption{Efficiency estimation method used in measuring the efficiency of
a spectrum analyzer.}
\label{fig:pace_daq_efficiency_estimation_method}
\end{center}
\end{figure}

For a single noise spectrum, the mean divided by standard deviation is
an estimator for the number of independent samples averaged. The ideal
acquisition time is then estimated using this number with the spectrum
resolution bandwidth information. We also apply a correction coming from
the spectral estimation method applied by the spectrum analyzer.  See Fig.~\ref{fig:pace_daq_efficiency_estimation_method} for the efficiency estimation
method we use.

\begin{figure}[h]
\begin{center}
\includegraphics[width=.75\textwidth]{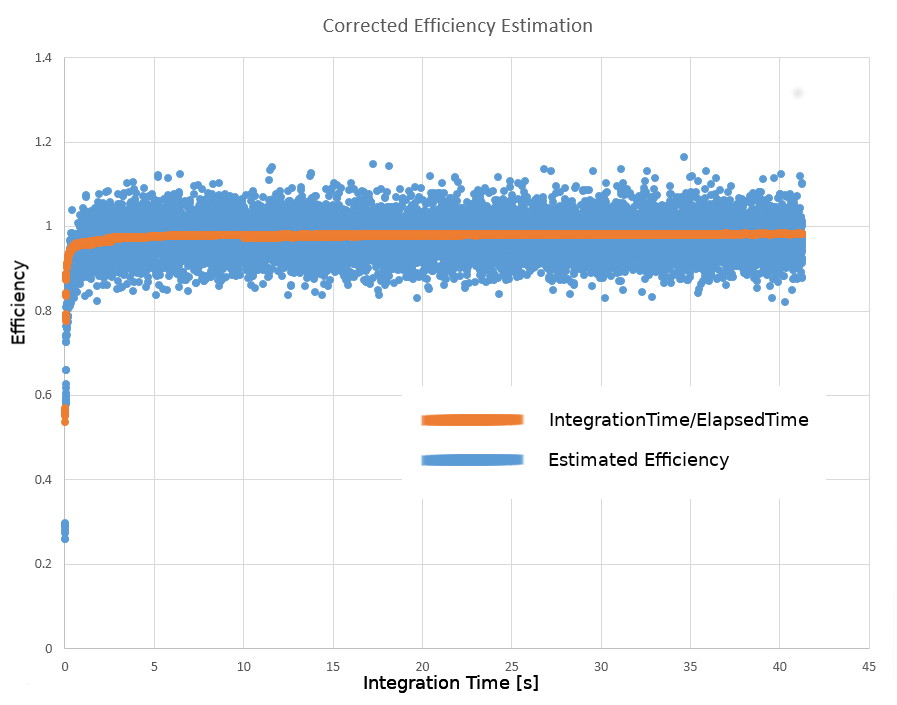}
\caption{Efficiency measurement with correction included.}
\label{fig:pace_daq_efficiency_measurement}
\end{center}
\end{figure}

The spectrum analyzer, with the minimum sweep time, collects the
samples, computes the FFT, fetches the FFT result to update the display
and make the data ready for output. When we use the trace average
functionality, the spectrum analyzer still updates the display for every
spectrum acquired, thus the bottleneck coming from fetching and
displaying the result dominates the overall efficiency. In this case, we
achieved efficiencies that are only up to 50\%. FSV series spectrum
analyzers are also capable of fast averaging the captured spectra using
more than minimum sweep time. The only downside of this method is that
the display is not updated at the intermediately acquired spectrum, which
isn't a problem for our experiment at all. Using this method, we were
able to achieve close to spectrum acquisition efficiencies, see figure
\ref{fig:pace_daq_efficiency_measurement}.

Currently, we are using two main monitoring systems: web-based and
desktop based. The web-based monitoring system is based on an
open-source web application for general purpose monitoring integrations
called Grafana. The desktop software is written in-home for having more real-time access to physics and cavity data, and also for future
flexibility. Apart from monitoring systems, we have a non-stop alert
management system that continuously monitors safety-critical aspects of
the whole system.

\begin{figure}[h]
\begin{center}
\includegraphics[width=.75\textwidth]{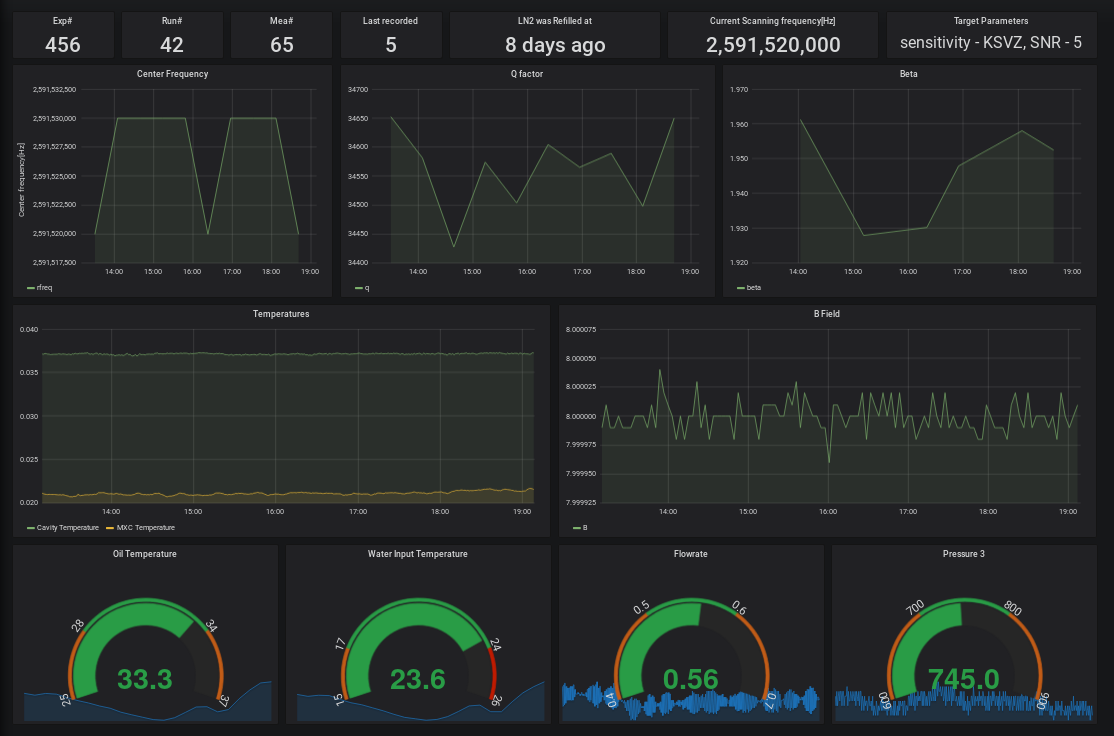}
\caption{A view of PACE's Grafana dashboard.}
\label{fig:pace_daq_grafana}
\end{center}
\end{figure}

Grafana is an open-source web application for monitoring variables
fetched from databases. It provides a very natural interface that
separates presetation from the data itself. The data panels are arranged
in a dashboard using its drag and drop interface. Each panel is
connected to it's source data via simple SQL queries. Figure
\ref{fig:pace_daq_grafana} shows a view of our web dashboard.

\begin{figure}[h]
\begin{center}
\includegraphics[width=.75\textwidth]{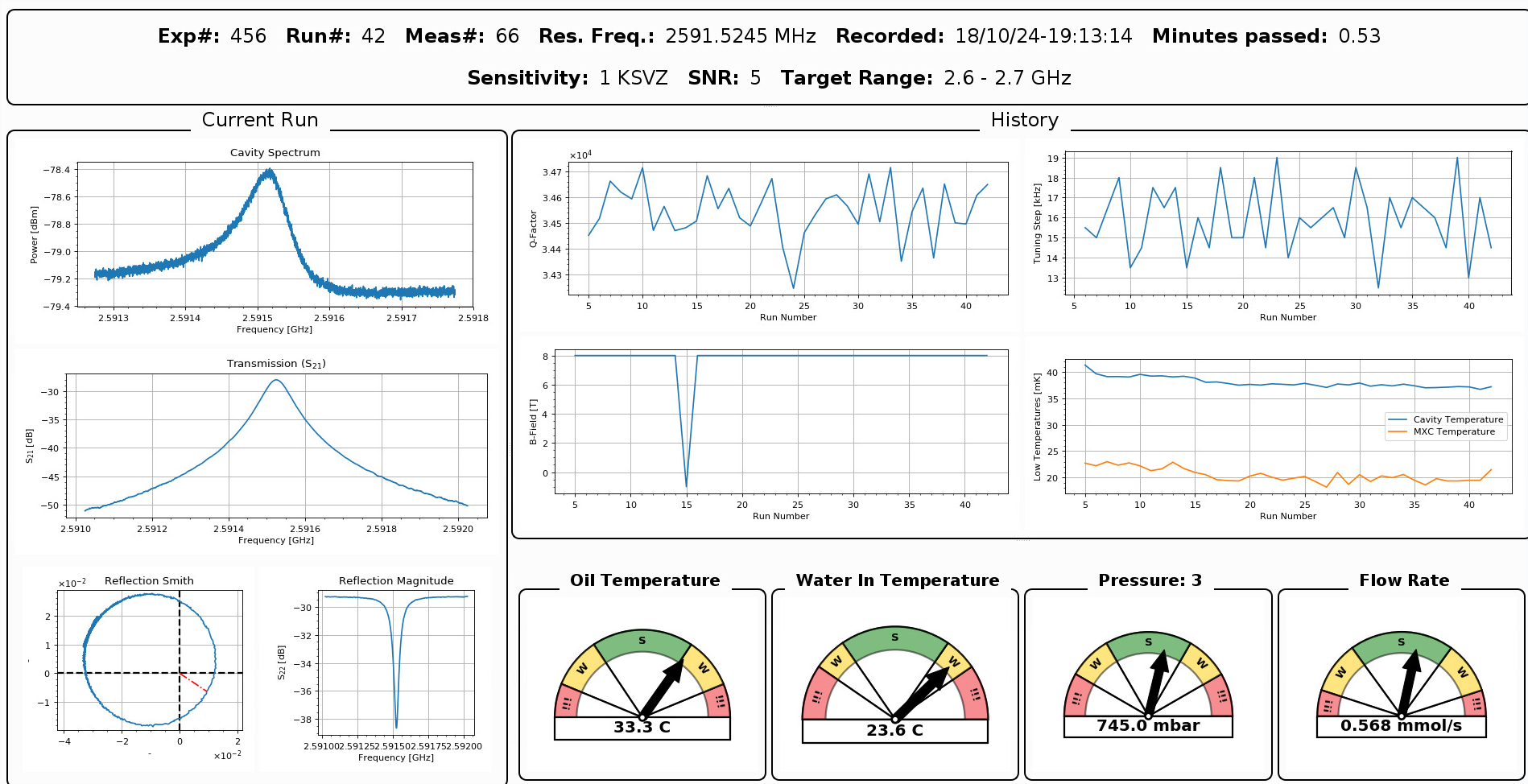}
\caption{A view of the Qt based desktop monitor application.}
\label{fig:pace_daq_desktop_monitor}
\end{center}
\end{figure}

The desktop application has access both to binary data files and to the
database. We mainly access the desktop monitor by accessing the daq
computer via VNC remote monitoring software. Desktop monitoring software was 
developed in house using Python3 with Qt5 library. It currently provides
us with a hackable platform to implement new visualizations of our data as it
is more flexible in terms of plotting techniques. The plotting uses the
widely available matplotlib library.  See Figure
\ref{fig:pace_daq_desktop_monitor} for a view of our desktop application monitor.

Alert program runs on the main computer continuously. It acts on with
response to the change in any of the observed system parameters such as
temperatures in the dilution unit. There are mainly three levels of
critical conditions for the parameters. When the first condition is
triggered, the software sends an alert over Telegram, but doesn't
perform any action. If the level two trigger is tripped, an alert
message is continuously sent. After the level three gets triggered, the
last alert message is sent asking for an action to be performed. If the
action isn't performed within a specified time the alert software
performs the action by itself in order to protect the experimental
fixture. Currently, that action is mainly ramping down the magnet.


\subsection{Analysis}

The upper limit to the sensitivity of a haloscope experiment is solely
determined by the experimental apparatus including the data recording
instruments. The job of analysis is to extract the information contained
within without reducing this intrinsic sensitivity any further.

In the PACE experiment, we are currently developing a set of tools that
would allow us to build further analyses and improve our system. The
summer KSVZ run of our experiment was successfully completed, leaving us
with good quality data. From here we will give a brief overview of our data, the
software we used and the overall analysis procedure.

\begin{figure}[h]
\begin{center}
\includegraphics[width=.95\textwidth]{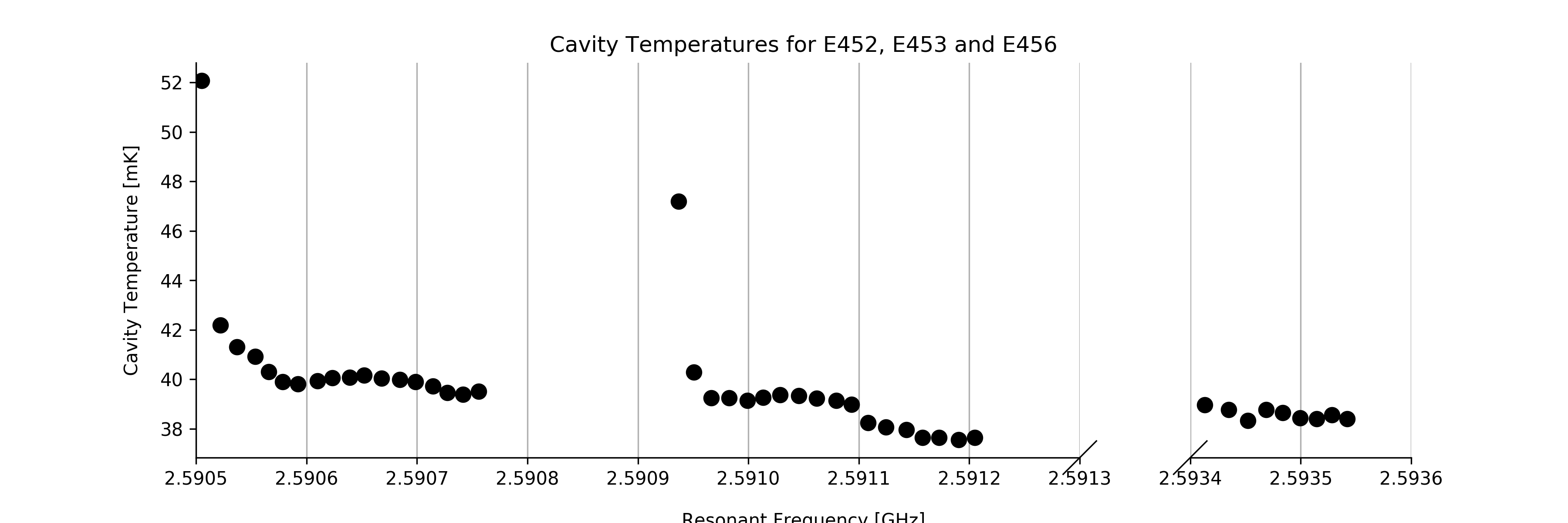}
\caption{Temperature vs. resonant frequency of the cavity from our KSVZ sensitive runs.}
\label{fig:pace_analysis_temperature_vs_rfreq}
\end{center}
\end{figure}

We categorize our data into three main groups:
supplementary, auxiliary and physics data. \emph{Supplementary data}
consists of all the data that we do not actively need for analysis, but
necessary for system diagnosis. Examples of this are temperature and
pressure levels at various levels of the dilution refrigerator unit. We
refer all the cavity parameter measurements as \emph{auxiliary data},
including coupling coefficient ($\beta$) and quality factor of the
cavity. Finally, we refer to the recorded spectrum of the RF signal
coming from the cavity as the \emph{physics data}. During PACE's 2018
Summer KSVZ run, we collected 3870 heavily averaged spectra, along with
approximately 720 hours of supplementary and auxiliary data. Each
spectrum we acquire is centered at the resonance frequency of the
cavity.  See Figure \ref{fig:pace_analysis_temperature_vs_rfreq} for the cavity
measurement from our latest KSVZ run.

For the implementation of the analysis, we mainly used the Python 3.6
programming language with its CPython implementation. For investigating
different analysis implementations and strategies, we needed a flexible
tool that is fast to experiment with. Having a concise syntax with
support to multi-paradigm programming and a vast amount of libraries for
scientific computing, python is an excellent choice for our analysis
needs.

Being a dynamically typed language and relying on automatic garbage
collection makes python programs run slower when compared to programs
written in more traditional languages such as C, C++ or FORTRAN. We
mitigate this problem by relying heavily on numerical libraries such as
numpy, scipy and pandas, which are written in $C$ under the hood. As of
now, even without the slightest consideration of computational
efficiency, our analysis program runs under 10 minutes for a dataset of
2 months. Apart from numerical libraries, we use matplotlib for
visualizations. Internally we make use of jupyter and jupyter-lab for
producing documented analysis notes.

As for programming style, we make use of object-oriented style for
mainly data retrieval and persistence and functional paradigm for
composing parts of the analysis. We try to document important bits of
functionality whenever possible, and also continuously improve the
structure of our codebase. It is crucial that the analysis code is
transparent, documented and logically true. For this, a structured and
systematic approach, as well as a very organized codebase is necessary.
One of the goals of our analysis software project is to reach the level
of maturity of more established analysis projects including
astropy from the astrophysics community and gwpy from gravitational wave community.

Any haloscope analysis ultimately aims to locate a persistent signal
among power spectrum bins populated with noise coming from thermal
fluctuations inside the cavity. The signal's frequency distribution will
be mainly determined by the velocity dispersion of the dark matter
axions in our galactic halo. Our goal is to reliably detect or reject
the existence of such a signal in our experiment's sensitive frequency
range with a sound statistical confidence level. Our analysis depends on
the chosen coupling model which we set to KSVZ. After
choosing the model we choose a final SNR of 5 which can be seen as the
conventional \(5 \sigma\) discovery threshold that is used in many other
HEP experiments. Our analysis mainly follows the recipe outlined in
Ref.~\cite{PRL_118_061302_2017}.

The main goal of the analysis is to combine the acquired spectra and
auxiliary data to construct a final spectrum from which we can search
for signal excesses. For constructing this spectrum, our process is
divided into several steps consisting of preprocessing, baseline
removal, scaling, vertical combination, horizontal rebinning and
lineshape integration.

\begin{figure}[h]
\begin{center}
\includegraphics[width=.85\textwidth]{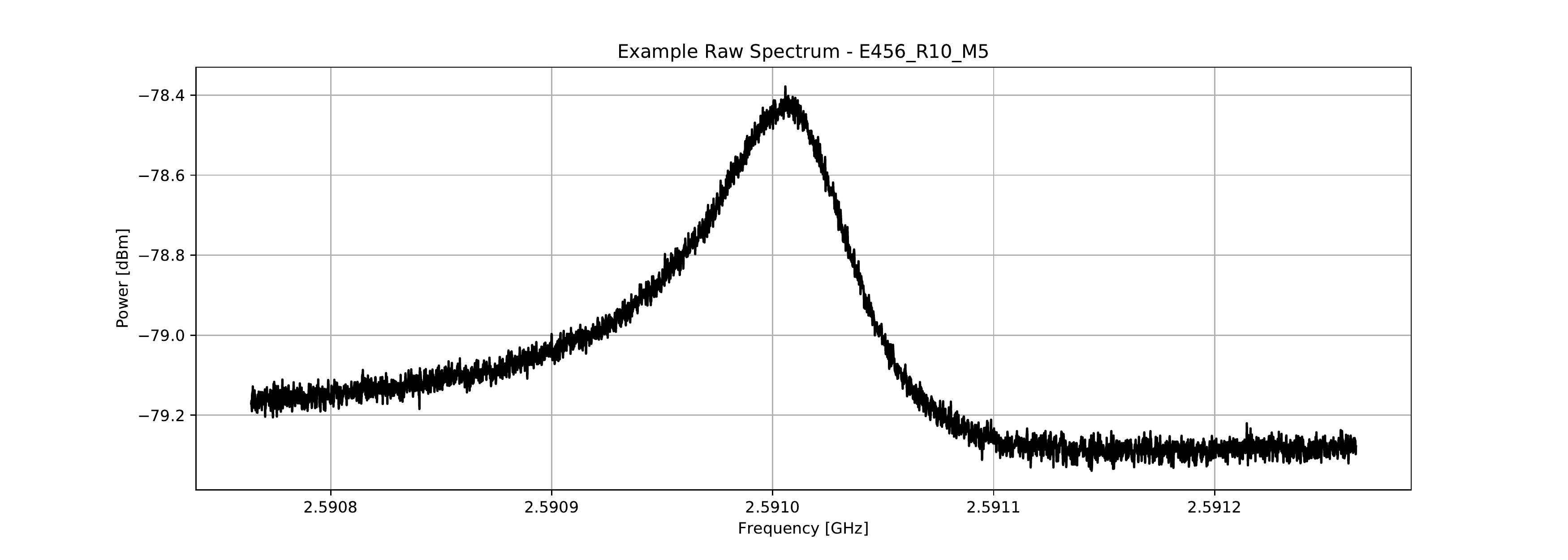}
\caption{Example raw spectrum from one of the KSVZ sensitive experiments(E456 Run 10 Measurement 5).}
\label{fig:pace_analysis_example_raw}
\end{center}
\end{figure}

For PACE experiment's KSVZ run, at each tuning step we collect 90
individual spectra that are averages of 60000 spectra themselves. An example of a raw spectrum is shown in Figure \ref{fig:pace_analysis_example_raw}.
As a first step, we average these 90 spectra with each other. We did not see
any change in the results if we remove the baseline prior to this step
instead. For all our measurements, we observe an intrinsic baseline in
our spectra. To remove this, we first obtain an average baseline by
averaging all the spectrum between different frequency tunings. We
divide each raw spectrum with this average baseline and apply a 5th
order Savitzky Golay with an appropriate window length to obtain a fit
for the intermediate baseline in each spectrum. Then we divide each raw
spectrum with its respective baseline estimate and subtract one to
obtain a spectrum where each bin is drawn from a Gaussian distribution
with 0 mean and \(1/\sqrt{N}\) standard deviation. When obtaining the
baseline, it is likely that we will also capture some part of the signal
if it exists. Thus, this step is the most crucial step for the
maximization of SNR in our analysis.

The baseline is due to the fact that the cavity, as seen from the signal
output port, is only perfectly matched to the coaxial transmission line
at the resonance. Since the noise power is proportional to the real part
of the impedance(that is, the resistance) seen from the input port of
the cavity, it is natural to have a frequency dependent noise spectrum.
The exact shape of the baseline depends mainly on the scattering of
noise-waves along the path from the cavity to the first amplifier. We
can also see a similar noise spectrum when we repeat the measurement in
noise temperature, and vary its shape by changing the length of the
coax connecting the cavity to the first amplifier. Our baseline removal
currently does not rely on this functional form of the lineshape as
described above, but will be in the future. A similar approach can be
found in Ref.~\cite{PRD_64_092003_2001}.

\begin{figure}[h]
\centering
\begin{subfigure}{.5\textwidth}
  \centering
  \includegraphics[width=1\linewidth]{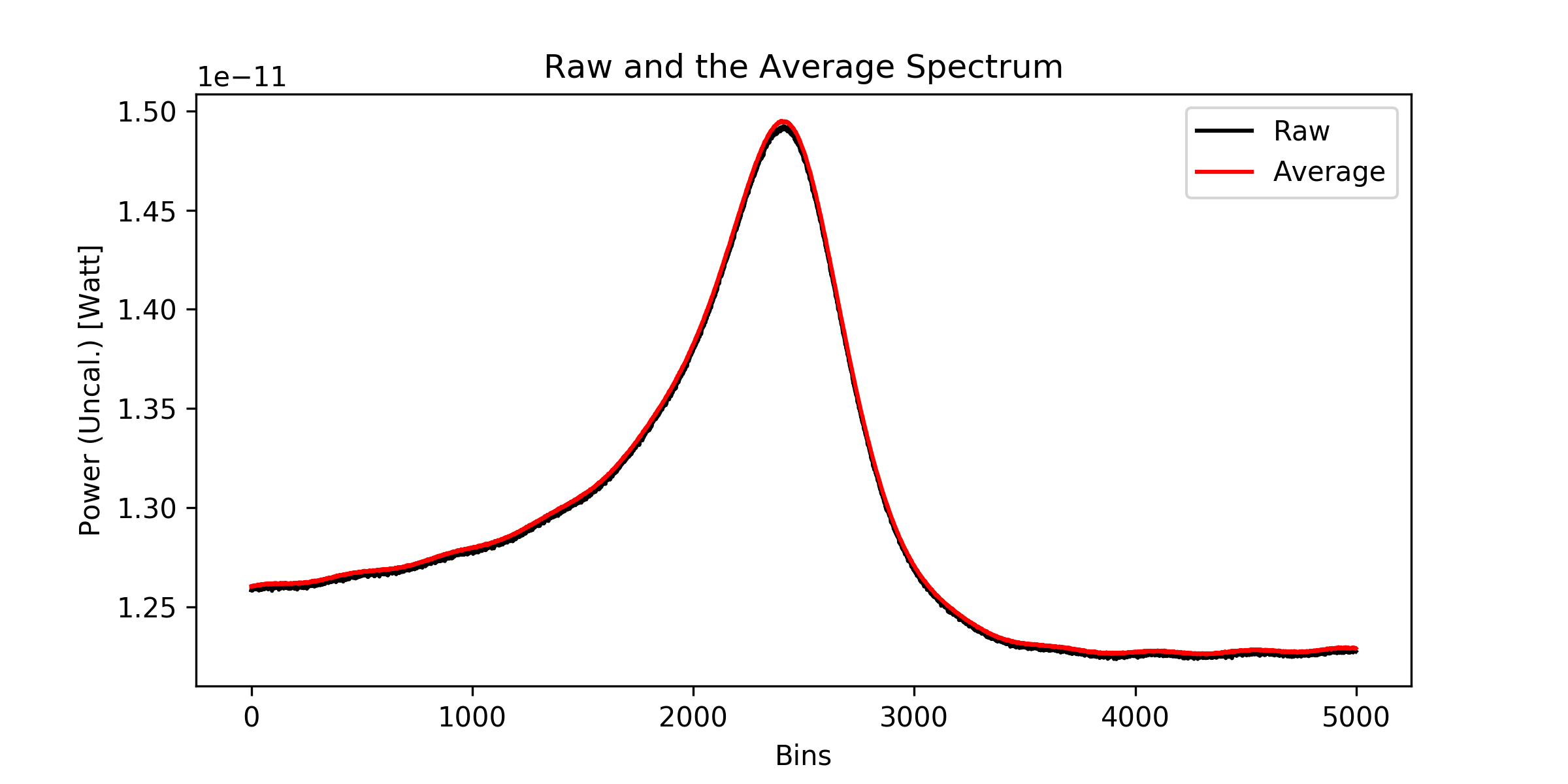}
  \caption{}
  \label{fig:pace_analysis_raw_and_average}
\end{subfigure}%
\begin{subfigure}{.5\textwidth}
  \centering
  \includegraphics[width=1\linewidth]{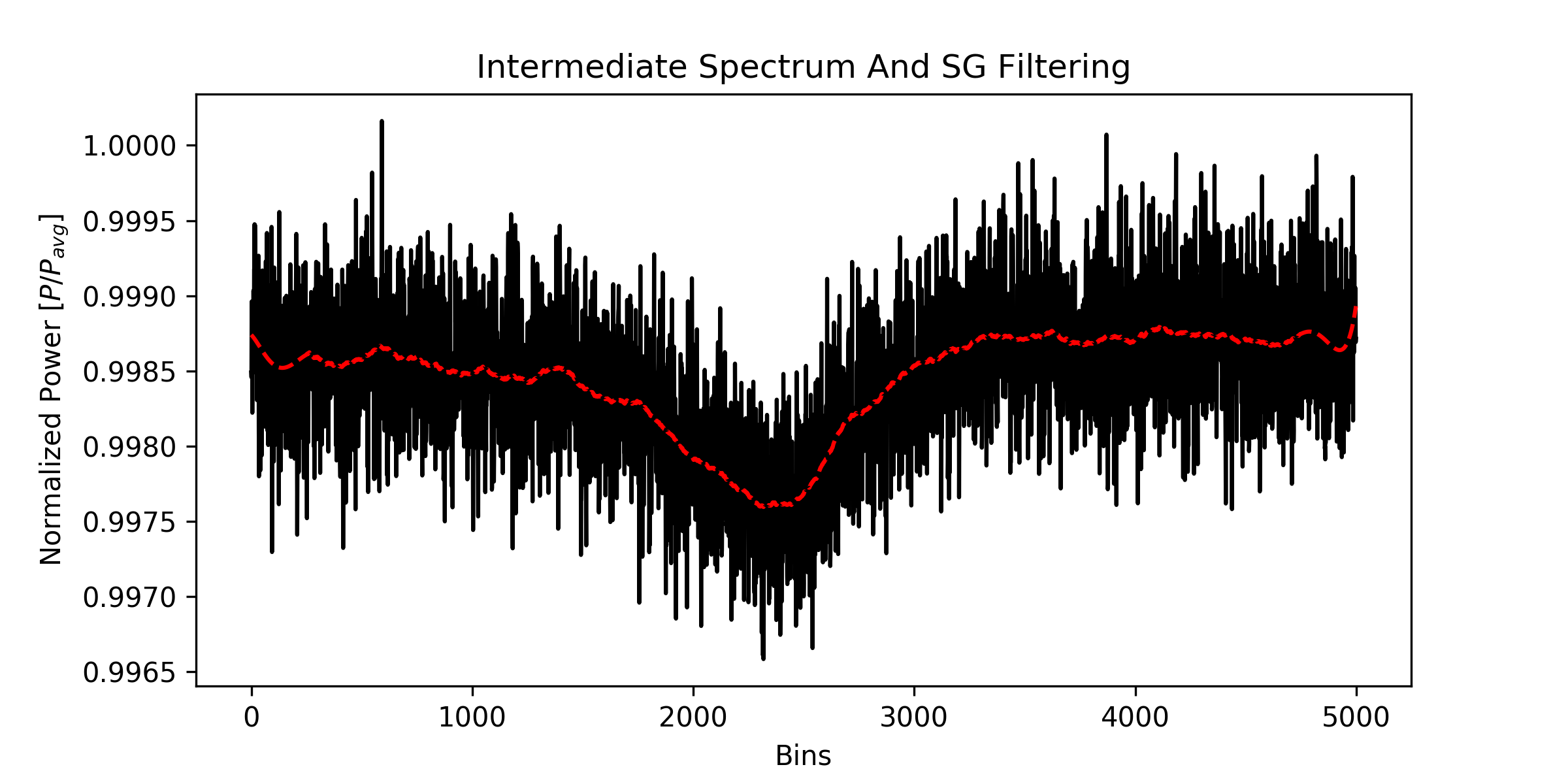}
  \caption{}
  \label{fig:pace_analysis_intermediate}
\end{subfigure}

\begin{subfigure}{.5\textwidth}
  \centering
  \includegraphics[width=1\linewidth]{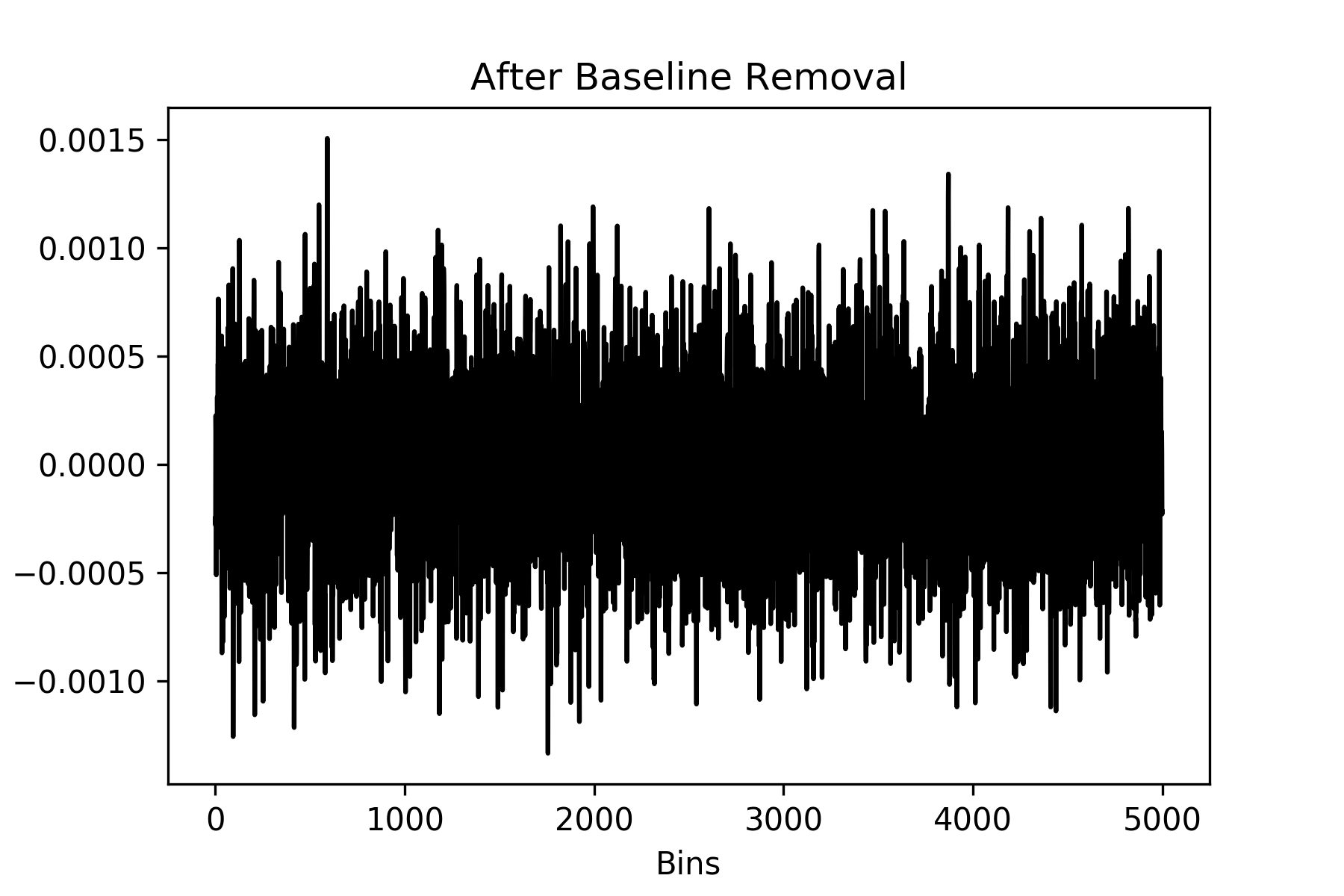}
  \caption{}
  \label{fig:pace_analysis_processed}
\end{subfigure}%
\begin{subfigure}{.5\textwidth}
  \centering
  \includegraphics[width=1\linewidth]{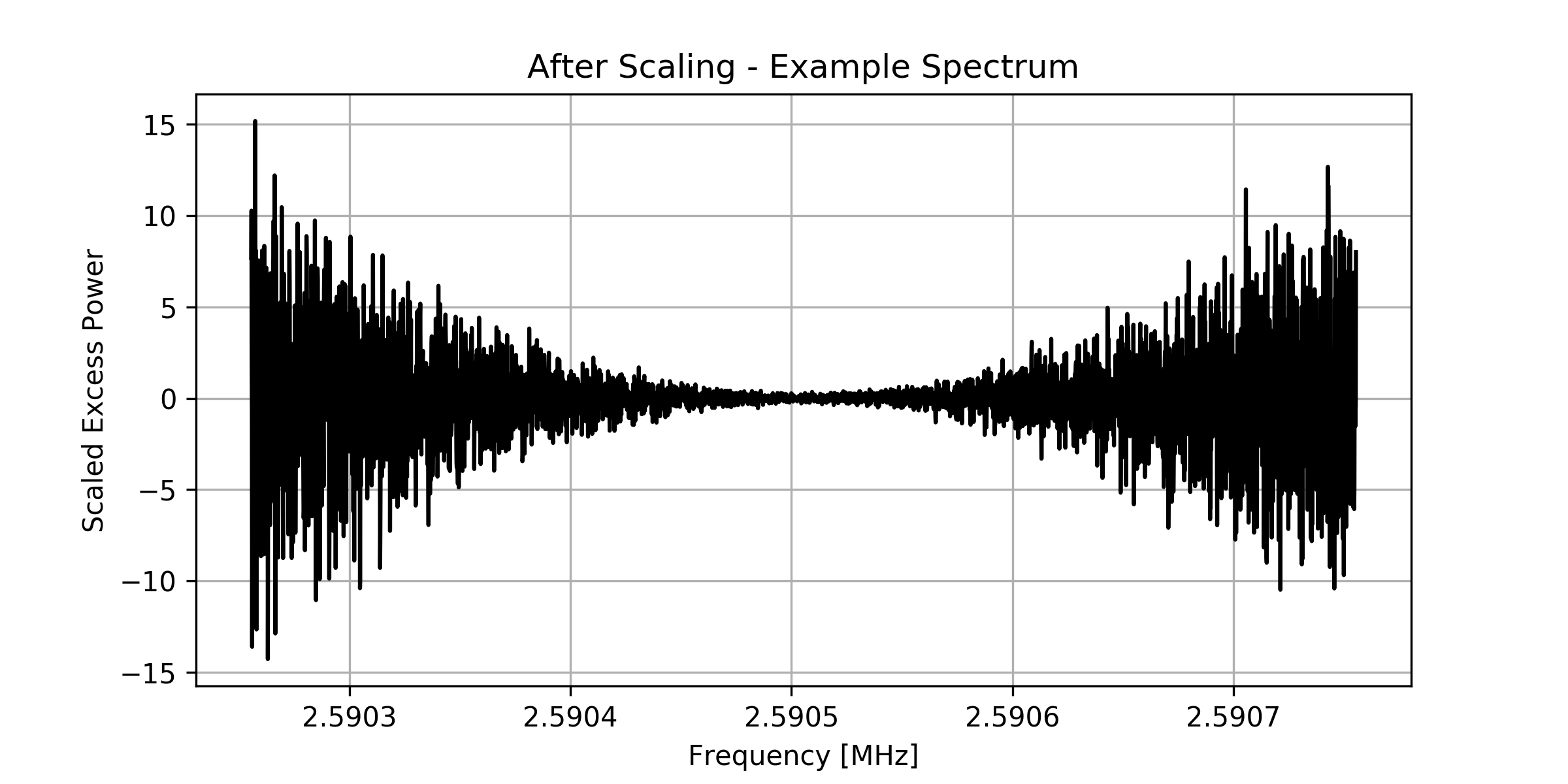}
  \caption{}
  \label{fig:pace_analysis_after_scaling}
\end{subfigure}
\caption{(a) An example raw spectrum in black and the bin-by-bin averaged
spectrum in red.  (b) After dividing the raw spectrum with the average. 
(c) After baseline removal. (d) Scaling to normalize bins to conversion power.}
\label{fig:pace_analysis_baseline_removal}
\end{figure}

Before proceeding for the combinations, we need to correct for the fact
that the signals appearing at a certain offset from the cavity center
frequency will be attenuated due to the obvious finite resonance width.
The single bin conversion power is given as:

\begin{align}
    \mathrm{P}_s(\beta, Q_L, \delta \nu, \Delta\nu_c) = A g_{\gamma} B_0^2 V \left(\frac{\beta}{1 + \beta} C \frac{Q_L}{1+(2\delta \nu/\Delta \nu_c)^2} \right)
\end{align}

where \(A\) simply encapsulates all the fixed parameters and physical
constants, \(g_{\gamma}\) is model dependent axion-photon coupling,
\(B_0\) is magnetic field, \(V\) is cavity volume, \(\beta\) is coupling
coefficient of the antenna to the cavity, \(C\) is the form factor of
the TM$_{010}$ mode, \(Q_L\) is the loaded quality factor, \(\delta \nu\) is
the offset from the resonant frequency of the cavity, and
\(\Delta \nu_c\) is the bandwidth of the cavity. The Last term is the usual
Lorentzian spectral shape for the resonance. It is obvious that this
power diminishes at the edges of each spectrum, thus scaling we do
simply corrects this factor and restores the noise power. Let the k'th
bin of the i'th normalized spectrum be denoted by \(\delta_i[k]\) and
let \(\nu_i[k]\) correspond to the frequency of the k'th bin in the i'th
spectrum. We create a scaled spectrum for each baseline removed spectrum
by~\cite{PRL_118_061302_2017}:

\begin{align}
    \delta_i^{(s)}[k] = 
        \frac{k_BT_{sys}\Delta f}
        {\mathrm{P}_S(\beta_i, Q_Li, \nu_{0i} - \nu_i[k], \Delta \nu_{ci})},
            \delta_i[k]
\end{align}

where \(\Delta f\) is bin frequency spacing, and \(k_B\) is Boltzmann
constant. Also, \(\beta_i\), \(Q_Li\), \(\nu_{0i}\) and
\(\Delta \nu_{ci}\) are the coupling coefficient, q-factor, resonant
frequency and bandwidth of the cavity. All parameters are measured when
the i'th spectrum is collected. See Fig.~\ref{fig:pace_analysis_processed}
and \ref{fig:pace_analysis_after_scaling} for before scaling and after
respectively.

\begin{figure}[h]
\centering
\begin{subfigure}{.5\textwidth}
  \centering
  \includegraphics[width=1\linewidth]{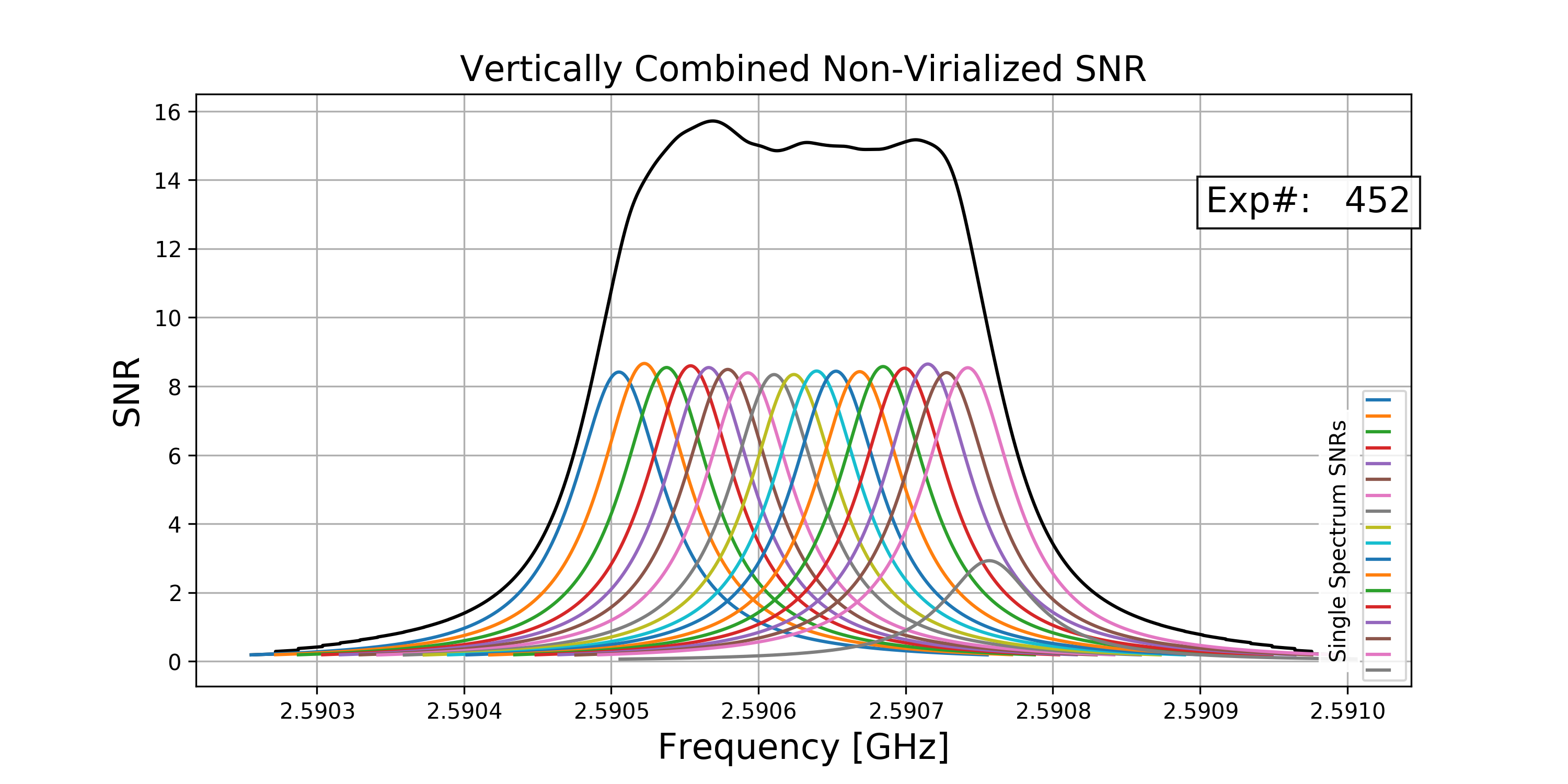}
  \caption{}
  \label{fig:pace_analysis_e452_snr_vc}
\end{subfigure}%
\begin{subfigure}{.5\textwidth}
  \centering
  \includegraphics[width=1\linewidth]{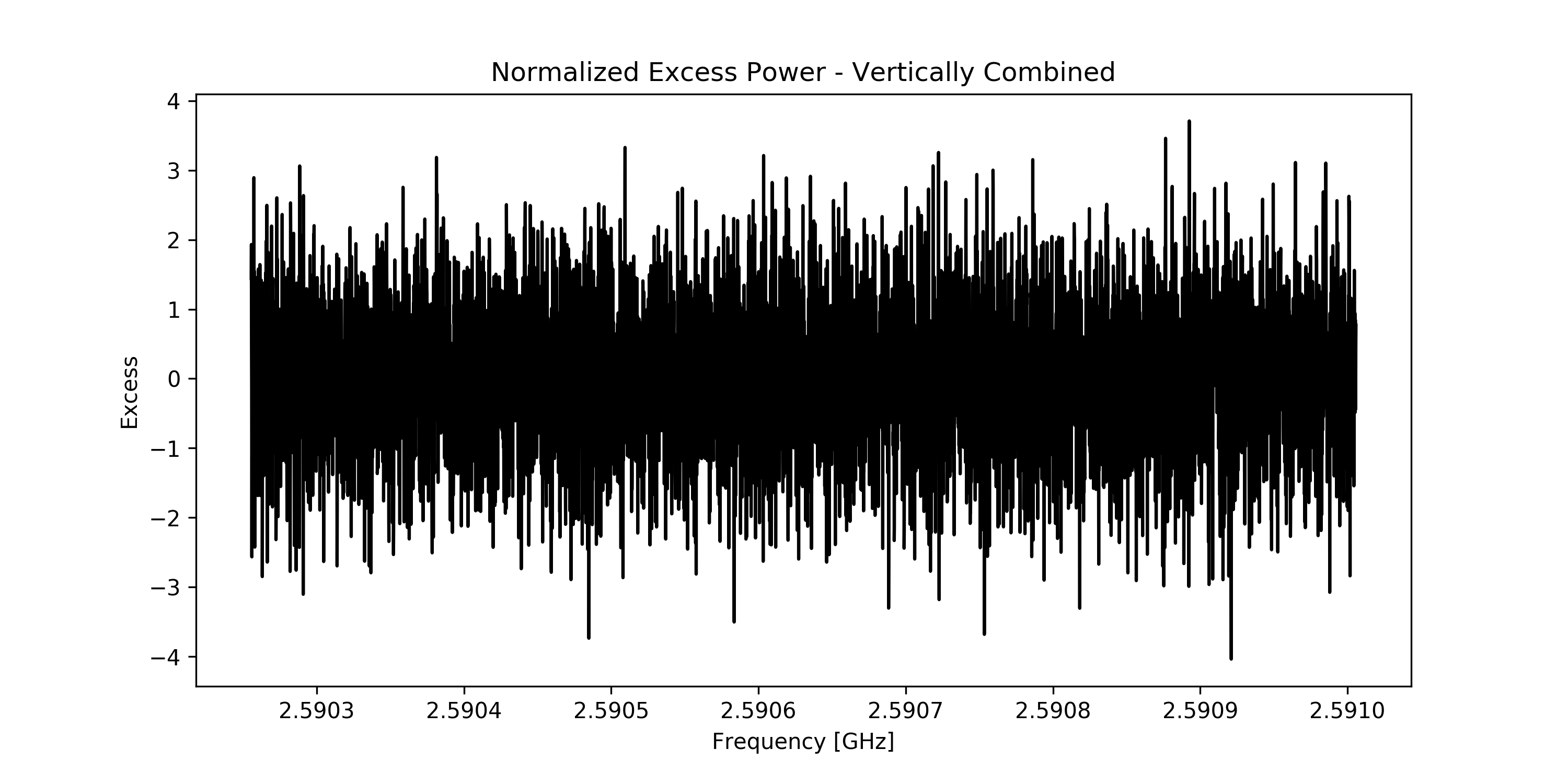}
  \caption{}
  \label{fig:pace_analysis_e452_excess_vc}
\end{subfigure}
\caption{(a) Non-virialized SNR for the vertically combined spectrum of E452.
(b) Normalized excess spectrum after vertical combination.}
\label{fig:pace_analysis_vertical_combination}
\end{figure}

Since we collect spectrum for every tuning step, we need a way to
combine these. The process at which we carefully combine overlapping
frequency bins from consecutive bins is called vertical combination. We
currently add the aligned powers with RMS which gives the maximum
likelihood in case of no correlations.  Fig.~\ref{fig:pace_analysis_e452_snr_vc} shows the resulting SNR and the excess spectrum after the vertical combination.

Our resolution bandwidth is approximately 20 times smaller than the
axion bandwidth at our frequencies, which is approximately
\(2\,\mathrm{kHz}\). Having a higher resolution allows us to tailor our
analysis for different dispersion models easily. We perform a horizontal
combination of consecutive bins in order to integrate this dispersed
signals. We divide this combination process into two steps following the
recipe from HAYSTAC~\cite{PRL_118_061302_2017} which builds upon the single step
done in ADMX~\cite{PRD_64_092003_2001}. The first step is referred to as
\emph{horizontal rebinning} and is basically a root-mean-square
aggregation of non-overlapping consecutive bins. For the current state of
our analysis, we only use 2 bins for rebinning. 

\begin{figure}[h]
\centering
\begin{subfigure}{.5\textwidth}
  \centering
  \includegraphics[width=1\linewidth]{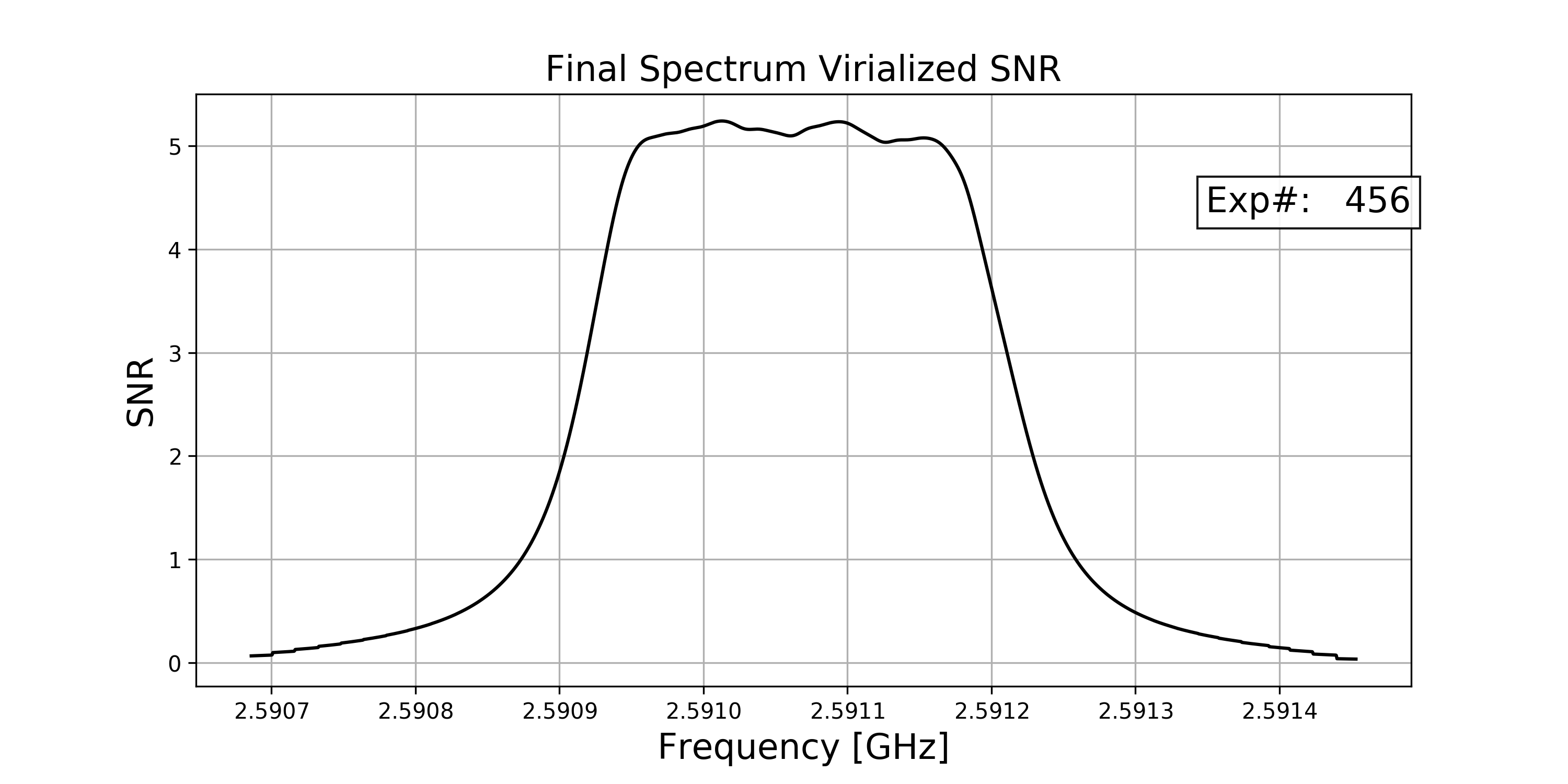}
  \caption{}
  \label{fig:pace_analysis_e456_snr}
\end{subfigure}%
\begin{subfigure}{.5\textwidth}
  \centering
  \includegraphics[width=1\linewidth]{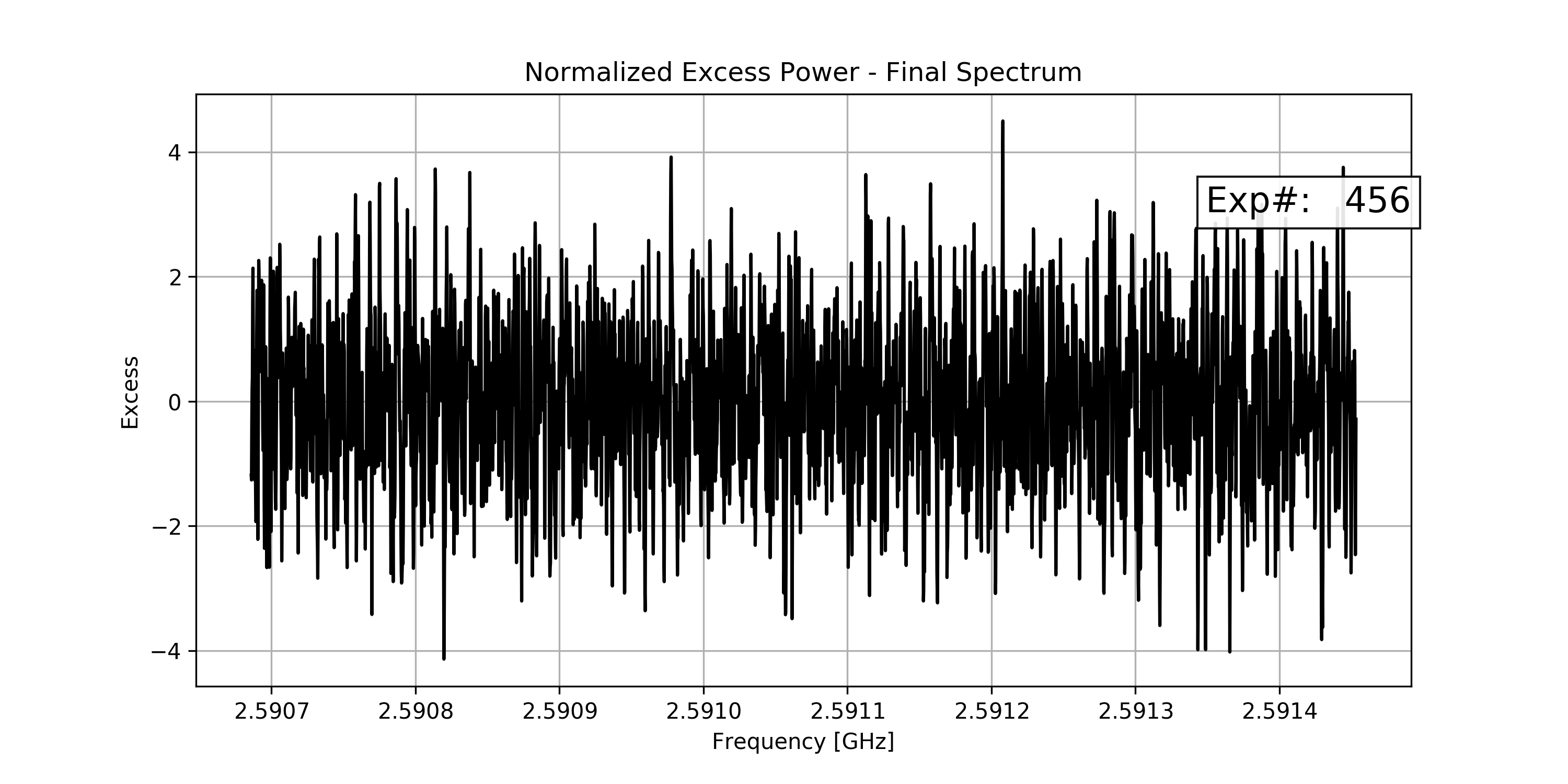}
  \caption{}
  \label{fig:pace_analysis_e456_excess}
\end{subfigure}
\caption{(a) Virialized SNR for the final spectrum. Note that this estimated
design parameter matches with our design goal. (b) Normalized excess for our
final spectrum}
\label{fig:pace_analysis_final_spectrum}
\end{figure}

The second step of horizontal combination incorporates the axion signal
dispersion, which we refer to as lineshape integration. Our analysis currently
uses the Maxwellian model for the velocity distribution given
as\textasciitilde{}:

\begin{align}
    f(\nu; \nu_a, \langle\beta^2\rangle) = \frac{2(\nu - \nu_a)}{\sqrt{\pi}}
    \left(\frac{3}{\nu_a \langle\beta^2\rangle}\right)^{3/2}
    \mathrm{exp}\left(-\frac{3(\nu-\nu_a)}{\nu_a\langle\beta^2\rangle}\right)
\end{align}

where \(\nu\) is the frequency variable, \(\nu_a\) is the axion
frequency,
\(\langle \beta^2 \rangle = \langle v^2 \rangle/c^2 \approx 9\times 10^{-9}\).
This lineshape is convolved with the horizontally rebinned spectrum with
6-bin overlaps which yielded the highest SNR of all reasonable bin
choices. The result of this operation gives us the final excess power
spectrum along with the virialized SNR distribution of the final
spectrum (see Fig.~\ref{fig:pace_analysis_e456_snr}).




In summary, we have successfully developed a stack of tools for processing our
data.  With the insights we gained in this process, it will be easier to
incorporate different analysis strategies. After completion of the horizontal
combination process, we are left with a final SNR spectrum along with normalized
excess power spectrum. Further processing that is flagging bins that exceed a
certain threshold, will be performed on this final excess power spectrum using
the final SNR spectrum to determine the threshold values per bin.  However,
before proceeding further, our analysis group is pursuing the correction of
certain systematics such as the attenuations due to the transfer function of the
spectrum analyzer , bin correlations introduced by the baseline removal filter.
Furthermore, we are working on incorporating Monte Carlo simulations as a
testing target for our analysis stack.

%% file: 1.3.3/main.tex
\subsection{Introduction}
\label{paragraph:capp8tb_introduction}
The CAPP-8TB is an experiment to search the axion dark matter in a mass range of 6.62$-$7.04\,$\mu$eV with a microwave resonant cavity under a magnetic field of 8\,T. In this experiment, the activity of axion dark matter search will be done in two stages: searching the axion dark matter at QCD axion sensitivity; reaching the experimental sensitivity near KSVZ model~\cite{PRL_43_103_1979,NP_B166_493_1980} predictions.

The experimental technique is based on P. Sikivie's method~\cite{PRL_51_1415_1983}, and to search unknown mass of axion dark matter, a mass range of 6.62$-$7.04\,$\mu$eV is scanned with a frequency tuning mechanism of the experiment. The performance of the experiment can be estimated by its scan rate which tells that how fast the experiment scans a target frequency range with a given experimental sensitivity. The scan rate ($df/dt$) for a given signal-to-noise ratio (SNR) is given by:
\begin{eqnarray}
\label{eq:capp8tb_scanrate}
\frac{df}{dt} & \propto & \frac{B^{4}V^{2}C^{2}_{010}Q_{L}}{T^{2}}
\end{eqnarray}
where $B$ is an external magnetic field, $V$, $C_{010}$, and $Q_{L}$ are a volume, form factor, and loaded quality factor of a microwave resonant cavity, respectively, and $T$ is a system noise temperature. Therefore, the critical components are following: a powerful refrigerator is necessary to minimize the cavity temperature included in the system noise temperature, and it is described in Section~\ref{paragraph:capp8tb_refrigerator}; a strong magnet to provide an external magnetic field is described in Section~\ref{paragraph:capp8tb_magnet}; a tunable microwave resonant cavity system is described in Section~\ref{paragraph:capp8tb_cavity}; radio-frequency (RF) receiver chain is described in Section~\ref{paragraph:capp8tb_rfchain}; data acquisition, system control and monitoring system are described in Section~\ref{paragraph:capp8tb_daq}; timeline and cost estimations are described in Section~\ref{paragraph:capp8tb_timeline}; prospects of the experiment are described in Section~\ref{paragraph:capp8tb_prospects}.

\subsection{Refrigerator}
\label{paragraph:capp8tb_refrigerator}
From the Eq.~(\ref{eq:capp8tb_scanrate}), the scan rate of the experiment is inversely proportional to the system noise temperature squared, in other words, a lower system noise temperature gives a faster scan rate, therefore, the performance of the experiments can be increased by reducing the system noise temperature.

The system noise power is contributed by a thermal noise of the resonant cavity and an RF noise temperature from the components in the RF receiver chain. Since the thermal noise of the cavity is proportional to the physical temperature of the cavity, a colder resonant cavity is better for the experimental sensitivity. To make the cavity cold, we employ BlueFors dilution refrigerator, BF\-LD400~\cite{web:BlueFors_LD}, which is able to maintain the temperature of the coldest stage to be less than 10\,mK without any loads. The dilution refrigerator consists of several temperature stages, and its cooling power is summarized in Table~\ref{table:capp8tb_refrigerator}.

\begin{table}[h]
\centering
\begin{tabular}{|c|c|}
\hline
Base temperature & 8\,mK \\
Cooling power @ 20\,mK & 15\,$\mu$W \\
Cooling power @ 100\,mK & 450\,$\mu$W \\
Cooling power @ 120\,mK & 650\,$\mu$W \\
Cool-down time to base & 22 hours \\
\hline
\end{tabular}
\caption{\label{table:capp8tb_refrigerator}Base temperature, cooling power, and cool-down time to base of BlueFors BF-LD400~\cite{web:BlueFors_LD}.}
\end{table}

\subsection{Magnet}
\label{paragraph:capp8tb_magnet}
\begin{figure}[h]
\begin{center}
\includegraphics[width=.6\textwidth]{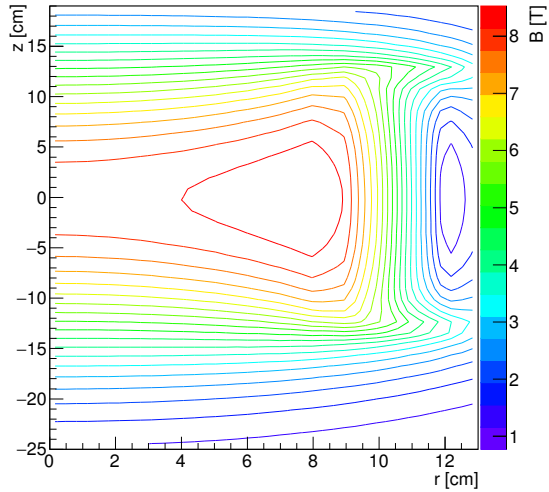}
\caption{Magnetic field map. Horizontal is radial ($r$) direction and vertical is axial ($z$).}
\label{fig:capp8tb_magnet_fieldmap}
\end{center}
\end{figure}

As shown in Eq.~(\ref{eq:capp8tb_scanrate}), the magnetic field is the most effective parameter to increase the sensitivity. We use a superconducting magnet made by American Magnetics Inc.~\cite{web:AMI}, which is capable up to 8\,T at a current of 96.56\,A. The inner diameter of clear bore of the magnet is 165\,mm, and this limits the size of the resonant cavity since it is placed inside the clear bore. Detailed specifications of the magnet is summarized in Table~\ref{table:capp8tb_magnet}, and the magnetic field map is shown in Figure~\ref{fig:capp8tb_magnet_fieldmap}.

\begin{table}[h]
\centering
\begin{tabular}{|c|c|}
\hline
Type & Cryogen-free compensated solenoid \\
\hline
Rated central field at 4.2\,K & 8\,T \\
\hline
Rated operating current & 96.56\,A \\
\hline
Field to current ratio & 0.0828\,T/A \\
\hline
Homogeneity & $\pm$ 0.1\,\% \\
\hline
Inductance & 53\,H \\
\hline
Diameter of clear bore & 165\,mm \\
\hline
Total magnet resistance & 1423\,$\Omega$ \\
\hline
Overall length & 480\,mm \\
\hline
Maximum outside diameter & 325\,mm \\
\hline
Weight & 52.7\,kg \\
\hline
\end{tabular}
\caption{\label{table:capp8tb_magnet}Detailed specifications of the magnet.}
\end{table}

\subsection{Resonant cavity and tuning system}
\label{paragraph:capp8tb_cavity}
\begin{figure}[h]
\begin{center}
\includegraphics[width=.45\textwidth]{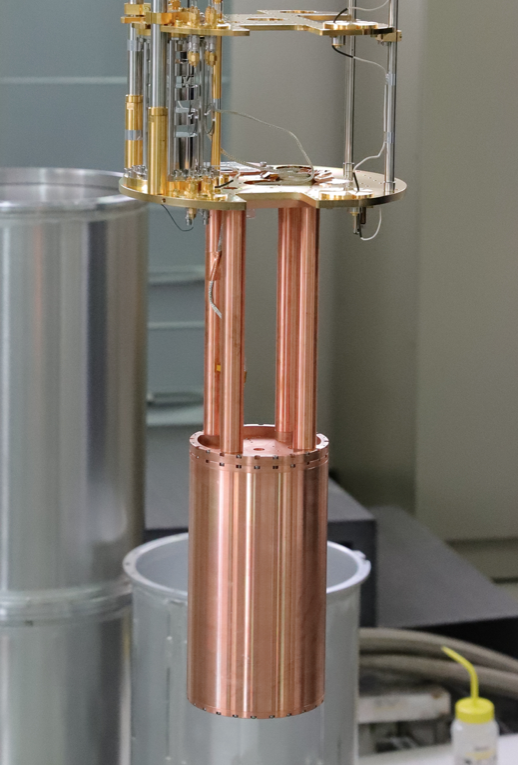}
\caption{Microwave resonant cavity mounted on the refrigerator.}
\label{fig:capp8tb_cavity}
\end{center}
\end{figure}

To amplify signal from an axion-to-photon conversion, a cylindrical microwave resonant cavity is used as a detector. As mentioned in Section~\ref{paragraph:capp8tb_magnet}, the cavity is located in the clear bore of the magnet, and the dimension of the clear bore determines the size of the cavity. The inner diameter and height of the cavity are 134\,mm and 236\,mm, respectively, and it fits to the clear bore to maximize the volume of the cavity. The cavity consists of two end-caps and a barrel to form a close cylindrical shape, and the parts are all made of pure copper as shown in Figure~\ref{fig:capp8tb_cavity}.

The cavity is mounted by hanging on the coldest stage in the refrigerator with four polls of supporting structure which are made of pure copper as well. This supporting structure maintains that cavity center is at the center of the magnet, therefore, it takes full advantage of the magnetic field. There is gradient of the magnetic field in side the cavity volume, and the average magnetic field inside the cavity volume is about 7.3\,T.

\begin{figure}[h]
\begin{center}
\includegraphics[width=.7\textwidth]{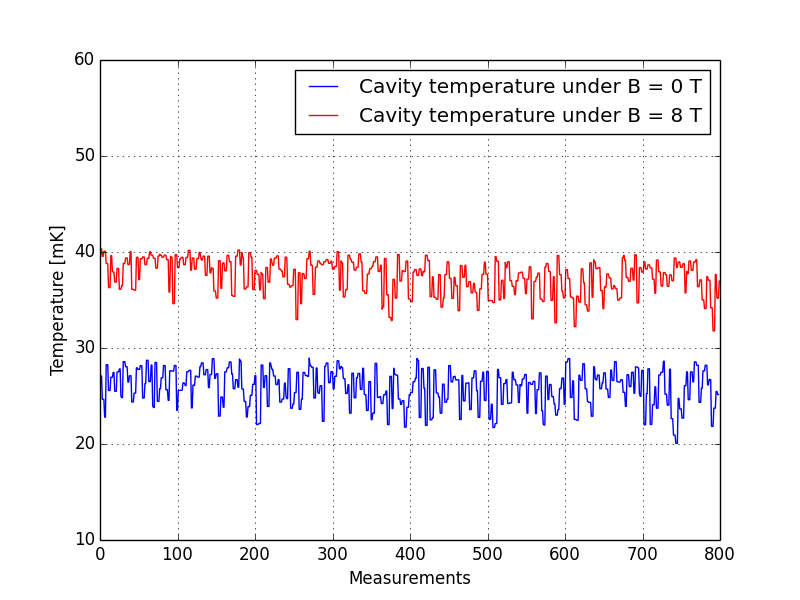}
\caption{Measured temperatures of the cavity without (blue) and with (red) an external magnetic field.}
\label{fig:capp8tb_temperature}
\end{center}
\end{figure}

The cavity temperature is measured as around 25\,mK without any external magnetic fields. Under a magnetic field of 8\,T, its temperature goes up due to the presence of the Eddy current with a small vibration, and the temperature is measured as about 40\,mK as shown in Figure~\ref{fig:capp8tb_temperature}.

\begin{figure}[h]
\begin{center}
\includegraphics[width=.9\textwidth]{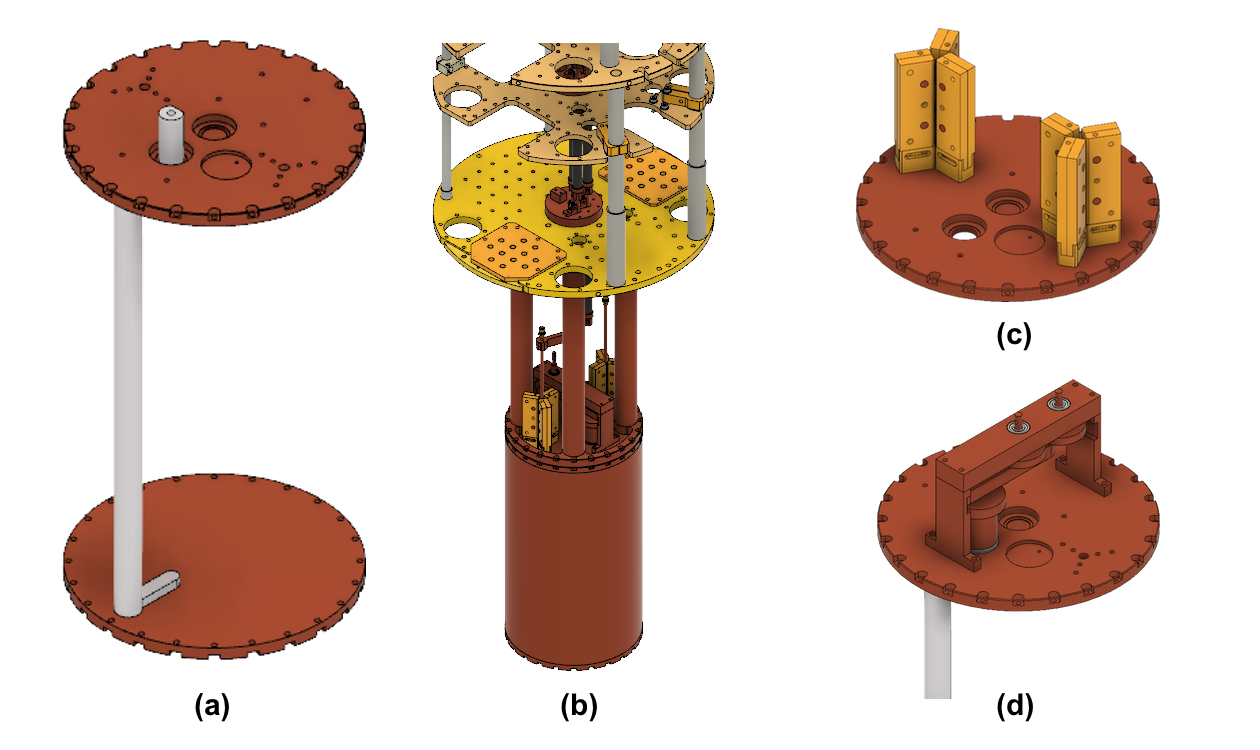}
\caption{Frequency and coupling tuning structure. (a) Alumina tuning rod inside the cavity. Cavity wall is not shown for an illustration. (b) Full configuration of the tuning system with the resonant cavity mounted on the lowest temperature stage. (c) Antenna guides for the coupling tuning. (d) Locomotive frequency tuning structure.}
\label{fig:capp8tb_tuning}
\end{center}
\end{figure}

The resonant frequency of the cavity is tuned by a rotating tuning rod inside the cavity volume as shown in Figure~\ref{fig:capp8tb_tuning} (a). In addition, the antennas are controlled to tune the coupling with microwave signals as well. Full configuration of the frequency and antenna tuning system are depicted in Figure~\ref{fig:capp8tb_tuning} (b). Stepping motors are used to drive the tuning system, and they are mounted outside the refrigerator at room temperature. 

There are two antennas, one is for detecting the resonant microwave signal from the cavity, and the other is for injecting microwave signals to measure the properties of the cavity. The former one is directly related to the strength of detected signals, and determines the loaded quality factor shown in Eq.~\ref{eq:capp8tb_scanrate}, for example, a critical coupling ($Q_{L}=Q_{0}/2$ where $Q_{0}$ is a unloaded quality factor) can be found by tuning the antenna properly. The coupling coefficient is $\beta=Q_{0}/Q_{L}-1$, thus, $\beta=1$ at the critical coupling. In the experiment, we use $\beta=2$ by tuning the antenna. Another antenna in the system is weakly coupled to the cavity not to disturb the microwave inside the cavity. Those antennas are guided by a structure shown in Figure~\ref{fig:capp8tb_tuning} (c). There are cryogenic bearings in the structure, and they keep the antennas stable.

Since there is an offset between the axis of the driving force from the stepping motor which is at the center of the refrigerator and the axis of the tuning rod, the driving force should be properly transferred. We employ a locomotive design for this purpose as shown in Figure~\ref{fig:capp8tb_tuning} (d). To minimize the frictional heats during the tuning, cryogenic bearings are included in the structure.

\begin{figure}[h]
\begin{center}
\includegraphics[width=.7\textwidth]{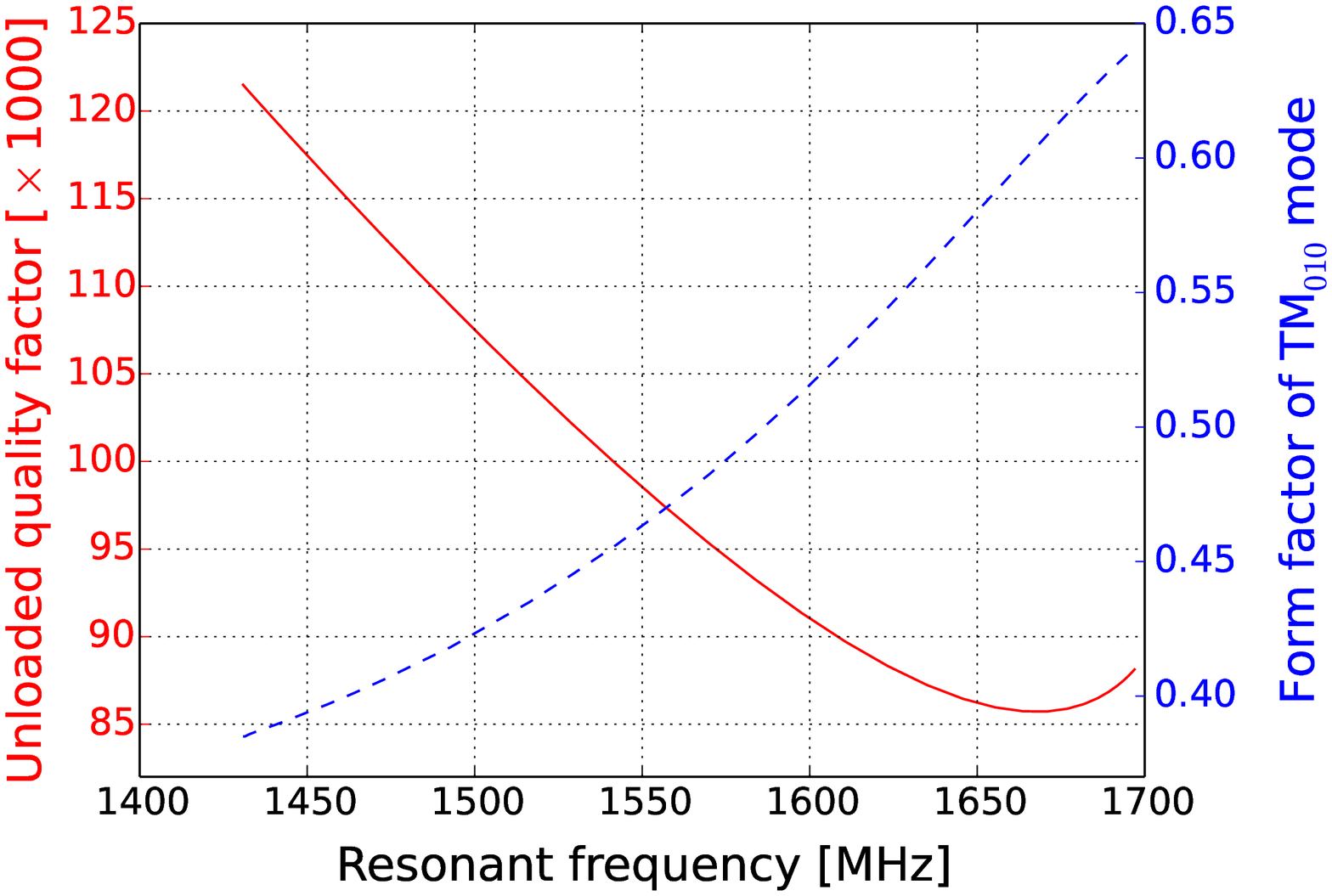}
\caption{Unloaded quality factor (red) and form factor (blue) of TM$_{010}$ mode of the cavity as a function of resonant frequency of the cavity.}
\label{fig:capp8tb_quality_factor}
\end{center}
\end{figure}

The tuning rod changes the geometry inside the cavity, therefore, it distorts the electromagnetic fields and changes the resonant frequency, quality factor, and form factor of the cavity. We use a dielectric tuning rod which is made of pure alumina. Figure~\ref{fig:capp8tb_quality_factor} shows the changes of the unloaded quality factor and form factor of TM$_{010}$ mode of the cavity as a function of resonant frequency.

The stepping motor for the frequency tuning is capable to move by $2.88\times10^{-5}$ degree, therefore, it is roughly able to move every 24\,Hz. Since the fluctuation of the resonant frequency is larger by an order of magnitude, the frequency tuning system gives an enough resolution for the experiment.

\begin{figure}[h]
\begin{center}
\includegraphics[width=.75\textwidth]{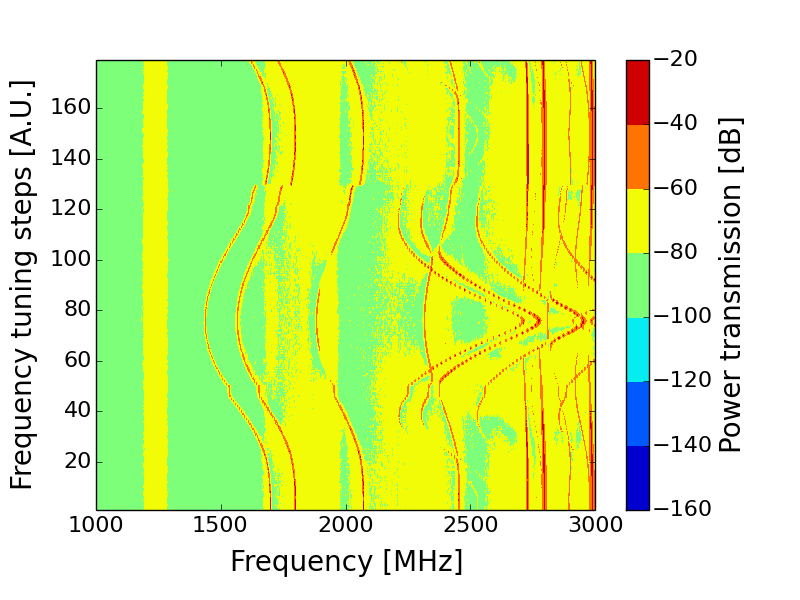}
\caption{Frequency mode map of the cavity. The leftmost curve which touches 1400\,MHz is TM$_{010}$ mode, and later curves are higher modes.}
\label{fig:capp8tb_frequency_map}
\end{center}
\end{figure}

One important issue in the frequency tuning system is so-called ``mode-crossing'' that a target mode to measure (TM$_{010}$ in this case) is overlapped by another mode such as TE modes at a certain frequency. As those modes are close to each other, their resonant shape in a frequency domain is distorted, therefore, it is not possible to measure the parameters of the cavity precisely. This makes a blinded region to search axion dark matter, and we lose sensitivity in this region. The cavity used in this experiment does not have such a mode-crossing as shown in Figure~\ref{fig:capp8tb_frequency_map}, therefore, there is no loss of sensitivity in the designed frequency region.
We also measure the temperature of the alumina tuning rod, and it is measured to be less than 150\,mK.

\subsection{RF receiver chain}
\label{paragraph:capp8tb_rfchain}
\begin{figure}[h]
\begin{center}
\includegraphics[width=.95\textwidth]{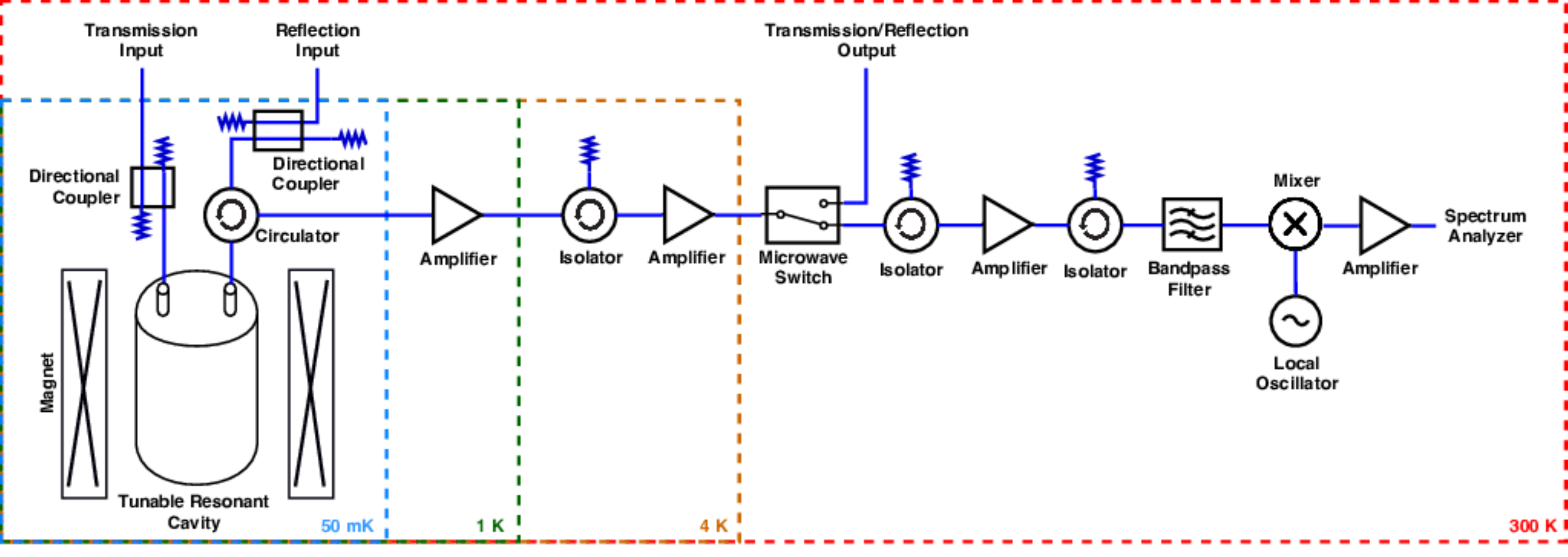}
\caption{RF receiver chain of the experiment. Different temperature stages are also shown. The magnet is thermally linked to the 4\,K stage.}
\label{fig:capp8tb_rfchain}
\end{center}
\end{figure}

Signals picked up by the resonant cavity system described in previous section are transmitted and processed through a radio-frequency receiver chain shown in Fig.~\ref{fig:capp8tb_rfchain}.

Since the axion signal is expected extremely small ($\mathcal{O}(10^{-22}\,\textrm{W})$ at the upper QCD axion band), the signal has to be amplified to visible level. For this purpose, we employ four radio-frequency amplifiers: two cryogenic low noise amplifiers based on HEMT (high-electron-mobility-transistor) technology, and two room temperature amplifiers.

To block the reflection from upstream, we add isolators and circulators in front of the amplifiers. The amplified signal up to the third amplifier is down-converted to a lower frequency by mixer to make the signal easy to handle. Finally, the signal is taken by a spectrum analyzer, and it transforms time-domain data into frequency-domain. 

In addition, a network analyzer is used to measure the properties of the cavity such as a resonant frequency, a quality factor, and a coupling coefficient of an antenna. To isolate the cavity from this auxiliary stream, we put two directional couplers in transmission and reflection input lines from the network analyzer. The transmission and reflection outputs are amplified by the cryogenic amplifiers, and are picked up by the network analyzer. The paths to network analyzer and spectrum analyzer are altered by radio-frequency switches at room temperature.

The system noise temperature does not affect to the signal power, however, the scan rate is inversely proportional to the system noise temperature squared as shown in Eq.~\ref{eq:capp8tb_scanrate}, and the system noise temperature consists of the physical temperature of the cavity as mentioned in Section~\ref{paragraph:capp8tb_cavity} and noise temperature by radio-frequency components in the receiver chain. In other words, the noise power is contributed by a thermal noise from the cavity itself, and the Johnson-Nyquist noise from the RF components in the receiver chain of the experiment. Therefore, a lower noise temperature of components in the chain is preferred to maximize the overall performance and sensitivity of the experiment. 

\begin{figure}[h]
\begin{center}
\includegraphics[width=.7\textwidth]{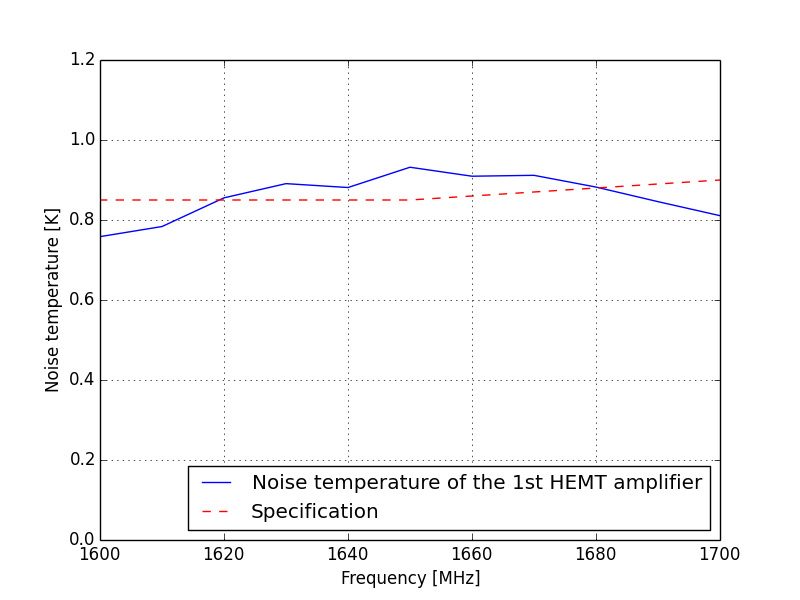}
\caption{Measured noise temperature temperature (blue solid) of the first HEMT amplifier in the RF chain. Specification of the amplifier is shown together (red dashed).}
\label{fig:capp8tb_noise_temperature}
\end{center}
\end{figure}

In a cascade RF chain, it is well-known that the first amplifier is the most significant source of the noise contribution, and the noise contributions from later components are negligible if the gain of the first amplifier is large enough. Therefore, precise measurement of the noise temperature of the first amplifier is essential to understand the RF receiver chain. To take the advantage of which the noise contribution from the first amplifier is dominant, we employ a high electron mobility transistor (HEMT) based amplifier that gives a low noise temperature. From the specification, the noise temperature of the amplifier is less than 1\,K in the frequency range from 1.4\,GHz to 1.7\,GHz, and its gain is about 33\,dB in the same frequency range. We measure the noise temperature of the first amplifier in various configurations for cross checks, and Figure~\ref{fig:capp8tb_noise_temperature} shows the agreements of the measurements, and they are measured to be below 1\,K which is consistent with the specification.

Considering the physical temperatures of the resonant cavity and tuning rod, and the noise temperature of the amplifier, the total system noise temperature is expected below 1.5\,K. And the total gain of the system is expected around 140\,dB.

\subsection{Data acquisition system}
\label{paragraph:capp8tb_daq}
To take the power spectra from the chain, we use a commercial spectrum analyzer, Rhode \& Schwarz FSV 4~\cite{web:rhode_schwarz_fsv4}, and a network analyzer, Rhode \& Schwarz ZND~\cite{web:rhode_schwarz_znd}, is used to take auxiliary data. They are connected to a data acquisition computer via GPIB and Ethernet protocols. As a part of auxiliary data, we also record temperatures and magnetic fields by using temperature controllers, LakeShore Model 372~\cite{web:lakeshore_model372}, and a magnet controller, American Magnetics Inc. Model 430~\cite{web:ami_model430}. Those are also connected to the data acquisition computer via GPIB and RS-232 protocols. All those devices connected to the computer are remotely controlled and monitored by a data acquisition software, CULDAQ, and more details of the software is discussed in Section~\ref{sec:control_system}.

\subsection{Timeline and cost estimations}
\label{paragraph:capp8tb_timeline}
The team for the experiment has been established on January 2017, and the manpower is summarized in Table~\ref{table:capp8tb_manpower}. 
The team has been developing the experiment. The development will be done in the end of 2018, and commissioning and physics runs will start after the development. About three months of data taking will be done. After the data taking, the experiment will be upgraded with a SQUID (superconducting quantum interference device) based amplifier, which will significantly reduce the noise temperature of the receiver chain. We expect that the upgrade will be done in the end of 2019, and the commissioning and physics runs will start in the beginning of 2020.

\begin{table}[h]
\centering
\begin{tabular}{c|c|c|c|c|c}
\hline
Name & Institute & Position & FTE, 2017 & FTE, 2018 & FTE, 2019 \\
\hline
Saebyeok Ahn & KAIST & Graduate student & 30\% & 75\% & 75\% \\
Jihoon Choi & IBS/CAPP & Research Fellow & 50\% & 40\% & 45\% \\
Byeong Rok Ko & IBS/CAPP & Research Fellow & 60\% & 50\% & 40\% \\
Soohyung Lee & IBS/CAPP & Research Fellow & 80\% & 90\% & 50\% \\
\hline
\end{tabular}
\caption{\label{table:capp8tb_manpower}Manpower of the experiment (alphabetical order).}
\end{table}

\subsection{Prospects}
\label{paragraph:capp8tb_prospects}
\begin{figure}[h]
\begin{center}
\includegraphics[width=1.0\textwidth]{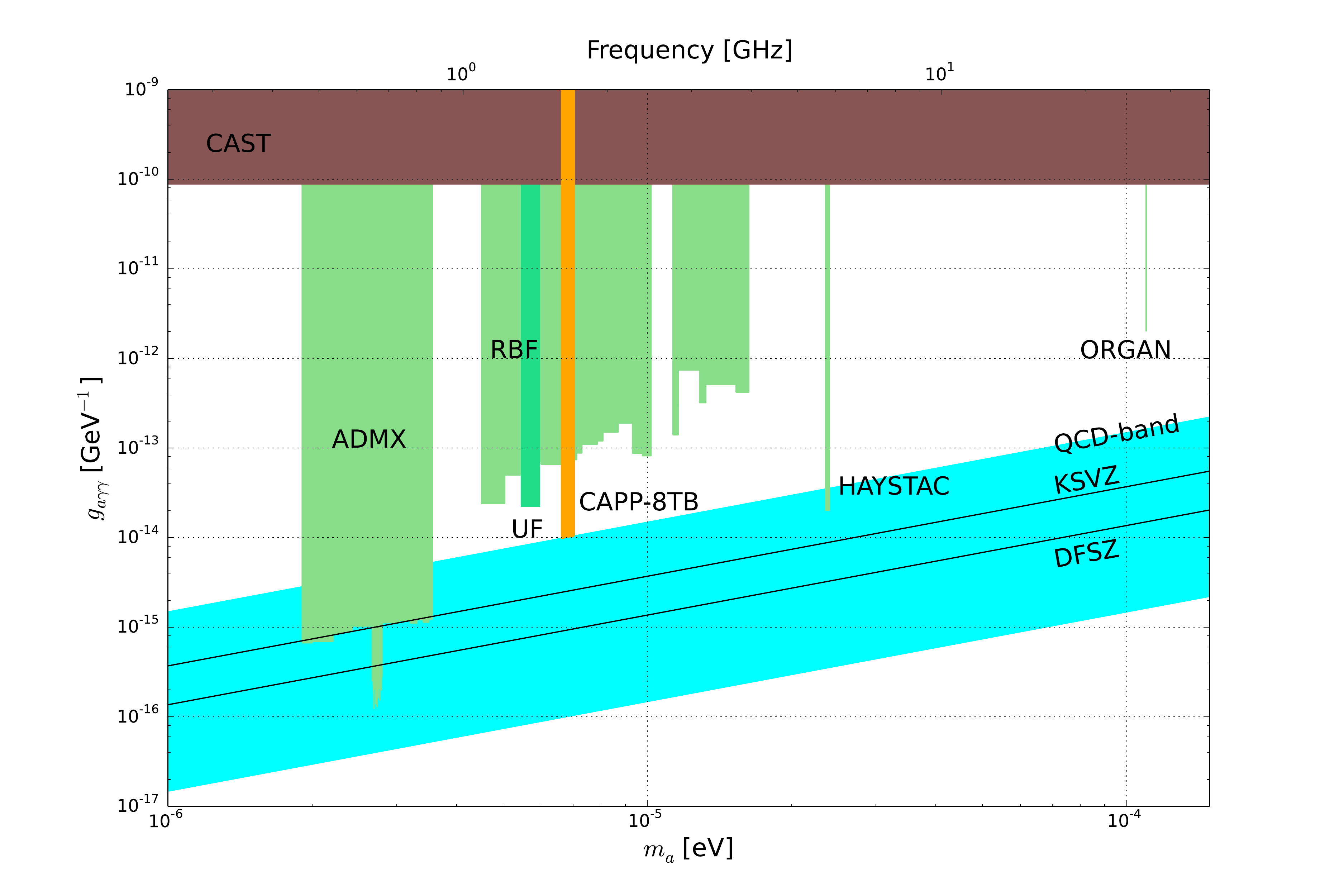}
\caption{Experimental sensitivity of axion dark matter search with results from experiments done~\cite{PRD_64_092003_2001,AJL_571_L27_2002,PRL_104_041301_2010,PRL_120_151301_2018,PRD_40_3153_2001,PRD_42_1297_1990,PRL_118_061302_2017,PDU_18_67_2017} and projection of the first stage of the CAPP-8TB experiment. QCD axion band~\cite{PRD_52_3132_1995} (light-blue) and model predictions of KSVZ~\cite{PRL_43_103_1979,NP_B166_493_1980} and DFSZ~\cite{PL_104B_199_1981,SJNP_31_260_1980} are shown toghether.}
\label{fig:capp8tb_prospects}
\end{center}
\end{figure}

In the first stage of the experiment with HEMT based amplifiers, we will reach the sensitivity of QCD axion band as shown in Figure~\ref{fig:capp8tb_prospects}. In the current experimental configuration described earlier with a data acquisition efficiency of $\sim50$\,\%, this can be done in three months including rescan of possible problematic frequency regions due to technical issues.

In the second stage of the experiment, which is an upgrade with a SQUID amplifier, we will search the axion dark matter up to the sensitivity near KSVZ. We expect a half year of data taking to reach this sensitivity.

%% file: 1.3.4/main.tex
\subsection{Introduction}
\begin{figure}[h]
\centering
\includegraphics[width=1.\textwidth]{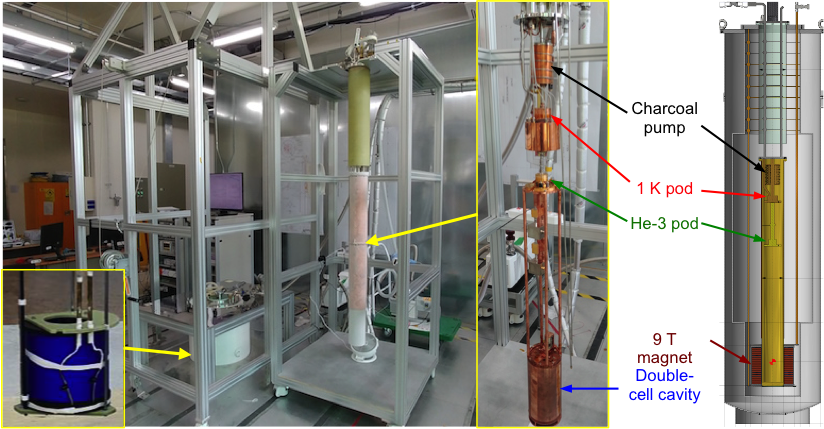}
\caption{Major equipment for the CAPP-9T MC experiment: Janis He-3 cyrogenics system, 9\,T SC magnet, and multiple-cell cavities.
The system is installed at one of the low vibration pad in the CAPP experimental area.}
\label{fig:janis_system}
\end{figure}

Three years of the Young Scientist (YS) program at CAPP have developed reliable phase-matching mechanisms for various types of multiple-cavity system, as discussed in detail in Sec.~\ref{sec:pizza_cavity}. 
Experimental demonstrations of their feasibility convince us that these approaches are promising for axion dark matter search in higher mass regions. 
In particular, the multiple-cell (pizza) cavity design turns out to be advantageous in many aspects: it provides larger detection volume, simpler experimental setup and easier phase-matching mechanism than the conventional array of multiple cavities~\cite{bib:pizza_cavity}. 
The major research effort for the remaining term of the YS program focuses on setup and operation of an axion search experiment exploiting the pizza cavity design to probe high and wide ranges of axion mass. 
The experiment, named CAPP-9T MC, where MC is an abbreviation of multiple-cell cavity, utilizes a He-3 cryogenic system currently available at the center. 
This system is equipped with a 9\,T superconducting (SC) solenoid magnet with a 5” diameter clear bore. 
The cryogenic system and the SC magnet, comprising the major equipment of the experiment, are shown in Fig~\ref{fig:janis_system}.

\subsection{Janis He-3 refrigerator}
A standard model HE-3-SSV cryostat, supplied by Janis Inc., was customized for axion research at IBS/CAPP. The major components of the He-3 insert, including the charcoal sorption pump, 1\,K pot and He-3 pot, are located inside the inner vacuum can (IVC). The charcoal sorption pump is adopted for multiple applications - depending on its temperature, it releases the He-3 gas for condensation; reduces the saturated vapor pressure of the condensed liquid He-3 to cool down the system; and holds all the evaporated He-3 gas.  The 1\,K pot is employed to condense the He-3 gas by pumping on liquid He-4 filling in the pot using an external pump. The He-3 pot is made of gold plated OFHC copper and it is located at the bottom of the insert. The charcoal pump evaporates the condensed He-3 liquid to cool the He-3 plate down to sub-K level. Using 15 NTP liter of He-3 gas contained in the system, the base temperature is maintained at 300\,mK for $> 150$ hours with free load. The stainless steel IVC fits the 5.0 clear bore of the SC magnet. 

\subsection{9\,T superconducting magnet}
\begin{figure}[b]
\centering
\includegraphics[width=1.0\textwidth]{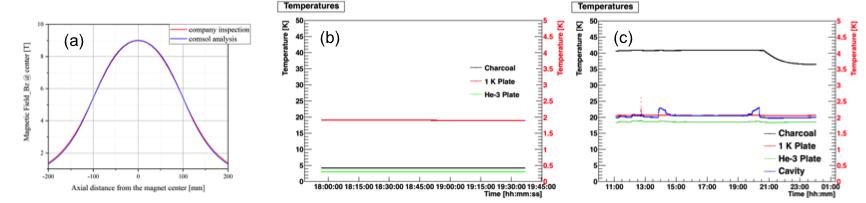}
\caption{(a) Magnetic field distribution along the z axis at the center of the magnet. 
(b) and (C) Operational temperatures of the major components of the He-3 system: charcoal pump (black), 1\,K pod (red), He-3 pod (green) and cavity (blue).}
\label{fig:janis_test_operation}
\end{figure}

A custom-made solenoid magnet with 5.0" diameter clear bore, manufactured by Cryomagnetics Inc., was designed to meet the requirements of the Janis system in accordance with our application. The SC magnet was fabricated using NbTi wire to generate the central magnetic field of 9\,T at a liquid helium (LHe) reservoir. A copper matrix is introduced in the wire, which acts as a form of quench protection along with diodes. A capability of operation in persistent mode, once a current is placed, eliminates power consumption to maintain the current. This feature reduces the heat load on the 4\,K cold mass and provides the stability of the magnetic field. The magnet design was simulated based on its mechanical structure and the field distribution is verified to be consistent with the company specification (see Fig.~\ref{fig:janis_test_operation} (a)). The maximum magnetic field measured at the center of the magnet at 4.2\,K read 9.0\,T at 81.0\,A, which is also consistent with both the spec and simulation. The magnet is placed on the bottom of the Janis cryostat, 29" apart from the He-3 plate, in order to minimize the magnetic field effect on the RF devices.

\subsection{Cavity and tuning system}
Multiple-cell cavities with 110\,mm inner diameter and 220\,mm inner height are designed to fit into the IVC. The dimension of cavities is determined through a simulation study to maximize the sensitivity taking the magnetic field distribution into account. The averaged field strength inside the cavities is estimated to be 7.8\,T. A new tuning mechanism depicted in Sec.~\ref{sec:pizza_cavity} is employed such that a single piezoelectric actuator placed under the cavity rotates frequency-tuning rods, joined together by a holder, simultaneously. A linear actuator is installed on top of the cavity to position a single pickup antenna at the center of the cavity. A typical loaded quality factor of the cavities with the tuning rods inside is measured to be about 20,000 at 4\,K. Degradation of about 30\% is assumed to be due to non-lossless dielectric tuning rods. Depending on the cell multiplicity, the frequency range covered by a series of multiple-cell cavities is up to 7\,GHz with the corresponding mass up to about 30\,$\mu$eV. 

\subsection{Test operation}
The simplicity and stability of the cryogenic system was tested at the LHe temperature (4.2 K) with the SC magnet running in persistent mode. 
Manipulation of the charcoal and 1\,K pot temperatures enables us to maintain the HE-3 plate at 270 mK for about 200 hours (shown in Fig.~\ref{fig:janis_test_operation} (b)), which fulfills the company specs. 
The cryogen (LHe) consumption was estimated to be about 20\,L/day. 
Test operation satisfying the desired specifications confirms that the system is well understood. 
The cavity is suspended from the He-3 plate to the middle of the magnet bore via four long copper rods as seen in Fig.~\ref{fig:janis_test_operation}. 
The entire load with major RF components weights about 10\,kg. 
With such a load, the He-3 system has a limited capability of maintaining the cavity temperature about 600\,mK for less than 2 hours. As this condition is not applicable to axion search business, we determine to operate the system in continuous mode with the He-3 gas in condensed state. 
By maintaining the 1\,K pot temperature as low as possible, we are able to achieve the He-pot and cavity temperatures of 1.8 and 2.0\,K, respectively (see Fig.~\ref{fig:janis_test_operation} (c)). 
We also observe the Eddy current effect on cavity temperature while ramping the magnet up and down. 

\begin{table}[b]
\begin{center}
\begin{tabular*}{0.85\textwidth}{@{\extracolsep{\fill}}c|ccc}
\hline
Cell multiplicity &2 & 4 & 8 \\
\hline
\raisebox{3\height}{Geometry}
& \raisebox{-.07\height}{\includegraphics[width=0.18\textwidth]{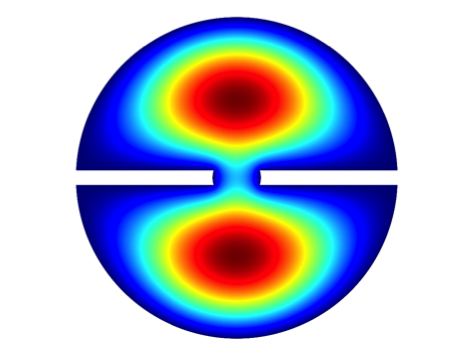}} 
& \raisebox{-.07\height}{\includegraphics[width=0.18\textwidth]{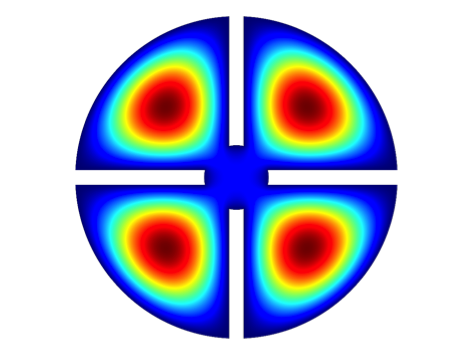}}
& \raisebox{-.07\height}{\includegraphics[width=0.18\textwidth]{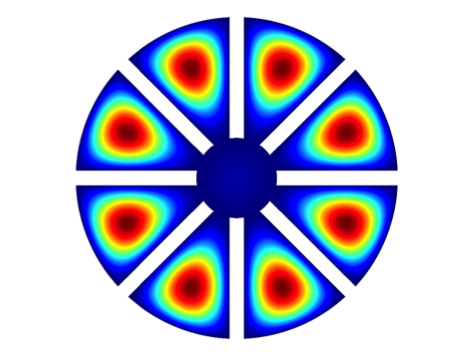}} \\
\hline
$f_{010}$ [GHz] & [2.8, 3.3] & [3.8, 4,5] & [5.7,7.0] \\
\hline
$Q_0$ & 60,000 & 51,000 & 51,000 \\
\hline
$C_{0101}$ & 0.45 & 0.45 & 0.40 \\
\hline
$B_{\rm avg}$ [T] & \multicolumn{3}{c}{7.8} \\
\hline
$V$ [L] & 2.0 & 1.9 & 1.7 \\
\hline
$P_{\rm sig}$ [10$^{-21}\,$W] & 0.51 & 0.56 & 0.68 \\
\hline
$T_{\rm sys}$ [K] & 2.1 + 2.0 & 2.1 + 3.0 & 2.1 + 4.0 \\
\hline
SNR & \multicolumn{3}{c}{5} \\
\hline
DAQ efficiency & \multicolumn{3}{c}{0.5} \\
\hline
$df/dt$ [GHz/year] & 5.4 & 4.8 & 5.0 \\
\hline
Scan time [month] & 1.1 & 1.8 & 2.9 \\
\hline
\end{tabular*}
\end{center}
\caption{Cavity scheme and experimental parameters for each stage of the CAPP-9T MC experiment.}
\label{tab:capp_mc_prospect}
\end{table}

\subsection{Prospects}
\begin{figure}[t]
\centering
\includegraphics[width=0.8\textwidth]{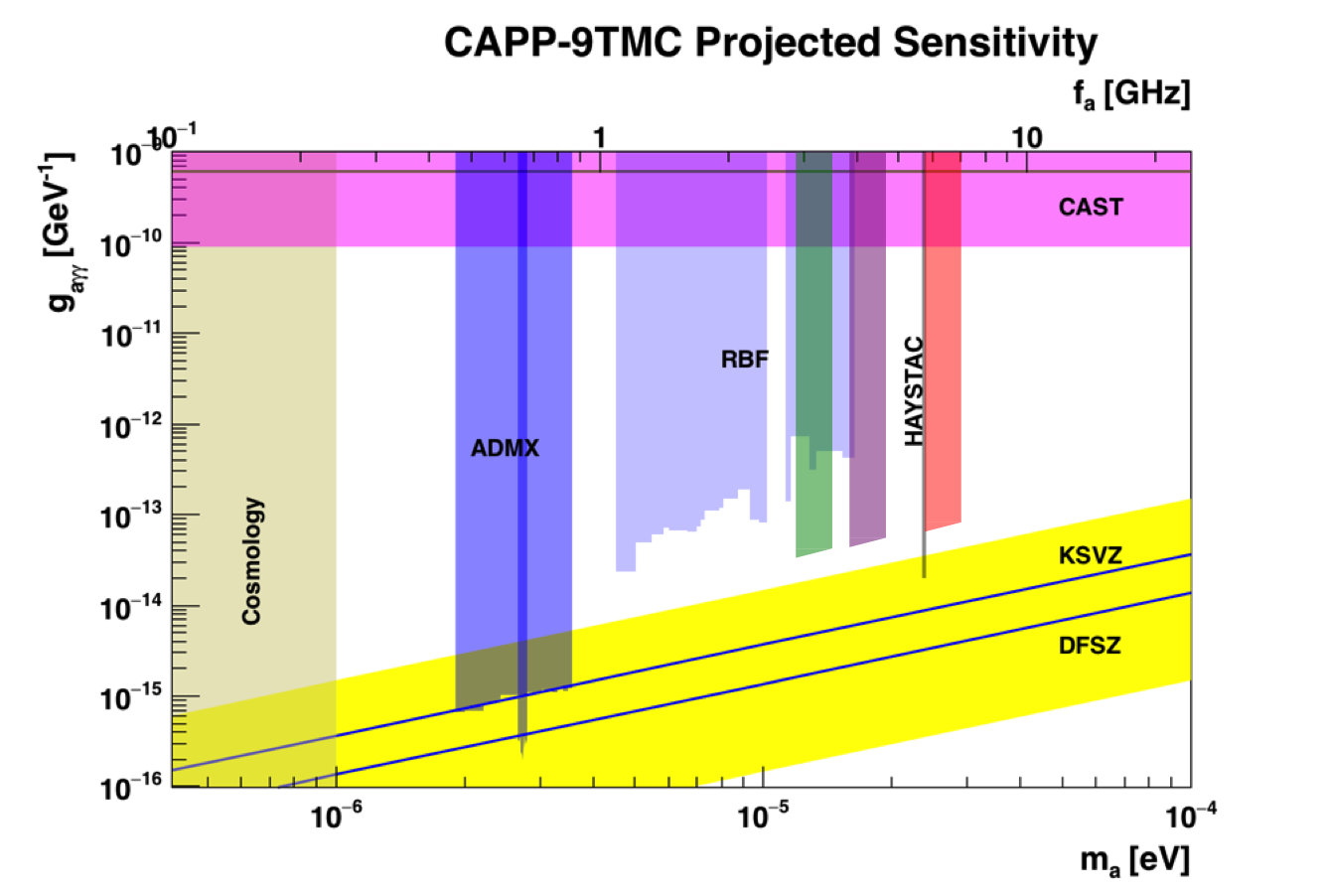}
\caption{Projected sensitivities of CAPP-9T MC in the parameter space of the axion-to-photon coupling vs. axion mass. 
The expected coverage regions by the first, second, and third stages are represented by the areas in green, purple, and red, respectively.}
\label{fig:capp_mc_prospect}
\end{figure}

CAPP-9T MC is a proof-of-concept experiment for the pizza cavity design. 
The objective of the experiment lies in demonstration of this concept to verify the capability of extending the search range for axion dark matter by a factor of two to three.
Therefore, the main focus is not on experimental design to be sensitive to the QCD axions, but on reliable operation of the tuning mechanism to show its feasibility at high frequency regions. 
The experiment consists of multiple stages depending on the cell multiplicity. 
The first stage will employ a double-cell cavity whose frequency coverage is 2.8 to 3.3\,GHz. 
A quadruple-cell cavity will replace the double-cell cavity in the second stage in order to cover a frequency range between 3.8 and 4.5\,GHz. 
At the last stage, an octuple-cell cavity will be considered expecting a frequency coverage of 5.8 to 7.0\,GHz. 
A series of low noise high-electron-mobility transistors (HEMT), whose typical noise temperatures are $1-3$\,K depending on frequency, is considered for the first stage amplification. Under a conservative assumption that the noise of the entire readout chain be 2 times larger than the amplifier noise, the total system noise temperature is estimated to be $4-6$\,K. 
Using conventional theoretical parameters and targeting at a sensitivity of 10$\times$KSVZ, the scan rate is evaluated to be approximately 5\,GHz/year. 
The corresponding time scale to scan the desired frequency range at each stage is estimated to be 1 to 3 months. 
Table~\ref{tab:capp_mc_prospect} and Fig.~\ref{fig:capp_mc_prospect} summarize the experimental parameters for each stage and projected sensitivity of the CAPP-9T MC project. 
It is expected large portions of the parameter space to be explored by the experiment.

%% file: 1.3.6/main.tex
\begin{figure}[b]
\centering
\includegraphics[width=0.7\textwidth]{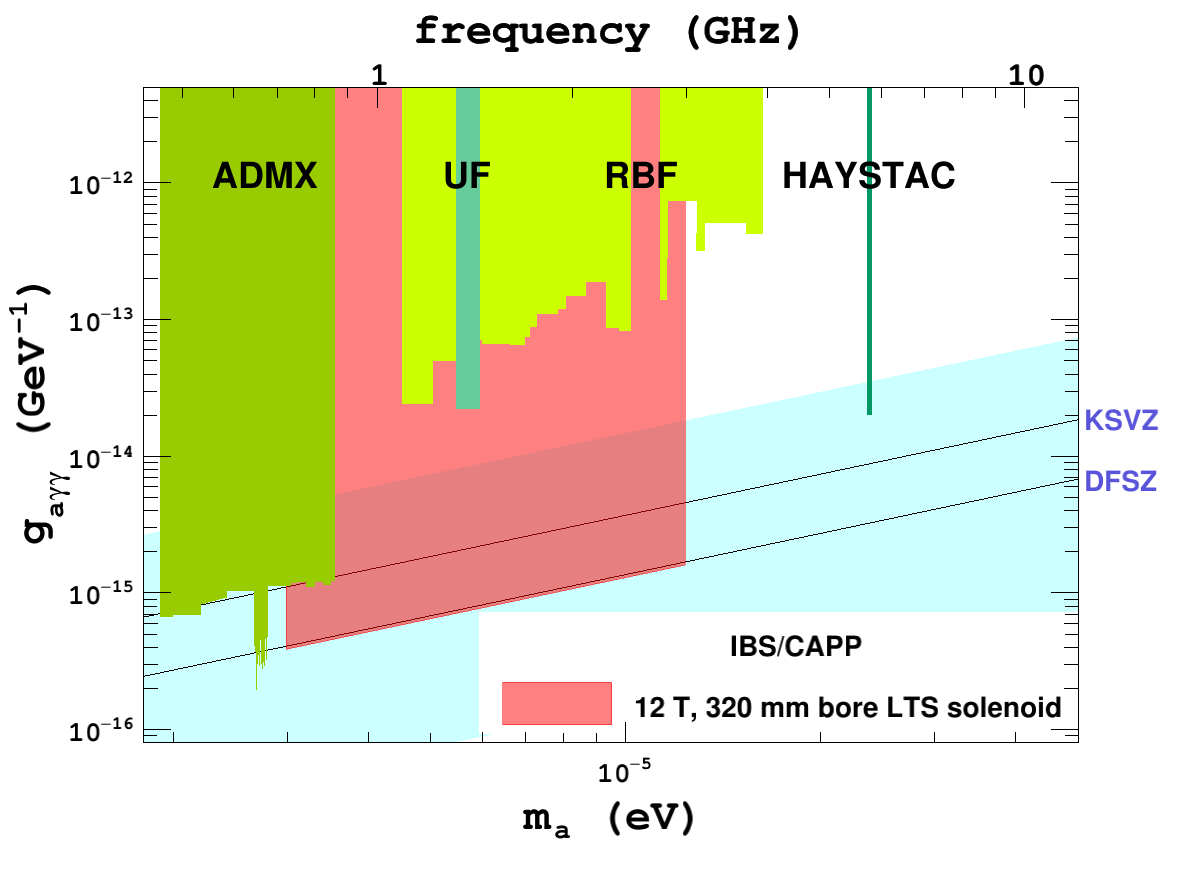}
\caption{Axion parameter space. The CAPP-12TB experiment would discover or exclude the axion in the red colored region.}
\label{fig:CAPP-12TB-1}
\end{figure}

CAPP-12TB is an axion haloscope search experiment utilizing a solenoid with the magnetic field of 12\,T and the Big bore of 320\,mm, a dilution refrigerator with the cooling power of 1\,mW at 100\,mK, a tunable cavity with the volume of 30\,L, and quantum-noise-limited amplifiers, i.e., SQUIDs. The primary goal of the CAPP-12TB experiment would become a DFSZ axion 
definitive experiment in the axion frequency range between 0.7 and 3\,GHz as shown in Fig.~\ref{fig:CAPP-12TB-1}., which would be unprecedented in the world for the next ten years or forever. 

\subsection{Magnet}
\begin{figure}[h]
\centering
\includegraphics[width=0.7\textwidth]{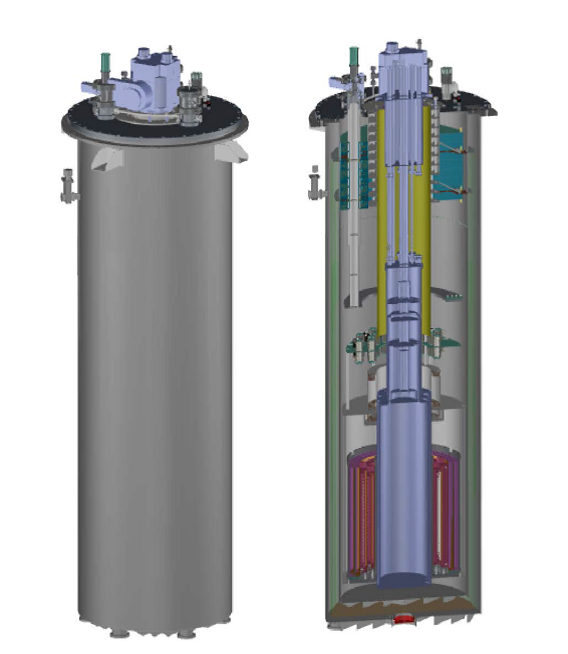}
\caption{Outline model of the magnet system, including the dilution refrigerator insert (violet), for visual aid only. The total height of the system is about 3.5\,m.}
\label{fig:CAPP-12TB-2}
\end{figure}

\begin{figure}[h]
\centering
\includegraphics[width=0.9\textwidth]{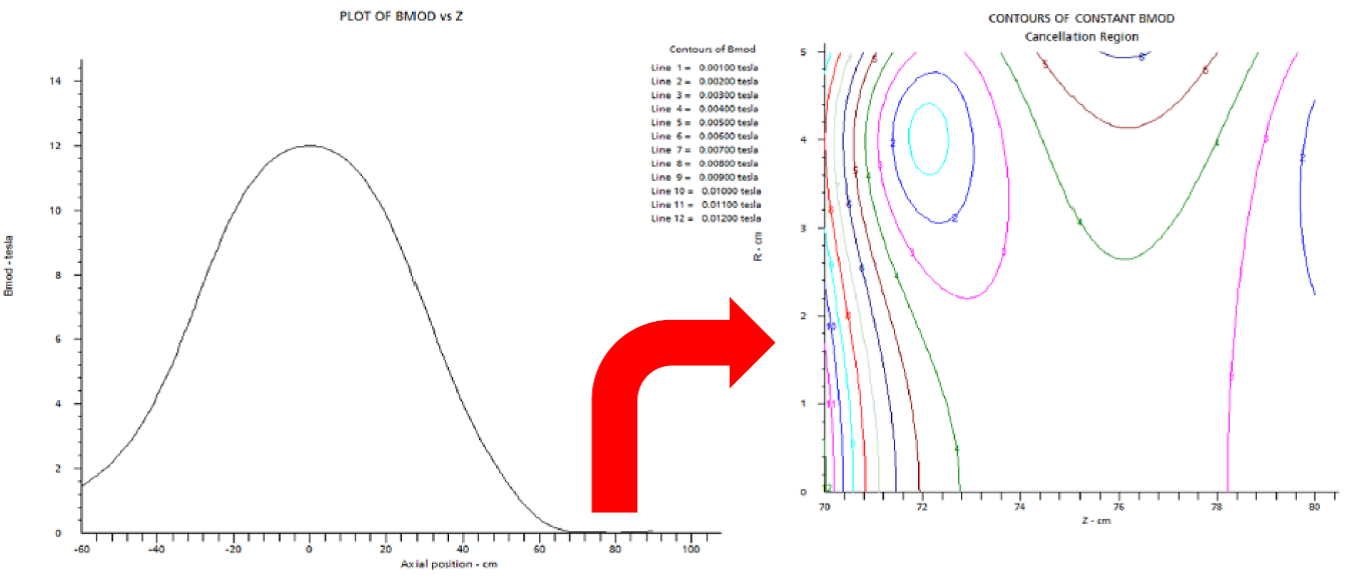}
\caption{Designed magnetic field at axial position (left) and that in the cancelation region which is about 80 cm away from the magnet center or the axial position of 0 (right).}
\label{fig:CAPP-12TB-3}
\end{figure}

\begin{figure}[h]
\centering
\includegraphics[width=0.9\textwidth]{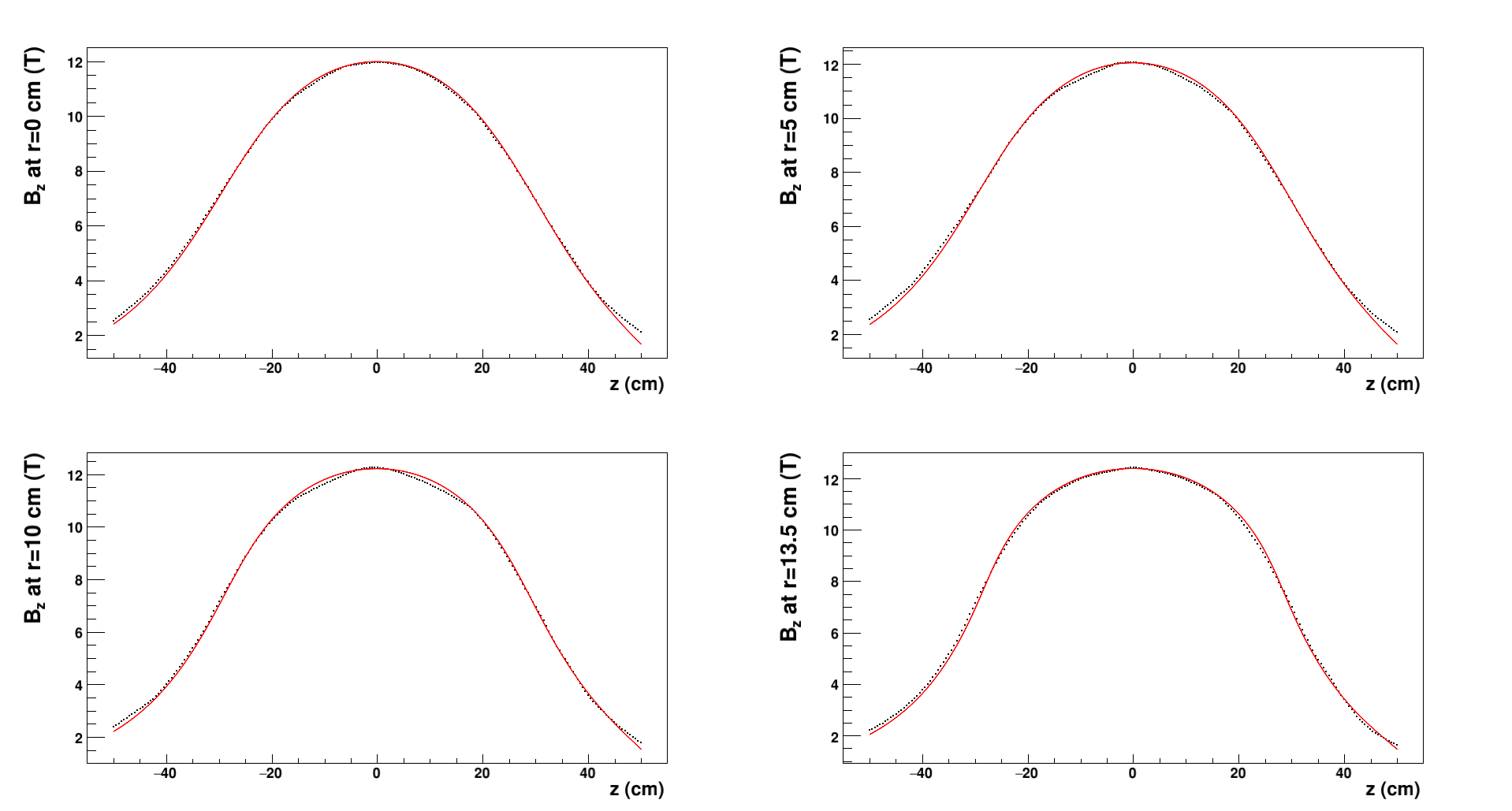}
\caption{Magnetic field in the axial position. Red lines are provided by the manufacturer and black dots are reproduced by our FEM solution.}
\label{fig:CAPP-12TB-4}
\end{figure}

\begin{figure}[h]
\centering
\includegraphics[width=0.9\textwidth]{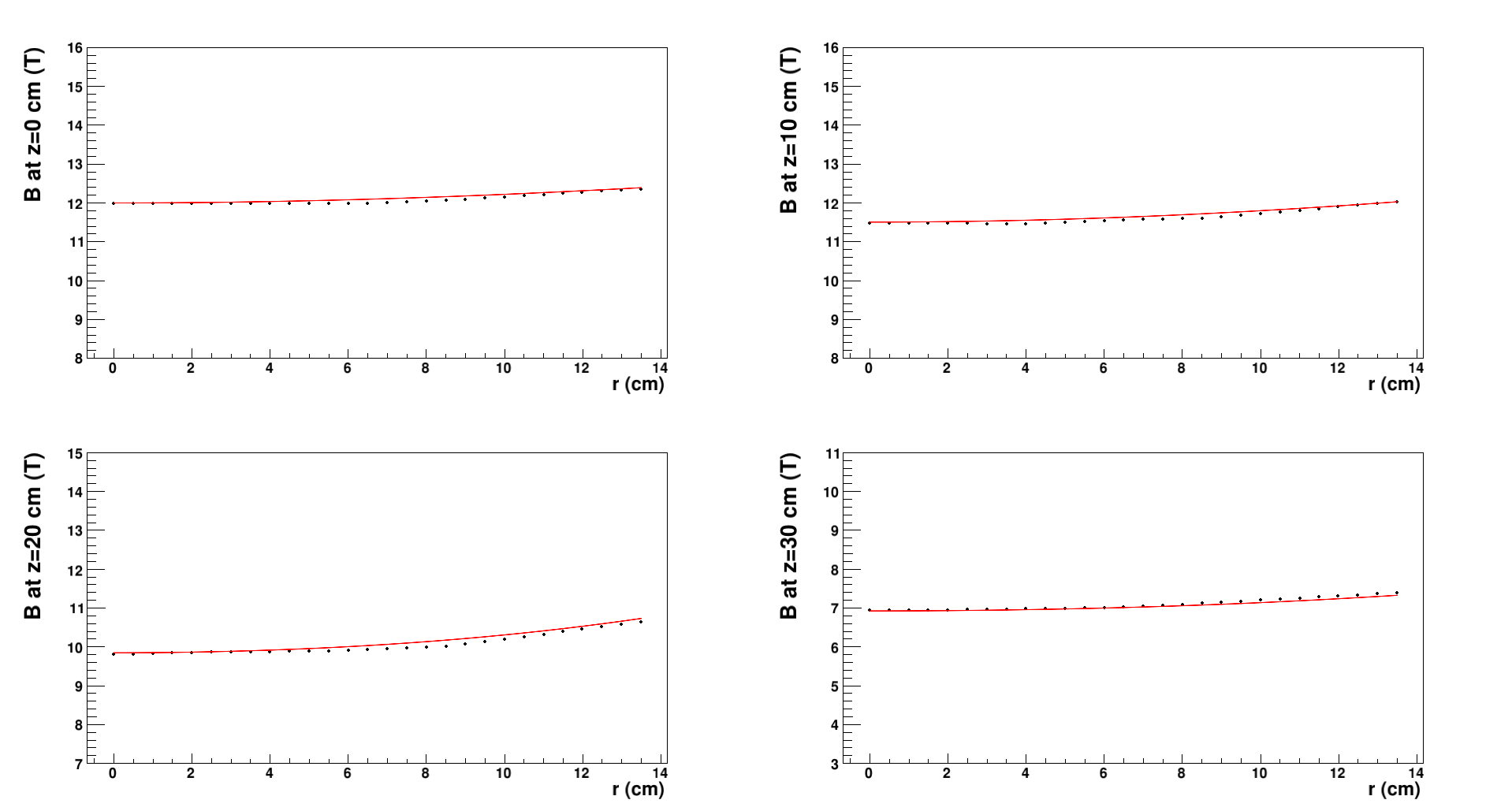}
\caption{The same as Fig.~\ref{fig:CAPP-12TB-4} except for magnetic field in the radial position.}
\label{fig:CAPP-12TB-5}
\end{figure}

The magnet system for the CAPP-12TB experiment provides not only the strong magnetic field region for the cavity, but also the cancelation region for the SQUID amplifier. In order to introduce the magnet system to IBS/CAPP, we went through the IBS and NFEC (National research Facilities \& Equipment Center) reviews in 2016 and NFEC approved the purchasing. The magnet system will be delivered at IBS/CAPP in 2020. Figure~\ref{fig:CAPP-12TB-2} shows the overall of the magnet system, including our dilution fridge insert which will be discussed later. Figure~\ref{fig:CAPP-12TB-3} shows the designed magnetic field along the axial position and that in the cancelation region which is about 80 cm away from the magnet center. Figures~\ref{fig:CAPP-12TB-4} and~\ref{fig:CAPP-12TB-5} also show the magnetic field along the axial and radial positions, respectively, together with those from our Finite Element Method (FEM) solutions. Our FEM solutions show good agreement with the solutions from the magnet manufacturer and they will be used to calculate the form factors of the cavity modes which are practically everything to predict the axion signal power in axion haloscope search experiments.

\subsection{Dilution fridge}
\begin{figure}[h]
\centering
\includegraphics[width=0.9\textwidth]{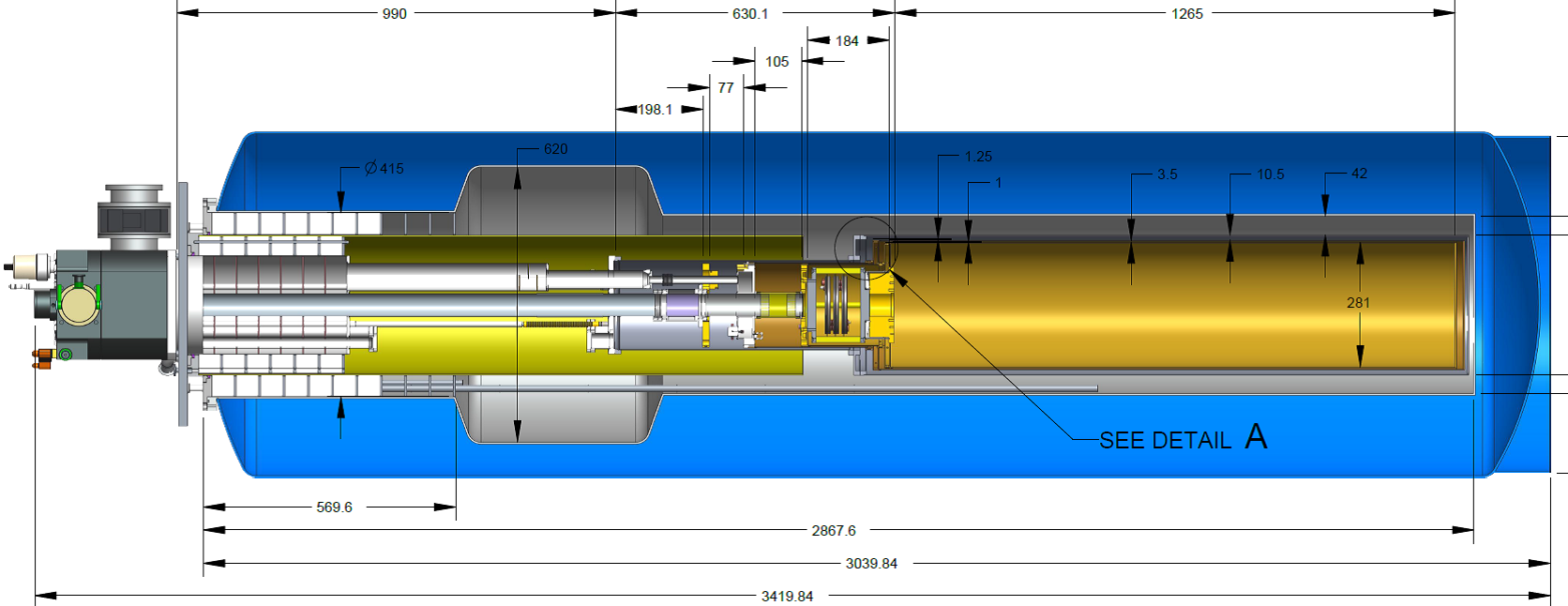}
\caption{Dilution fridge system for the CAPP-12TB experiment. The region just below the mixing chamber plate corresponds to the cancelation region of the magnet. The total height is about 3.5\,m.}
\label{fig:CAPP-12TB-6}
\end{figure}

Though the magnet amplifies the axion signal by providing high magnetic field, the matter in axion haloscope searches is the Signal to Noise Ratio (SNR), where the noise is a sum of temperatures from the cavity and the receiver chain. Our receiver chain would be a typical heterodyne receiver chain whose first amplifier is a SQUID, thus the SQUID noise dominates the receiver chain noise. Because the SQUID noise temperature is known to be proportional to its physical temperature, the SQUID itself should be cooled down to the quantum noise limit, i.e., ~50\,mK for the microwave signal whose frequency is 1\,GHz. The first phase of the CAPP-12TB experiment would run the frequency range around 1\,GHz. Therefore, with help from the dilution fridge, we expect the total noise would be ~100\,mK, 50\,mK from the cavity and the rest from the receiver chain. The design of the dilution fridge insert, inner vacuum chamber, and two radiation shields for the dilution fridge system has been being updated and finalized in order to be compatible with the magnet system, especially to fit the cancelation region of the magnet. Figure~\ref{fig:CAPP-12TB-6}. shows the overall of the dilution fridge system for the CAPP-12TB experiment. The system will be installed in IBS/CAPP around 2018 November. 

\subsection{Cavity}
\begin{figure}[b]
\centering
\includegraphics[width=0.7\textwidth]{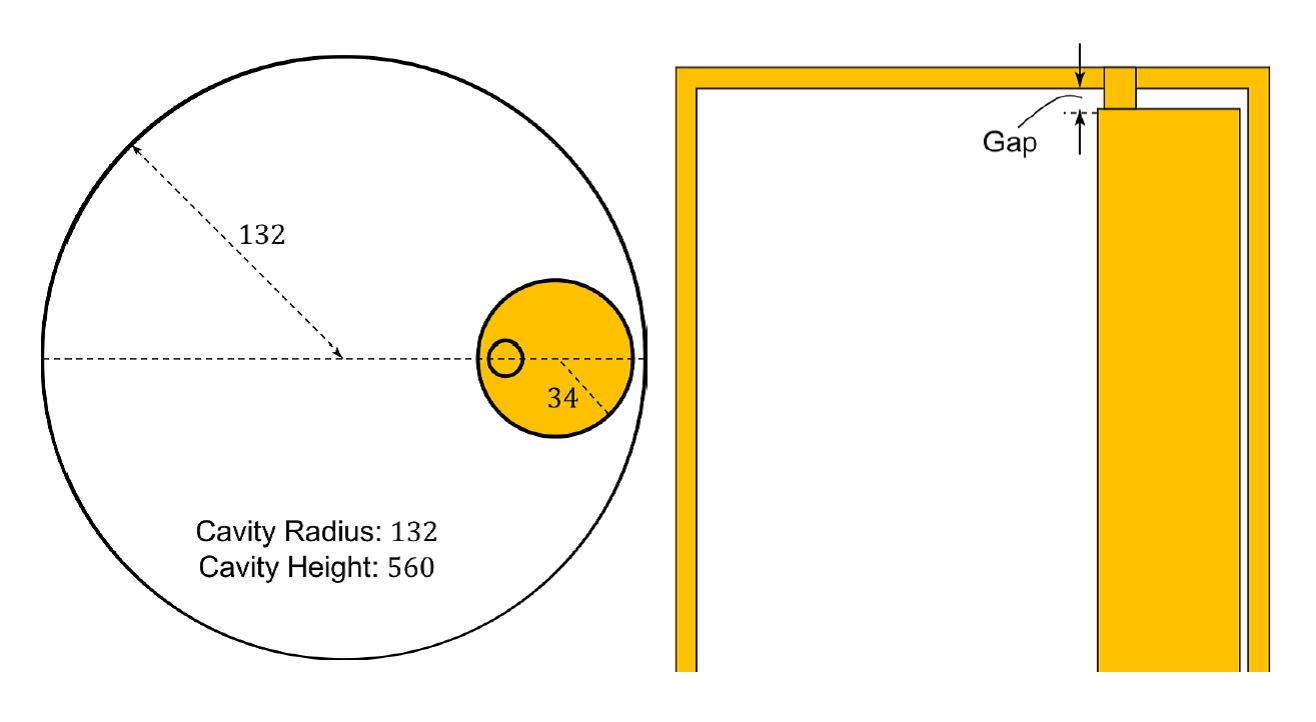}
\caption{Simplified design of the cavity where left shows top view and right shows lateral view (top half only). All dimensions are in\,mm.}
\label{fig:CAPP-12TB-7}
\end{figure}
\begin{figure}[t]
\centering
\includegraphics[width=0.95\textwidth]{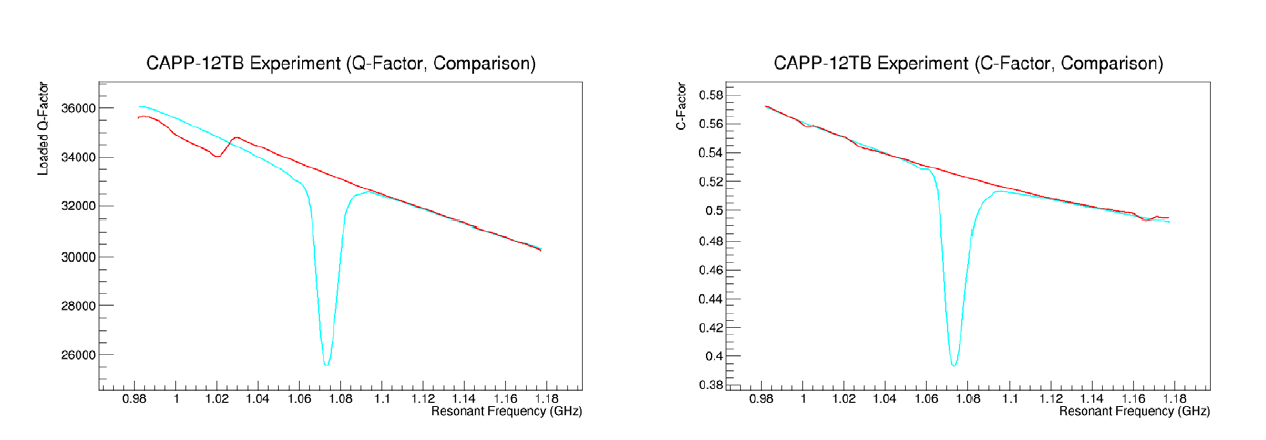}
\caption{Quality factor (left) and form factor (right) of the cavity mode as a function of the resonant frequency. Cyan and red lines are those without and with the additional dielectric rod.}
\label{fig:CAPP-12TB-8}
\end{figure}
\begin{figure}[t]
\centering
\includegraphics[width=0.9\textwidth]{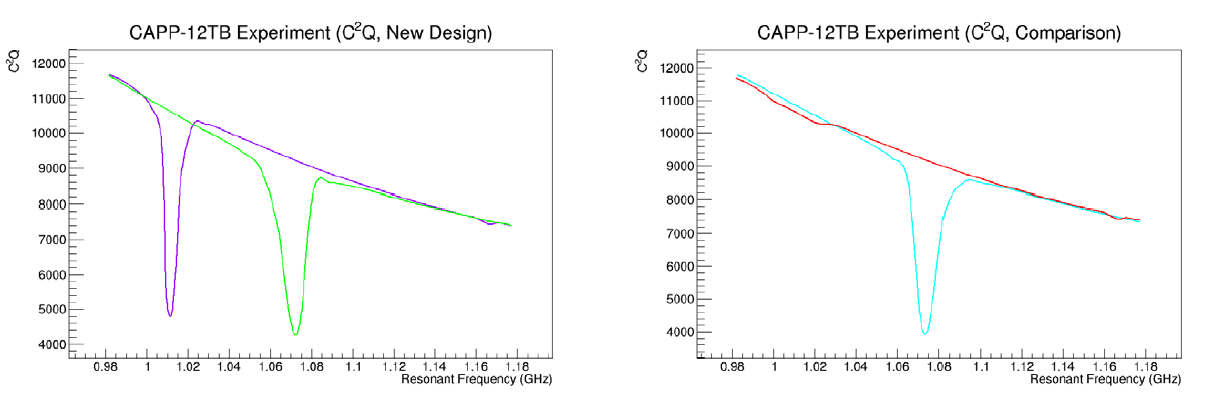}
\caption{Product of quality factor and form factor squared, which is proportional to the experimental sensitivity. Left plot shows those with the dielectric rod where the purple line is obtained from the first half rotation of the tuning rod and the green line is from the rest half rotation of it. The region between 1 and 1.03\,GHz would be covered by the green line and the rest of the region by the purple line, which corresponds to red line in right plot. Cyan line in right plot is obtained without the additional dielectric rod.}
\label{fig:CAPP-12TB-9}
\end{figure}
\begin{figure}[hp]
\centering
\includegraphics[width=0.8\textwidth]{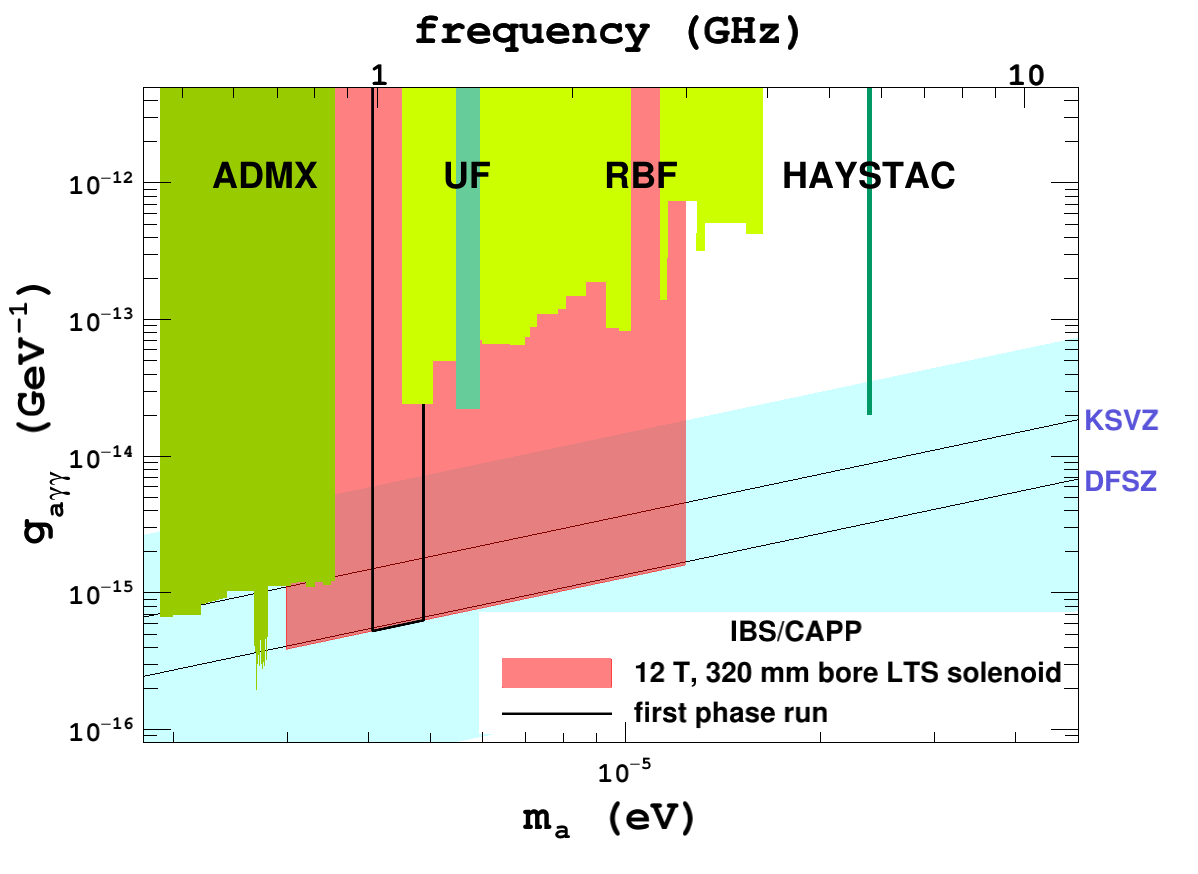}
\caption{The region with black line is the target of the first phase run of the CAPP-12TB experiment.}
\label{fig:CAPP-12TB-10}
\end{figure}

Thanks to the huge magnetic field volume providing from the magnet, we have designed to realize a tunable cavity with the volume of about 30\,L which is one of the significant advantages of the CAPP-12TB experiment. The first phase of the experiment has been designed with the copper tuning mechanism to run the frequency range around 1\,GHz (see Fig.~\ref{fig:CAPP-12TB-7}), then the frequency range will be extended up to 2\,GHz without any significant modifications of the tuning mechanism. Introducing the conductor tuning mechanism usually does not lose the form factor or the axion signal power in the absence of dielectric constant in the relevant form factor relation. However, the conductor tuning mechanism may introduce unwanted mode crossings where we lose the experimental sensitivity completely. Due to the gap between the two conductors, one is the cavity end cap and the other is the tuning rod end (see right plot in Fig.~\ref{fig:CAPP-12TB-7}), we cannot avoid the capacitive effect which usually results in hybrid modes that originate from the mode crossings. The hybrid modes, then, degrade both the cavity quality factor and the form factor of the cavity mode for axion haloscope searches (see cyan lines in Fig.~\ref{fig:CAPP-12TB-8}), thus equivalently degrade the experimental sensitivity (see cyan line of right plot in Fig.~\ref{fig:CAPP-12TB-9}). Through thorough simulation study, our first-phase tunable cavity fully avoids both mode crossings and hybrid modes by breaking the symmetry of the cavity volume geometry with an additional dielectric rod in the cavity (see red lines in Figs.~\ref{fig:CAPP-12TB-8} and~\ref{fig:CAPP-12TB-8}. As shown in left plot in Fig.~\ref{fig:CAPP-12TB-9}, drop-off in experimental sensitivity depends on rotational direction of the tuning rod with broken symmetry of the cavity volume geometry, thus such drop-off can be avoided by the full rotation of the tuning rod. Our first phase cavity design without such cases is a significant achievement. With such an achievement, the first phase run would search for axion dark matter in the axion parameter space indicated by the black line in Fig.~\ref{fig:CAPP-12TB-10}.

\subsection{Receiver chain}
Everything is ready to go with a lot of experience from the CAPP-8TB experiment, except for the SQUID matching to the CAPP-12TB experiment. 
Therefore, The SQUID acquirement and operation will be the main effort until the magnet delivery, utilizing the cryogen-free solenoid for the CAPP-8TB experiment, for the time being.
In the light of the CAPP-8TB experiment, the expected DAQ efficiency is about 50\% which is mainly limited by a spectrum analyzer. 
We are planning to improve this DAQ efficiency by employing a digitizer. 

\subsection{Plan}
\begin{figure}[hp]
\centering
\includegraphics[width=0.8\textwidth]{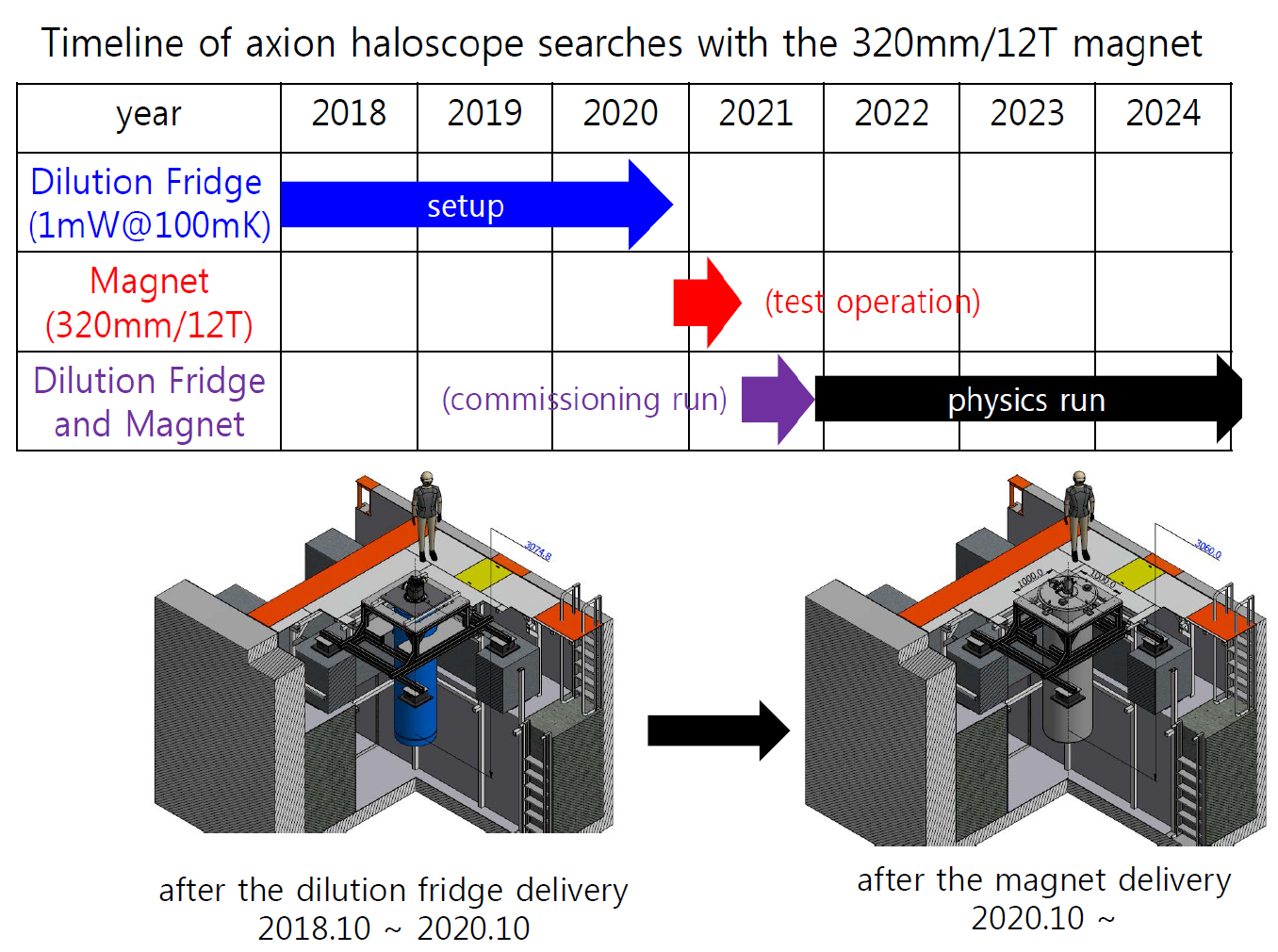}
\caption{Time line of the CAPP-12TB experiment.}
\label{fig:CAPP-12TB-11}
\end{figure}

The table in Fig.~\ref{fig:CAPP-12TB-11} shows the timeline of the CAPP-12TB experiment. 

\begin{table}[hp]
\begin{tabular*}{0.85\textwidth}{@{\extracolsep{\fill}}cccc}
\hline
Phase & Year	 & Search range [GHz]	& $a\gamma\gamma$ coupling \\
\hline
1st phase	& 2021     & 0.98$\sim$1.18	& DFSZ \\
2nd phase	& 2022	& 1.16$\sim$1.46	& DFSZ \\
3rd phase	& 2023	& 1.35$\sim$1.62	& DFSZ \\
4th phase	& 2024	& 1.56$\sim$1.70	& DFSZ \\
5th phase	& 2025	& 1.69$\sim$1.87	& DFSZ \\
\hline
\end{tabular*}
\end{table}

We will prepare everything for the experiment with the dilution fridge only (see the bottom left figure in Fig.~\ref{fig:CAPP-12TB-11}) until the magnet delivery. Once the magnet is delivered, then we will do our jobs, axion dark matter searches (see the bottom right figure in Fig.~\ref{fig:CAPP-12TB-11}). Table below lists the short term plan of the CAPP-12TB experiment, which can be realized without any serious difficulties because we plan to change the cavity volume geometry only without modifying the tuning mechanism employed for the 1st phase physics run.

After the 5th phase physics run, we will modify the tuning mechanism to search for the frequency range beyond that of the 5th phase physics run or below that of the 1st phase physics run, which belongs to the long term plan of the CAPP-12TB experiment.

%% file: 1.3.7/main.tex
25\,T magnet is designed for Axion research to get higher conversion power and advantage of scanning time. Goal of developing magnet is 25\,T center magnetic field and 100\,mm bore size. To get the high magnetic field, no insulation magnet structure is considered because of mechanical stress issue. Winding bore diameter of designed magnet is 105\,mm using 12\,mm width Superpower HTS conductor. Outer diameter of magnet is 200\,mm and overall length is 341\,mm. Detail specification are summarized in Table~\ref{tab:CAPP-25T-1}.

\begin{figure}[h]
\centering
\includegraphics[width=0.8\textwidth]{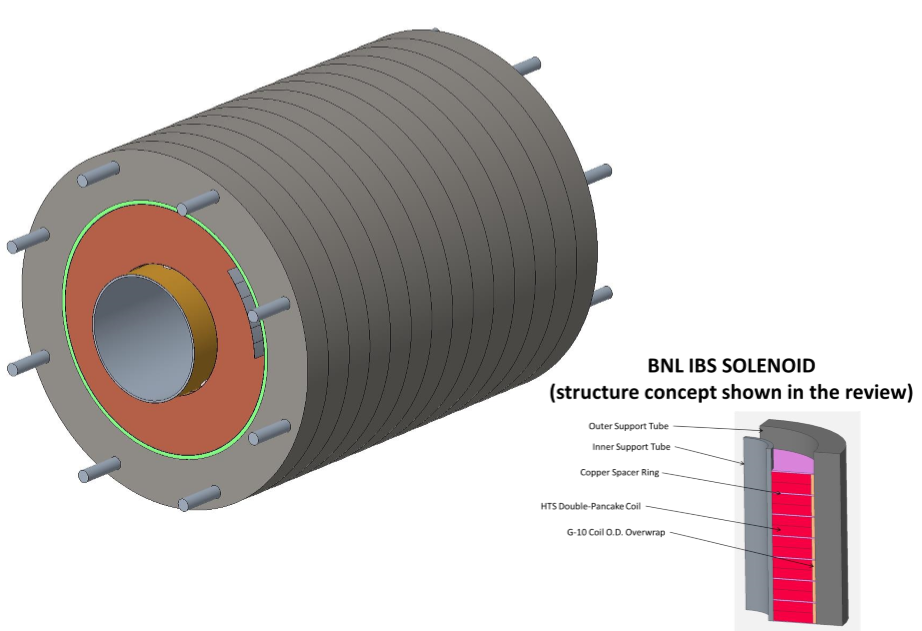}
\caption{\label{fig:CAPP-25T-1}25\,T magnet conceptual design}
\end{figure}

\begin{table}[b]
\centering
\caption{\label{tab:CAPP-25T-1}Superconducting magnet specifications}
\begin{tabular}{cc}
\hline
Specifications & Value \\
\hline
Center magnetic field & 25\,T \\
Clear bore & 100\,mm \\
Winding bore & 105\,mm \\
Operating current & 458\,A \\
Operating temperature & 4\,K, liquid helium \\ 
Superconductor width & 12\,mm \\
Superconductor thickness & 75 $\mu$m \\
Number of turns & 297 (Single) \\
Total conductor length & 8.3\,km \\
Magnet length (height) & 341\,mm \\
\hline
\end{tabular}
\end{table}

Estimated stress of magnet is 200\,MPa, thus BNL tested three different thickness conductor and 75$\,\mu$m conductor is selected which has 300\,A of minimum critical current at 77\,K.  Operating temperature is 4.2\,K to maximize magnetic field, so current density is 500\,A/mm\textsuperscript{2} at 4\,K (Operating current is 450\,A @ 25\,T) and current margin is 50 \% (Expected critical current is 600 A)

\begin{table}[t]
\centering
\caption{\label{tab:CAPP-25T-2}Specifications of superconductor}
\begin{tabular}{cc}
\hline
Specifications & Value \\
\hline
Critical current, self-field & Min 300\,A \\
Critical current, @ 8\,T, 4K & 675\,A \\
Dimension (width, thickness) & 12\,mm, 75 $\mu$m \\
Conductor type & YBCO, 2\textsuperscript{nd} generation HTS \\
N value & 20 \\
Substrate & Hasteloy \\
\hline
\end{tabular}
\end{table}

Center field of designed magnet is 25\,T at 450\,A and current density of magnet is 500\,A/mm\textsuperscript{2}, Stress of conductor is important because of stability of structure. Maximum hoop stress of superconductor, which they selected for magnet, is 620\,MPa at 25\,T. 

\begin{figure}[b]
\centering
\includegraphics[width=0.9\textwidth]{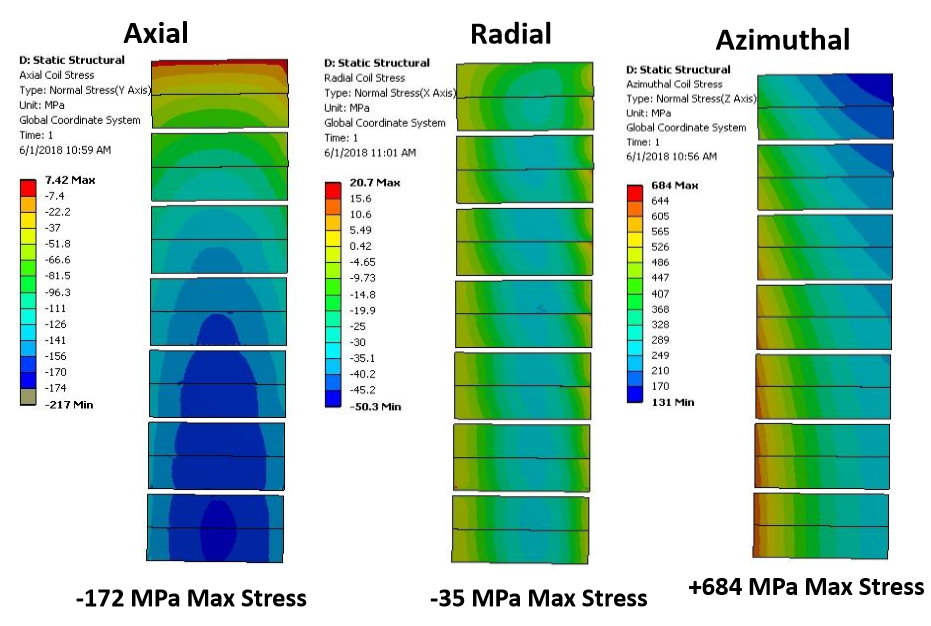}
\caption{\label{fig:CAPP-25T-2}Stress analysis of 25\,T magnet}
\end{figure}

Double pancake coils are formed from two single pancake coils with an internal splice spanning almost all of the inner surface of the coils. Fourteen double-pancake coils are installed on a tight-fitting tube having a 100 mm inner diameter and 1 mm wall thickness with fiberglass insulation over it.  The insulation between two single pancakes and between double pancakes is 0.25 mm thick and consists of two Nomex® sheets. The double pancake will be over-wrapped with fiberglass epoxy which will be accurately machined to the desired outer diameter with a nominal thickness of 3 mm. The primary structure to contain the large hoop stresses over each double pancake will be 40 mm thick outer support rings made of high Strength 7075-T651 aluminum which has a yield strength of 500 MPa. Aluminum structure with higher thermal contraction than the coil is chosen to overcome a gap of 0.13 mm between the coil and tube to allow for assembly tolerances. Stainless Steel inner and outer end plates and axial tie rods with thermal contraction similar to the coil form the axial structure. Mechanical structure analysis is performed with ANSYS using 2-D axi-symmetric model of1/4 of the structure. Lorentz forces from Maxwell are mapped to the ANSYS static structural model where appropriate boundary conditions, material properties, contacts, and thermal conditions are applied. All contacts are assumed to be frictionless except G10 overwrap which is bonded to the O.D. of the double-coil pancakes. 

\begin{figure}[t]
\centering
\includegraphics[width=0.41\textwidth]{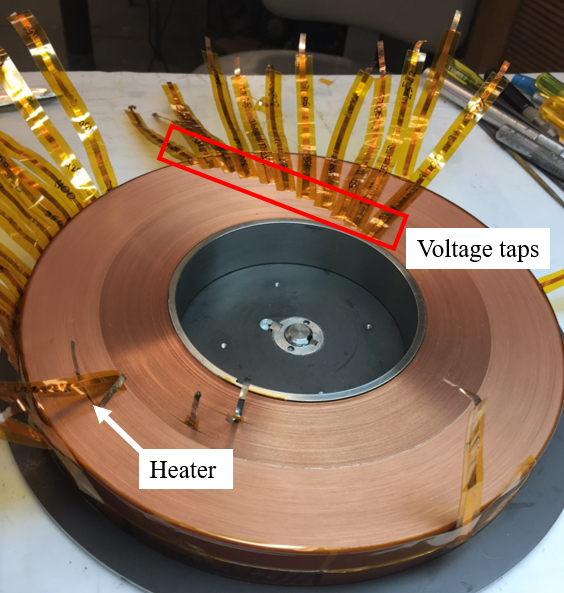}
\includegraphics[width=0.5\textwidth]{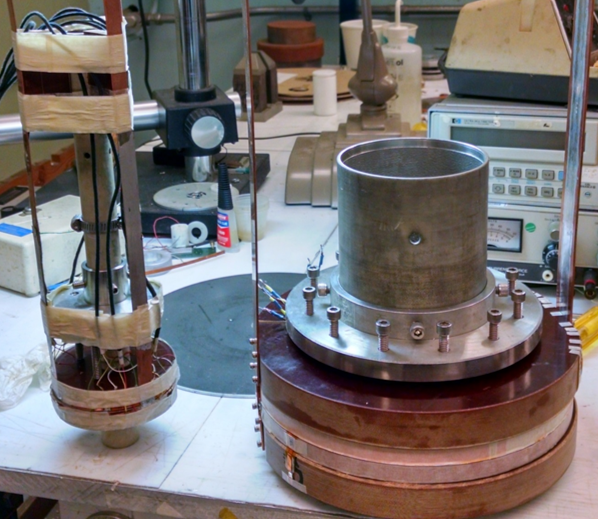}
\caption{\label{fig:CAPP-25T-3}Test coil and experimental setup}
\end{figure}

\begin{figure}[b]
\centering
\includegraphics[width=0.8\textwidth]{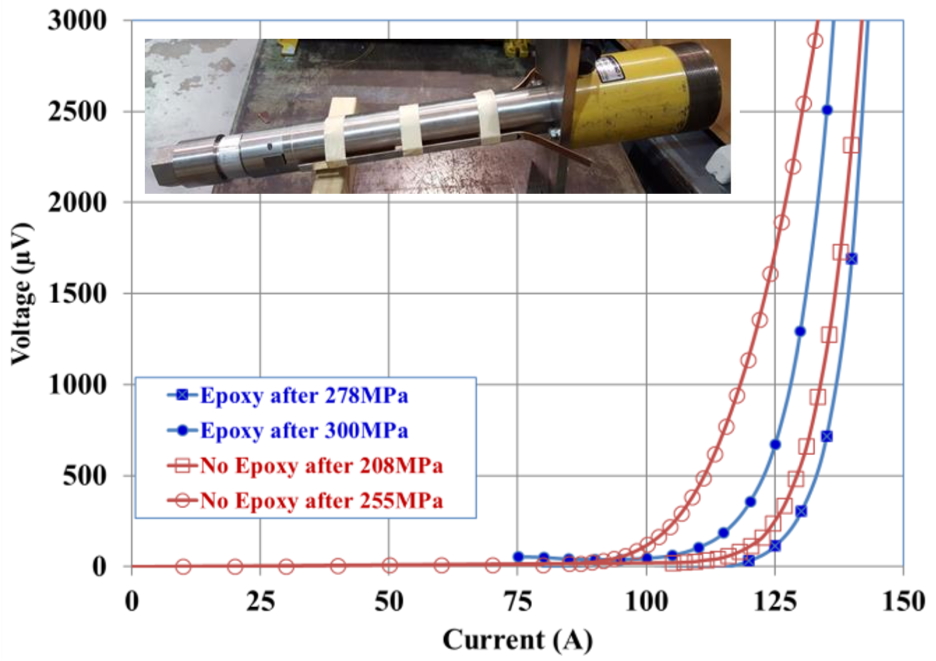}
\caption{\label{fig:CAPP-25T-9} V-I measurements of a prototype coil.}
\end{figure}

Mechanical properties of the conductor are based on the measurements at SuperPower on the wide face of conductor. The influence of loading narrow face of conductor was obtained through measurements at BNL with a fixture specifically designed and built for this purpose. Figure~\ref{fig:CAPP-25T-9} shows the V-I measurements on a small coil made with 40\,$\mu$m copper and 50\,$\mu$m Hastelloy. The actual conductor used has even lower copper (20\,$\mu$m).
The hoop stresses are all well within the margins of the superconductor, all of them more than 50\%. The weakest point is at the ends of the magnet, where the margins are still more than 50\% as estimated by ANSYS simulations.

The magnet is reliable and that it can operate for long periods at a time reliably, it can withstand numerous quenches safely, and, in addition, its operating margins are large. Tests have shown performance degradation more than 50\% beyond estimated loading from ANSYS simulations. The use of epoxy to the wound coils shows that the compressive loading limits increase and it was adopted in the design.

Before manufacturing actual coil for 25\,T magnet, model HTS coil fabricated (double pancake, no insulation) to study quench behavior and performance of no insulation magnet. 12\,mm wide, 120$\,\mu$m thickness HTS conductor were used winding coil. Detail parameters are summarized in Table~\ref{tab:CAPP-25T-3}. This magnet has several splices in coil and single pancake coils were jointed together for double pancake coil.

\begin{table}[h]
\centering
\caption{\label{tab:CAPP-25T-3}Test coil design parameters}
\begin{tabular}{cc}
\hline
Test coil parameters & Value \\
\hline
Winding diameter & 100\,mm \\
Outer diameter & 220 \\
Number of turns & 971 /double pan cake \\
HTS conductor type & 115$\,\mu$m thickness, 12\,mm width \\
Total conductor length & 505\,m \\
Critical current & \makecell{830 A @ 4\,K\\ 60\,A @ 77 K} \\
\hline
\end{tabular}
\end{table}

\begin{figure}[b]
\centering
\includegraphics[width=0.8\textwidth]{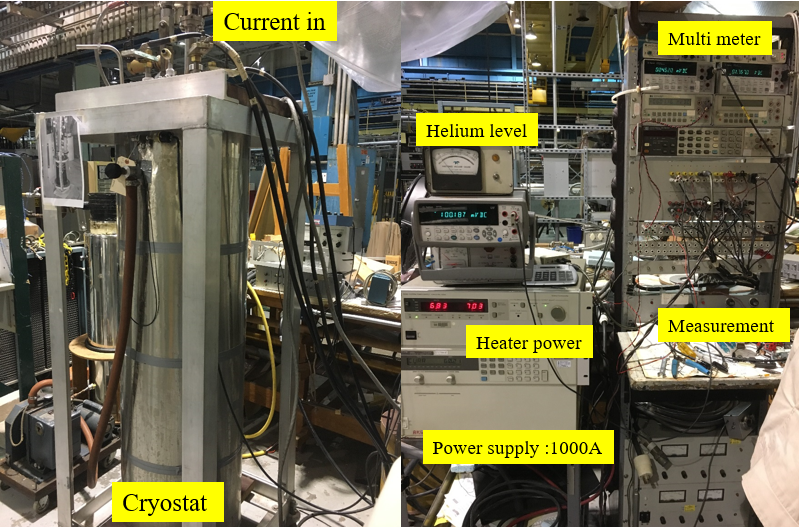}
\caption{\label{fig:CAPP-25T-4}Experimental setup}
\end{figure}

\begin{figure}[t]
\centering
\includegraphics[width=0.8\textwidth]{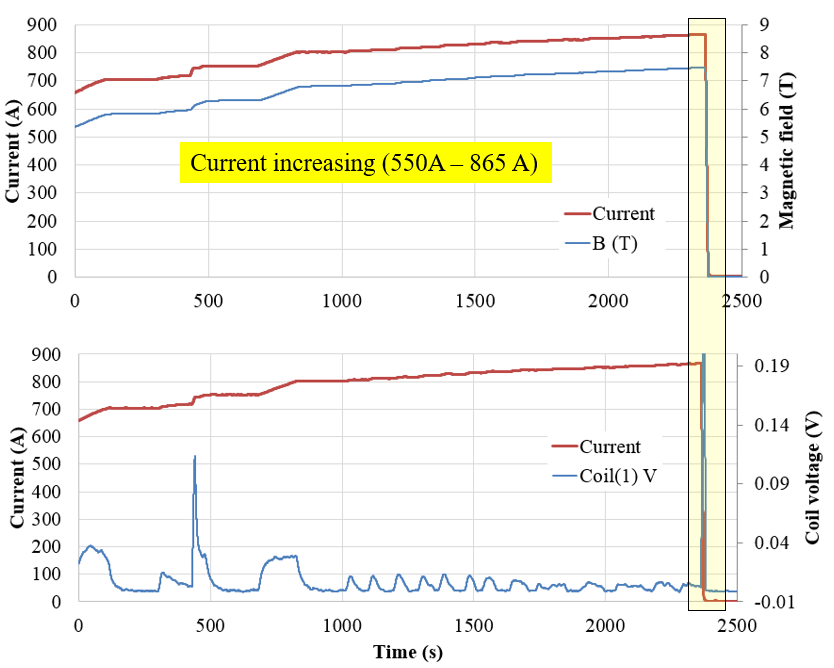}
\caption{\label{fig:CAPP-25T-5}Critical current measurement results of model coil}
\end{figure}

Critical current measured in 4\,K liquid helium. Fig.~\ref{fig:CAPP-25T-4}. Shows the measurement system including current source, cryostat and DAQ systems. Test coil is installed in the cryostat and cool down using liquid helium. Current applied to superconducting coil and voltages, current, temperatures are monitored. Critical current of sample coil was 860\,A and test results are shown in Fig.~\ref{fig:CAPP-25T-5}. Over 860 A, voltages were suddenly dropped because of quench and liquid helium evaporated shortly. Magnet voltages increased by current discharging from the coil. The local defects were simulated with three stainless steel heaters at 600\,A. the coil kept operating and didn’t runaway (quench) despite a significant local defect simulated with the heater.  The coil turned only partially resistive (40 mV across the coil) with 30\,W. The coil recovered immediately after the heater was turned off. No damage in coil performance was observed following such experiments even after the thermal runaway (quench). This demonstrates the tolerance against significant local disturbances or defects even in a large no-insulation coil operating at high current. 

\begin{figure}[h]
\centering
\includegraphics[width=0.9\textwidth]{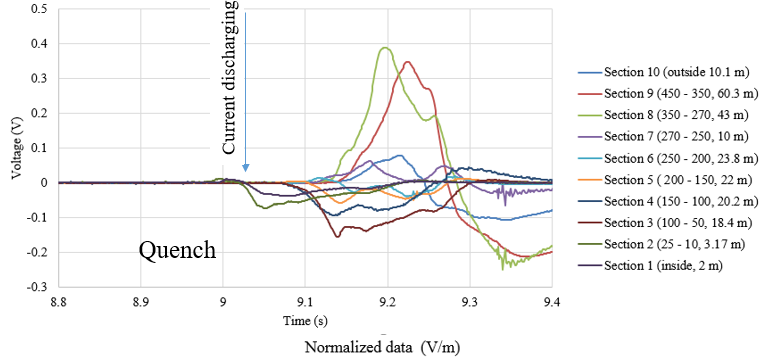}
\caption{\label{fig:CAPP-25T-6}Quench characteristic of model coil}
\end{figure}

Over critical current, HTS coil occurs quench, Temperature raised and liquid helium evaporated shortly. Fig.~\ref{fig:CAPP-25T-6} shows several sections of HTS coil voltage behavior. After quench, all section voltage were increased out most sections voltages were larger than inner. Outer section is longer than inner. 

Tested HTS magnet does not have insulation between superconductors, so magnetic field is not proportional to applied current. 

Conductor provider is Superpower, and first 1\,km conductor arrived BNL from Superpower Company at May 2018. 1\,km conductor has 4 splices. Minimum critical current at 77\,K is 300\,A, arrived conductors performance is quite higher than specs, 380\,A, 400\,A, 420\,A, 500\,A respectively. 

\begin{figure}[h]
\centering
\includegraphics[width=0.8\textwidth]{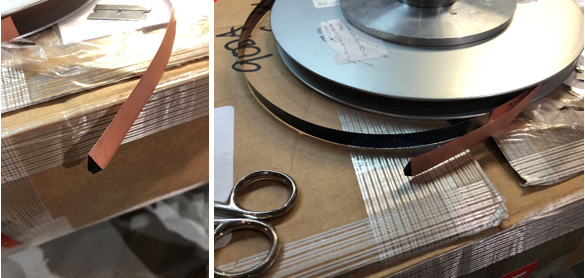}
\caption{\label{fig:CAPP-25T-7}First conductor delivery from Superpower (1 km)}
\end{figure}

\begin{figure}[h]
\centering
\includegraphics[width=0.8\textwidth]{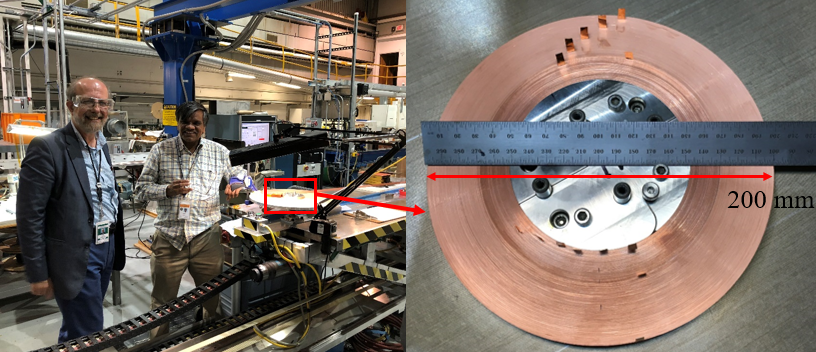}
\caption{\label{fig:CAPP-25T-8}First single HTS coil (no insulation)}
\end{figure}

First HTS coil was fabricated 4th June, Bending diameter is 105\,mm without inner bobbin. Before winding, first turn soldered with indium to make joint with other coil. Total length of single coil is 260\,m so, 2 splice of conductors were used and made joint at 227 turn. Total number of turn is 630 and outer diameter is 200\,mm. Picture of 1st coil is shown in Fig.~\ref{fig:CAPP-25T-8}.

Single coil was tested at liquid nitrogen first. Critical current was 80\,A. After that, BNL fabricated three more coil with same dimension. Totally four coils were fabricated until now. Test results are summarized at Table~\ref{tab:CAPP-25T-4}. Double pancake coil was assembled with two of single pancake coil and preparing 4\,K test now. 

\begin{table}[h]
\centering
\caption{\label{tab:CAPP-25T-4}Fabricated coils performance}
\begin{tabular}{| l | l | l | l | l |}
\hline
 & Coil 1 & Coil2 & Coil3 & Coil4 \\
 \hline
Number of turns & 630 & 630 & 629 & 624 \\
 \hline
Critical current @ 77\,K & 80\,A & 70\,A & 73\,A & 70\,A \\
 \hline
Soldering turn \# & 227 & 611 & 346 & 423 \\
 \hline
Contact resistance & 9.1 $\Omega$ & 8.3 $\Omega$ & 8.4 $\Omega$ & 8.5 $\Omega$ \\
 \hline
Time constant & 78.4 & 97.1 & 92.6 & 95.2 \\
 \hline
\end{tabular}
\end{table}

Final design has 105\,mm winding diameter and 200\,mm outer diameter with 630 turns of 12\,mm superconductor. Totally 14 double pancake coil requires for 25\,T full magnet. BNL made 4 single coil (2 for double pancake coil) and first double pancake coil is ready to test. They need totally 10\,km conductor to complete project, at this moment they bought 5\,km and 2\,km received. Four single coils tested in liquid nitrogen and coil \#2 and \#4 is used for first double pancake coil because of similar critical current. 

\begin{figure}[h]
\centering
\includegraphics[width=0.95\textwidth]{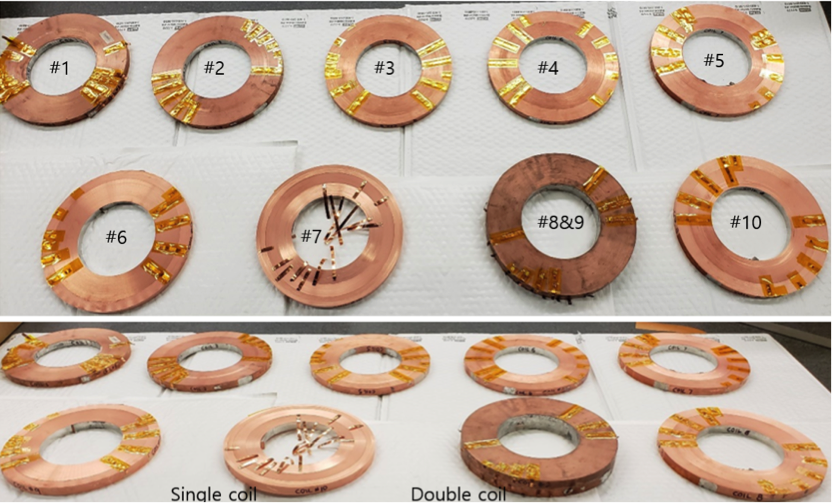}
\caption{\label{fig:CAPP-25T-10}Manufacturing process (10 HTS coils).}
\end{figure}

\begin{figure}[h]
\centering
\includegraphics[width=0.95\textwidth]{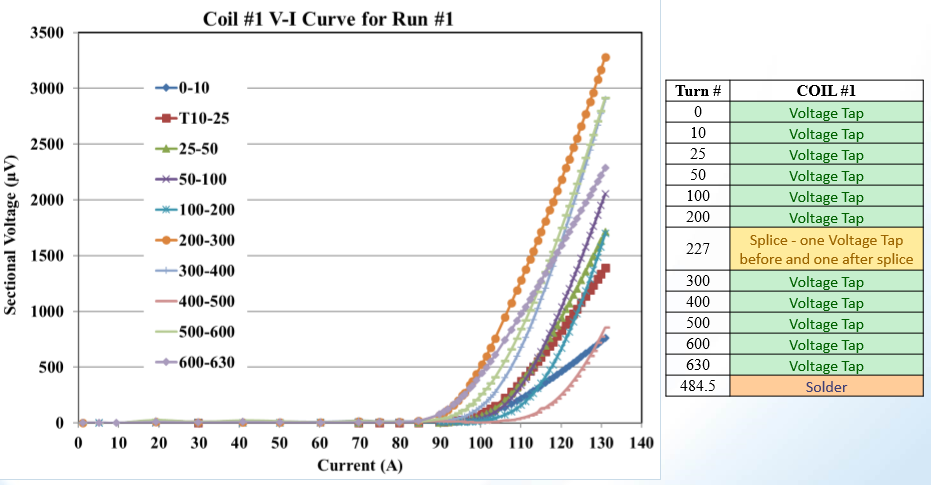}
\caption{\label{fig:CAPP-25T-11}Critical current measurement results (First single HTS coil)}
\end{figure}

BNL fabricated 10 single pancake coils using HTS conductor. 6 coils critical currents are measured in the liquid nitrogen (77\,K). Figure~\ref{fig:CAPP-25T-10} shows 8 single pancake coils and 1 double pancake coil (totally 10 single pancake coils are fabricated). Experiment results of 6 coils are shown in Fig.~\ref{fig:CAPP-25T-11} and critical currents were 60$\sim$80\,A. Four more coils critical current will be measured in liquid nitrogen too.

%% file: 1.4.1/main.tex
\subsection{Introduction}

A Superconducting QUantum Interferometer Devices or SQUID is the most sensitive magnetic sensor that takes advantage of two quantum phenomena of superconductor, Josephson effect and flux quantization.
SQUIDs evidence macroscopic quantum effects as quantum tunnelling, macroscopic wave function and quantum interference.
They are used to measure multiply physical quantities, which are conceivable to transmute to a magnetic flux.
SQUID sensors have the lowest physically possible intrinsic noise and, with some modifications, they may reach extremely wide frequency bands.
At IBS/CAPP we use SQUIDs in three projects: 1) axion dark matter search; 2) proton beam positioning monitor; and 3) Axion Resonant InterAction Detection Experiment (ARIADNE).
In current section we describe development of microwave SQUID amplifiers for axion search experiments. 

In the experiments for dark matter QCD axion searches, a tiny microwave signals from a low-temperature high-Q resonant cavity should be detected using the highest sensitivity.
Noise level of the amplifier defines the measurement time if the sensitivity is set. 
In general, the measurement time is proportional to the square of the noise level. 
An axion search experiment requires a frequency scan in a wide frequency range from hundreds of megahertz to tens of gigahertz.
Wide band signal amplifiers provide an advantage because of a time saving coming from no needs of frequency adjustment and amplifiers replacement.
The best commercial wide-band low-noise cryogenic semiconductor amplifiers (High Electron Mobility Transistor or HEMT) have the lowest noise temperature above 1.0\,K in the frequency range of our interest, $1-10$ GHz.
For our setup it takes about one month to scan a 1\,MHz frequency range at the KSVZ axion model limit. 
A SQUID-based microwave amplifier with a noise temperature of 0.2\,K will allow us to do such a scan in one day.  

It is already known, SQUIDs are able to outperform the best semiconductor microwave amplifiers with significantly lower noise temperature close to the standard quantum limit~\cite{bib:SQLN}
\begin{equation}
T_{\rm{SQL}} = \hbar\omega/k_B,
\label{eq:sql}
\end{equation}
where $\hbar$ is the Planck constant, $\omega$ is the circular frequency, and $k_B$ is the Boltzmann constant.
$T_{\rm{SQL}}$ is $\sim50$\,mK at 1\,GHz.

There are two types of microwave SQUID-based amplifiers with some modifications: MSA (Microstrip SQUID Amplifier) and JPA (Josephson Parametric Amplifier). 
In general both of them have narrow frequency resonant bands, although some new modifications may have non-resonant wide bandwidths. 
In the case of a resonant MSA its resonance frequency can be tuned in about $\pm$ 20\% range using GaAs cryogenic varactors. 

\begin{figure}[t]
\centering
\includegraphics[width=1.0\textwidth]{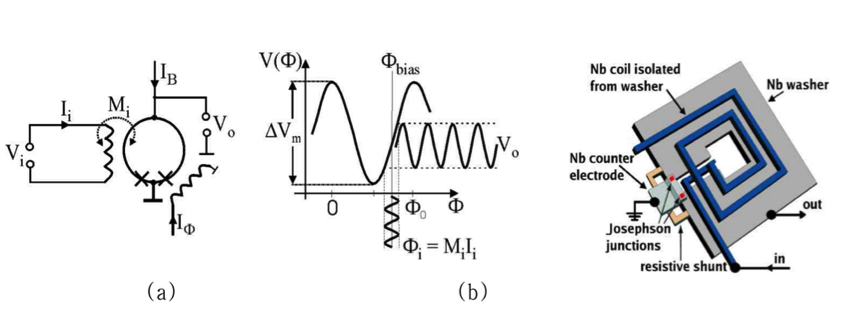}
\caption{Principle of signal detection using an SQUID. (a) Schematic diagram of the SQUID. Input signal ($V_i$) is coupled into the SQUID via input coil with a mutual inductance $M_i$ with the SQUID. (b) Change of the SQUID output voltage versus the input flux. The input signal ($\Phi_i$) is converted into the output voltage signal ($V_o$). 
Basic structure of the microstrip SQUID amplifier (MSA) for measuring input signal~\cite{bib:MSA, bib:MSA_ADMX}. The tight-coupling of input coil is useful when the parasitic capacitance between the input coil and the SQUID washer is not very high, which is the case for the frequencies below about 10\,MHz. At higher frequencies this parasitic capacitance can be used to form a resonate circuit together with the input coil inductance.}
\label{fig:squid_principle}
\end{figure}

Usually SQUID sensors are designed for a low-frequency measurement, say, below 10\,MHz. 
To connect a SQUID with a source, an input coil is ordinarily introduced (Fig.~\ref{fig:squid_principle}), which has a high inductive coupling with the SQUID washer. 
At higher frequencies the parasitic capacitance between the input coil and the SQUID washer increases its influence and reduces the SQUID gain.
In 1998 M. M\"{u}ck and J. Clark proposed a design with a microstrip input coil, where the input coil forms a microstrip transmission line with the SQUID loop (Fig.~\ref{fig:squid_principle})~\cite{bib:MSA}. 
In this design, the input coil is open and the input signal is applied between one of its ends and one terminal of the SQUID washer. 
So, the parasitic capacitance is used to form a microstrip resonant circuit. 
Such MSAs may reach a power gain up to 25\,dB and a noise temperature near to the quantum limit.
Some of MSA's noise limitations come from an energy dissipation in Josephson junction resistive shuts, which causes heating effect that can increase Johnson noise. 
To minimize this effect the shunts should be very well thermalized at working temperatures below 1\,K. 

Although microstrip SQUID amplifiers concept and design have been proposed back in 1998, such amplifiers have not become commercially available even for 20 years. 
A company ``ezSQUID" claims commercial manufacturing of MSAs, although our preliminary tests of five non-commercial and two commercial MSA samples showed that they do not comply with our specifications in terms of frequency band, gain, stability and reliability. 
This is why the SQUID team has concentrated many efforts on development of new types of low noise SQUID-based microwave amplifiers since 2014.
This work was conducted in collaboration with KRISS in 2014$-$2016 and later with IPHT in 2017$-$2018.    

A review of the current status of MSAs including noise temperature, input and output impedance matching, and so forth, is discussed in Ref.~\cite{bib:MSA_review}. 
Mostly all known high gain MSAs have a narrow resonant frequency bandwidth.
This is why an experiment requires serial replacement of SQUID preamplifiers in order to scan a required frequency range.
This procedure is usually very complex and time consuming because every time a large mass of hardware should be warmed up to room temperature and then cooled down below 100\,mK. 
This problem can be resolved by designing a wideband microwave SQUID amplifier. 
We developed a few versions of wideband microwave SQUID amplifiers for $0.5-5$\,GHz range in 2017$-$2018 in cooperation with the Leibniz Institute of Photonic Technology (IPHT) in Jena, Germany.

\begin{figure}[t]
\centering
\includegraphics[width=0.9\textwidth]{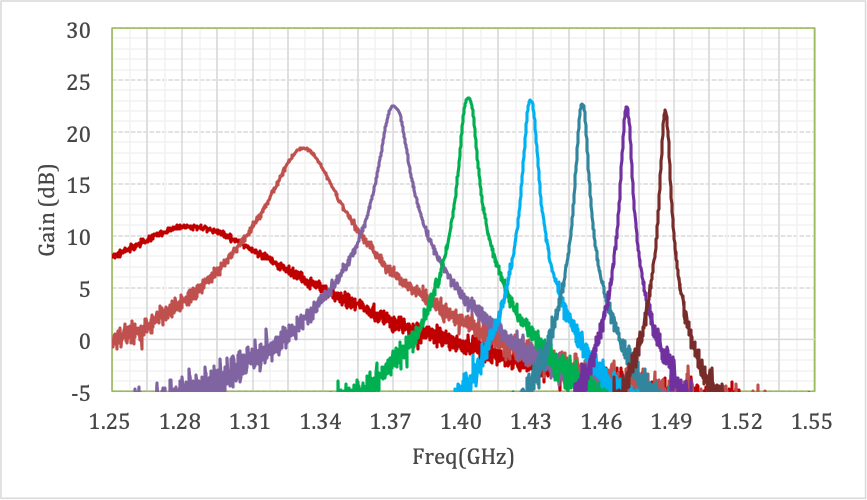}
\caption{Gain vs. frequency for the resonant MSA developed in UC Berkeley. The resonant frequency can be tuned using a varactor connected between the SQUID input coil to the main ground. Eight plots corresponds to different DC voltage on the varactor from 5\,V (red) to 12\,V (brown) with 1\,V step.}
\label{fig:berkeley_msa}
\end{figure}

IBS/CAPP also collaborated with John Clarke’s group at UC Berkeley. In 2017 and 2018 we tested one sample of the resonant MSAs designed and built by Sean O'Kelly. Although this MSA has a very narrow bandwidth, less than 10\,MHz, its resonant frequency can be tuned over a pretty wide frequency range from 1.3 to 1.5\,GHz using an embedded series of varactors controlled by an external DC voltage. The gain vs. frequency plot in Fig.~\ref{fig:berkeley_msa} shows the resonant frequencies at different DC voltage on a varactor that is connected between the output lead of the SQUID input coil and the ground. This resonant MSA has a small working bias current only about 5\,$\mu$A that makes it easy to thermalize at a mK temperature. The gain exceeds 20\,dB, meaning that this MSA can reach very low noise temperatures below 0.1\,K. This MSA will be tested by IBS/CAPP axion search experiments in 2019.

Josephson parametric amplifier's operation is based on non-linear properties of the Josephson tunnel junctions imbedded in a quarter-wave coplanar waveguide (CPW) resonator.
CPW central conductor is formed by a long array of non-shunted Josephson junctions (JJ).
An externally applied pumping microwave signal changes the kinetic inductance of the JJ array. 
When the pumping signal is close to the CPW resonance frequency, it becomes mixing with the input signal and creates side-band frequencies. 
The input signal amplification is a result of the mixing between the pumping and the input signals via the Josephson junction non-liner inductance modulation. 
The JPA can achieve a nearly quantum limited noise. 

The JJ array can be replaced with an array of SQUIDs. 
Each SQUID is formed by two Josephson junctions embedded in a small loop. 
An externally applied DC magnetic field changes the critical current of each SQUID in the array which in turn changes the CPW resonance frequency. 
This allows to make a JPA with tunable resonance frequency by a factor of about two, say, from 4 to 8\,GHz~\cite{bib:JPA}. 
One sample of such JPA with central frequency 8\,GHz was tested at IBS/CAPP in 2018 at 30\,mK.
There are a few different types of JPA with a power gain up to 30 dB within a narrow frequency range. 
There are no commercial JPAs on the market.

\subsection{Development MSA at KRISS}
Lack of commercial SQUID-based amplifiers for 1$-$10 GHz frequencies forced IBS/CAPP to start ``in house" MSA development in collaboration with Korean Research Institute of Standards and Science (KRISS, Korea) in 2014$-$2016 and later with Leibniz Institute of Photonic Technology (IPHT, Germany) in 2017$-$2018. 
KRISS has had complete technology to produce superconducting structures and tunnel Josephson junction, so called tri-layer technology (Fig.~\ref{fig:kriss_msa_structure}), with a minimum size of $2\times2\,\mu$m$^2$. 
Two superconducting thin films are separated by a dielectric layer of thickness $d$ and dielectric constant $\epsilon$. 
The multi-turn input coil has a linewidth $w$. 
Assuming that two superconducting films are much thicker than the London penetration depth of the superconductor $\lambda$ and $w\gg d$, the capacitance and inductance per unit length of the microstrip are respectively given as 
\begin{align}
C_0&=\epsilon\epsilon_0/d\; {\rm[F/m]}\\
L_0&=\frac{\mu_0d}{\omega}(1+2\frac{\lambda}{d})\; {\rm[H/m]},
\end{align}
where, $\epsilon_0$ and $\mu_0$ are permittivity and permeability in vacuum. 
The velocity of an electromagnetic wave $c^*$ in the microstrip is then given by
\begin{equation}
c^*=\frac{1}{2}\frac{1}{L_0C_0}=\frac{c}{[\epsilon(1+2\frac{\lambda}{d})]^{1/2}},
\end{equation}
where, $c$ is the speed of light in vacuum. Finally the characteristic line impedance of the microstrip is given by
\begin{equation}
Z_0=\frac{1}{2}(L_0/C_0)=\frac{d}{\omega}[\mu_0(1+\frac{2\lambda}{d})/\epsilon\epsilon_0]^{1/2}.
\end{equation}

When the microstrip has length $l$, and its ends are open with source resistance $(R_i>Z_0)$, it becomes a half-wave resonator with its wavelength given by the relation, $l=\lambda/2$. 
And the quality factor of the resonator is $Q=\pi R_i/2Z_0$. Since $Z_0$ can be made to be smaller than $R_i$, $Q$ can be larger than 1. 
Then, the resonance frequency $f_0(L_0)$ of the microstrip is given by,
\begin{equation}
f_0(L_0) = c^*/\lambda = c^*/2l = c/2l[\epsilon(1+2\frac{\lambda}{d})]^{1/2}.
\label{eq:resonant_frequency}
\end{equation}

For our band the resonance frequencies became smaller than one get from Eq.~(\ref{eq:resonant_frequency}) because of parasitic inductance. 
The MSA gain experimentally is given by:
\begin{equation}
T_N\sim f_0T/V_{\Phi},
\end{equation}
where, $T$ is the bath temperature. To have lower noise temperature, larger $V_{\Phi}$ is needed as in the amplifier gain. Physically, cooling the MSA device to as low temperature as possible is needed to achieve lower system noise temperature~\cite{bib:hot_electron_effect}.

\noindent\begin{minipage}[t]{0.48\textwidth}
\centering
\includegraphics[width=1.0\textwidth]{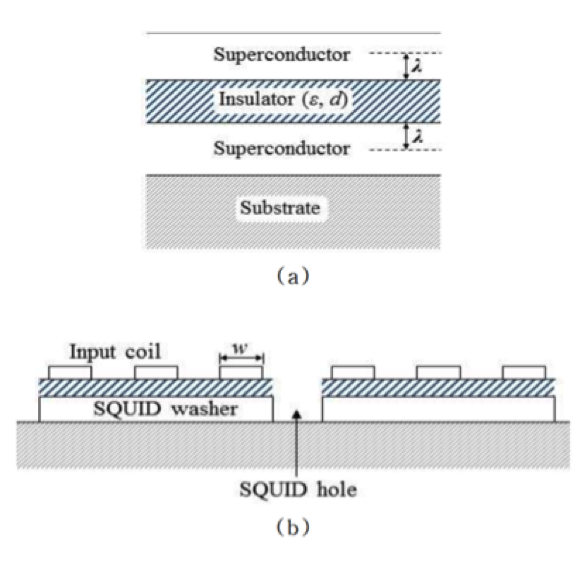}
\captionof{figure}{Superconducting structures of the MSA produced in KRISS: Cross-sectional view of the microstrip line. (a) Side view showing the vertical structure in which two superconductors are separated by an insulator, and (b) end view showing layout of the input coil on the SQUID washer.}
\label{fig:kriss_msa_structure}
\end{minipage}
\hspace{0.02\textwidth}
\begin{minipage}[t]{0.48\textwidth}
\centering
\includegraphics[width=1.0\textwidth]{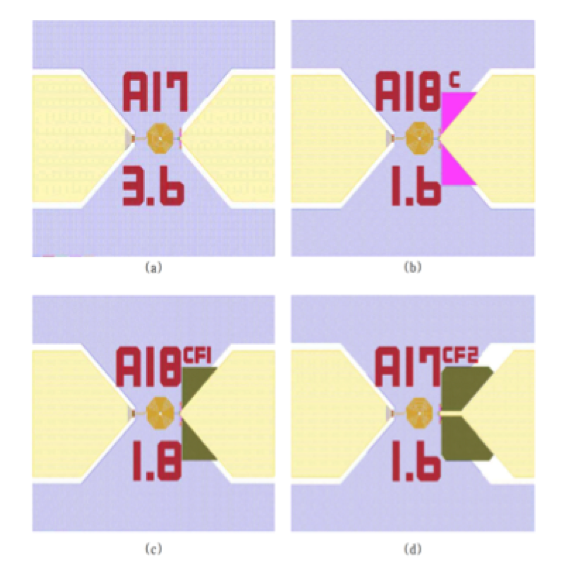}
\captionof{figure}{SQUID designs: (a) SQUID without cooling fin, (b) SQUID with small cooling fins made of 35\,nm Pd, (c) smaller cooling fin CF1 with 500\,nm thick Pd, and (d) largest cooling fin CF2 with 500\,nm Pd.}
\label{fig:kriss_squid_design}
\end{minipage}

In KRISS, SQUID chips were fabricated using optical lithography and multilayer thin film process. 
As a superconductor niobium (Nb) thin films are used with critical temperature about 9\,K. Such SQUIDs can operate at 4.2\,K, the temperature of liquid helium (LHe). 
Fabricated Nb Josephson junctions have low sub-gap leakage current and can be used to make low-noise SQUIDs. 
Especially, Nb is refractory metal and physically very stable against repeated thermal cycles between room temperature and low temperature. 
Furthermore, multilayer integration using semiconductor fabrication process is possible. 
The fabrication process consists of 5 depositions (Nb/AlO$_x$/Nb, SiO$_2$-1, SiO$_2$-2, Pd and Nb), 1 dry etching (Josephson junction), 4 lift-off processes, and 5 photo-lithography processes. 
A few designs of the MSA have been developed and tested in a collaboration with KRISS. 
The MSAs have a magnetometer loop configuration. 
To understand shunt thermalization some of the design have cooling fins to sink the Johnson heat from the shunt resistors into a silicon chip substrate (Fig.~\ref{fig:kriss_squid_design}).

\begin{figure}[h]
\centering
\includegraphics[width=1.0\textwidth]{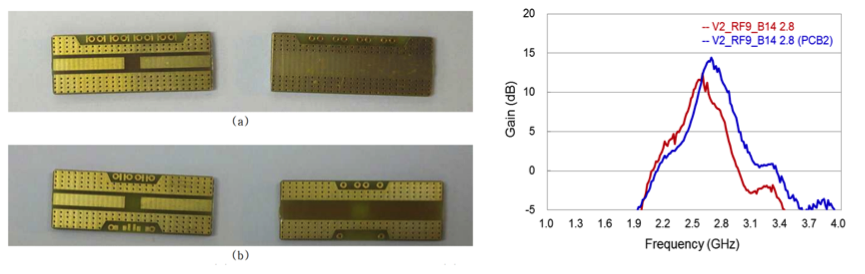}
\caption{(Left) Two PCB types - (a) Design used until Dec. 2016 and (b) new design from Dec. 2016. (Right) Gain curves depending on the PCB type.}
\label{fig:kriss_pcb}
\end{figure}

To connect the MSA in electronic circuit, special PCBs were developed. 
The MSAs gain curves show some ripples or satellite resonance peaks, possibly due to reflection or impedance mismatching of the microwave signal at the connectors, wiring pads and wires. 
We designed a new PCB board having pads for resistors or inductors in the bias current, which may result in reflection of microwave signal at the bias current pads and reducing the loss. Figure~\ref{fig:kriss_pcb} compares the PCB pattern used before and new design for the further measurements. 
For one SQUID, we measured the gain in one PCB and remounted on the other PCB and repeated the gain measurement. 
Figure~\ref{fig:kriss_pcb} also shows the gain difference between them. 
We can see that the peak gain and bandwidth increases slightly by using the new PCB design. 
In some other SQUIDs, the new PCB showed similar behaviours, that is, improved the gain and bandwidth.

The entire design of the amplifier with bounded MSA chip and PCBs with SMA connectors is shown in Fig.~\ref{fig:kriss_msa_amp}. 
Figure~\ref{fig:kriss_msa_gain} shows the frequency dependences of gain for 4 different amplifiers. 
As usual for MSA they have a narrow band typically 150\,MHz and a maximum power amplification of 14$-$17\,dB.

\noindent\begin{minipage}[h]{0.48\textwidth}
\centering
\includegraphics[width=1.0\textwidth]{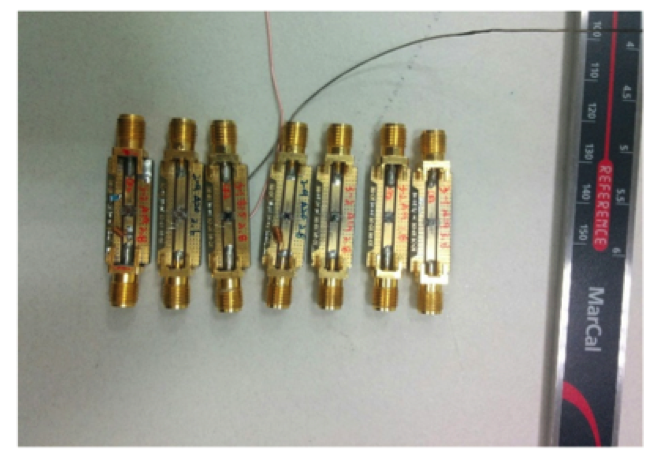}
\captionof{figure}{The MSA amplifiers developed in a collaboration with KRISS.}
\label{fig:kriss_msa_amp}
\end{minipage}
\hspace{0.02\textwidth}
\begin{minipage}[h]{0.48\textwidth}
\centering
\includegraphics[width=1.0\textwidth]{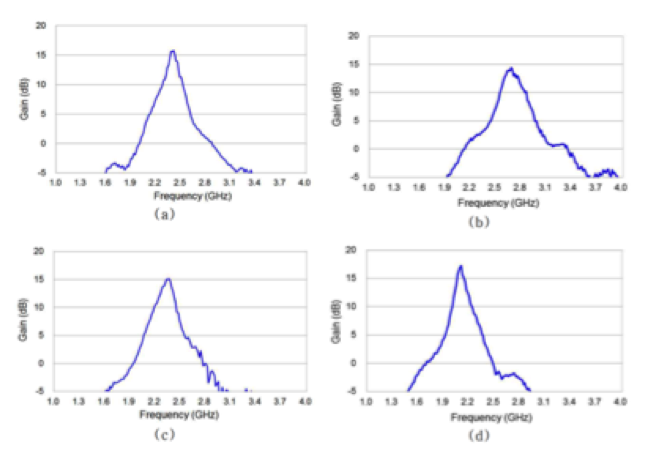}
\captionof{figure}{Some gain curves having gain about 15\,dB for amplifiers: (a) V2\_RF9\_A19 2.8, (b) V2\_RF9\_B14 2.8 (PCB2), (c) V2\_RF9\_B15 2.6 (PCB2), and (d) V2\_RF9\_A20 2.8 (PCB2).}
\label{fig:kriss_msa_gain}
\end{minipage}

During three years of development, 2014$-$2016, KRISS has designed and manufactured a variety of MSAs for IBS/CAPP based on a single-loop SQUID configuration. 
More than a hundred of them have been tested at IBS/CAPP in 2017$-$2018 using a dunk probe immersed in a LHe transport dewar at $T = 4.2$\,K. 
These MSAs have about 100$-$150\,MHz bandwidth and resonant frequencies from 2.2 to 2.45\,GHz. Our current inventory includes 25 functional MSAs with gain up to about 19\,dB.
Table~\ref{tab:kriss_msa_gain} lists the working frequency and gain for five of them.

\begin{table}[h]
\centering
\begin{tabular}{|c|c|c|}
\hline
Chip & Frequency [GHz] & Gain [dB] \\
\hline
3-2 A 19 2.8 & 2.40 & 18.7 \\
3-1 B 15 2.8 & 2.30 & 15.5 \\
3-2 B 15 2.8 & 2.20 & 14.1 \\
3-2 A 19 2.8 & 2.45 & 13.6 \\
3-2 B 15 2.8 & 2.20 & 12.2 \\
\hline

\end{tabular}
\caption{Central frequency and typical gain of various MSA chips from KRISS.}
\label{tab:kriss_msa_gain}
\end{table}

\subsection{Development wideband MSAs at IPHT}
IPHT has advanced technology with potential of making a submicron size tunnel Josephson junctions with a small parasitic capacitance. 
Because of that, their technology was promising to create a high performance MSA and in 2017 we started to design a prototype MSA in a collaboration with them. 
To get a prototype quickly we used existing masks of the IPHT lab and neglect some of the features which can be implemented later, for instance thermalization cooling fins. 

The main parameters of SQUIDs that determine their characteristics are SQUID loop inductance, $L_0$, a critical current, $I_0$, a capacitance of Josephson junctions, $C_{\rm JJ}$, and shunt resistors, $R_{\rm SH}$.  When a SQUID is biased with proper dc current, its output voltage changes periodically with increasing input magnetic flux through a SQUID loop with a period of the flux quantum, $\Phi_0 \approx 2.07\times10^{-15}$\,Wb. This periodic function is called a voltage-flux characteristic or a $V(\Phi)$ curve. 
For amplification of a small signal the crucial parameter is the steepness of $V(\Phi)$ curve called transfer function $dV/d\Phi$ at the working point. 
The power amplification is proportional to $(dV/d\Phi)^2$. 
The maximal value of the transfer function is given by $dV/d\Phi \approx R_{SH}/2L_0$ and could be large proportionally to $R_{\rm SH}$. 
However, having stable non-hysteretic $V(\Phi)$ curve characteristic requires the McCumber-Stewart parameter $\beta_C = 2\pi I_0R_{\rm SH}^2C_{\rm JJ}/\Phi_0 < 1$, which limits the maximum value of $R_{\rm SH}$. Moreover, as $L_0$ determines the coupling of the input signal to the SQUID, the most elegant way to increase the transfer function is to decrease the Josephson junction capacitance $C_{\rm JJ}$. 

Accordingly, the frequency characteristics of a SQUID should also be improved. Initially, the dc voltage across the SQUID, $V_{\rm DC}$, is a mean value of the voltage pulses train with the Josephson frequency $f_{\rm J}$: $V_{\rm DC} = f_{\rm J} \Phi_0$. 
The maximum of this frequency corresponds to the maximum $V_0 \approx I_0R_{\rm SH}$ and should be as high as possible. 
The other characteristic frequencies are determined by the time constants $R_{\rm SH}C_{\rm JJ} $and $L_0/2R_{\rm SH}$, which also should be as high as possible. 
These frequencies limit the maximum frequency of the signal that can be amplified by a SQUID. 
All these frequencies are approximately equal if the SQUID is optimized: $\beta_C = 1$ and $\beta_L = 2I_0L_0/\Phi_0 = 1$. 
The latter condition allows gaining a large $V_0$ and corresponding $dV/d\Phi$, if the critical current $I_0$ is high for a SQUID with low inductance. 
Hence, for suitable amplification at high frequency the SQUID should be designed with the smallest possible tunnel junction capacitance, $C_{\rm JJ}$, with reasonably low SQUID loop inductance, $L_0$, and maximal transfer function, $dV/d\Phi$, at the working point.  

\begin{figure}[h]
\floatbox[{\capbeside\thisfloatsetup{capbesideposition={right,top},capbesidewidth=0.5\textwidth}}]{figure}[\FBwidth]
{\caption{The scanning electron microscope image of two niobium cross-type Josephson tunnel junctions with an area $0.6\times0.6\,\mu$m$^2$. Each Josephson junction has a capacitance of about 0.04\,pF. Josephson junctions fabricated using conventional window-type technology have a typical capacitance of 0.4$-$0.8\,pF.}
\label{fig:ipht_semi}}
{\includegraphics[width=0.45\textwidth]{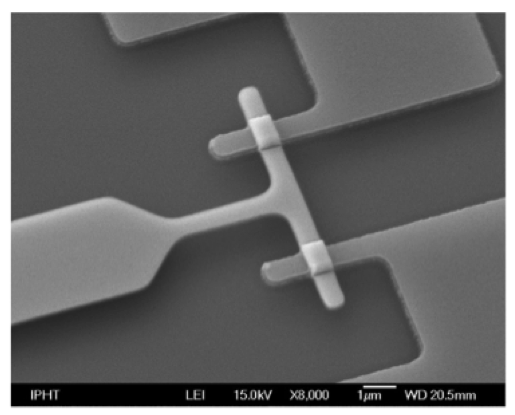}}
\end{figure}

Sub-micron size Josephson junctions with very low capacitance were designed at IPHT and used for low-frequency SQUID sensors applications. 
These cross-type Josephson junctions have a very low capacitance of about 0.04\,pF. This is more than 10 times smaller than the capacitance of Josephson junctions fabricated using conventional window-type technology. 
Fig~\ref{fig:ipht_semi} shows the scanning electron microscope image of two niobium cross-type Josephson junctions with an area $0.6\times0.6\,\mu$m$^2$.  
These types of cross-type Josephson junctions were used for fabrication of new families of SQUID current sensors. 
As we experimentally found out, some of these new SQUID current sensors work very well as microwave both resonant and wideband amplifiers. 

We chose SQUID current sensors with the total SQUID loop inductance of 30\,pH to be designed as a wideband microwave SQUID amplifier (MSA). 
The SQUID loop is built with two washers connected in parallel forming a gradiometer configuration. 
There were three MSA designs with 6-turn, 7-turn and 12-turn input coils. 
The input coil with 12 turns on each washer has a total input coil inductance 11.5\,nH and mutual inductance to the SQUID loop of about 0.6\,nH. 
Due to a relatively short input coil length (8.2\,mm) the estimated microstrip resonance should be at about 5$-$7\,GHz. 
The critical current of each Josephson junction and the shunt resistor were $I_0 \approx 35 \,\mu$A and $R_{\rm SH} \approx 8$\,Ohm respectively. 

\begin{figure}[h]
\floatbox[{\capbeside\thisfloatsetup{capbesideposition={left,top},capbesidewidth=0.5\textwidth}}]{figure}[\FBwidth]
{\caption{Voltage-flux characteristics of the investigated SQUID current sensor. The curves are plotted for the SQUID bias current ranging from 50\,$\mu$A (bottom) to 70\,$\mu$A (top) with 2\,$\mu$A steps. Flux is produced by a current through a one-turn coil placed above one of the SQUID loops. The voltage-flux characteristics has asymmetric shape with the left side steeper than the right side and the steepest transfer function exceeds 2 mV/ $\Phi_0$.}
\label{fig:ipht_vphi}}
{\includegraphics[width=0.5\textwidth]{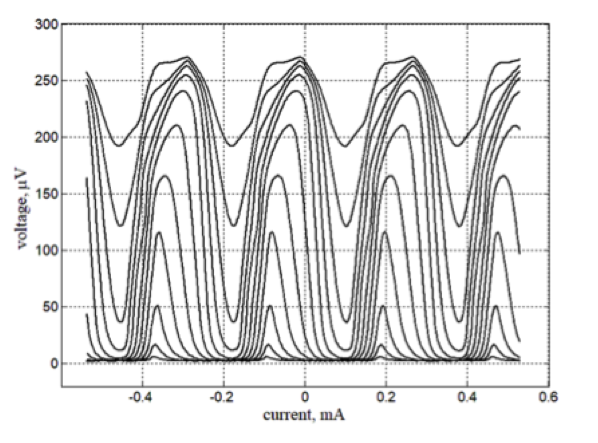}}
\end{figure}

A 1-turn coil was placed above one of the SQUID washers to apply DC bias flux. 
An example of the voltage-flux characteristics is shown in Fig~\ref{fig:ipht_vphi} for SQUID bias current changing between 50 $\,\mu$A and 70$\,\mu$A with 2$\,\mu$A steps. 
The maximal voltage swing of $V(\Phi)$ curve is 250$\,\mu$V and the steepest transfer function $dV/d\Phi$ exceeds 2\,mV/$\Phi_0$ at I$_{\rm BIAS}$ = 62$\,\mu$A and V$_{\rm WP} = 140\,\mu$V, where V$_{\rm WP}$ is the working point voltage above zero line.

\begin{figure}[h]
\centering
\includegraphics[width=0.8\textwidth]{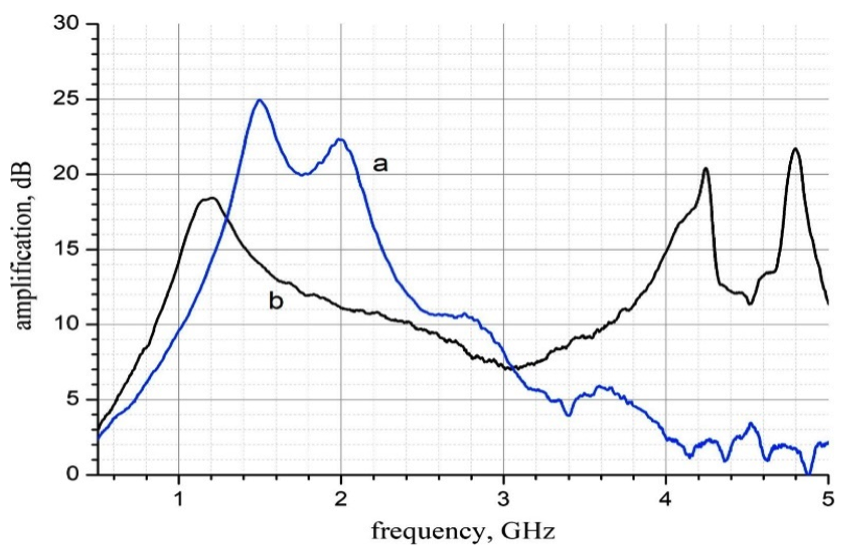}
\caption{Gain vs. frequency plots of the SQUID-based microwave amplifier at two working points positioned on gentle (a) and steep (b) sides of the $V(\Phi)$ curve approximately at 140\,$\mu$V above a zero line~\cite{bib:andrei}.}
\label{fig:ipht_gain_freq}
\end{figure}

We used SQUIDs with 30\,pH loop inductance and 12-turn input coil for preliminary tests as a microwave SQUID amplifier at temperature of 4.2\,K inside a liquid helium transport cryostat. 
The SQUID amplifier output was connected to a network analyzer input through $-$10\,dB cryogenic attenuator and two cascaded room temperature semiconductor amplifiers ZX60-V63+ with a frequency bandwidth of 0.05$-$6.0 GHz and $+$42\,dB gain at 1\,GHz. 
Figure~\ref{fig:ipht_gain_freq} demonstrates the power gain vs. frequency plot for two working points positioned on gentle (a) and steep (b) sides of the $V(\Phi)$ curve. 
Both working points were set at voltage approximately $V_{\rm WP} = 140\,\mu$V that corresponds to about a 67.6\,GHz Josephson frequency. 
Shapes of the SQUID $V(\Phi)$ curves and the working point positions were controlled using a slightly modified commercial direct read-out electronic. 
It allows to check the actual shape of $V(\Phi)$ curves at any bias current and to set a working point by adjusting dc bias flux. 
The highest amplifications at any DC bias current can be achieved by accurate adjustments of the working point position on $V(\Phi)$ curves. 

In September 2017 we developed our first new microwave SQUID amplifiers with high gain in a wide frequency range. 
A single amplifier can be used in a very wide frequency range of approximately from 1 to 3\,GHz with gain above 25\,dB. 
Such amplifiers outperform world analogues by bandwidth and amplification factor. 
Compare with analogues, the amplifier does not have pronounced resonances and that indicates a high potential noise characteristics. 
We tested ten different noncommercial MSA samples from IPHT and six commercial MSA samples from Supracon, spin-off Company from IPHT. 
Four commercial MSA chips were redesigned and assembled specifically for axion search systems at IBS/CAPP. 

After first preliminary successful tests of MSAs in 2017 we have made a decision to modify existing IPHT SQUID-amplifier topology specifically for 1$-$5\,GHz range applications. 
IPHT is a government organization, so we needed to move our business connection to Supracon AG – the commercial spin-off company of IPHT. 
As a result, all further MSAs development and commercial connections went through this company. 

\begin{figure}[h]
\centering
\includegraphics[width=0.9\textwidth]{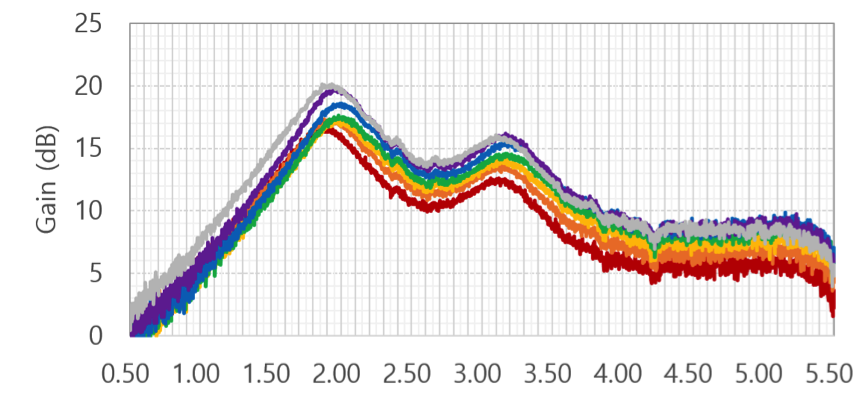}
\caption{Gain vs. frequency for new IPHT MSAs with 6-turn input coil: Seven colored plots correspond to different SQUID bias currents from 84 (red) to 96 (gray)\,$\mu$A with 2\,$\mu$A step, $T=4.2$\,K.}
\label{fig:ipht_gain_freq_6turn}
\end{figure}

In mid of 2018 the Supracon/IPHT experts designed a few new modifications of their SQUID amplifiers orienting toward better performances at GHz range. 
As the results of such efforts, IBS/CAPP was able to buy a few new amplifiers with 6-turn and 12-turn input coils and with embedded the flux-bias coil. 
The new MSA design allowed us significantly approve performance and reliability of such very unique cryogenic amplifiers. 
Figure~\ref{fig:ipht_gain_freq_6turn} shows gain vs. frequency plot demonstration above 10\,dB gain in 0.8$-$3.6\,GHz band.    

\begin{figure}[h]
\centering
\includegraphics[width=0.9\textwidth]{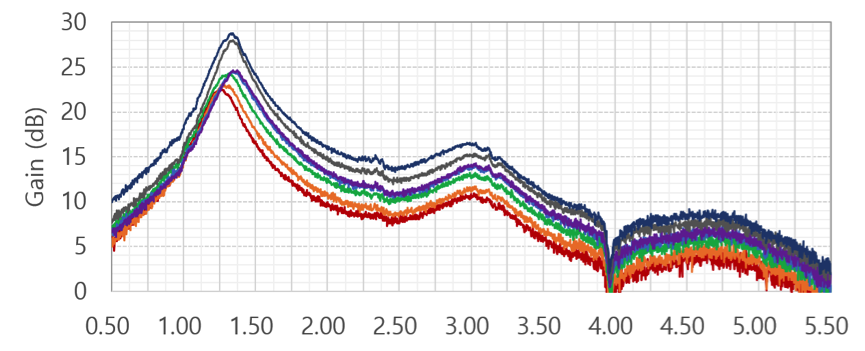}
\caption{Gain vs. frequency for new IPHT MSAs with 12-turn input coil: seven plots correspond to bias currents from 80 (red) to 102 (dark blue) \,$\mu$A, $T = 4.2$\,K. 
As one can see, this MSA demonstrates more than 15\,dB gain in 0.8$-$2.0\,GHz range with 29\,dB gain at about 1.3\,GHz.}
\label{fig:ipht_gain_freq_12turn}
\end{figure}

\subsection{SQUID thermalization modeling}

The first wideband MSA design was made using old photomasks for a quick prototyping. In spite of that, the MSA shows the record amplification factor at wide frequency band. 
An implementation of this MSA in CAPP’s experiments promises to save a lot of manpower and budget because of easier frequency adjustment and reduction of amplification stages. 
At the next design iteration we stress on the MSA noise performances. 
As it is already known physical temperature of the MSA and its components limits amplifier’s noise characteristics. 
The major source of the heat in the MSA is its resistive shunts. 
The shunts are important elements of the MSA as they linearize VAC of the Josephson junctions and make them non-hysteretic. 
The downside of the shunts is their heat dissipation in a range of 1$-$100\,nW. 
It is a significant dissipation for a microscopic device operating at temperature below 1\,K.

\begin{figure}[h]
\floatbox[{\capbeside\thisfloatsetup{capbesideposition={right,top},capbesidewidth=0.5\textwidth}}]{figure}[\FBwidth]
{\caption{Thermal model of the MSA shuts thermalization. $P_{sh}$ – heating power; $T_e^{sh}$ and $C_e^{sh}$ - electron temperature and electron heat capacitance of the shunts; $T_{ph}^{sh}$ and $C_{ph}^{sh}$ - phonon temperature and lattice heat capacitance of the shunts; $T_{ph}^{sub}$ and $C_{ph}^{sub}$ - phonon temperature and phonon heat capacitance of the substrate; $T_{ph}^{pad}$ and $C_{ph}^{pad}$ - phonon temperature and phonon heat capacitance of the contact pads; $T_e^{pad}$ and $C_e^{pad}$ - electron temperature and electron heat capacitance of the contact pads; $T_e^b$ - electron temperature of the temperature bath (MXC). $G_{e-ph}^{sh}$, $G_K^{sh-sub}$, $G_K^{sub-pad}$, $G_{e-ph}^{pad}$, $G_e^{pad-b}$  - shunt’s electron-phonon thermal conductance, Kapitza conductance between shunts and substrate, Kapitza conductance between substrate and contact pads, pad’s electron-phonon thermal conductance, thermal conductance of the bonding wires between pads and MXC, correspondently.}
\label{fig:squid_thermal_model}}
{\includegraphics[width=0.5\textwidth]{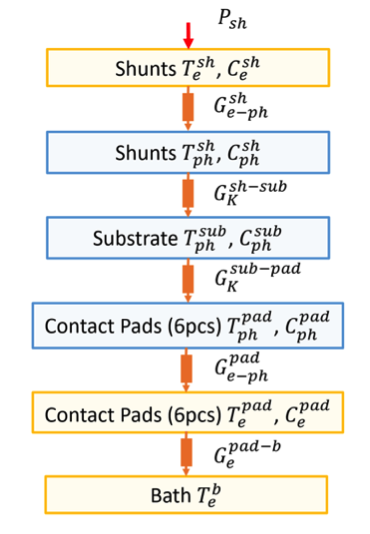}}
\end{figure}

To understand thermalization processes and estimate temperature of the MSA parts we created a thermal model of the MSA. 
The model taking into account heat dissipation in the Josephson junction shunts. 
At low temperature we need to take into consideration specific effect as a low electron-phonon coupling in normal metals and acoustic mismatch on the material interfaces. 
The Johnson dissipation in the shunts P$_{sh}$ heats its electron system with a heat capacitance $C_e^{sh}$ up to temperature $T_e^{sh}$. 
Thermalization of the heating power dissipating in the shunts to the lattice is going through electron-lattice thermal conductance in the shunts volume $G_{e-ph}^{sh}$ and heat the shunt phonon system with heat capacitance $C_{ph}^{sh}$ to a temperature $T_{ph}^{sh}$.  
Further the heat passes to substrate with heat capacitance of $C_{ph}^{sub}$ through a Kapitza resistance of the shunt-substrate interface $G_K^{sh-sub}$ and heat it up to temperature of $T_{ph}^{sub}$. 
The thermalization in the crystal Ge substrate is a very quick process and we can exclude it from our consideration. 
The MSA attached to the non-plated PCB surface and it thermalization through the PCB dielectric is very poor. 
We neglect PCB dielectric heating as it is slower process and it is not effect on the MSA thermalization in the steady state. 
All of thermalization goes through a Kapitza heat conductance $G_K^{sub-pad}$   to 6 contact pads with a common phonon heat capacitance $C_{ph}^{pad}$. The phonon temperature of the pads rises up to $T_{ph}^{pad}$. 
Then the phonon systems of the pads heats their electron systems with heat capacitance $C_{ph}^{pad}$ up to $T_e^{pad}$ through electro-phonon coupling $G_{e-ph}^{pad}$. 
The pads are connected to the MSA enclosure through 12 gold bonding wires (two for each of the pads) with a conductance $G_e^{pad-b}$. 
The MSA is well thermalized to the MXC with a temperature of $T_e^b$. 
The full thermal model (Fig.~\ref{fig:squid_thermal_model}) includes all of that effects. 

\begin{figure}[h]
\centering
\includegraphics[width=1.0\textwidth]{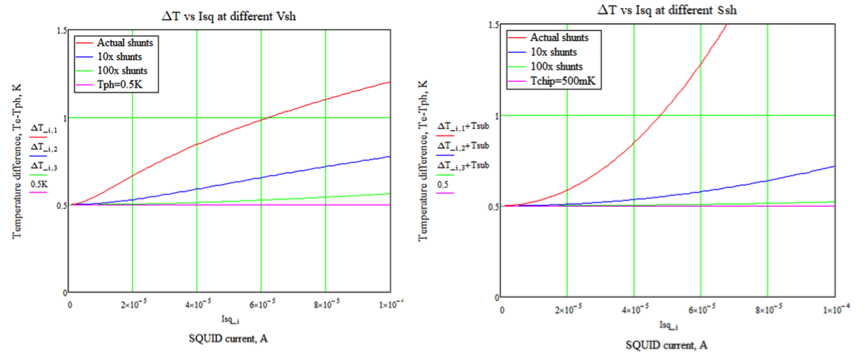}
\caption{Dependence of the temperature difference between electron and phonon systems of the shunt $\Delta T$ from SQUID bias current $I_{sq}$ at the different shunt volume $V_{sh}$. MXC temperature of the fridge is 500\,mK.}
\label{fig:squid_temp_cur}
\end{figure}

We estimated the influence of the each of thermal resistance on the shunts temperature for our current shunts. 
The main problem in thermalization came from poor heat conductance $G_{e-ph}^{sh}$ and $G_K^{sh-sub}$. 
Figure~\ref{fig:squid_temp_cur} shows the dependence of the overheating dues to that conductances. 
As our estimation showed, to reduce the shunt overheating to an acceptable level of 100\,mK, the area and volume (thickness) of the shunts should be increased about in 100 times. 
The current shunts have a small size of 120\,$\mu$m$^2$ and the increasing is not a practical difficulty.  

As the shunts overheating is a main source of the MSA noise, the next iteration will allow us to achieve the lowest possible noise of our MSA. 
Therefore, the modelling determined the limitations associated with low-temperature effects on the cooling of MSA and allows us to provide recommendation for a future MSA design, particularly for shunt area and dimensions of thermalization fins.

\subsection{Test facilities}
To test the MSA characteristics we developed a special testing facilities. 
The facilities were installed in two dry dilution fridges BF5 and BF4 and on a special designed 4\,K probe for a quick tests in a transport dewar. 
Test setups in the dilution fridges use a high frequency line for RF signals. The base temperatures of the fridges are below 20\,mK. 
Therefore we are conducting noise measurements and noise calibrations both at LHe $T = 4.2\,$K and in dilution refrigerators at the temperature below 100\,mK.

\begin{figure}[h]
\caption{Schematics of the SQUID voltmeter. 
The MSA is located at base temperature below 100 mK. Its output is connected to the SQUID input coil through the resistor $R_4 = 1\,\Omega$. 
This SQUID is located at temperature $T = 2\,$K. 
The SQUID feedback is connected to the reference resistor $R_5 = 10\,\Omega$ and compensates to zero the current flows through the SQUID input coil.}
\label{fig:squid_voltmeter}
{\includegraphics[width=0.9\textwidth]{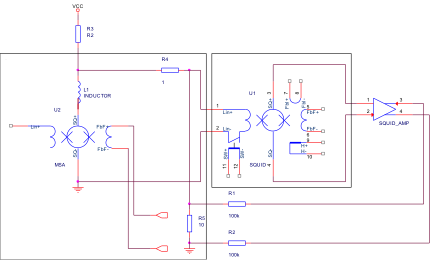}}
\end{figure}

\begin{figure}[h]
\centering
\includegraphics[width=0.6\textwidth]{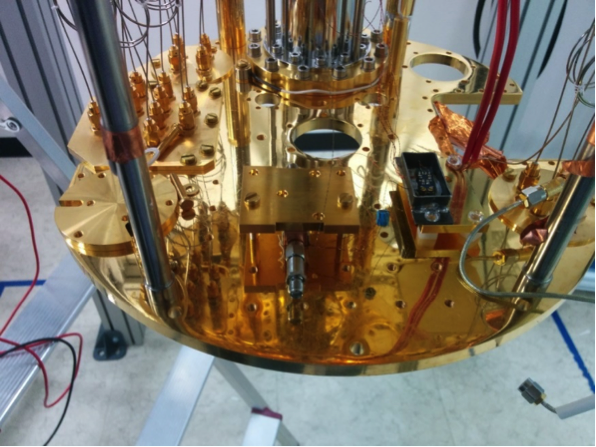}
\caption{Test facility in the BF4 fridge. On the mixer chamber plate a pico-voltmeter is installed (a grey box from the right side). 
The pico-voltmeter is placed in a lead shield (the cover is open), which is fixed on a thermally isolated plate. The plate is thermalized to the 4\,K fridge stage. 
This configuration allows minimizing connections between DUT (device under test) and the pico-voltmeter. }
\label{fig:squid_test_facility}
\end{figure}

A low frequency SQUID pico-voltmeter is installed in the BF4 fridge (Fig.~\ref{fig:squid_voltmeter}). 
The pico-voltmeter SQUIDs operates at the temperature about 2\,K and is mounted on MXC plate through a thermal isolated parts. 
The pico-voltmeter SQUID platforms thermalized to the 2\,K stage by means of two thick wires (red wires on the Fig.~\ref{fig:squid_test_facility}). 
The pico-voltmeter allows to measure temperature of shunt by measurements of their Johnson noise. 
The shunt overheating is a dominated source of noise in the MSA and the pico-voltmeter is an important tool for the MSA noise investigation at low frequency. 
It allows to estimate actual physical temperature of the shunt resistors. 

As it can be seen from the picture, the MSA is connected with a twisted pair wires with a length about 10\,cm. It has a specially designed filtered low frequency lines which are used for power and control signal (Fig.~\ref{fig:squid_test_facility}). 
The filters PCB is designed to provide a maximum thermalization of the filter components to the 4K fridge stage. The filter components tested to warranty their performance in the temperature range of mKs to 300\,K temperature range.

There are three key problems, which we are currently solving before MSAs will be installed inside BlueFors dilution refrigerators and used for routine axion search experiments with noise temperature well below 1\,K. 
First, we are working on development of experimental setups and procedures for accurate measurements of MSA noise temperature in temperature range from 4\,K down to 30\,mK. 
Second, we are developing an optimal schematic for microwave circuits and test all components at 4.2\,K before they will be installed inside dilution refrigerators. 
Third, we are designing the effective shields and filters to suppress both ambient DC field and all kinds of radio frequency interferences (RFI). 
Our final goal is significant noise temperature decreasing of the microwave amplifiers used for axion search experiments. 
One our option is finding MSAs with noise temperature significantly below 1\,K at frequency range 0.5$-$5.0\,GHz. 
Our 2018 inventory includes three types of functional MSAs, which were tested at 4.2\,K. 
Three IPHT MSAs were tested at below 100\,mK temperature and one of them were used for preliminary noise temperature measurements at 35\,mK inside BF5 dilution refrigerator. Noise temperature measurements in GHz frequency range inside dilution refrigerator is not a trivial technically and extremely time consuming procedure. 
One thermal cycle takes about one week. This is why we need to move some preliminary measurement to liquid helium probes and also to the second dilution refrigerator BF4. 
We are also developing a new experimental setup to make noise measurements at low frequencies, 1.0$-$10\,kHz, that will allow us to estimate physical temperature of Josephson junctions (JJ) shunt resistors at different working points on VФ curves and at different bias currents. 
Knowing JJ shunts temperature will allow us to develop improved thermalization technique for MSA silicon chips. Therefore, our current testing facilities allow:

\begin{itemize}
\item Measure noise temperature of all available amplifiers at 4.2\,K with LHe probe. 
\item Develop low frequency noise measurement setup using SQUID-based pV-meter.
\item Measure low frequency noise of amplifiers inside BF4 and BF5 at $T < 1\,$K.
\item Measure noise temperature of amplifiers at GHz range in BF4 at $T < 1\,$K.
\item Optimize working point and operation temperature of amplifiers. 
\end{itemize}

\subsection{Microwave SQUID amplifier shielding}
The MSA is very sensitive to magnetic field. Its operation in a close distance to a multi-Tesla magnet is a challenge and to make it possible we need to take a special measure. 
CAPP’s magnets have a special compensation coils which allow to reduce the field down to 20$-$100\,mT in the MXC area. 
The IPHT design of the MSA has a gradiometer configuration and allows to reduce the field sensitivity in hundredth times. 
The KRISS design of the MSA has magnetometer configuration and it is more sensitive to the magnetic field. 
Moreover, the JJ itself are sensitive to the high magnetic field and their properties can be dramatically change near to the magnet. 
For a further reduction of the magnetic field we started to develop a magnetic shield around MSA. We decide to combine a mu-metal shield together with a superconducting shield. 
The mu-metal shield should reduce the field down to an acceptable level to provide an environment during superconducting transition of the MSA. 
The superconducting shield provides a further field reduction and secure the environment against field change in the time of amplifier operation. 

\begin{figure}[h]
\centering
\includegraphics[width=1.0\textwidth]{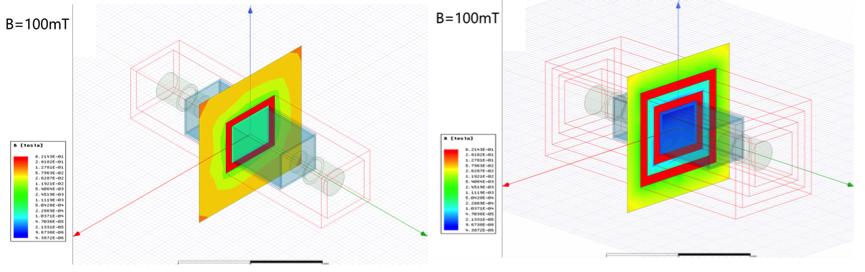}
\caption{ Modelling of the single-layer (left) and double-layer (right) mu-metal shields for the MSA placed in the magnet environment with a field of $B=100\,$mT. 
The shield residual field with a material thickness of 2\,mm is about 500\,$\mu$T and 10\,$\mu$T for the single-layer and double-layer shields respectively.}
\label{fig:squid_shielding}
\end{figure}

As our simulation showed, a simple one-layer mu-metal shield is insufficient to get an acceptable magnetic environment for the amplifier (Fig.~\ref{fig:squid_shielding}). 
The residual field of 2 mm thick one layer shield showed in Fig.~\ref{fig:squid_shielding} is about 500\,$\mu$T. 
The field is in an order of magnitude higher than the Earth field and is too big for the MSA operation. 
To get an acceptable field value we need to use at least a double-layer shield and test a MSA in the fridge with a running magnet. 
The simulation of the double layer shield showed that we can achieve the residual field in the MSA area about 10\,$\mu$T (Fig.~\ref{fig:squid_shielding}). 
Our experiments shows that with a superconducting lead shield the MSA can operates in Earth field of 70\,$\mu$T. 
Therefore, the combination of a two-layer mu-metal shield and a superconducting shield should be sufficient for MSA operation in the magnet vicinity with a field of 100\,mT. 

%% file: 1.4.2/main.tex
Haloscope axion dark matter search experiments conventionally employ cylindrical microwave resonant cavities immersed in a solenoid magnetic field. 
Exploring higher frequency regions requires a smaller cavity size as the frequency of the resonant mode of our interest, TM$_{010}$, scales inversely with the cavity radius. 
One of the intuitive ways to make a maximal use of a given magnet volume, and thereby to increase the experimental sensitivity, is to bundle multiple cavities together and combine the individual signals coherently~\cite{bib:multicavity_the}. 
An experimental attempt to implement this concept was made by ADMX in 2000~\cite{bib:multicavity_exp}, but its methodological advantages were not fully addressed because the reliability and increased complexity of operation were significant factors.
Herein, the YS program has been dedicated to develop cavity designs and tuning mechanisms that are applicable to axion search experiments in higher frequency regions. 
The methodological approaches are in two folds: a) designing a realistically feasible phase-matching mechanism for multiple-cavity systems~\cite{bib:multiple_cavity}; and b) developing a new concept of cavity design, as know as multiple-cell cavity, for more efficient detection~\cite{bib:pizza_cavity}.
Research efforts on these methods at CAPP are described in details in the following subsections.

\subsection{Phase-matching of multiple-cavity systems}
\label{sec:multi-cavity}

\subsubsection{Introduction}
As the first approach to effectively increase the axion sensitivity at high frequencies, the YS project performs an extensive study of the conceptual design of a phase-matching mechanism for an array of multiple cavities. 
We consider three possible configurations in designing the receiver chain for a multiple($N$)-cavity system as summarized in Table~\ref{tab:multicavity_config}. 
Configuration 1 comprises $N$ single-cavity experiments, consisting of a complete receiver chains per cavity, where the signals are statistically combined after all, eventually resulting in a $\sqrt{N}$ improvement in sensitivity. 
The other two configurations introduce a power combiner at an early stage of the receiver chain to build an $N$-cavity experiment - one with the first stage amplification taking place prior to the signal combination; and the other with the signal combination preceding the first stage amplification. 
Configuration 2 is characterized by $N$ amplifiers and a combiner, while configuration 3 is characterized by a signal combiner followed by a single amplifier. Assuming the axion signal from individual cavities is correlated while the noise from the system components is uncorrelated, configuration 2 gains an additional $\sqrt{N}$ improvement yielding the highest sensitivity. 
On the other hand, configuration 3 provides a slight low sensitivity de to insertion of the combiner before the amplifier, but it supplies the simplest structure in the receiver chain. 
As simpler design is significantly beneficial especially for large cavity multiplicities, configuration 3 is chosen as the final design.

\begin{table}[h]
\begin{center}
\begin{tabular*}{0.85\textwidth}{@{\extracolsep{\fill}}cccc}
\hline
Configuration &1 & 2 & 3 \\
\hline
\raisebox{1.5\height}{Schematic}
& \raisebox{-.07\height}{\includegraphics[width=0.15\textwidth, height=0.09\textwidth]{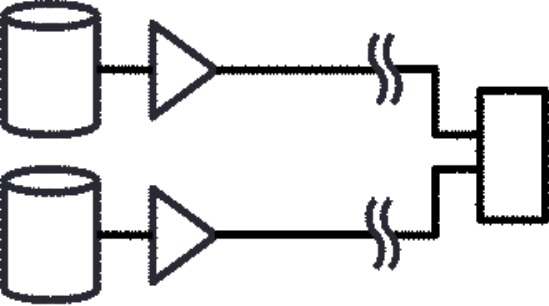}} 
& \raisebox{-.07\height}{\includegraphics[width=0.15\textwidth, height=0.09\textwidth]{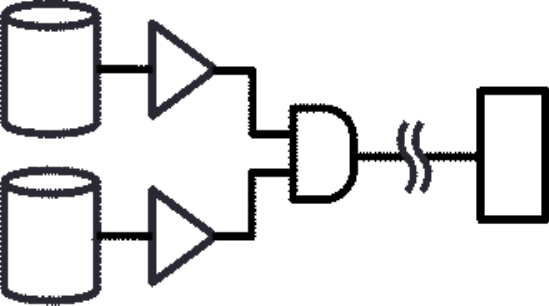}}
& \raisebox{-.07\height}{\includegraphics[width=0.15\textwidth, height=0.09\textwidth]{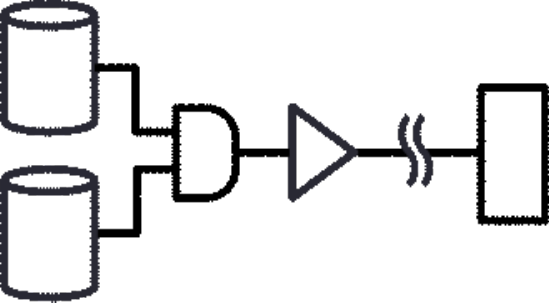}} \\
\multirow{2}{*}{Characteristics} & $N$ complete & $N$ amplifiers & 1 amplifier \\
& chains & 1 combiner & 1 combiner \\ 
Sensitivity & $\sqrt N\cdot\rm{SNR_{sgl}}$ & $N\cdot \rm{SNR_{sgl}}$ & $N\cdot\rm{SNR_{sgl}}$ \\
\multirow{2}{*}{Pros} & Accessibility to & \multirow{2}{*}{Highest sensitivity} & \multirow{2}{*}{Simplest design} \\
& individual cavities & & \\
\multirow{2}{*}{Cons} & Low sensitivity & \multirow{2}{*}{$N$ amplifiers} & \multirow{2}{*}{SNR$_3$ $\lesssim$ SNR$_2$\footnotemark} \\
& Complex design & & \\
\hline
\end{tabular*}
\end{center}
\caption{Possible configurations of the receiver chain for a multiple($N$)-cavity system. 
The cylinders, triangles, and D-shaped figures represent cavities, amplifiers, and combiners, respectively. 
$\rm{SNR_{sgl}}$ refers to the signal-to-noise ratio (SNR) of a single-cavity experiment.
The gain of the amplifiers is assumed to be large enough.}
\label{tab:multicavity_config}
\footnotetext{Configuration 3, where the combiner is placed prior to the first amplifier, degrades the SNR due to imperfection of the combiner. For instance, a system with a combiner with a noise figure of 0.5 and a amplifier with a gain of 12 and a noise figure of 6 yields a SNR reduction of $\sim10$\%.}
\end{table}

\subsubsection{Phase-matching and frequency matching tolerance}
\begin{figure}[h]
\includegraphics[width=1.0\textwidth]{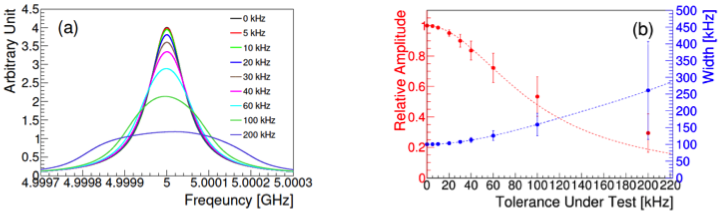}
\caption{(a) Combined power spectra averaged over 1,000 pseudo-experiments for several tolerances under test (TUT). 
The power amplitude of each cavity is normalized to unity. (b) Relative power amplitude in red and full width at half maximum in blue as a function of TUT. 
The error bars represent the statistical uncertainties. These distributions are fitted with the Lorentzian and forth order polynomial functions, respectively.}
\label{fig:multicavity_fmt}
\end{figure}

Due to the large de Broglie wavelength of the coherent axion field and relatively facile achievement of constructive interference in signal combination, the phase-matching of a multiple-cavity system is in practice equivalent to frequency tuning of individual cavities to the same resonant frequency, which is referred to as frequency-matching. 
Realistically, however, an ideal frequency-matching of multiple cavities is not possible mainly because of the machining tolerance of cavity fabrication (typically $< 50\,\mu$m) and nonzero step size of the turning system with a typical size of 1\,m$\degree$. 
Instead, a practical approach is to permit frequency mismatch up to a certain level at which the combined power is still sufficiently enough that the resulting sensitivity is not significantly degraded. We refer to the certain level as the frequency matching tolerance (FMT).
To determine FMT for multiple-cavity systems, a pseudo-experiment study is performed using a quadruple-cavity detector searching for 5\,GHz axion signal. 
The unloaded quality factors of all cavities are assumed to be the same as $Q_0 = 10^5$. 
Supposing $Q_a \gg Q_0$, the signal power spectrum follows the Lorentzian distribution with its mean of 5\,GHz and half width of 50\,kHz. Several values of frequency matching tolerance, tolerance under test (TUT), are considered, i.e., 0, 5, 10, 20, 30, 40, 60, 100, and 200\,kHz, where 0\,kHz corresponds to the ideal frequency-matching. 
The cavities in the array are randomly tuned to the target frequency, 5\,GHz, following a uniform distribution with its center at 5\,GHz and half-width of the TUT under consideration. Assuming each cavity is critically coupled, the individual power spectra are linearly summed up. 
The combined power spectrum is fitted with the Lorentzian function to get the amplitude and full-width at half maximum. 
The procedure is repeated 1,000 times over which the combined power spectra are averaged. 
Figure~\ref{fig:multicavity_fmt} shows the distributions of the averaged combined power spectra for different TUT values (a) and displays the normalized amplitudes and full widths at half maximum as a function of TUT (b). 
For the realistic approach, we put criteria that the relative amplitude of the combined power spectrum is greater than 0.95. 
From this we find that the FMT is 21\,kHz for a system consisting of four identical cavities with $Q_0 = 10^5$ seeking for a 5\,GHz axion signal. 
The FMT has a dependence on the cavity quality factor and target frequency, and can be generalized as FMT$(Q_0, f$) = 0.42\,GHz / $Q_0 \times f$\,[GHz].

\subsubsection{Tuning mechanism}
The basic principle of the tuning mechanism for a multiple-cavity system employs the same principle as for conventional single-cavity experiments. 
It relies on target frequency shift by rotating a single dielectric rod inside the cavity; frequency matching by finely tuning the individual cavity frequencies; and critical coupling by adjusting the depth of a single RF antenna into the cavity in a global manner. 
From the experimental point of view, frequency-matching is assured by achievement of minimal bandwidth of the combined reflection peak in the scattering parameter ($S$-parameter) sapce, yielding the maximal $Q$ value. 
Critical coupling is characterized by the minimum reflection coefficient in the $S$-parameter and by the constant resistance circle passing through the center of the Smith chart. 
Therefore, the tuning mechanism for a multiple-cavity system consists of three steps: 1) shifting the target frequency by simultaneously operating the rotational actuators; 2) achieving frequency-matching by finely manipulating the individual rotational actuators; and 3) achieving critical coupling of the system by globally adjusting the antenna depth using the linear actuator.

\subsubsection{Experimental demonstration}
The feasibility of the tuning mechanism for multiple-cavity systems is experimentally demonstrated using a double-cavity detector at room temperature. 
It is composed of two identical copper cavities with an inner diameter of 38.8\,mm whose corresponding resonant frequency is 5.92\,GHz and an unloaded quality factor of about 18,000. 
A single dielectric rod made of 95\% aluminum oxide (Al$_2$O$_3$) with 4\,mm diameter is introduced to each cavity and a piezoelectric rotator is installed under the cavity to rotate the rod for frequency tuning. With the tuning rod positioned at the center of the cavity, the resonant frequency decreased to 4.54\,GHz and $Q_0$ degrades to about 5,000 due to energy loss by the rod. 
A pair of RF antennae, each of which is coupled to each cavity, sustained by an aluminum holder attached to a linear piezoelectric actuator above the cavities. 
The double-cavity system is assembled by connecting the two antennae to a two-way power combiner that transmits signals to a network analyzer. 
Critical coupling of one cavity is made while the combiner input port, which the other cavity is connected to, is terminated with a 50$\,\Omega$ impedance terminator, and vice versa. 
Two cavities are configured to be critically coupled at slightly different resonant frequencies. 
The initial values of the loaded quality factor $Q_L$ and scattering parameter $S_{11}$ are measured. After the system is re-assembled with the both RF antennae being connected to the combiner, the initial system is represented by two reflection peaks with $S_{11} = −6\,$dB in the $S$-parameter spectrum and two small circles with half-unit radius on the Smith chart.
In order to match the frequency, one of the rotational actuators is finely manipulated until the combined reflection coefficient becomes minimized. 
Following that, the linear actuator is operated to adjust the antenna positions in a global manner to achieve critical coupling of the system until the reflection peak becomes the deepest and the large circle passes through the center of the Smith chart. 
The final $Q_L$ and $S_{11}$ values are measured. 
The consistency between initial and final values confirms that the turning mechanism with frequency-matching is successful.

\subsection{New concept of cavity design: multiple-cell cavity}

\subsubsection{Introduction}

\begin{figure}[h]
\centering
\includegraphics[width=0.8\textwidth]{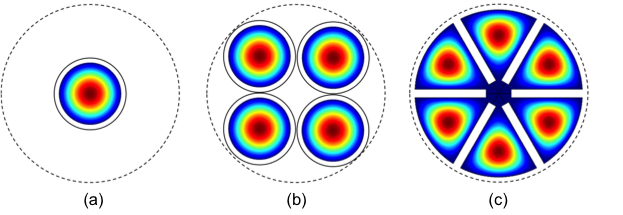}
\caption{Configurations of cavity detector design for high frequency axion search: (a) single cavity; (b) multiple-cavity system; and (c) multiple-cell cavity.}
\label{fig:cavity_configuration}
\end{figure}

The successful experimental demonstration, depicted in Sec.~\ref{sec:multi-cavity}, convinces us that phase-matching of an array of multiple cavities is certainly plausible and applicable to axion search business. 
The multiple-cavity design, however, is still inefficient in terms of volume usage, mainly due to unused volume and cavity wall thickness, as can be seen in Fig.~\ref{fig:cavity_configuration} (b). 
An alternative design is comprised of a single cylindrical cavity, fitting into the magnet bore, with metal partitions placed at equidistant intervals to make multiple identical cells (see Fig.~\ref{fig:cavity_configuration} (c)). 
This concept provides a more effective way to increase the detection volume while relying on the same frequency tuning mechanism as that of multiple-cavity systems, i.e. a single (pair) of dielectric (metal) tuning rod(s) in each cell. 
The YS, furthermore, brought in an innovative idea of introducing a narrow hollow gap at the center of the cavity, as shown in Fig.~\ref{fig:cavity_configuration} (c). 
In this design, all cells are spatially connected among others, which allows a single RF coupler to extract the signal out of the entire cavity volume. 
This simplifies the readout chain by not only reducing the number of pickup antennae but also eliminating the necessity of a power combiner, both of which could be bottlenecks for multiple-cavity systems especially when the cavity multiplicity is large. We refer to this cavity concept as ``pizza" cavity. 

\subsubsection{Characteristics}
\begin{figure}[h]
\centering
\includegraphics[width=0.8\textwidth]{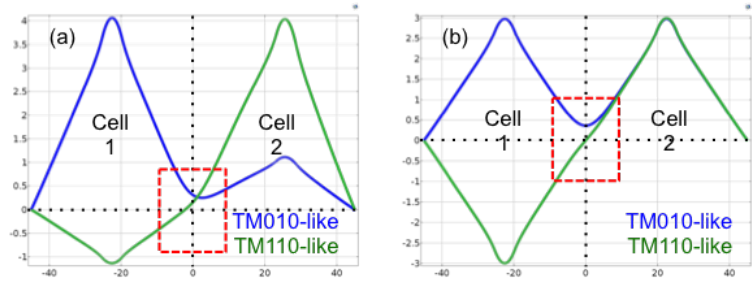}
\caption{Electric field profiles for a double-cell cavity: asymmetric configuration (a) and symmetric configuration (b). 
The field profiles of the TM$_{010}$-like and TM$_{110}$-like modes are represented by the blue and green solid lines respectively.}
\label{fig:pizza_E_profile}
\end{figure}

This pizza cavity design is characterized, based on simulation studies, by several features, which supply critical advantages. Introduction of a narrow hollow in the middle of the cavity breaks the frequency degeneracy with the lowest mode corresponding to the TM$_{010}$-like mode regardless of the cell multiplicity. Since the individual cells are spatially connected and thus the EM fields can mix, the relative tuning rod position in a cell affects the entire field distribution of the cavity and breaks the field symmetry, as can be seen in Fig.~\ref{fig:pizza_E_profile} (a). Therefore, the frequency tuning mechanism for the TM$_{010}$-like mode requires that the field distribution in individual cells be identical and that the overall field distribution be symmetric. We refer to this condition as “phase-matching”. It is noticeable that once phase-matching is successfully accomplished, the field strength at the center of the cavity becomes zero for the higher TM$_{110}$-like mode, while it remains non-zero for the lowest TM$_{010}$-like mode, as illustrated in Fig.~\ref{fig:pizza_E_profile} (b). This indicates that electrical coupling at the center of the cavity will be sensitive only to the lowest TM$_{010}$-like mode, not to any higher modes. 
This, in turn, provides an idea that a single monopole RF antenna in the middle of the cavity is sufficient enough to couple to the desired mode.

The features described above are in practice represented in the scattering parameter ($S$-parameter) and the Smith chart using a network analyzer. For a double-cell cavity, an asymmetric rod configuration and arbitrary coupling strength are represented by two reflection peaks in the S-parameter space and two circles in the Smith chart, as illustrated in Fig.~\ref{fig:pizza_tuning} (a). When the rods are symmetrically arranged, the higher resonant frequency peak and the corresponding circle fade away, as seen in Fig.~\ref{fig:pizza_tuning} (b). This implies that phase-matching of multiple-cell cavities is characterized by disappearance of any higher mode peaks in the scattering parameter space and corresponding constant resistance circles in the Smith chart.

\subsubsection{Tuning mechanism}
\begin{figure}[t]
\centering
\includegraphics[width=1.0\textwidth]{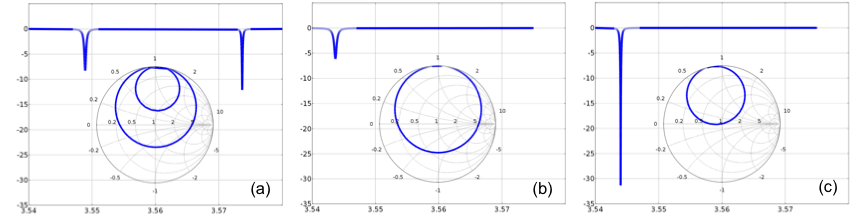}
\caption{Representation of the tuning mechanism in the $S$-parameter and Smith chat for a double-cell cavity: 
(a) initial state (two reflection peaks and two circles); (b) phase-matching (higher mode reflection peak and circle vanishing); and (c) critical coupling.}
\label{fig:pizza_tuning}
\end{figure}

The basic principle of the tuning mechanism is the same as that for conventional multiple-cavity systems. It consists of two steps: phase-matching followed by critical coupling. From an experimental point of view, phase-matching is assured by aligning the tuning rods until the higher mode peaks in the $S$-parameter space and, equivalently, the corresponding constant resistance circles in the Smith chart vanish. The coupling mechanism is simple enough that a single coaxial monopole antenna is placed at the center of the top (or bottom) end cap and its depth into the cavity changes the coupling strength to the lowest TM$_{010}$-like mode independently of the other higher modes. Figure~\ref{fig:pizza_tuning} shows the sequence of the tuning mechanism and characteristics represented in the $S$-parameter and Smith chart using a network analyzer.

\subsubsection{Experimental demonstration}
\begin{figure}[h]
	\centering
	\begin{subfigure}[]{0.45\textwidth}
		\centering
		\includegraphics[width=\linewidth]{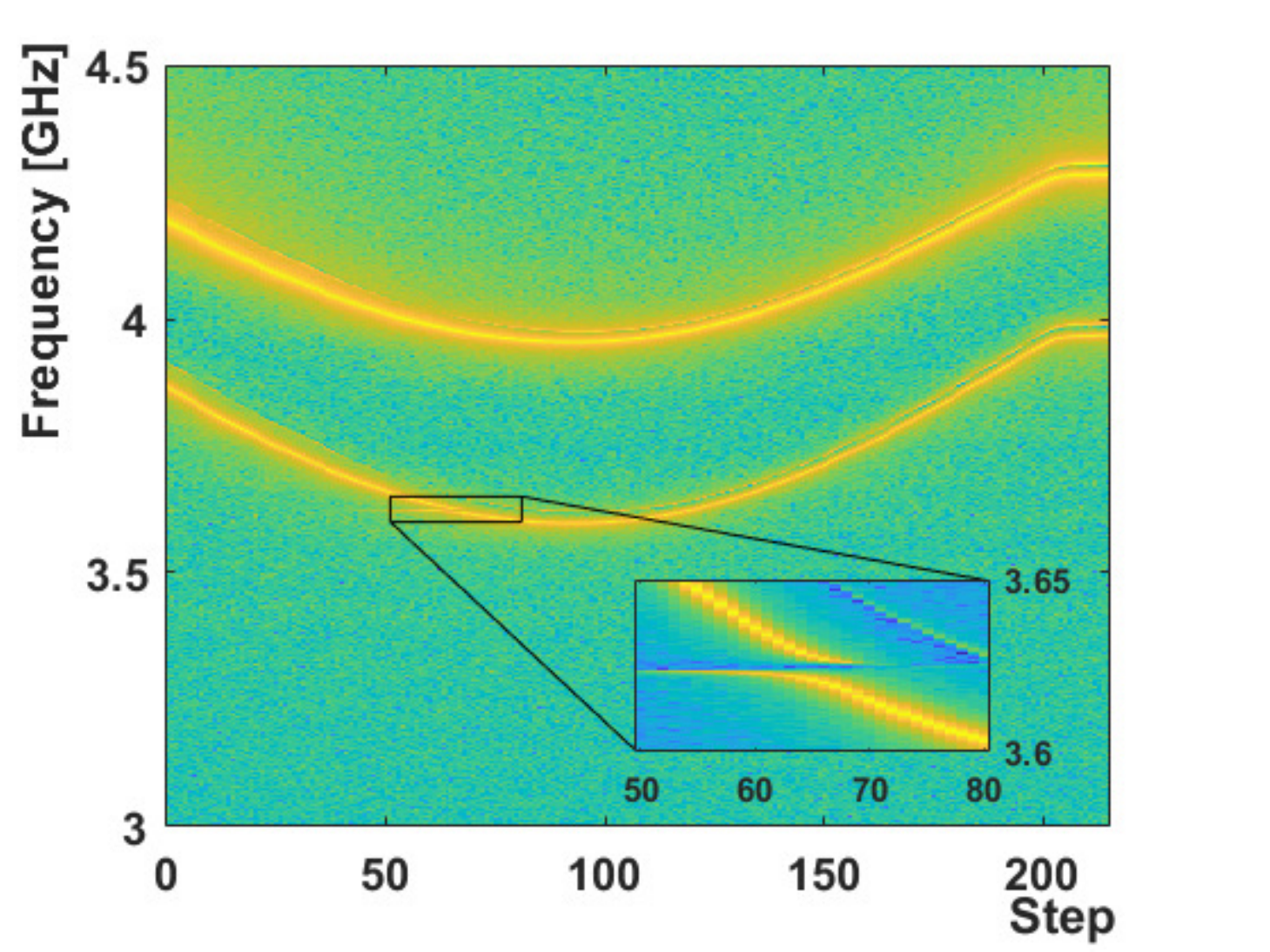}
		\subcaption{}
	\end{subfigure}\\
	\begin{subfigure}[]{0.45\textwidth}
		\centering
		\includegraphics[width=\linewidth]{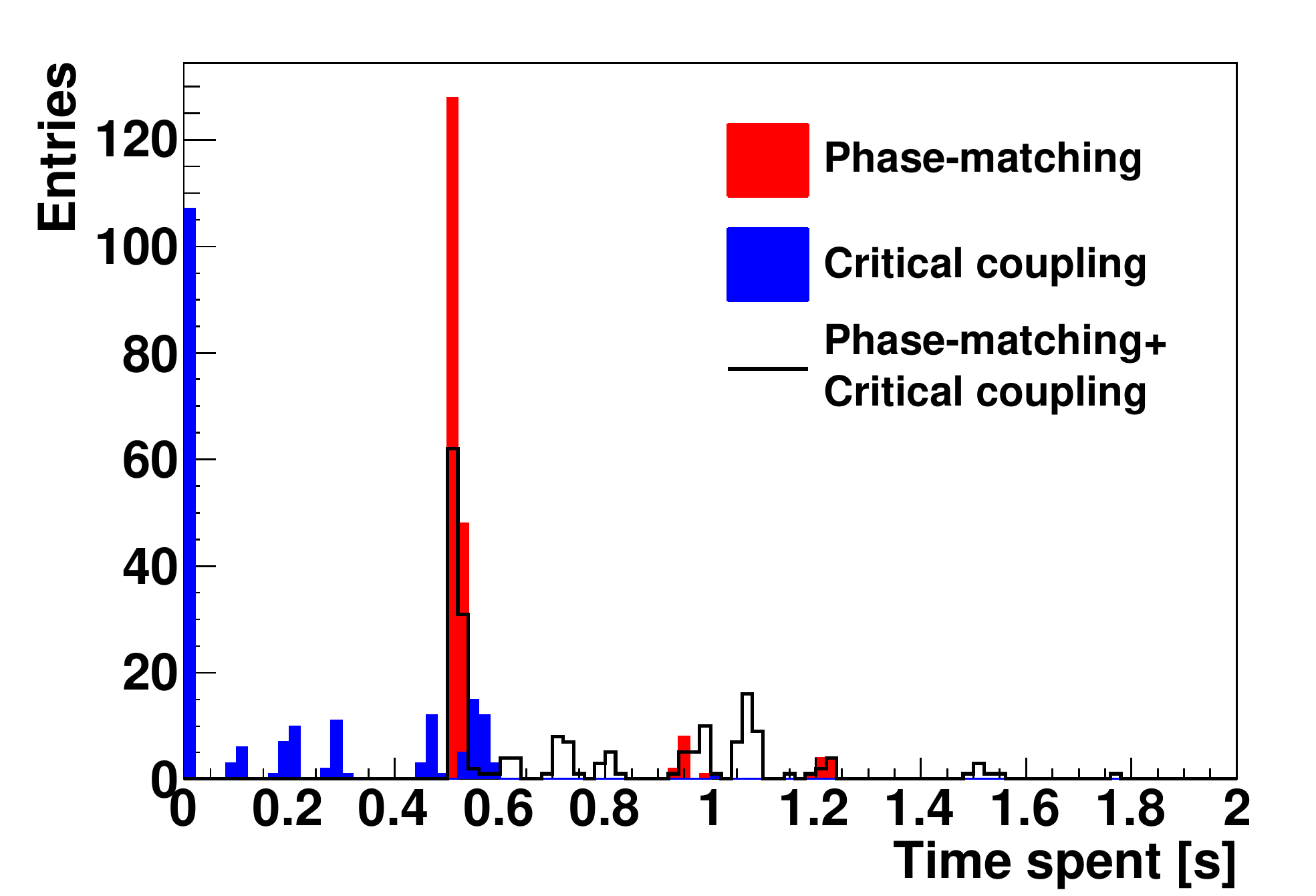}
		\subcaption{}
	\end{subfigure}
	\begin{subfigure}[]{0.45\textwidth}
		\centering
		\includegraphics[width=\linewidth]{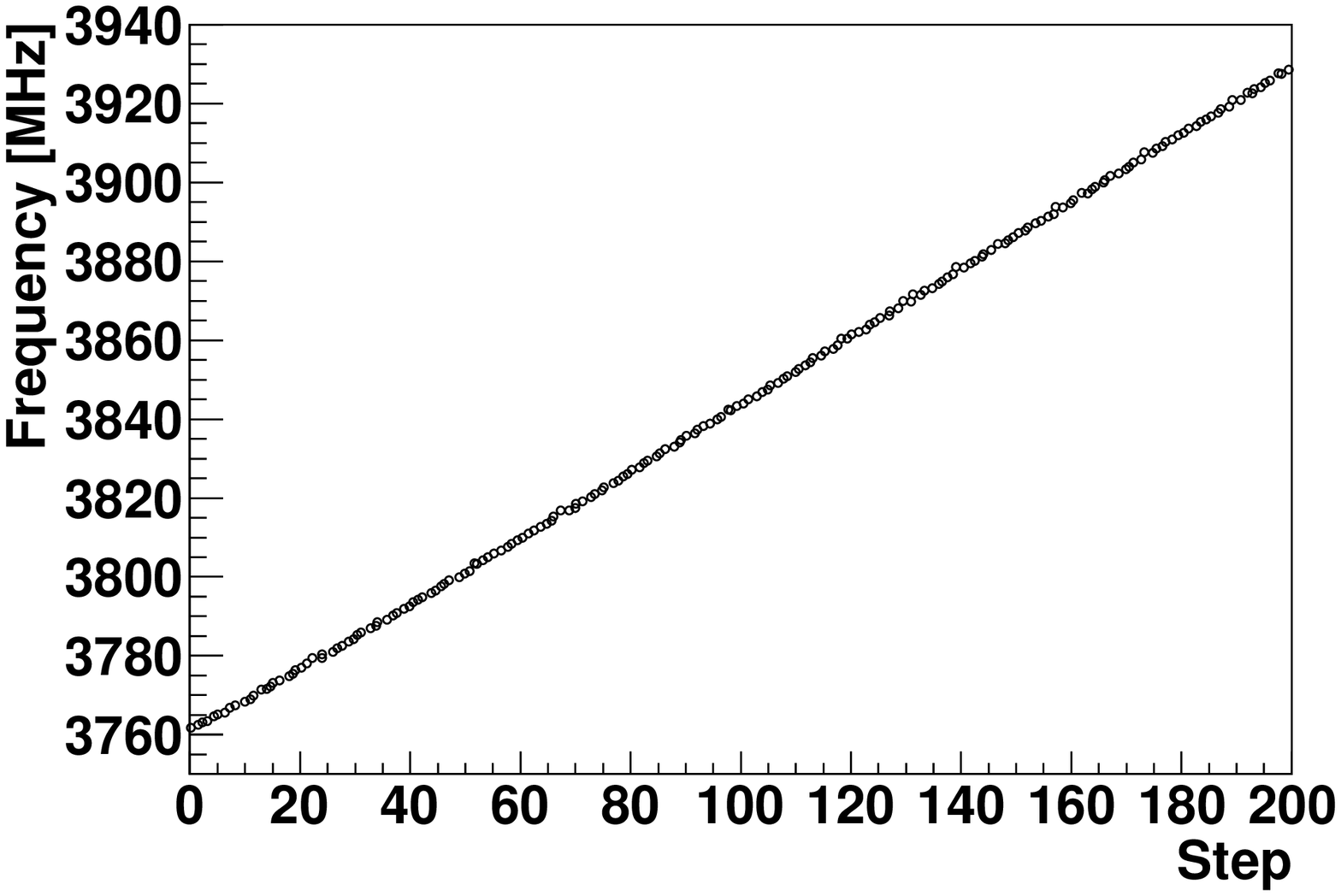}
		\subcaption{}
	\end{subfigure}
	\caption{(a) Mode map of the double-cell cavity described in the text. 
(b) Time durations for the tuning mechanism. 
The red and blue filled histograms represent the time require for phase-matching and critical coupling in seconds. 
The total time spent is represented by the empty black histogram. 
(c) Resonant frequency of the TM$_{010}$-like mode with step.}
	\label{fig:pizza_demo}
\end{figure}

Experimental feasibility of the multiple-cell design is verified at room temperature using a copper double-cell cavity with 90\,mm inner diameter and 100\,mm inner height. The concept of split cavity design, introduced in Ref.~\cite{bib:magnetoresistance}, is adopted to eliminate the contact resistance. A single alumina rod is introduced in each cell for frequency tuning and a single coaxial RF antenna is inserted through a hole at the top center of the cavity for critical coupling. 
This sequence of the tuning mechanism, described above, is repeated 200 times, with the target frequency being shifted by roughly 1\,MHz at every step. During the exercise, the time durations required to complete phase-matching and critical coupling are measured and the TM$_{010}$-like resonant frequency is recorded. The frequency mode map is drawn in Fig.~\ref{fig:pizza_demo} (a). It is remarkable that the criteria for phase-matching (critical coupling) are satisfied more than 90\% (50\%) of the time, and for the rest 10\% (50\%) of the time, less than 2 (1) seconds are required to complete the process (see Fig.~\ref{fig:pizza_demo} (b)). Taking into account the typical data acquisition (DAQ) time of an order of minutes or hours, this indicates that the dead time owing to the procedure for the tuning mechanism is negligible. A good linear behaviour of the target frequency with step, shown in Fig.~\ref{fig:pizza_demo} (c), implies the stability of the tuning mechanism.

\subsubsection{Advanced tuning system}
Through the study above, it is proven that the pizza cavity design is superior to the conventional multiple-cavity design in terms of detection volume, simplicity of the experimental setup, and facilitation of the phase-matching mechanism. 
Nonetheless, the reliability and complexity of operation with large cell multiplicities could still be limiting factors to the experimental sensitivity. 
Hereby, we exploit an advanced (simplified) tuning mechanism in which all the tuning rods are simultaneously turned by a single rotator. 
This approach would be valid if the sensitivity loss, which mainly comes from the form factor reduction due to cavity fabrication tolerance and misalignment of tuning rods, is not significant. 
As can be inferred from the description above, the precision of phase-matching of a system is evaluated by the coupling strength for the higher mode. 
A simulation study indicates that as long as, with the lowest mode being critically coupled, the higher mode coupling strength is maintained less than 0.1\,dB, the sensitivity (scan rate) remains more than 98\% of the ideal value. 

\begin{figure}[h]
	\centering
	\begin{subfigure}[]{0.45\textwidth}
		\centering
		\includegraphics[height=0.75\linewidth]{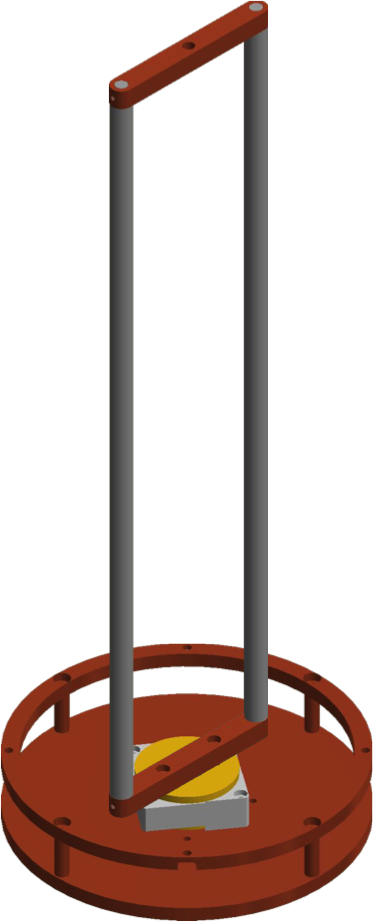}
		\subcaption{}
		\label{fig:pizza_new_tuning_a}
	\end{subfigure}
	\begin{subfigure}[]{0.45\textwidth}
		\centering
		\includegraphics[width=\linewidth, height=0.75\linewidth]{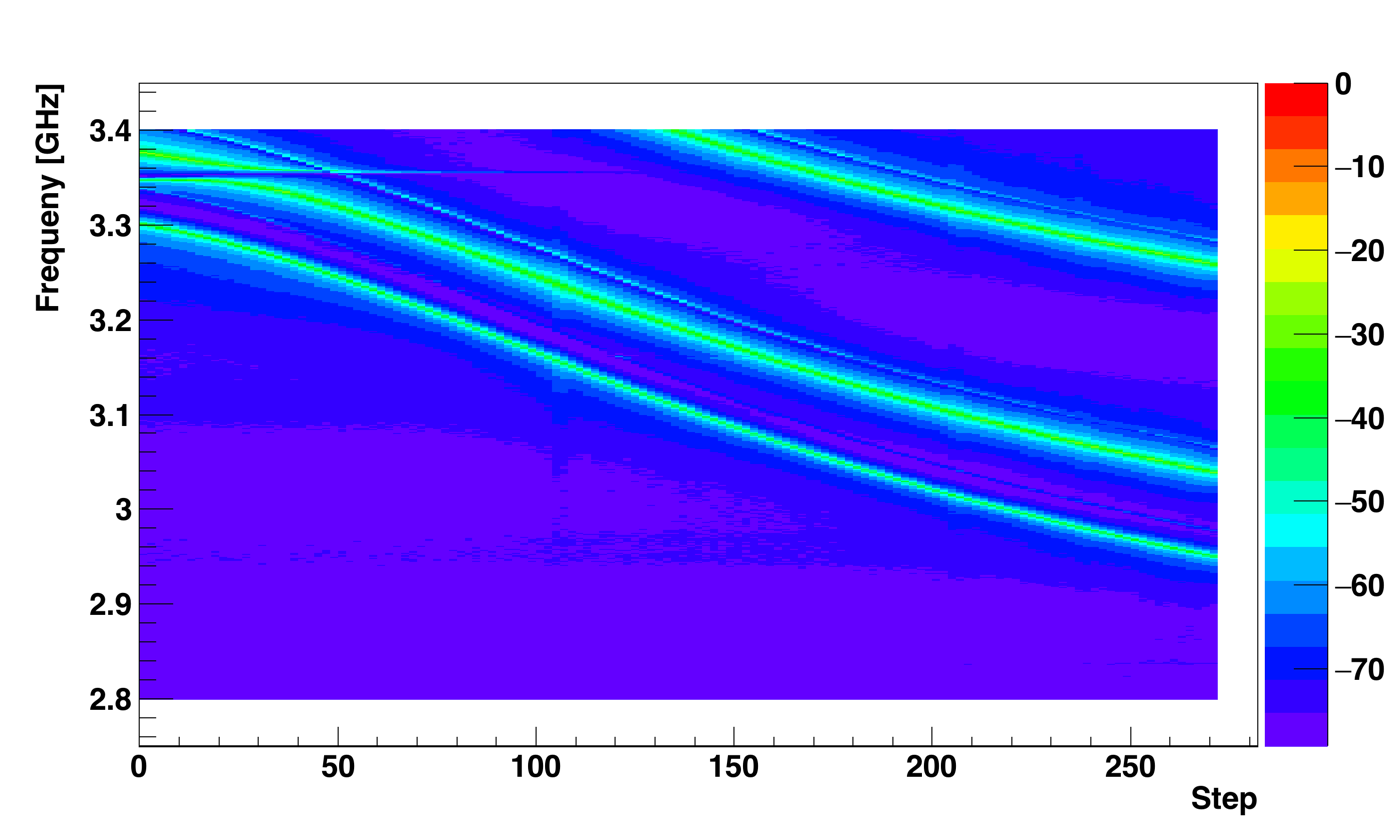}
		\subcaption{}
	\end{subfigure}
	\begin{subfigure}[]{0.45\textwidth}
		\centering
		\includegraphics[width=\linewidth]{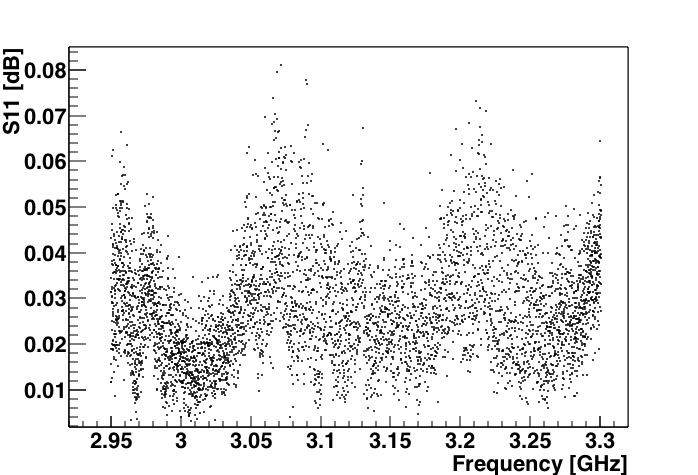}
		\subcaption{}
	\end{subfigure}
	\begin{subfigure}[]{0.45\textwidth}
		\centering
		\includegraphics[width=\linewidth]{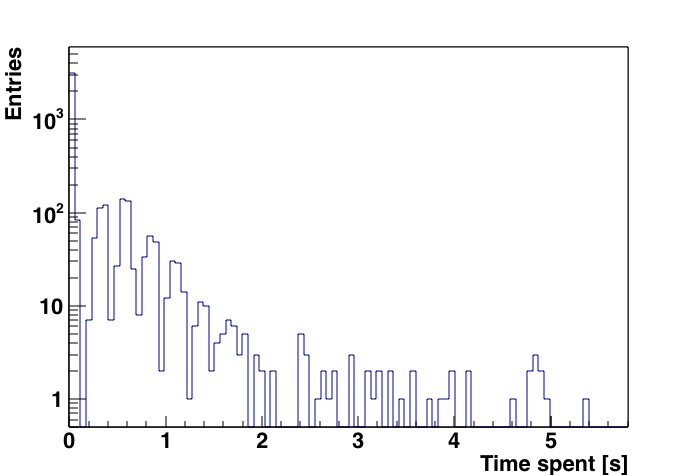}
		\subcaption{}
	\end{subfigure}
	\caption{(a) Design of the frequency tuning system using a single rotator. 
(b) Frequency mode map. 
(c) Coupling strength of the higher TM$_{110}$-like mode in dB. (d) Time duration for critical coupling.}
	\label{fig:pizza_new_tuning}
\end{figure}

An experimental demonstration was done using a copper double-cell cavity with 110\,mm inner diameter and 220 mm inner height and a pair of alumina tuning rods held by a copper holder attached on a single rotator, as seen in Fig.~\ref{fig:pizza_new_tuning}(a). 
The system is brought to cryogenic temperature (4\,K) and the frequency is tuned using the simplified tuning mechanism, including critical coupling of the lowest TM$_{010}$-like mode. 
Figure~\ref{fig:pizza_new_tuning}(b) shows the mode map and measurement results relevant to important variables. These verify that the tuning mechanism works appropriately at cryogenic temperature. 
In particular, Fig.~\ref{fig:pizza_new_tuning}(c) shows the coupling strength for the higher mode with the lowest mode critically coupled. 
It is remarkable that the coupling strength is maintained less than 0.1\,dB throughout the entire frequency range, indicating the sensitivity drop is less than 2\%. 
The time spent for the critical coupling of the lowest mode is ~0.3 seconds on average. 
This implies that the pizza cavity design is robust against the manufacturing tolerance and rod misalignment while multiple-cavity design would significantly suffer from them. 
This also convinces us that this design would enhance its applicability to axion dark matter search for higher frequency regions.

%% file: 1.4.3/main.tex
\subsection{Introduction}
\begin{figure}[h]
\centering
\includegraphics[width=1.0\textwidth]{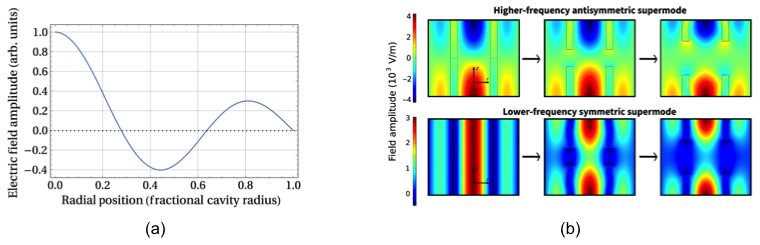}
\caption{(a) Electric field profile of the TM$_{030}$ mode as a function of radial distance from the center of the cavity. 
(b) Electric field distribution for the two supermodes discussed in Ref.~\cite{bib:TM030_australia}. 
The black sold lines represent of the boundaries of the half dielectric hollows.}
\label{fig:TM030_australia}
\end{figure}

As another approach in pushing towards higher axion mass regions, utilizing higher-order resonant modes, such as TM$_{020}$ or TM$_{030}$, could be beneficial. 
They allow us to access to higher frequency regions than the lowest TM mode with higher cavity quality factors without reduction of the detection volume.
However, high degrees of field variation give rise to out-of-phase electric field components, which, in the presence of an external static magnetic field, results in cancellation of $\left|E\cdot B\right|$ in the nominator of the form factor, defined as
\begin{equation}
C=\frac{\left|\int {\bf E_c}\cdot {\bf B_0} dV\right|^2}{\int \epsilon \left|{\bf E_c}\right|^2 dV \int \left| {\bf B_0}\right|^2 dV},
\label{eq:form_factor}
\end{equation}
where ${\bf E_c}$ is the electric field of the cavity resonant mode under consideration, ${\bf B_0}$ is the external magnetic field and $\epsilon$ is the dielectric constant inside the cavity volume.
The electric field profile for the TM$_{030}$ mode is shown in Fig.~\ref{fig:TM030_australia} (a). 
The cancellation effect becomes larger with increasing order of resonant modes. 
For instance, the form factor for the TM$_{020}$ and TM$_{030}$ modes is 0.13 and 0.05 respectively, while it is 0.69 for the TM$_{010}$ mode. 
This gives rise to significant reduction in experimental sensitivity and thus the higher modes have not been considered for axion research purposes.

Recently, there was an attempt to employ the concept of a Bragg resonator in different conditions to achieve reasonable sensitivities in axion dark matter haloscopes utilizing the higher-order resonant modes~\cite{bib:TM030_australia}. For the TM$_{030}$ resonant mode, a cylindrical dielectric hollow with optimal dimension is introduced at a place inside the cavity where the negatively oscillating field components are maximally suppressed. They also invented a tuning mechanism, in which two half hollows are taken apart along the axial direction, as illustrated in Fig.~\ref{fig:TM030_australia} (b). However, we find that this mechanism gives rise to significant degradation in the cavity quality due to field leakage while the dielectric hollows are being taken out of the cavity. Hereby, the YS refuted their faulty claims [2] and furthermore brought up a new idea to tune the frequency for this particular higher resonant mode while preserving the aforementioned advantages.

\subsection{Tuning mechanism}
\begin{figure}[h]
\centering
\includegraphics[width=1.0\textwidth]{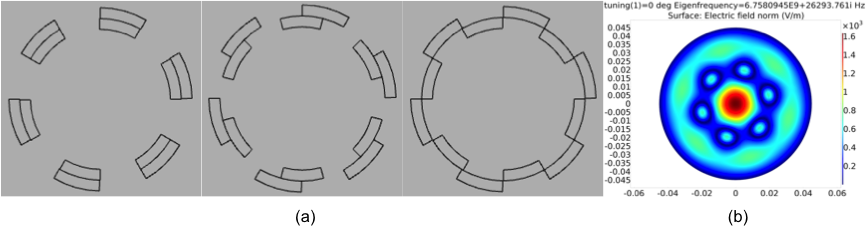}
\caption{(a) Illustration of the tuning mechanism described in the text. 
The inner layer of the segments is simultaneously rotated with respect to the fixed outer layer. 
(b) Electric field distribution of the TM$_{030}$ mode with the layer configuration corresponding to the first scheme in (a).}
\label{fig:TM030_tuning}
\end{figure}

According to a simulation study, we find that the TM$_{030}$ resonant frequency alters with the thickness of the cylindrical dielectric hollow. Based on this, we invent a new concept of tuning mechanism, in which we introduce double layers of concentrically segmented dielectric pieces and one layer of the segments rotates with respective to the other. In this way, the thickness of the dielectric segments effectively changes. The outer layer of dielectric segments is fixed in position, while the inner layer is rotated simultaneously by a single rotator outside the cavity. This concept of the tuning mechanism is illustrated in Fig.~\ref{fig:TM030_tuning} (a). An example of the electric filed distribution for the TM$_{030}$ mode is found in Fig.~\ref{fig:TM030_tuning} (b), where one can see the negative field components are suppressed. As a result, this design provides a more reliable tuning mechanism, which is also easy to manipulate.

\subsection{Simulation studies}
\begin{figure}[h]
\centering
\includegraphics[width=1.0\textwidth]{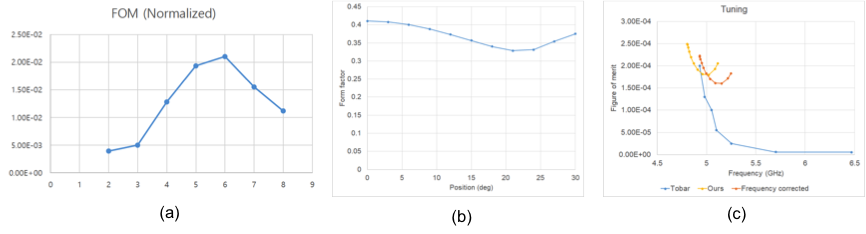}
\caption{(a) Figure of merit for different numbers of dielectric segments per layer. 
(b) Form factor distribution of the TM$_{030}$ mode as a function of rotation angle of the inner layer using 6 segments per layer as shown in Fig.~\ref{fig:TM030_tuning}. 
(c) Comparison of the tuning mechanisms in terms of the F.O.M. The yellow and blue lines represent our mechanism and the one in Ref.~\cite{bib:TM030_australia}, respectively. The F.O.M of our mechanism is rescaled to have the same starting frequency as the latter.}
\label{fig:TM030_simulation}
\end{figure}

The dimension of the tuning system is optimized based on COMSOL simulation studies in terms of three parameters: number of segments per layer, thickness of the double layers, and inner radius of the layer. The figure of merit (F.O.M.) is chosen such that the design enhances both the scan rate ($df/dt$) and the frequency tuning range $(\Delta f)$, i.e. F.O.M. = $df/dt \times \Delta f$. Assuming a copper cylindrical cavity with 90\,mm inner diameter and 100\,mm inner height and dielectric material with a dielectric constant of 10, we find six segments per layer is optimal to maximize the F.O.M., as shown in Fig.~\ref{fig:TM030_simulation} (a). It turns out that the optimal thickness of the layer pair is approximately $\lambda$/2, where $\lambda$ is the wavelength of the EM wave corresponding to the TM$_{030}$ mode of an empty cavity, and the radius of the layer approximately 0.37R, where R is the cavity radius. We observe that the frequency tuning range of the higher resonant mode is about 6\% with respect to the central frequency and the form factor is enhanced to be greater than 0.33 (compared to 0.05 for an empty cavity) over the entire tuning range using this design (see Fig.~\ref{fig:TM030_simulation} (b)). A comparison with the design introduced in Ref.~\cite{bib:TM030_australia} is also made to reveal that our design significantly enhances the scan rate, which is proportional to $C^2V^2G$, as shown in Fig.~\ref{fig:TM030_simulation} (c). This convinces us that this design is going to be another breakthrough in approaching higher frequency regions for cavity-based experiments.

\subsection{Design of the tuning system}
\begin{figure}[t]
\centering
\includegraphics[width=1.0\textwidth]{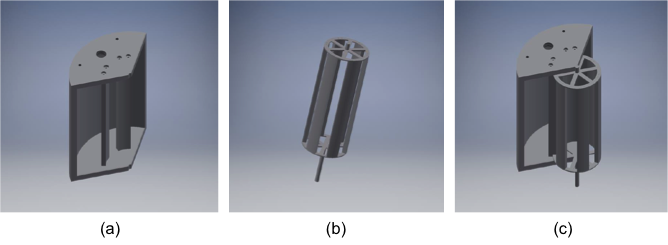}
\caption{Drawings of the tuning system described in the text. (a) Two outer segments are fixed at a copper cavity piece. 
(b) Structure of the inner layer. Six dielectric segments are supported by a pair of wheel-shaped structures. 
The extended piece on the bottom is attached to a rotator outside the cavity. 
(c) Combined view of (a) and (b).}
\label{fig:TM030_design}
\end{figure}

As mentioned earlier the tuning system consists of a double layer of dielectric segments with a low loss tangent and a high dielectric constant, such as sapphire or alumina, and a single rotator. Based on the split design, the cavity is assembled with three identical copper pieces, each of which houses two segments of the outer layer, as shown in Fig.~\ref{fig:TM030_design} (a). The six segments of the inner layer are supported by a pair of support structures made of dielectric material with a low dielectric constant, such as teflon as shown in Fig.~\ref{fig:TM030_design} (b). The support structure on the bottom has an extended rod, which is eventually attached to a single rotator outside the cavity. The overall structure can also be seen in Fig.~\ref{fig:TM030_design} (c).

%% file: 1.4.5/main.tex
\begin{figure}[h]
\centering
\includegraphics[width=0.5\textwidth]{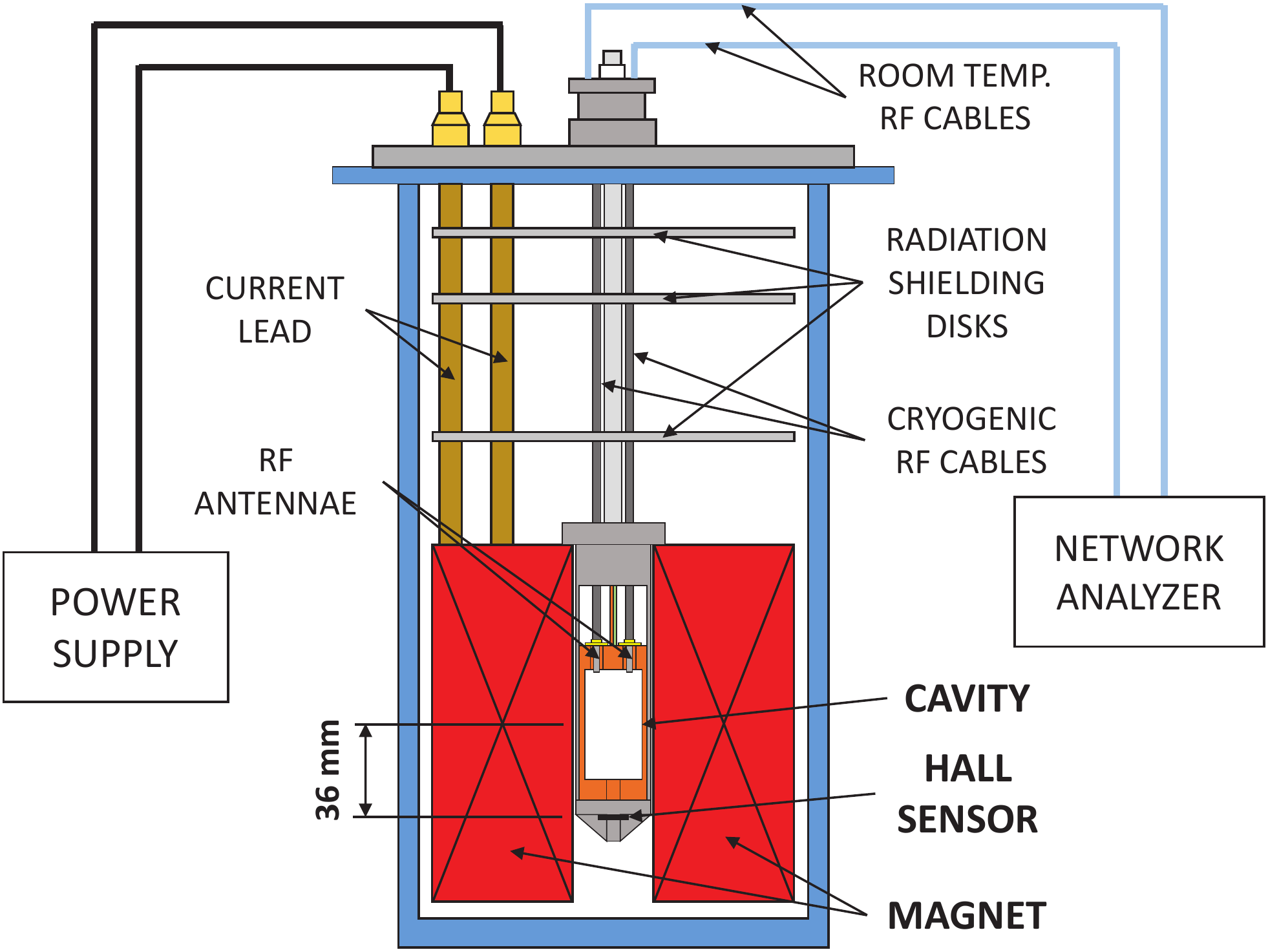}
\caption{Schematic view of the experimental setup. The major components include a copper cavity (orange) with a Hall sensor (black) and a HTS magnet (red) immersed in an LHe cryostat.}
\label{fig:magneto_setup}
\end{figure}

The YS project team has played a leading role in an experiment to study the magnetoresistance of copper at a higher frequency by measuring the surface resistance of a copper cavity under high magnetic fields at cryogenic temperature~\cite{bib:magnetoresistance}. A 26\,T high temperature superconducting (HTS) high field magnet manufactured by SuNAM Co. Ltd. was employed and a copper split-cavity with 18\,mm inner diameter and 50\,mm inner height are employed and the experiment was performed at a liquid helium reservoir. Due to unexpected electrical issues on a couple of double pancakes, the magnet could be energized up to 15\,T. The TM$_{010}$ resonant frequency and quality factor of the cavity were measured at the low temperature to be 12.9\,GHz and 33,600 respectively. The overall structure of the experimental setup is shown in Fig.~\ref{fig:magneto_setup}.

The experiment is performed while the magnet is being energized and de-energized. The cavity quality factor is measured consecutively though transmission signals between a pair of weekly coupled RF antenna. The measured quality factors are transformed into the electrical surface resistivity ($R_s$) and the fractional changes in $R_s$, as known as magnetoresistance, are obtained as a function of the magnetic field. 

\begin{figure}[h]
\centering
\includegraphics[width=0.9\textwidth]{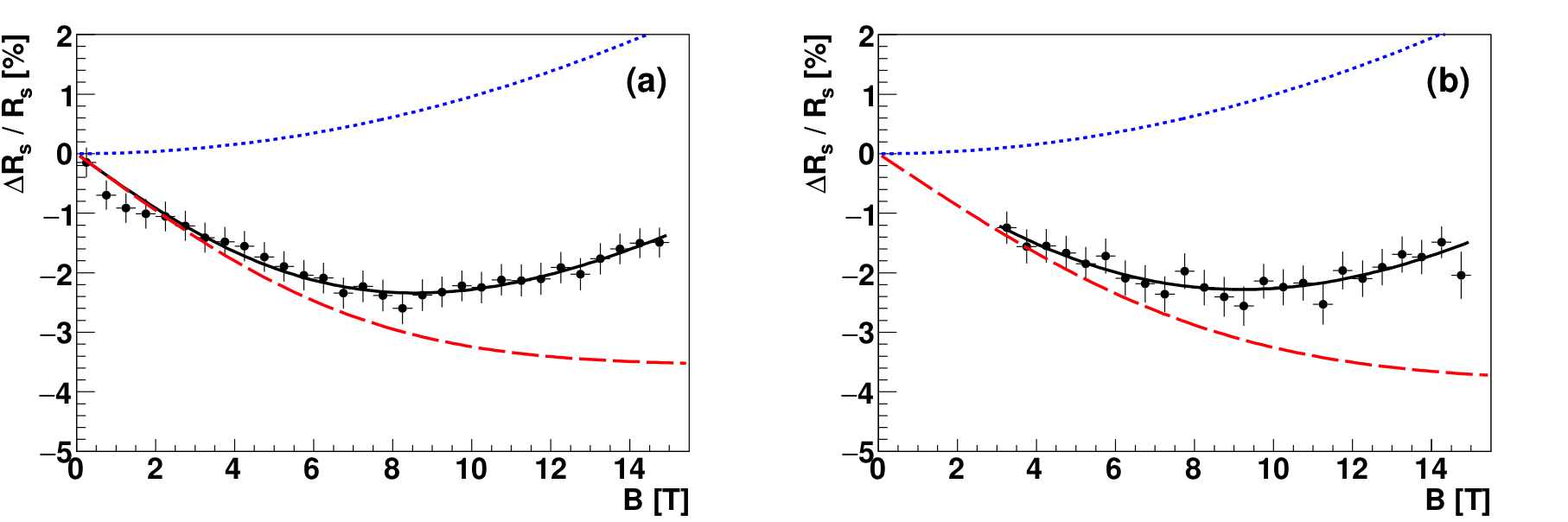}
\caption{Fractional change in surface resistivity while charging (a) and discharging (b) the HTS magnet. 
The data distributions (black points with an error bar) are modelled by a linear combination of the second order polynomial (dotted blue lines) to describe the nominal behaviour of the magnetoresistance, and the Error function (dashed red lines) to describe the abnormal behaviour of the phenomenon.}
\label{fig:magneto_result}
\end{figure}

Figure~\ref{fig:magneto_result} shows the experimental results of the magnetoresistance of copper at 12.9\,GHz. A parabolic dependence on magnetic field is clearly seen and about $2-3$\% of gain in the quality factor at around 9\,T is observed. The behaviour is consistently explained by the size effect - the relative size of the skin depth and cyclotron radius to the mean free path of electrons - at the given frequency and magnetic fields.  In addition, a small variation in the surface resistivity, equivalently the cavity quality factor, over a wide range of magnetic field at a high frequency indicates that the influence of strong magnetic fields on the cavity-based axion experiments exploring high mass regions will be tolerable.
Finally, this measurement is the first user-based physics result employing the multi-width no-insulation 2G HTS magnet technologies for the scientific purpose.

%% file: 1.4.6/main.tex
\subsection{Background}

\subsubsection{Purpose of high Q cavity}

As we discussed in the previous sections, improving cavity is one of the most important issue in the axion experiment. The purpose of making better cavity is increasing scan rate, because we can rule out larger parameter space with larger scan rate. It makes the experiment more efficient and faster. To make bigger scan rate, we have to understand the relationship between system parameters and scan rate. In the Brubaker's note~\cite{Brubaker_Thesis}, he shows that the several parameters related to cavity ($B$: Magnetic Field, $V$: Volume of Cavity, $C$: Geometric Factor of Cavity, $Q_L$: Loaded Q Factor of Cavity, $T_{sys}$: System Temperature) are the main parameters of {\bf scan rate}.
\begin{equation}
    \frac{d \nu}{dt} \propto B^4 V^2 C^2 Q_L T^{-2}_{sys}
\end{equation}

Now the main issue in Cavity R$\&$D to make larger scan rate is increasing $V^2C^2Q_L$, so we can make larger scan rate with increasing $Q_L$ which means that reducing surface impedance of cavity inner surface. $Q_L$ usually inversely proportional to the surface impedance of cavity.

\subsubsection{Electronic structure and surface impedance}
To understand the relationship between Q factor and surface impedance, we have to consider complex conductivity. Since we can deduce our main parameter which are impedance ($R_s + i X_s$), Q factor, and skin depth $\delta$ (penetration depth, $\lambda$) from the complex conductivity. We usually use the term "skin depth" to normal conductor, and "penetration depth" to superconductor, but we will use the term "penetration depth" to all situation. Complex conductivity and complex surface impedance also can contain the complex aspects of electronic property of conductor-superconductor transition. In this section we will calculate complex conductivity and complex impedance, and we will only discuss about the result of the calculation and the physical meanings of the formulas. You can see the detailed derivation of following formulas in the textbook~\cite{Lancaster_Textbook}.

Two fluid model is one of the most useful and simple tactics to understand superconducting material. We will divide total electron flow into two parts, normal electron flow and superconducting electron flow ($\overrightarrow{J_{tot}} = \overrightarrow{J_n} + \overrightarrow{J_s}$). In the case of normal conductor, electrons feel drag force because of collision with lattice. The equation of motion of the electrons inside conductor is,
\begin{equation} \label{eq:EOMnormal}
    m \frac{d \overrightarrow{v_n}}{dt} + m \frac{\overrightarrow{v_n}}{\tau} = -e \overrightarrow{E}.
\end{equation}
However inside superconductor electron pairs do not feel drag force which means that resistivity is zero. Thus the equation of motion of the electrons inside superconductor is,
\begin{equation} \label{eq:EOMsuper}
    2m \frac{d \overrightarrow{v_s}}{dt} = -2e \overrightarrow{E}.
\end{equation}

Now we can substitute the definition of current density ($\overrightarrow{J_{n,s}} = -n_{n,s}e\overrightarrow{v_{n,s}}$) in the Eq.\ref{eq:EOMnormal} and Eq.\ref{eq:EOMsuper}. Then the equations become,
\begin{equation}
    \frac{d \overrightarrow{J_n}}{dt} + \frac{\overrightarrow{J_n}}{\tau} = \frac{n_n e^2}{m} \overrightarrow{E}
\end{equation}
\begin{equation}
    \frac{d \overrightarrow{J_s}}{dt} = \frac{n_s e^2}{m} \overrightarrow{E}.
\end{equation}
Also we can rewrite upper two equations in frequency space.
\begin{equation}
    i \omega \overrightarrow{J_{n0}} + \frac{\overrightarrow{J_{n0}}}{\tau} = \frac{n_n e^2}{m} \overrightarrow{E_{0}}
\end{equation}
\begin{equation}
    i \omega \overrightarrow{J_{s0}} = \frac{n_s e^2}{m} \overrightarrow{E_{0}}.
\end{equation}
Ohm's law ($\overrightarrow{J_{tot}} = (\sigma_1 + i \sigma_2) \overrightarrow{E}$) lead the equations to {\bf complex conductivity} which are,
\begin{equation} \label{eq:sigma1}
    \sigma_1 = \frac{n_n e^2 \tau}{m \left( 1 + \omega^2 \tau^2 \right)}
\end{equation}
\begin{equation} \label{eq:sigma2}
    \sigma_2 = \frac{n_s e^2}{\omega m} + \frac{\omega n_n e^2 \tau^2}{m \left( 1 + \omega^2 \tau^2 \right)}
\end{equation}

Now we can substitute the conductivity (Eq.\ref{eq:sigma1}, Eq.\ref{eq:sigma2}) into the surface impedance and penetration depth formulas~\cite{Brubaker_Thesis}. The formulas are derived from plane wave and plane boundary between air and conductor. We assumed also the conduction current is larger than displacement current. ($\sigma \gg \omega \epsilon$)
\begin{equation} \label{eq:SurfaceImpedance}
    Z_s = \sqrt{\frac{i \omega \mu}{\sigma_1 + i\sigma_2}}
\end{equation}
\begin{equation} \label{eq:PenetrationDepth}
    \lambda = \frac{1}{\sqrt{\omega \mu \sigma_2}}
\end{equation}

Therefore we can obtain the surface impedance and the penetration depth of conductor and superconductor as below.
\begin{equation} \label{eq:SurfaceImpedanceN}
    Z_n = \sqrt{\frac{\omega \mu}{2 \sigma}} + i \sqrt{\frac{\omega \mu}{2 \sigma}}
\end{equation}
\begin{equation} \label{eq:PenetrationDepthN}
    \lambda_n = \sqrt{\frac{2}{\omega \mu \sigma}}
\end{equation}
\begin{equation} \label{eq:SurfaceImpedanceS}
    Z_s = \frac{\omega^2 \mu^2 \sigma_1 \lambda^3}{2} + i \omega \mu \lambda
\end{equation}
\begin{equation} \label{eq:PenetrationDepthS}
    \lambda_s = \lambda_L = \sqrt{\frac{m}{n_s e^2 \mu}}.
\end{equation}
In this result, there is no assumption that there is strong DC magnetic field. However, in the main experiment, we will apply the strong DC magnetic field on the cavity. We will discuss about magnetic property of the superconductor in the further research. 

\subsubsection{Superconducting cavity quality factor}
The quality factor of RF system is defined as below.
\begin{equation} \label{eq:Qfactor}
    Q = \omega \frac{U}{P_{loss}}
\end{equation}
As you see in Eq.\ref{eq:Qfactor}, Q factor is inversely proportional to the power loss. From the linearity of the power loss the inverse of Q factor can be decomposed as below. (Q$_m$: measured Q factor, Q$_c$: Q factor for the conducting surface, Q$_a$ Q factor for the antennas, Q$_{cont}$: Q factor for the contact area)
\begin{equation}
    \frac{1}{Q_{m}} = \frac{1}{Q_{c}} +\frac{1}{Q_{a}} + \frac{1}{Q_{cont}}
\end{equation}

Especially, Q factor for the conducting surface is inversely proportional to the surface resistance.
\begin{equation}
    Q_c = \frac{\omega \mu}{R_s} \frac{\int_{cavity} | \overrightarrow{H} |^2 dv}{\int_{conductor} | \overrightarrow{H} |^2 ds} \equiv \frac{\Gamma}{R_s}
\end{equation}
The other factors in Q$_c$ are related to the cavity mode, resonant frequency and the geometry. To analyze experimental data, we have to obtain Q factor value for certain cavity mode and material by simulation. Based on the result, we can calculate the surface resistance in the experimental situation comparing the two Q factors.

We can also see the penetration depth change through temperature with changing resonant frequency of a cavity mode. When the penetration depth becomes shorter, the resonant frequency of the cavity becomes smaller. This effect is also related to the imaginary part of the surface impedance, because, as you see in the Eq.\ref{eq:SurfaceImpedanceS}, imaginary part of impedance is proportional to the penetration depth.
\begin{equation}
    \frac{\Delta \omega}{\omega} = \frac{\Delta X_s}{2 \omega \mu} \frac{\int_{conductor} | \overrightarrow{H} |^2 ds}{\int_{cavity} | \overrightarrow{H} |^2 dv}
\end{equation}
\begin{equation}
    R_s + i \Delta X_s = \frac{\Gamma}{Q_c} + 2i \Gamma \frac{\Delta \omega}{\omega}
\end{equation}

\subsubsection{Antenna loss and microwave measurement}
We can measure the unloaded Q factor with network analyzer with below formulas.
\begin{equation}
    Q_0 = Q_L (1+\beta_1+\beta_2)
\end{equation}
\begin{equation}
    \beta_i = \left| \frac{1-S_{ii}}{1+S_{ii}} \right|
\end{equation}

\subsection{Method}

\subsubsection{Physical property measurement system (PPMS)}

PPMS is the commercial measurement system which provides the cryogenic system to measure the electronic and magnetic property of the sample. Its working temperature is from 2 K to room temperature, and it can apply maximum 9 T magnetic field on the sample.

\begin{figure}[h]
\centering
\includegraphics[width=0.8\textwidth]{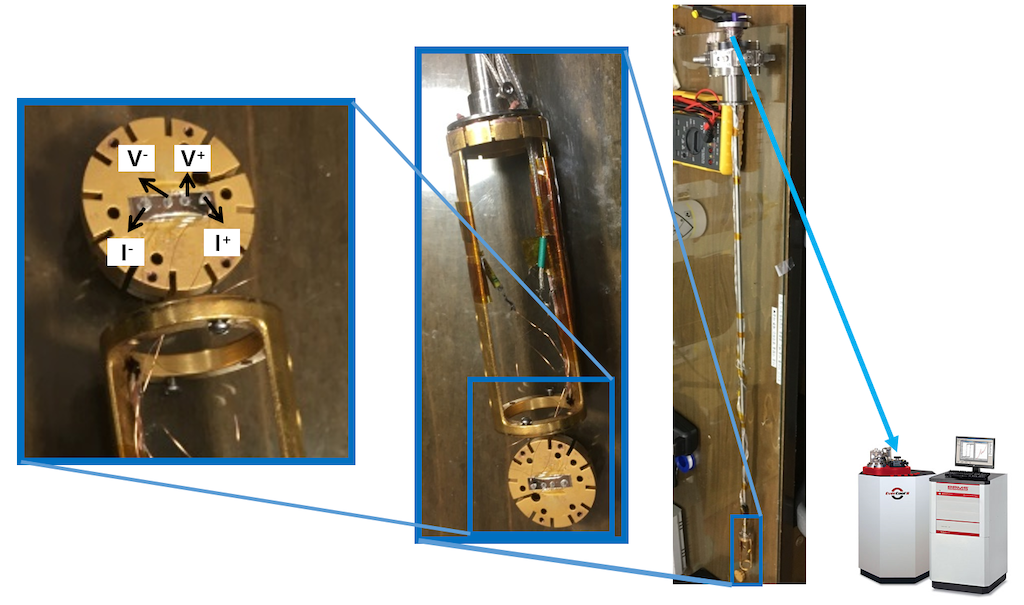}
\caption{\label{fig:PPMSmethod} The typical procedure to use PPMS.}
\end{figure}

As you can see in the Fig.~\ref{fig:PPMSmethod}, we put the four probes on the superconducting film to conduct 4-probe measurement. Through the coaxial cable, we apply DC current into the sample and measure the voltage between the two outer probes.

\subsubsection{Cavity preparation}

The split cavity is the first cavity which we used for testing magnetron sputtering method with NbTi and RF measurement. The geometry of the cavity is same as what we used in the main experiments. The split cavity is designed for using the TM$_{010}$ mode without contact problem. We used stainless steel as substrate material. The cavity inner diameter is 88 mm, and the inner height is 150 mm. The resonant frequency of TM$_{010}$ mode is 2.56 GHz. However its inner surface was not electro-polished, so the surface roughness was high.

\begin{figure}[h]
\centering
\includegraphics[width=0.8\textwidth]{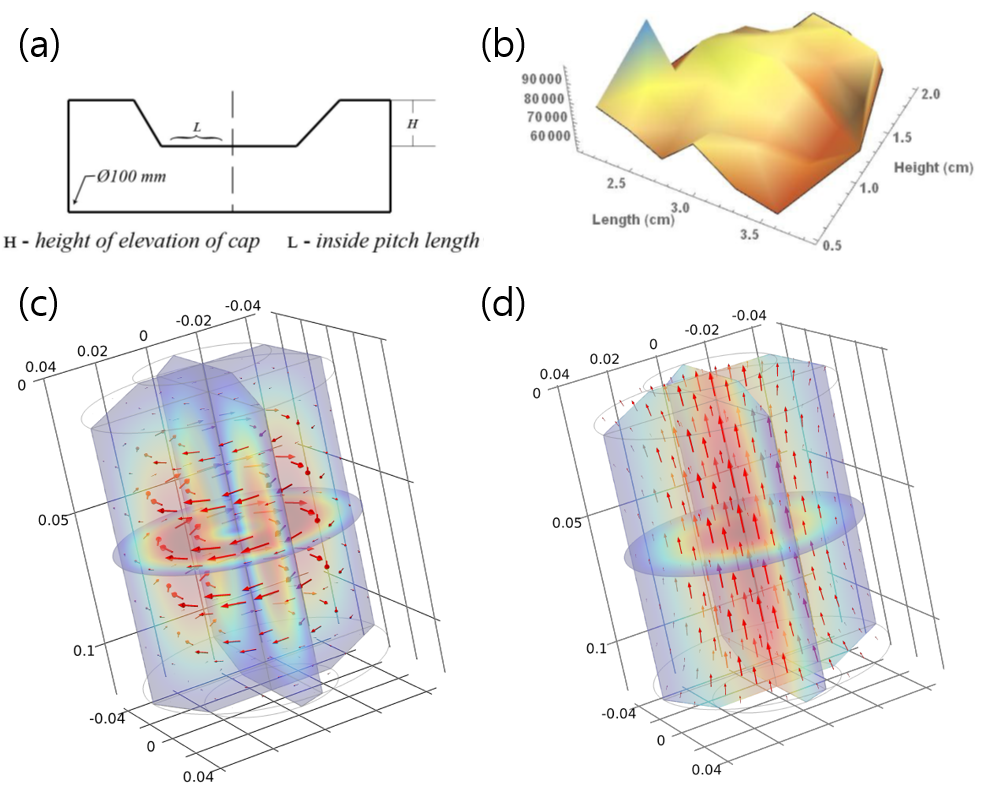}
\caption{\label{fig:FrustumCavitySimulation}~\cite{Frustum_Cavity_KUSP} (a) The schematic of frustum cap. H and L are the parameters for the COMSOL simulation, (b) The simulation result in the parameter space. It exhibits a peak at $L=2.200$cm and $H=1.100$cm. (c) TE$_{011}$ mode in the optimized condition. The arrows show the vectors of electric field. (d) TM$_{010}$ mode in the optimized condition. The surface current flows same as electric field direction, but at the contact line between frustum cap and hollow cylinder the heavy electric loss is produced.}
\end{figure}

The frustum cavity was designed to avoid high surface roughness and degenerate problem between TE$_{011}$ and TM$_{111}$ modes. First, the electro-polished stainless steel hollow cylinder was designed to make NbTi film uniform and clean. Its height is 100 mm and inner diameter is 88 mm However, to avoid contact problem, we had to use TE$_{011}$ mode which degenerate to the TM$_{111}$ mode. If surface current flows through contact line between two geometrically separated bulk, electric loss becomes large. It makes Q factor small. For this reason, the two modes should be separated.

The solution for separating degenerated modes was the frustum cap design. Since degenerated two modes can be separated by geometric distortion of cavity. If we change the geometry of the cavity, field distribution and Q factor will be changed. To optimize the condition, COMSOL simulation had been conducted. With two parameters which are a height of elevation of cap and an inside pitch length, we calculated Q factor. The optimal parameter was $L=2.200$ cm and $H=1.100$ cm. The Q factor of frustum cavity with OFHC copper and the resonant frequency were $Q \sim 45814$ and $f_{TE_{011}} \sim 4.365$ GHz.

\begin{figure}[t]
\centering
\includegraphics[width=0.6\textwidth]{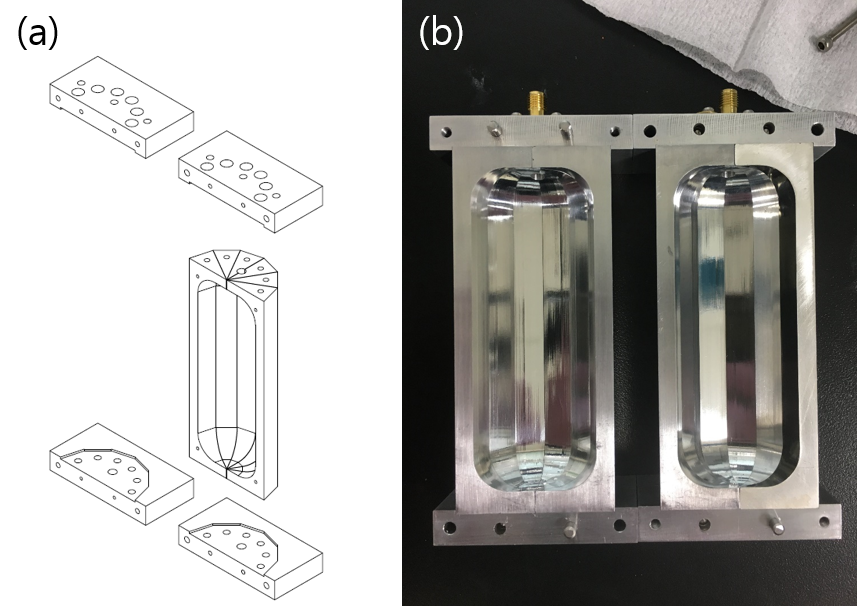}
\caption{\label{fig:PolygonCavity} (a) The schematic of the polygon cavity, (b) The picture of the polygon cavity.}
\end{figure}

The polygon cavity is designed for making RF resonant cavity with YBCO tape, because, to reduce contact loss and field distortion, the tapes should be carefully attached on the cavity inner surface. If we attach the tapes on the curved inner surface, the space between tape and substrate surface make large loss. To avoid the problem we designed polygon shape cavity. Its Q factor with OFHC copper is $Q \sim 16256$, and the resonant frequency is $f_{TM_{010}} \sim 6.854$ GHz. We used commercial 12 mm width AMSC YBCO tape.

\subsubsection{Magnetron sputtering}

\begin{figure}[h]
\centering
\includegraphics[width=0.8\textwidth]{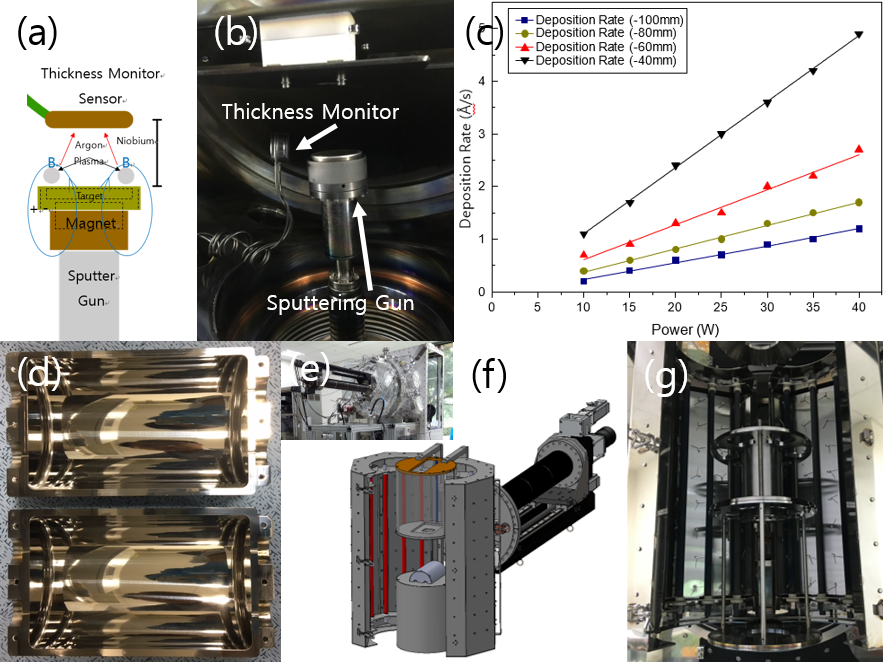}
\caption{\label{fig:MagnetronSputtering} (a) The schematic of the magnetron sputtering gun, (b) The picture of thickness monitor test setup, (c) The result of deposition rate with varying the height of the sputtering gun the experimental setup is same as Fig.~\ref{fig:MagnetronSputtering}-(b), (d) The split cavity after sputtering NbTi film, (e) The picture of outside of the sputtering chamber, (f) The schematic of the setup for sputtering NbTi film on the inner surface of the split cavity, (g) The picture of the setup for sputtering NbTi film on the inner surface of the stainless steel hollow cylinder.}
\end{figure}

Magnetron sputtering is one of the physical deposition method which use argon plasma and a permanent magnet. When we apply strong electric field between two electrodes in the certain amount of argon gas, the argon plasma is produced. The plasma trapped by magnetic field which is induced by permanent magnet contains a lot of argon nucleus, and they hit the target material. The atoms in the target material gets energy and they become to move freely. Finally, the atoms goes to the substrate and they form a film on the surface. Fig.~\ref{fig:MagnetronSputtering}-(a) shows whole process of the sputtering.

Before the main sputtering process, we did thickness monitor test for the system. [Fig.~\ref{fig:MagnetronSputtering}-(b)] We install the thickness monitor around the sputtering gun and we read deposition rate on the control panel and change the position of the gun and the power which is given to the gun. Thus we could define the deposition rate in the certain condition. [Fig.~\ref{fig:MagnetronSputtering}-(c)]

In this experiment we use $5$ mTorr argon gas with $10^{-8}$ Torr base pressure, and we deposit the NbTi film with $1$ $\mu$m thickness. For the split cavity we maintain the temperature of the cavity at 200 $^{\circ}$C, and we deposit NbTi from the equidistant 2 points during 30 minutes 8 times. In the case of the hollow cylinder, we had sputtered NbTi 35 minutes with moving the gun from the top to the bottom in both direction. The substrate temperature was 200 $^{\circ}$C. The sputtering system can rotate with the motor, so the process have been done efficiently. [Fig.~\ref{fig:MagnetronSputtering}-(f)]

\subsubsection{Cryogenics and data acquisition}

To understand RF property of the conducting surface, we have to vary the temperature and apply and receive RF signal with the cavity. Our system is based on the BlueFors LD400. It has a pulse tube and a dilution unit which can cool down the fridge to 10 mK. We have installed our cavities on the mixing chamber of LD400 system and we communicate with the cavity through RF chain inside of the cryogenics. The system can monitor the pressure of each point and temperature of plates and additional cavities.

\begin{figure}[h]
\centering
\includegraphics[width=0.8\textwidth]{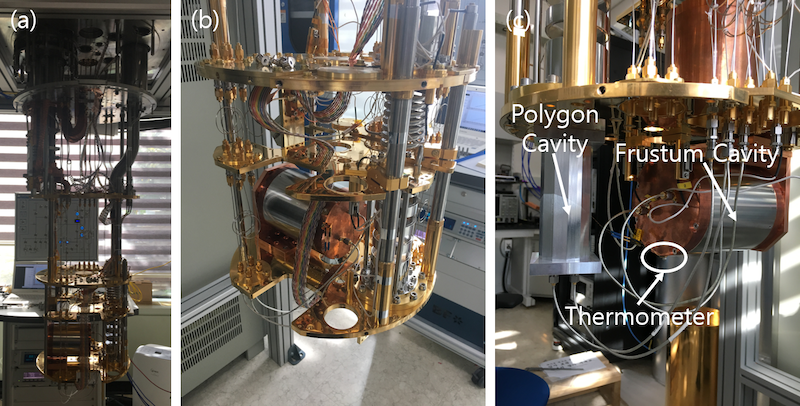}
\caption{\label{fig:Cryogenics} (a) Picture of the inside of LD400 system, (b) Picture of the installed frustum cavity on the mixing plate, (c) The picture of the installed frustum cavity and the polygon cavity under the mixing plate.}
\end{figure}

To collect RF data, we use NIVISA package for CentOS with GPIB connection with the network analyzer Keysight E5063A. Using VISA command we can give RF signal to the cavity and collect data which are loaded Q factor and S11/S22 reflection coefficient. In case of temperature data, we can collect them from the LD400 system. In our center we collect all the pressure and the temperature data continuously in the server computer, so each computer can download it in real time. All the following results are obtained from these systems.

\subsection{Result}

\subsubsection{Physical property measurement system (PPMS)}

Before the deposition of superconducting film on the cavity wall, we had made the samples of NbTi film on stainless steel substrate. From the SEM measurement we could define the thickness of the film and atomic ratio of Nb atoms and Ti atoms. The number ratio between two elements are about 6:4.

\begin{figure}[h]
\centering
\includegraphics[width=0.8\textwidth]{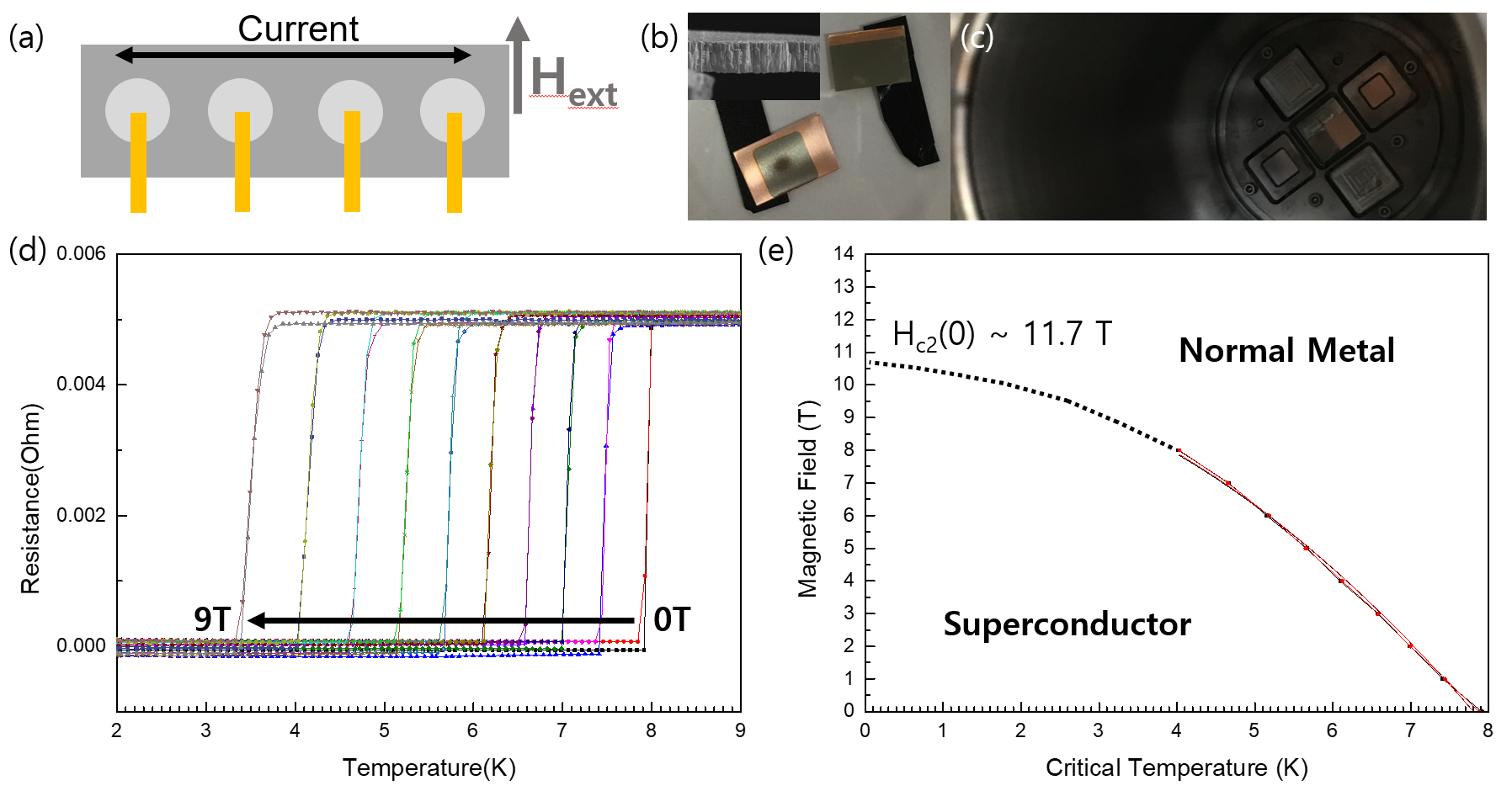}
\caption{\label{fig:PPMSresult} (a) The schematic of 4-point measurement circuit for the NbTi film samples, (b) NbTi film samples on the copper substrate and SEM picture of NbTi film segment, (c) As the picture shows, we put small copper and stainless steel substrates inside the sputtering chamber, (d) The result of PPMS measurement for NbTi film on the stainless steel substrate, (e) The critical temperature plot for each magnetic field strength. The boundary divide normal metal and superconductor states.}
\end{figure}

DC electronic property of the samples was measured by PPMS system. From the R-T curve we could define the critical temperature of the sample. [Fig.~\ref{fig:PPMSresult}-(d)] Measuring critical temperature with varying the magnetic field give phase boundary. [Fig.~\ref{fig:PPMSresult}-(e)] To make sure the direction of magnetic field is same as main experiment, we applied the magnetic field in-plane direction. The critical temperature without magnetic field was around 8\,K, and H$_{c2}$ was about 11.7\,T.

\subsubsection{Split cavity}

\begin{figure}[h]
\centering
\includegraphics[width=0.8\textwidth]{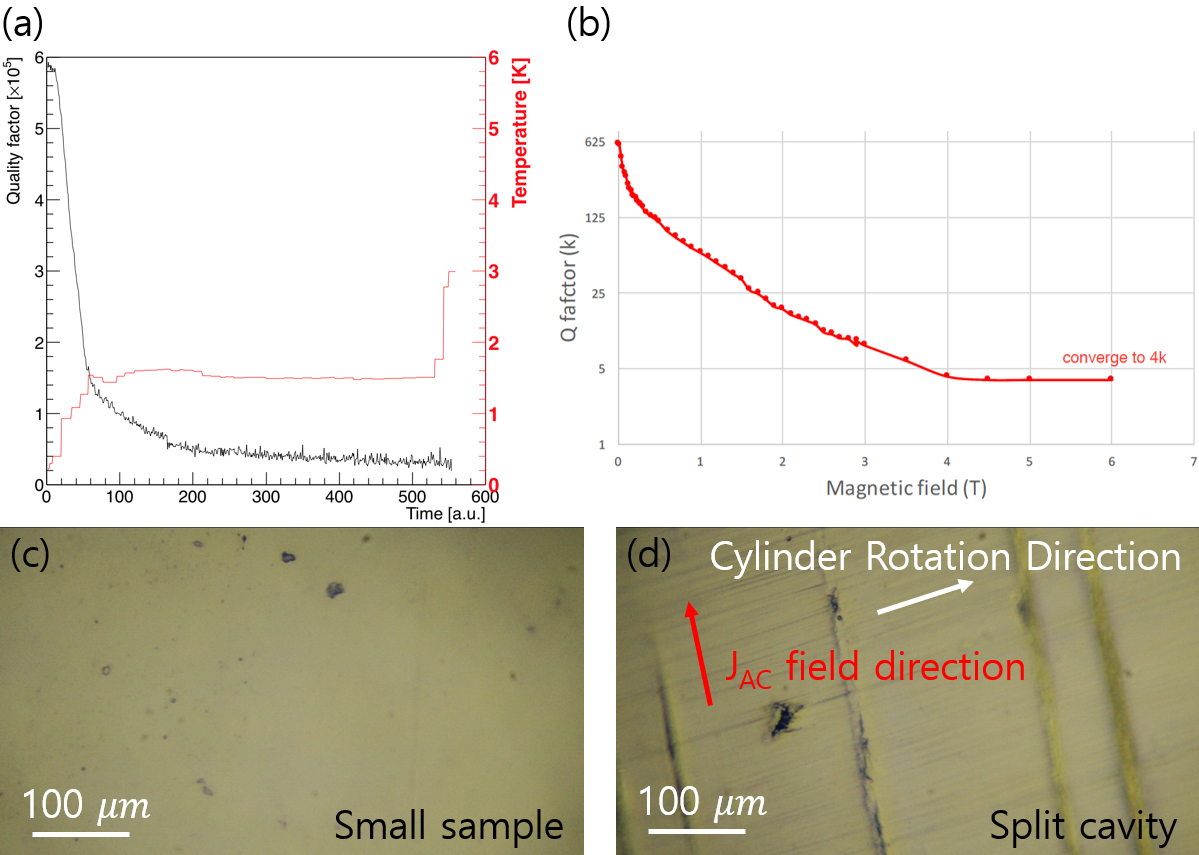}
\caption{\label{fig:SplitCavityResult} (a) Q factor and Temperature data in time series. The transition temperature is around 1\,K, (b) Q factor of the split cavity with varying the applying magnetic field. Q factor become 1/5 at 0.5\,T, and at 4\,T superconductivity vanished, (c) The surface picture of electro-polished stainless steel sample, (d) The surface picture of split cavity inner wall.}
\end{figure}

The data from the split cavity measurement show that the transition temperature of superconducting film is 1K, and H$_{c2}$ as 4 T. However this result is have no consistency with the PPMS DC measurement data from the small sample. We have repeatedly confirmed the transition temperature and H$_{c2}$ of NbTi films are around 8\,K and 11\,T.

These symptoms shows that the surface quality on the split cavity was not good. As you see in the Fig.~\ref{fig:SplitCavityResult}-(c) and (d), the surface condition of split cavity is much worse than the electro-polished small sample. Thus we can conclude that we can improve Q factor electro-polishing the cavity surface.

\subsubsection{Frustum cavity}

\begin{figure}[h]
\centering
\includegraphics[width=0.9\textwidth]{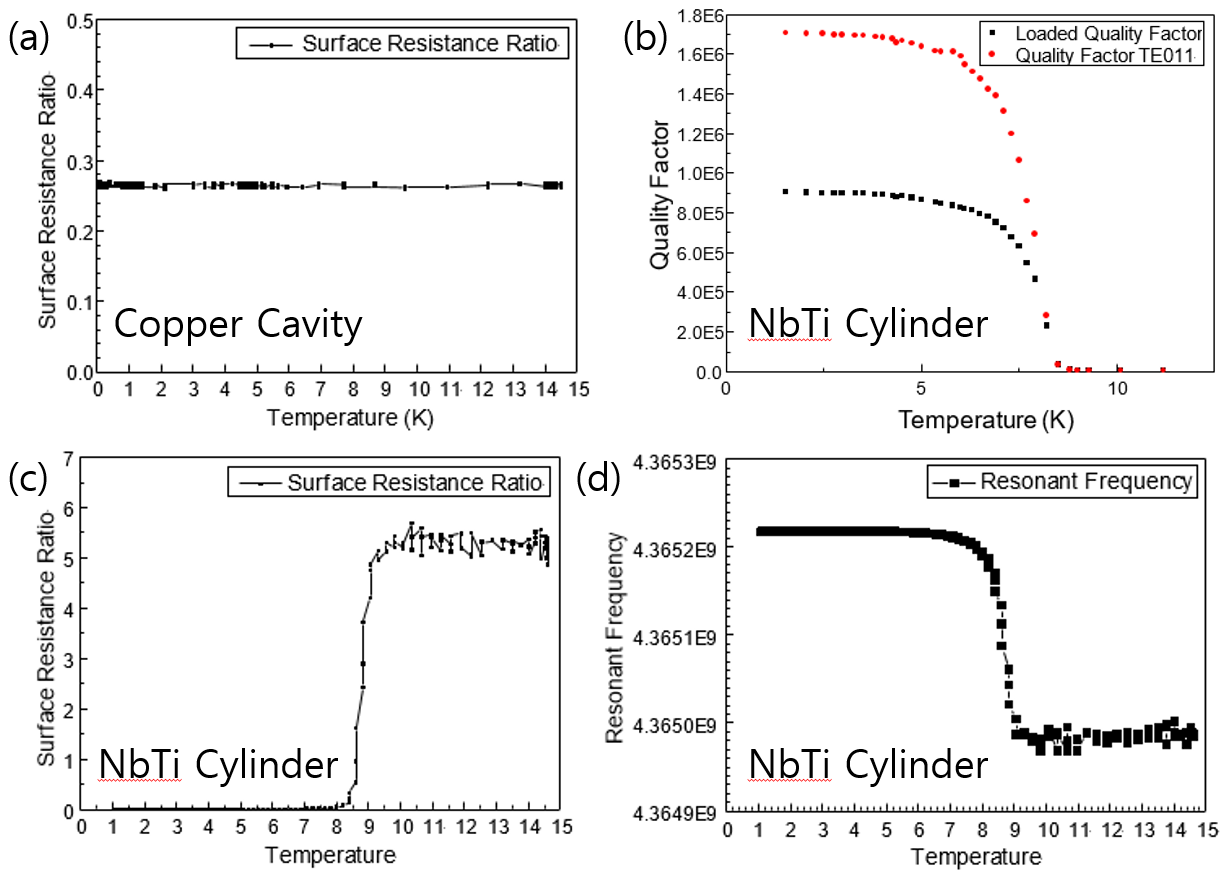}
\caption{\label{fig:FrustumCavityResult} (a) The temperature dependent surface resistance ratio between the frustum cavity with the copper hollow cylinder and the known surface resistance of OFHC copper, (b) The temperature dependent loaded and unloaded Q factor data of the frustum cavity with NbTi deposited stainless steel hollow cylinder, (c) Calculated surface resistance ratio between the frustum cavity in (b) and the known value of OFHC copper, (d) The resonant frequency data with varying temperature for the cavity in (b).}
\end{figure}

First we measured Q factor of the frustum cavity with OFHC copper cylinder to verify the copper surface resistance at low temperature. [Fig.~\ref{fig:FrustumCavityResult}-(a)] We could calculate the surface resistance ratio at low temperature by calculating the ratio between simulated Q factor (Q $\sim$ 45814) and measured Q factor, because Q factor is inversely proportional to the surface resistance of the surface material.

Based on the copper cylinder data and calculation, we could calculated the portion of electric loss only from the hollow cylinder with subtracting the loss from the copper frustum caps. Since the Q factor is inversely proportional to the total electric loss of the cavity, we can calculate the Q factor for only cylinder. The formula is,
\begin{equation}
    \frac{1}{Q_{cavity}} = \frac{1}{Q_{caps}} + \frac{1}{Q_{cylinder}}.
\end{equation}
From the data of the copper cylinder case we can calculate $Q_{caps}$ with electric loss ratio between caps and cylinder. ($1.11:12.52$, COMSOL) With $Q_{caps}$ we can calculate $Q_{cylinder}$ from the data of the NbTi deposited stainless steel cylinder case.

In the frustum cavity, the surface condition was improved by electro-polishing, so the result was better than the former case. The transition temperature is appeared at around 8 K which have consistency with former DC measurement results, and the surface resistance is improved 178.6 times at low temperature. From the number we can calculate the Q factor of TM$_{010}$ mode with split cavity as 8,200,000 which is much bigger than the result from the split cavity. (Q$_{split}$ $\sim$ 625,000)

The resonant frequency data shows another aspect of the conductor-superconductor phase transition. As we discussed in the background, a superconductor has almost constant RF penetration depth after phase transition. The penetration depth of NbTi is known as 240\,nm which is much smaller than the skin depth of usual conductors($1\sim2\,\mu$m)~\cite{CERN_FCC}. Decreasing length of RF penetration makes cavity dimension effectively small, so the resonant frequency increase at the transition temperature. [Fig.~\ref{fig:FrustumCavityResult}-(d)]

\subsubsection{Polygon cavity}

\begin{figure}[t]
\centering
\includegraphics[width=0.8\textwidth]{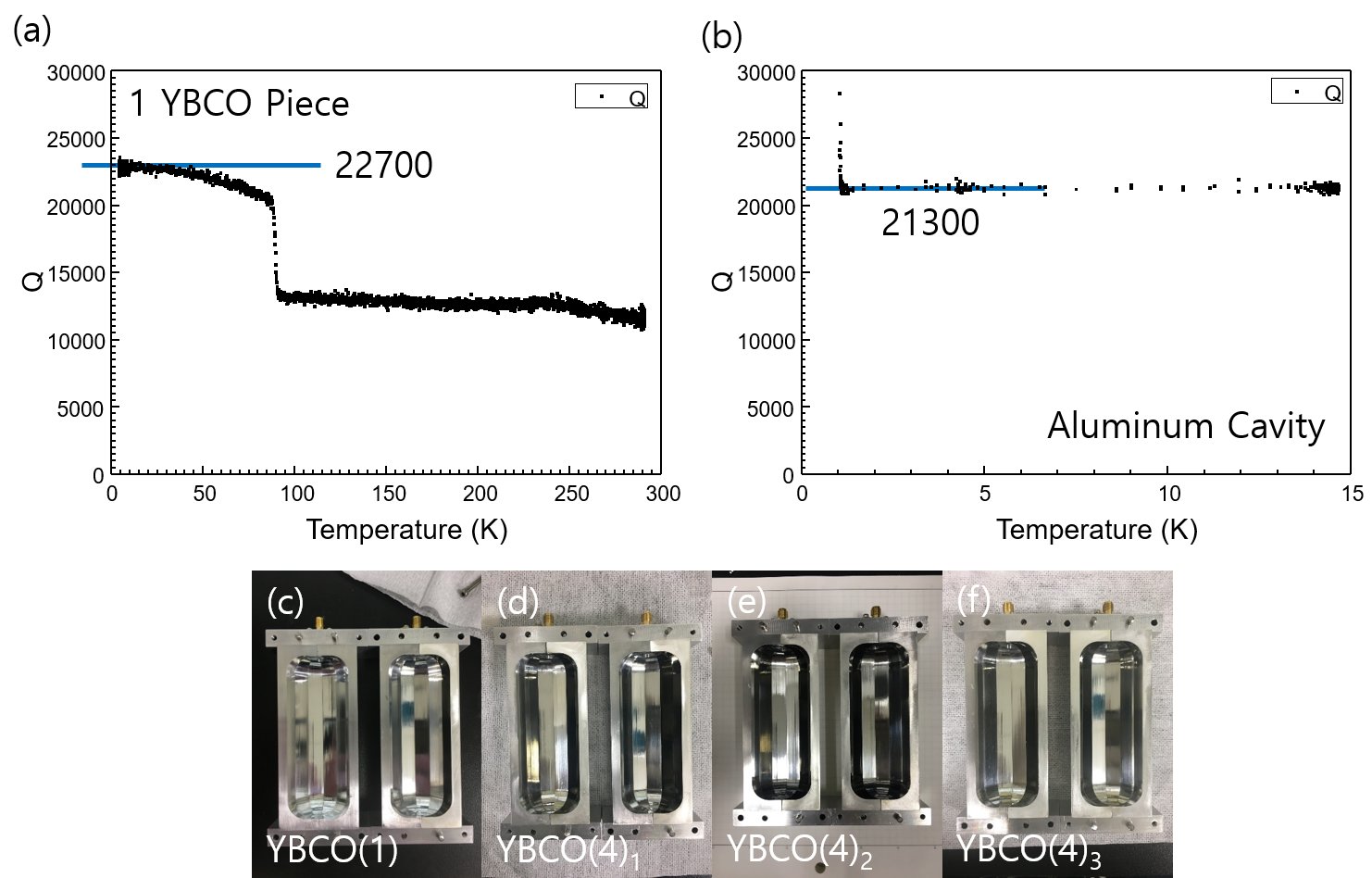}
\caption{\label{fig:PolygonCavityResult} (a) The Q factor data of the polygon cavity with one YBCO tape at the one side. The value is larger than Aluminum Caivity, (b) The Q factor data of the polygon cavity without YBCO tape, (c) The polygon cavity with one YBCO tape, (d) The polygon cavity with four YBCO tapes. Each two tapes are facing each other, (e) The polygon cavity with four YBCO tapes. The four tapes are fully connected, (f) The polygon cavity with four YBCO tapes. We polished the two sides of YBCO tape and sputtered silver in 2\,$\mu$m thickness.}
\end{figure}

Same as the frustum cavity measurement we obtained Q factor data with varying temperature. The transition temperature was around 90\,K, and it have consistency with known value. At low temperature, the Q factor of the polygon cavity with YBCO tape was higher than the full aluminum cavity. [Fig.~\ref{fig:PolygonCavityResult}-(a),(b)] It shows YBCO film has much smaller surface resistance than aluminum at 4.2\,K. The film quality was also very uniform. The surface resistance of the polygon cavity with four YBCO tapes and just one tape show the similar value.

However it seems that the main loss of the cavity is originated from the two sides of the YBCO tape. Below 1\,K, the polygon cavity with YBCO tapes show relatively smaller Q factor than the full aluminum cavity. To overcome this problem, we treated the sides of tapes in another way. After polishing the two sides of each film, we sputtered silver. As you can see in Table~\ref{table:PolygonCavityResult}, the Q factor value increased.

\begin{table}[h]
\centering
\begin{tabular}{|c|c|c|c|c|c|}
\hline
              & Aluminum & YBCO(1) & YBCO(4)$_{1}$ & YBCO(4)$_{2}$ & YBCO(4)$_{3}$ \\ \hline
Q$_0$ (4.2 K) & 21,300 & 22,700 & 28,300 & 28,400 & 32,300 \\ \hline
R$_s$ ratio (4.2 K) & 1 & 0.260 & 0.258 & 0.250 & -0.021 \\ \hline
Q$_0$ (0.5 K) & 3,000,000 & 870,000 & - & 175,400 & 361,200 \\ \hline
R$_s$ ratio (0.5 K) & 0.007 & 0.210 & - & 0.350 & 0.162 \\ \hline
\end{tabular}
\caption{\label{table:PolygonCavityResult} The Q factor and the surface resistance result for the polygon cavities. From the 4.2\,K surface resistance results of ``Aluminum", ``YBCO(1)", ``YBCO(4)$_1$", and ``YBCO(4)$_2$", we can notice that the tape quality is well controlled. From the case of ``YBCO(4)$_3$", we can notice that the two sides of tape gives large loss, because the surface resistance almost vanishes at 4.2\,K. The 0.5\,K results may not accurate, because at low temperature the Q factor fluctuation was large.}
\end{table}

\subsection{Discussion}

From the entire work, we studied basic theoretical and experimental approach to make superconducting cavity for axion search. From measuring Q factor and resonant frequency of the resonant cavity, we could see the surface resistance and penetration depth change through changing temperature. The result shows that the critical temperature of NbTi was around 8\,K and YBCO was about 90\,K. In the frustum cavity result, we also noticed that penetration depth of NbTi drastically decrease at critical temperature and it reached to the certain value. This observation matches the formal understanding of S-wave superconductor. At low temperature the penetration depth remains same.

The experimental technique to obtain uniform film and reduced contact loss also have been studied. In the case of sputtering NbTi film on metal surface, electro-polishing is necessary. Since, without electro-polishing, the surface of substrate have so many deep scratches. We could overcome the limitation of the first split cavity result by electro-polishing the inner surface of the stainless steel cylinder. Good quality of polishing was also important in the polygon cavity. We could obtain larger Q factor by polishing the two sides of the YBCO tape.

The remain tasks are related to the magnetic property of the superconductor. We have to reduce the electric loss induced by vortices when we apply high magnetic field on the superconducting cavity. Type II superconductor, such as NbTi and YBCO, form magnetic vortices inside the material when magnetic field is applied. If RF field penetrate into the superconductor, the vortices move due to surface current and changing magnetic field distribution makes an electric loss.

Therefore, to make efficient cavity, we have to make vortices stationary. To make vortex stable, vortex pinning and vortex direction are important. Pinned vortex do not move fast, so they do not produce heat. The vortex which direction is parallel to the surface current direction does not move due to Lorentz force law. We are planning to conduct Q factor measurement of the cavities in the strong magnetic field.

%% file: 1.4.7/main.tex
The signal from axion haloscope experiment is RF, and two  DAQ methods are possible in general: 
frequency domain DAQ and time domain DAQ. In
the case of frequency domain data taking, spectrum analyzer is widely used equipment which employs
the sweep analysis to take spectra. However, the prevalent problem of DAQ with spectrum analyzer is
high dead-time ratio, which is the ratio of off-DAQ time and on-DAQ time. The previous study on
dead-time ratio was found to be about 70\% 
when single spectrum is individually collected. This ratio was also found to get much lower
when the spectra are averaged via on-line function of spectrum analyzer, but it is not a preferred
way because of the degradation in data quality, and removing all the timing information. 

Time domain DAQ can be a breakthrough to solve this dead-time issue because of its capability of
parallel data taking with multi-channels. In this study, the preliminary study on single-channel DAQ
was done to analyze the dead-time by testing the KeySight digitizer M9211~\cite{WEB_M9211}. For further
understanding of time domain RF signal, Off-line analysis was done additionally by transforming time
domain data into frequency domain using fast Fourier transform (FFT).

\subsection{Digitization DAQ method in axion experiments}
In axion signal processing using the digitizer, there are three important parameters: intermediate
frequency $(f_i)$,  sampling rate $(f_s)$, and frequency resolution $(\Delta f)$ of FFT.
Intermediate frequency is the down-converted frequency of axion signal coming from the cavity, and
is determined by the RF setup of the experiment while there could be some constraints coming from
other electronics components such as driving frequency of the amplifier. Once $f_i$ is fixed, it
imposes the constraint on $f_s$ in which $f_s$ should be higher than $2f_i$, called Nyquist
frequency to ensure that there is no aliasing among reflected frequency bands. $(\Delta f)$ is not
related to neither the experimental setup nor the signal-to-noise ration (SNR), but it has its own
importance in signal analysis after FFT.

The digitizer used in this test is Keysight M9211A PXI-H 10-bit UWB High-Speed IF Digitizer. 
It is a one-slot 3U PXI (PCI eXtensions for Instrumentation)-Hybrid~\cite{WEB_PXI} single-channel Ultra-wideband IF digitizer able to capture
signals at up to 3\,GHz and running at up to 4\,GS/s, significantly reducing data acquisition and
testing times. The M9211A UWB IF Digitizer comes with on-board memory of 64\,MBytes. 
The data transfer rate is maximum 100\,MB/s, and Spurious-free dynamic range (SFDR) is 42\,dB at 1\,GHz. 
Owing to its state-of-art sampling speed, the RF signal from axion experiment, slower than 2\,GHz,  can be recorded without any down-conversion process which adds additional noise.
However, in general, it is preferred to down-convert the RF signal to an IF level, and digitize.

\subsubsection{Dead-time measurement of digitizer DAQ for axion experiment}

\begin{figure}[b]
\centering 
\includegraphics[width=0.5\textwidth]{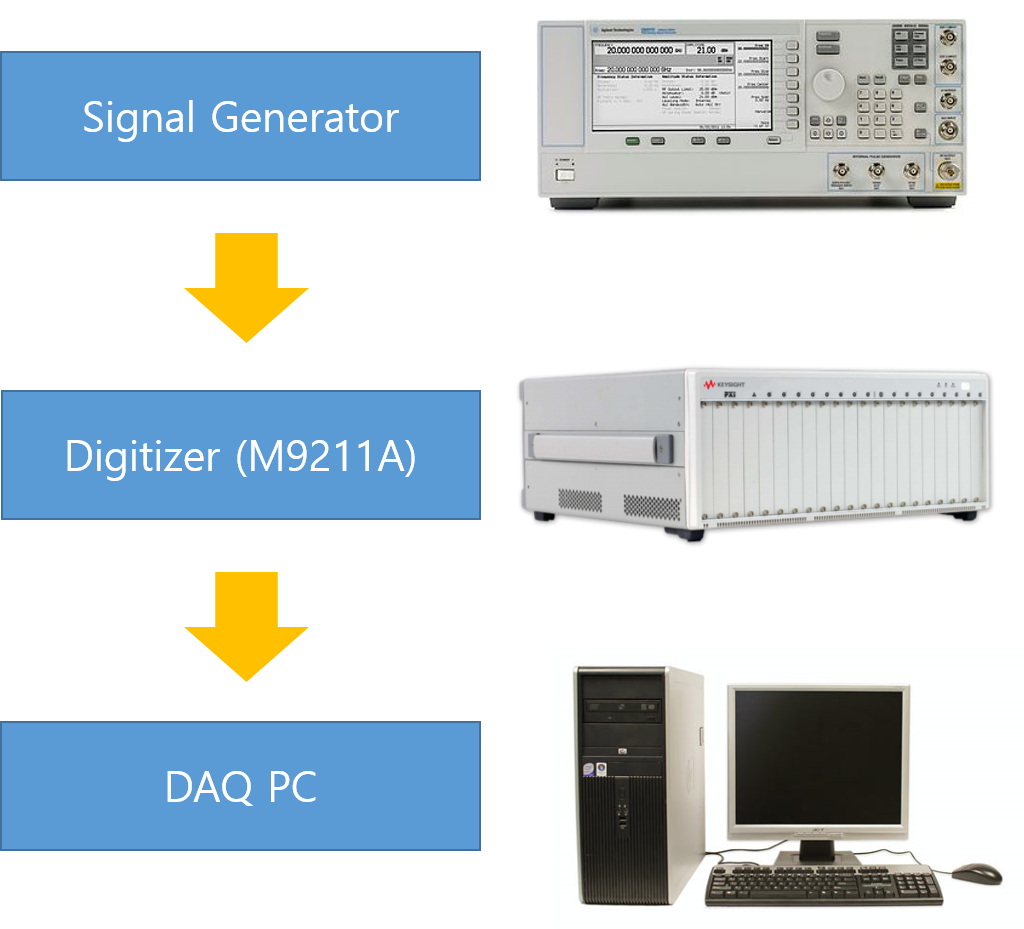}
\caption{ Experimental setup for testing the performance of the digitizer.
}\label{fig:muec_daq:digitizerSetup}
\end{figure}

The schematic of the experimental setup is shown in Fig.~\ref{fig:muec_daq:digitizerSetup}, where signal generator is
optional in measuring dead-time. The sequence of single DAQ iteration is as follows: 

\begin{enumerate}
\item Sampling: 50M ADC samples $(N_{tot})$ of the RF signal are recorded on the memory board of M9211A whose capacity is
64M samples. 
\item Transferring: Data is transferred to DAQ PC via PCIe cables (8-lanes). 
\item Recording: Data written in ROOT or binary format on the hard disk (NVMe(Non-Volatile Memory Express~\cite{WEB_NVMe}) and HDD).
\item Analysis: Offline FFT analysis.
\end{enumerate}

\begin{figure}[t]
\centering
\includegraphics[width=0.45\textwidth]{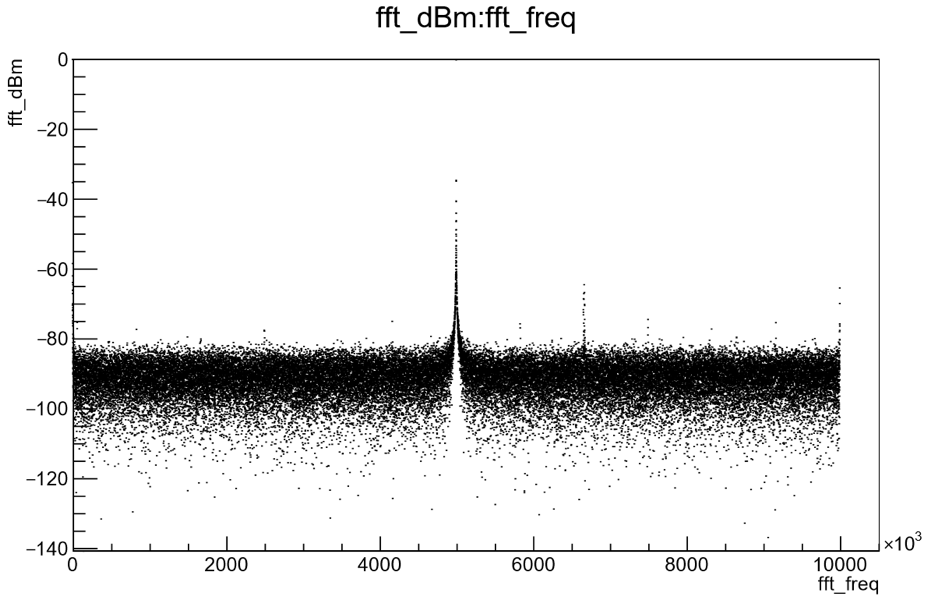}
\includegraphics[width=0.52\textwidth]{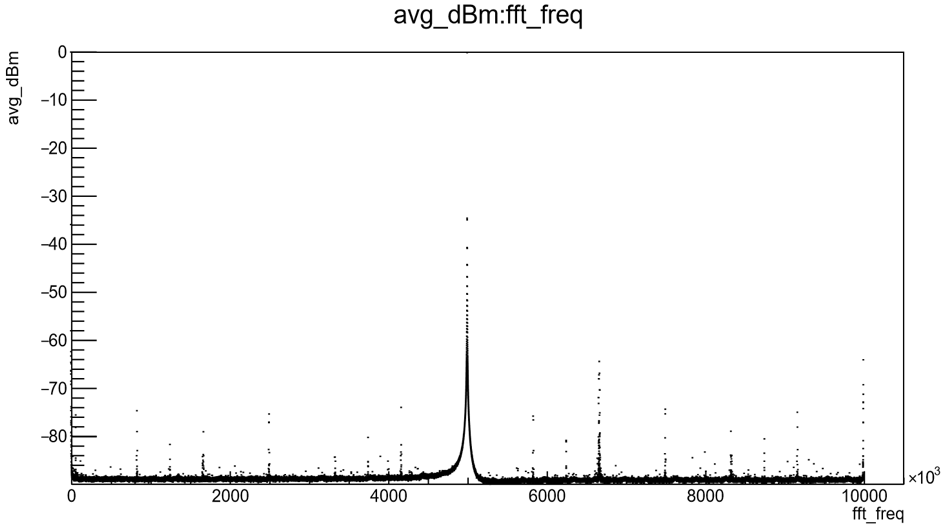}
\caption{FFT result of digitizer data, (Left) without averaging, (Right) with 500 spectrum averaging. Note that there are data points at 0\,dBm around the center of distributions, which are not actually visible.}
\label{fig:muec_daq:digitizer_fft_result}
\end{figure}

This fast sampling and offline FFT analysis scheme is basically working as expected. Figure \ref{fig:muec_daq:digitizer_fft_result} shows analyses data set with 5\,MHz test signal with 0\,dBm.  

The time taken in Sampling corresponds to live-time
$(t_l)$, and the time taken in Transferring and Recording to dead-time $(t_d)$, respectively. Since $N_s$ is
directly related to the SNR of the experiment, it is natural to set figure of merit (FOM) as the
number of spectra per single DAQ iteration:

\begin{equation}\label{eq:muec_daq:FOM}
\textrm{FOM}=\frac{N_s}{t_l +t_d}.
\end{equation}

Eq.~\ref{eq:muec_daq:FOM} can be rewritten by putting the relations between variables:
\begin{equation}
\textrm{FOM}=\frac{\Delta f}{1+f_s \frac{t_D}{N_{tot}}}=\frac{1}{N}\frac{f_s}{1+f_st_d/N_{tot}}.
\end{equation}

One aspect of the above equation is that lowering $N$ is the efficient way to get higher FOM, and
higher $f_s$ also makes better FOM but not as good as lowering $N$. Since the only unknown parameter
in Eq.~\ref{eq:muec_daq:FOM} is $t_D$, we focused on 
measuring $t_D$ in the experiment with 8/16/64 bits data type, respectively.

\subsubsection{Noise level measurement with FFT analysis}

The time domain data obtained is analyzed by applying FFT in fftw~\cite{IEEE_93_216_2005} library. The magnitude of FFT
output in Volt unit was normalized be dividing with $N/2$, which retains the sinusoidal signal level
from signal generator, but noise level decreases with $\sqrt{N}$. Power in Watt is subsequently
obtained considering $50 \Omega$ termination. To investigate the property of signal and noise
individually, two measurements were conducted (1) when signal is coming from signal generator, and
(2) quasi-flat noise is coming from signal generator with absolute power.

\subsection{Feasibility of using digitizer DAQ for axion experiment}
\subsubsection{Dead-time measurement} 

The results of dead-time measurement, which is the average of ten iterations, is shown in Table~\ref{tab:muec_daq:daqtime}. $f_s$ was given as $20 \textrm{ MHz}$, so the measurement time in table is consistent
with $N_{tot}/f_s$. The transfer time, which is time to take for sending ADC data to DAQ PC via PCIe
cable, shows tendency to increase with data size. However, measured transfer speed $(<200 \textrm{
MB/s})$ itself was not consistent with the expected data bandwidth of 8-lanes PCIe cable $(4
\textrm{ GB/s})$, and this requires further study. Writing time in ROOT format was much higher than
that in binary format, while the measured time is also not consistent with specification of writing
speed of hard disk. For example, the sequential writing speed of NVMe is higher than 2\,GB/s in
specification, but measured one is even lower than 1\,GB/s.

\begin{table}[t]
\centering
\caption{DAQ time breakdown in single iteration with 50 MSa. All measured times are averaged with ten iterations.}
\label{tab:muec_daq:daqtime}
\begin{tabular}{cccccc}
\hline \hline  
Data type & Data size & Measurement & Transfer & \multicolumn{2}{c}{Writing (sec)} \\ \cline{5-6}
(bit) & (MByte) & (sec) & (sec) & ROOT (NVMe) & Binary (NVMe/HDD) \\ \hline
8   & 50 & 2.5  & 1.2 & 3.7                & 0.1/0.3          \\
16  & 100 & 2.5  & 1.7 & 4.4                & 0.2/0.4          \\
64  & 400 & 2.5  & 2.4 & 4.3                & 0.4/0.7       \\   \hline \hline   
\end{tabular}
\end{table} 

\begin{figure}[b]
\centering
\includegraphics[width=0.7\textwidth]{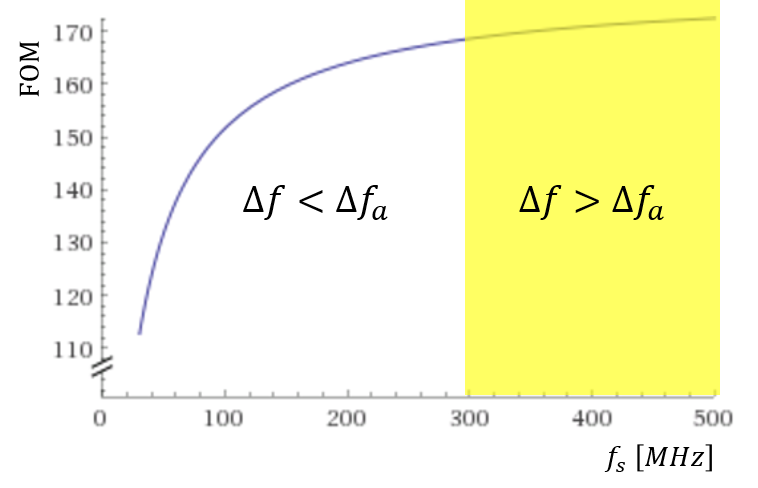}
\caption{FOM vs. $f_s$ when $t_d$, $N$, and $N_{tot}$ are 2.8 sec, $10^5$, and $5\times 10^7$, respectively. $\Delta f_a$ is assumed to be 3 kHz. 
}\label{fig:muec_daq:FOMvsFs}
\end{figure}

\begin{figure}[h]
\centering
\includegraphics[width=0.7\textwidth]{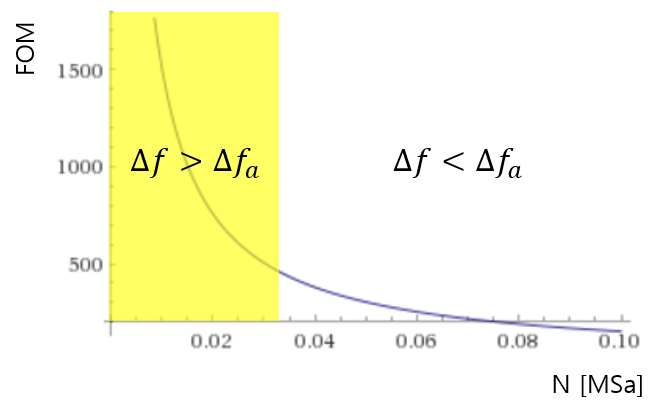}
\caption{FOM vs. $N$ when $t_d$, $f_s$, and $N_{tot}$ are 2.8 sec, $100 \textrm{ MHz}$, and $5\times 10^7$, respectively. $\Delta f_a$ is assumed to be 3 kHz. 
}\label{fig:muec_daq:FOMvsN}
\end{figure}

\begin{figure}[h]
\centering
\includegraphics[width=0.7\textwidth]{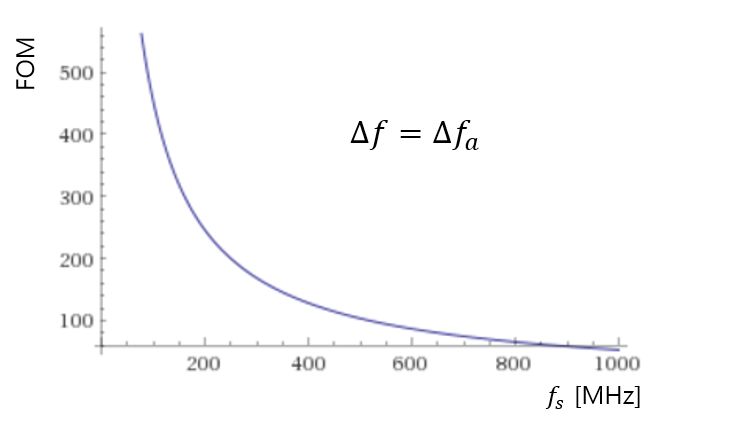}
\caption{FOM vs. $f_s$ when $\Delta f$ is fixed to $3 \textrm{ kHz}$, equivalently to $f_s/N=3 \textrm{ kHz}$.
}\label{fig:muec_daq:FOMboundary}
\end{figure}

FOM for 64 bit data type is also analyzed in terms of $f_s$ and $N$. Figure \ref{fig:muec_daq:FOMvsFs} shows the
relation between FOM and $f_s$ where N is set to $10^5$ which corresponds to 500 spectra in single
iteration. $t_d$ is assumed to be same regardless of $f_s$, and the prohibited region of $f_s$ from
the constraint $\Delta f < \Delta f_a \sim 3 kHz$ is shaded in yellow. The maximum value of FOM at
$\Delta f = 3kHz$ is 169. In Figure \ref{fig:muec_daq:FOMvsN}, we kept all parameters same but set $f_s$ to 
100\,MHz. Here, we get maximum FOM of 455. In Figure \ref{fig:muec_daq:FOMboundary}, $\Delta f$ is fixed to
$\Delta f_a$ (3\,kHz) to see the relation between FOM and $f_s$ with the constraint on frequency
resolution. The graph shows the tendency that FOM increases with lower sampling rate.

\subsubsection{Noise level measurement with FFT}

\begin{figure}[h]
\centering
\includegraphics[width=0.47\textwidth]{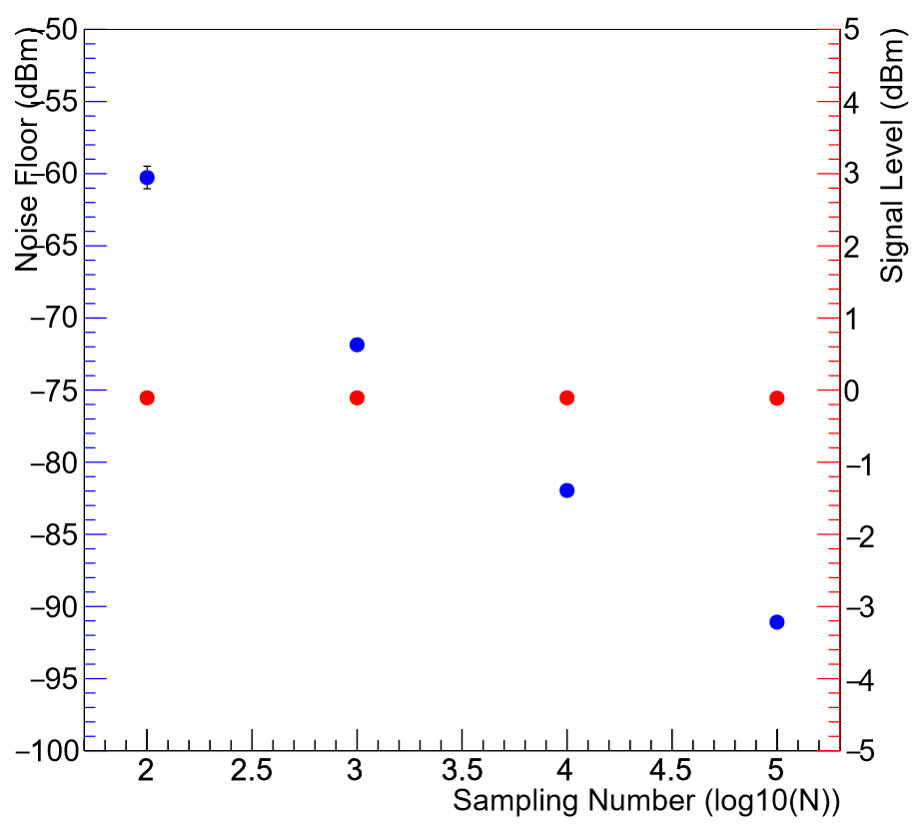}
\includegraphics[width=0.45\textwidth]{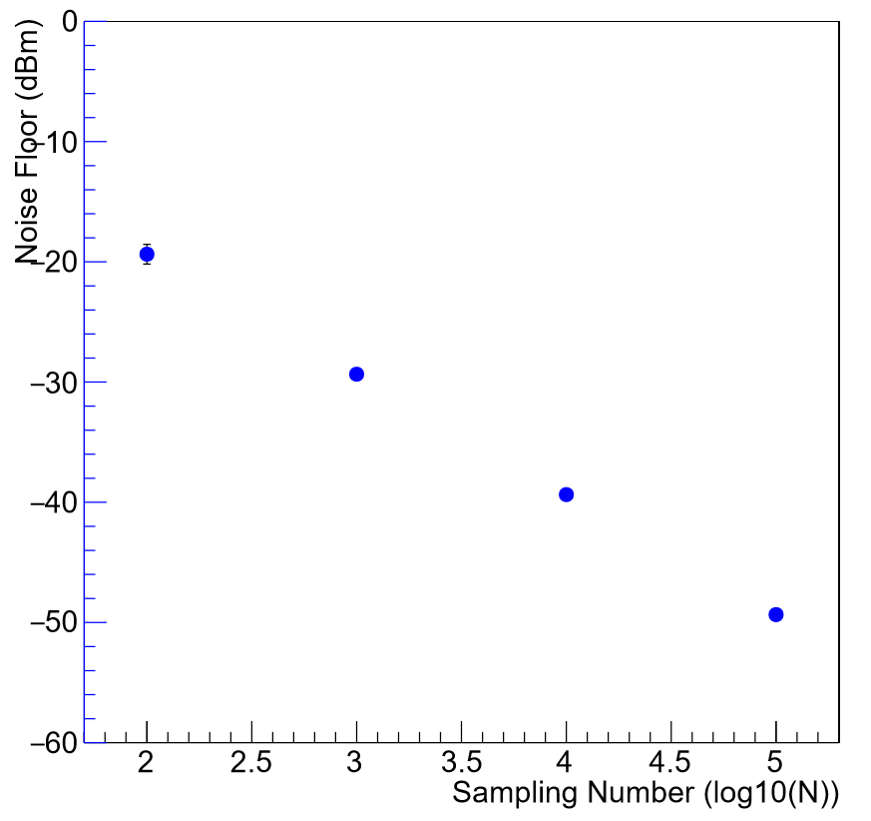}
\caption{(Left) Measurement of noise/signal level in terms of sampling number when 0 dBm sine wave is given to digitizer. Blue dots are noise levels, and red dots are signal levels, respectively. (Right) The measurement of noise level (blue dots) in terms of sampling number when 0 dBm flat noise is given to digitizer.
}\label{fig:muec_daq:SineWave}
\end{figure}

The obtained time domain data was transformed into frequency domain for the analysis of amplitude.
To test if the signal amplitude is achieved as expected, 0\,dBm power from signal generator was given
to digitizer. Since the normalization with $N/2$ retains signal level, the signal levels with
various $N$ ($10^2, 10^3, 10^4, 10^5$) were analyzed. In the left plot of Fig.~\ref{fig:muec_daq:SineWave}, we
can see the signal level is constant irrespective of the sampling number, while the noise level
decreases by 10\,dBm with a larger sampling number by a factor of ten. This behavior of noise level
is predicted because increasing sampling number by a factor ten means that noise power is dissipated
into ten times more bins. Considering the definition of $\textrm{dBm} (=10\log{P}$\,[mW]),
dBm value should go down with 10\,dB when the power is less by a factor of ten. In the right figure
of Fig.~\ref{fig:muec_daq:SineWave}, absolute noise power of 0\,dBm was given instead, and we can clearly see
that the noise level follows the equation of $0\,\textrm{dBm}-10\log{N}$.

\subsubsection{Deadtime measurement of a slower digitizer and comparison}
\begin{figure}
\centering
\includegraphics[width=0.7\textwidth]{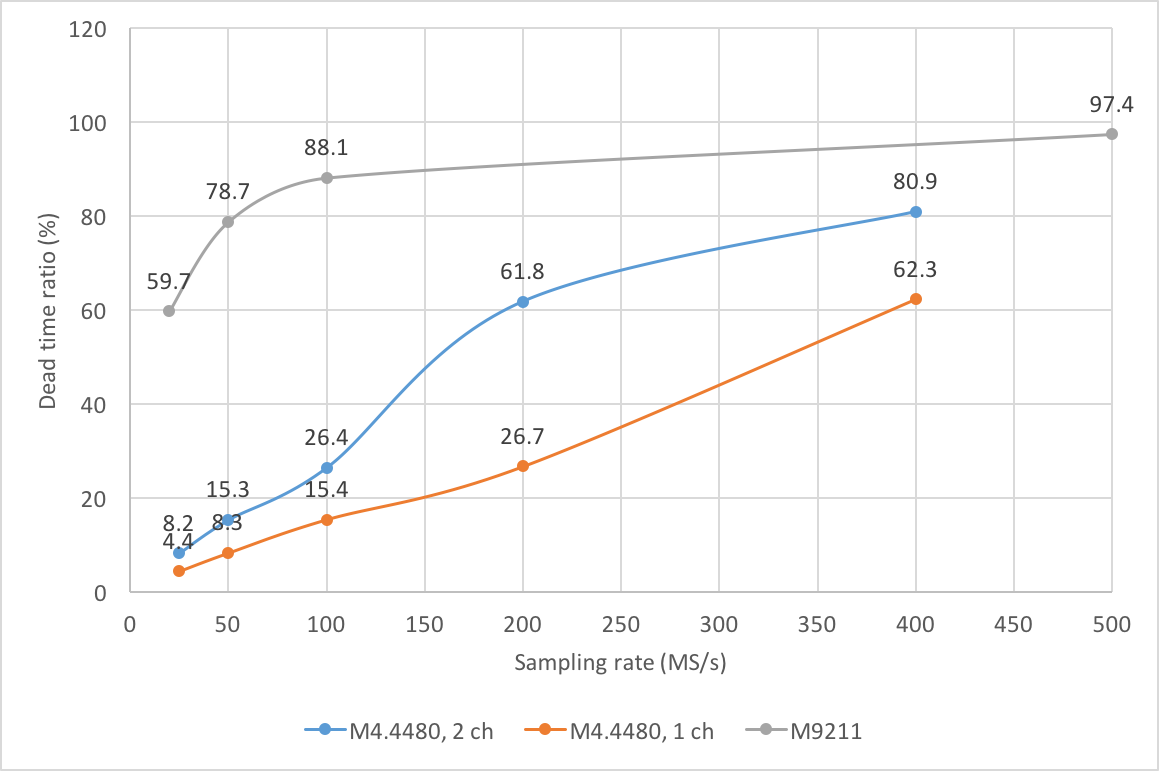}
\caption{Deadtime comparison for M4x.4480 and M9211 digitizers.}
\label{fig:muec_daq:deadtime_comparison}
\end{figure}

We also tested another digitizer module with a slower sampling speed. 
The Spectrum M4x.4480-x4~\cite{WEB_M4x4480x4} is PXI type 400\,MS/s digitizer
with two input channels, 14 bit resolution, and on-board memory for 2\,G samples, with PXIe x4 Gen 2 Interface interface. 
A similar dead-time measurement have been tested, where the result is shown and compared with M9211 digitizer in
Figure \ref{fig:muec_daq:deadtime_comparison}. It was evidently shown that using lower sampling rate will
improve the dead-time. 

While the dead-time issue can be mitigated by using lower sampling rate with specific choice of IF frequency 
and RF signal down conversion, it was also found that the storage is an issue. 
When using 100\,MS/s and taking data for 1 month with 2\,bytes per sampling, the total data size is around 264\,TByte. 
It is not feasible to store all this time domain data and applying offline FFT.

\subsection{Prosepct of digitizer DAQ for axion experiment}

In this study, we estimated dead-time between spectrum with various setups. After measuring the
dead-time, there are two ways to get better FOM, namely higher $f_s$ and lower $N$. It seems clear
that lowering $N$ has much more influence in increasing FOM once targeted $\Delta f$ is fixed.
Meanwhile, the measured dead-time (1--2 sec) was comparable to the live-time, so it should be
reduced by using better equipment or exploiting better DAQ technique. One recommended way is
parallel data taking by operating multiple digitizers, or ADC boards which are under development
currently. However, before applying such techniques, basic investigation should precede to be aware
of all necessary specifications of ADC: sampling rate, frequency resolution, voltage range, and ADC
bit resolution which may require at least 14 bits to interpret axion signal properly. Regarding
determining sampling rate, it is also important to get a highly down-converted intermediate
frequency (IF). The lower IF, we can have more flexibility in setting up the frequency resolution or
the sampling rate over the Nyquist frequency $(2f_s)$. Another thing to be developed further is data
analysis plan using the FFT, which is not covered in this writing.

\subsection{FPGA based realtime DAQ system development for axion experiment}

\subsubsection{Motivation of custom DAQ system for axion experiment}

The data of axion experiment is not conventional compared to various HEP experiment which records
the waveform of charge or voltage signal when triggered. In the axion experiment, the RF signal is
continuously monitored or recorded without trigger and the resulting power spectrum becomes the
final data set. Due to this difference, a very fast digitization technique and pipelined Fast
Fourier Transformation are required in the DAQ system. 

Typical DAQ equipment in axion experiment is a spectrum analyzer. While it is robust and proven
equipment, it is not designed to meet the requirement of CAPP axion experiment, in respect to the
Resolution bandwidth (RBW), center frequency, and dead-time. In many cases with small RBW, the DAQ
dead-time becomes very big, which limits the total DAQ time (and more Helium consumption). 

In other hands, while a state-of-art FADC chips runs at a few tens of Giga sampling rate, those
chips are mostly requiring trigger. This does not meet with axion experiment. A few Giga sampling
rate ADC with streaming the data to storage device is required in case of taking waveform data in
axion experiment. By taking time-domain data sample for a long DAQ time and applying offline FFT, it
enables much smaller RBW analysis. Even though most of axion halo model predicts a few kHz width of
axion signal, it is still possible that the width of axion signal is below the proposed RBW level,
and in this case, the conventional spectrum analysis techniques does not work. 

As a resolution of all these DAQ issues in axion experiment, a custom DAQ equipment based on FPGA
technology is developed in IBS/CAPP. The development got benefits from the experience of the
developments of COMET trigger system which is FPGA-development intensive. 

\subsubsection{Construction of FPGA based DAQ for axion experiment}

One of main bottleneck of axion data taking is that a spectrum measurement is slow in most of measurement devices, 
and only provides averaged spectrum in the end of data acquisition. This limits the data taking time and efficiency, 
which eventually results in more DAQ time to achieve target sensitivity, more Helium consumption, i.e. research cost. 
In order to overcome this bottleneck and increase the data taking efficiency to 100\%, 
and eventually to enable time-domain data taking for an exotic axion measurement, 
a custom axion DAQ board is developed based on FPGA technology. 
The concept is designed in CAPP, and fabricated in external company along with FPGA logic design. 
The board is delivered at June 2018 and being tested. The deployment of this system to CAPP-PACE or other experiments 
are expected to be possible in 2019. 

\begin{figure}
\centering
\includegraphics[width=0.8\textwidth]{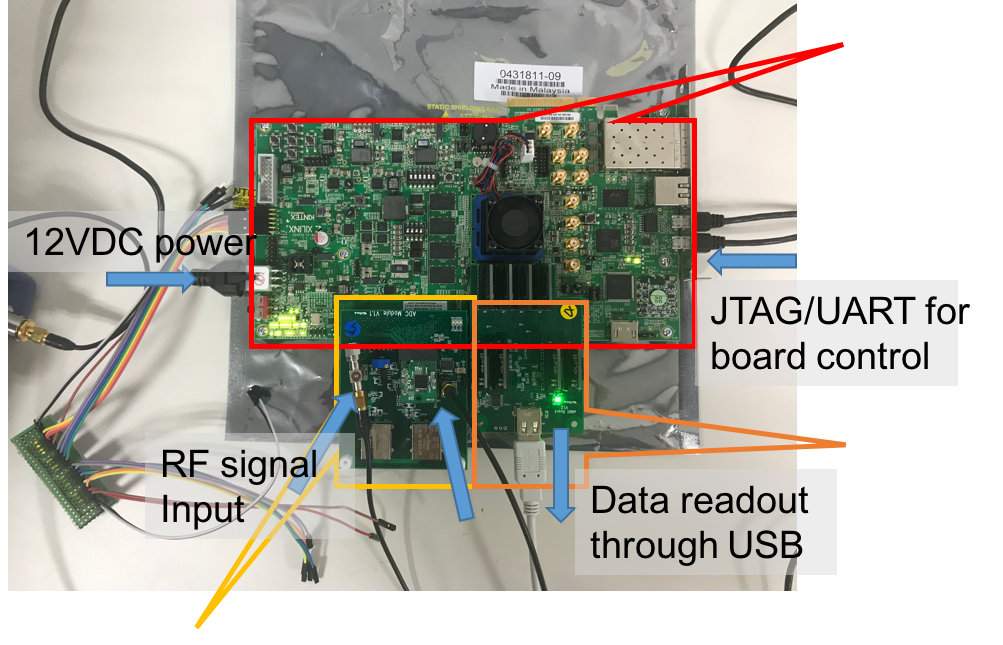}
\caption{The FPGA based custom axion DAQ system developed in CAPP.
}\label{fig:muec_daq:realtimedaq}
\end{figure}

The developed custom DAQ system is shown in the Fig.~\ref{fig:muec_daq:realtimedaq}. 
In order to reduce the FPGA board design complexity and possible errors, 
an off-the-self FPGA evaluation board (Xilinx KCU105~\cite{WEB_KCU105}) is employed. 
An ADC board is equipped with Texas Instrument ADS4149~\cite{WEB_ADS4149} 14\,bit 250\,MS/s ADC, 
and an eMMC (Embedded Multi Media Card) storage (up to 1TByte) 
board with USB interface are developed with close communication of IBS/CAPP with external company.

\begin{figure}
\centering
\includegraphics[width=0.9\textwidth]{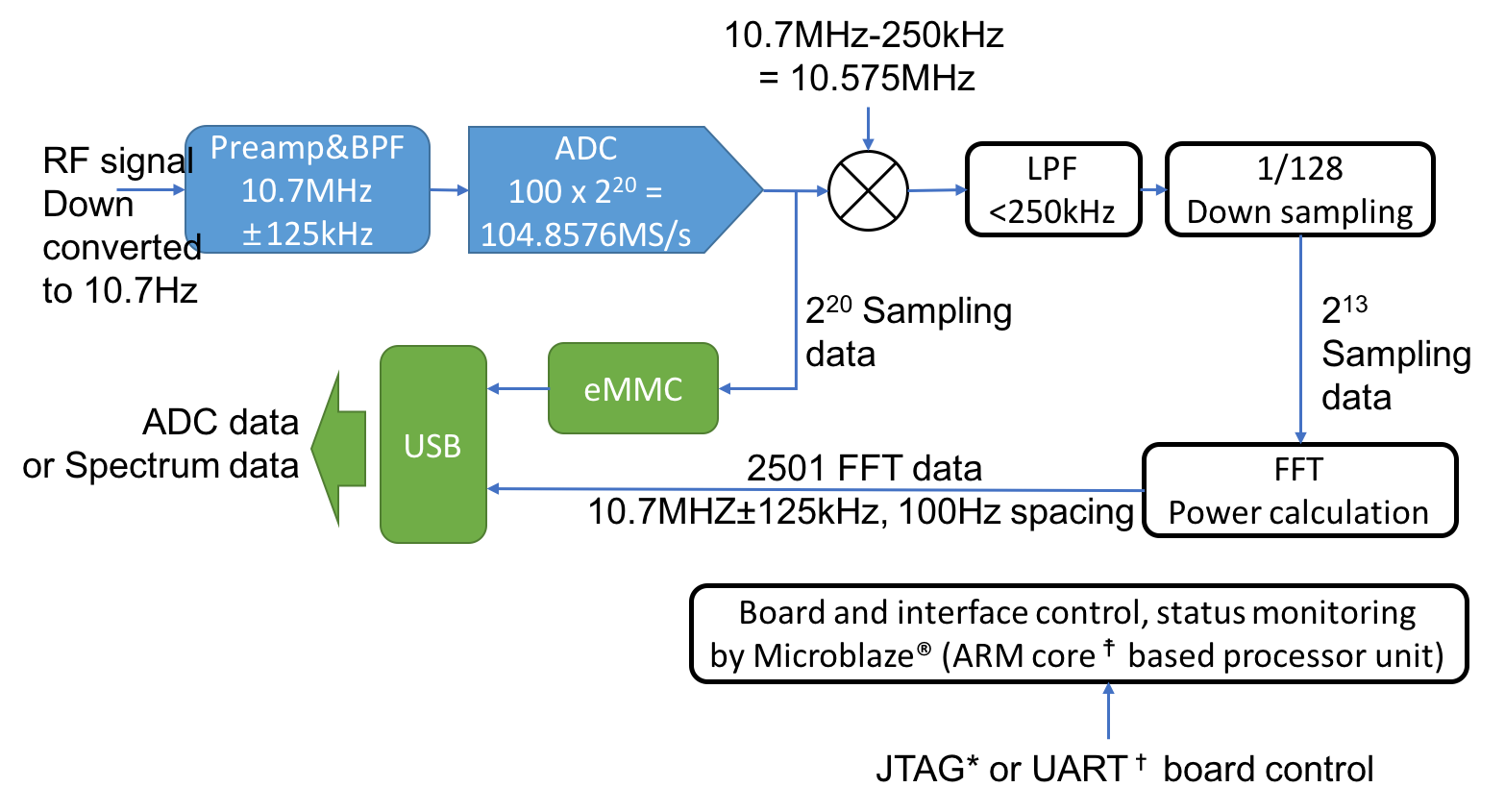}
\caption{The system diagram of FPGA based custom DAQ system in CAPP. JTAG is a chip test protocol by Joint Test Action Group, defined in IEEE-1149~\cite{IEEE_1149}, and  UART means Universal asynchronous receiver/transmitter, such as RS-232. For Microblaze, see \cite{WEB_Microblaze}.
}\label{fig:muec_daq:realtimedaq1}
\end{figure}

Figure~\ref{fig:muec_daq:realtimedaq1} shows the conceptual drawing of the custom DAQ system. 
After the pre-amplification of the RF signal from the axion experiment, 
the signal first digitized. The subsequent down-conversion directly to Baseband 
and band-pass filtering is purely digital, 
therefore they are immune to additional noise. 
The FFT process with down-sampled data performed in realtime inside FPGA, therefore, 
in principal, there is no dead time. The system is also featured with Processor core unit which cooperates the system interfaces. 

This custom FPGA based axion DAQ board is constructed at May 2018, and is being test and upgrade to meet the requirement of CAPP axion experiments. The installation of this system is expected at the early of 2019.

%% file: 1.4.8/main.tex
\subsection{CULTASK Data Acquisition Software (CULDAQ)}
CULTASK data acquisition software, or CULDAQ \cite{JPhysConfSer_898_032035_2017}, is a software framework that provides the following features for the axion dark matter experiments:
\begin{itemize}
	\item Data acquisition from experiment apparatus
	\item Standard data format
	\item Experiment control
	\item Monitoring of the experiments.
\end{itemize}
CULDAQ is a home-grown and an object-oriented software. Though it is intended to be used on Linux operating system, it can be extended to other operating systems such as Windows and Mac OS X. It is mostly written in Python, but other languages are also used for purposes.

\begin{figure}[h]
\begin{center}
\includegraphics[width=.85\textwidth]{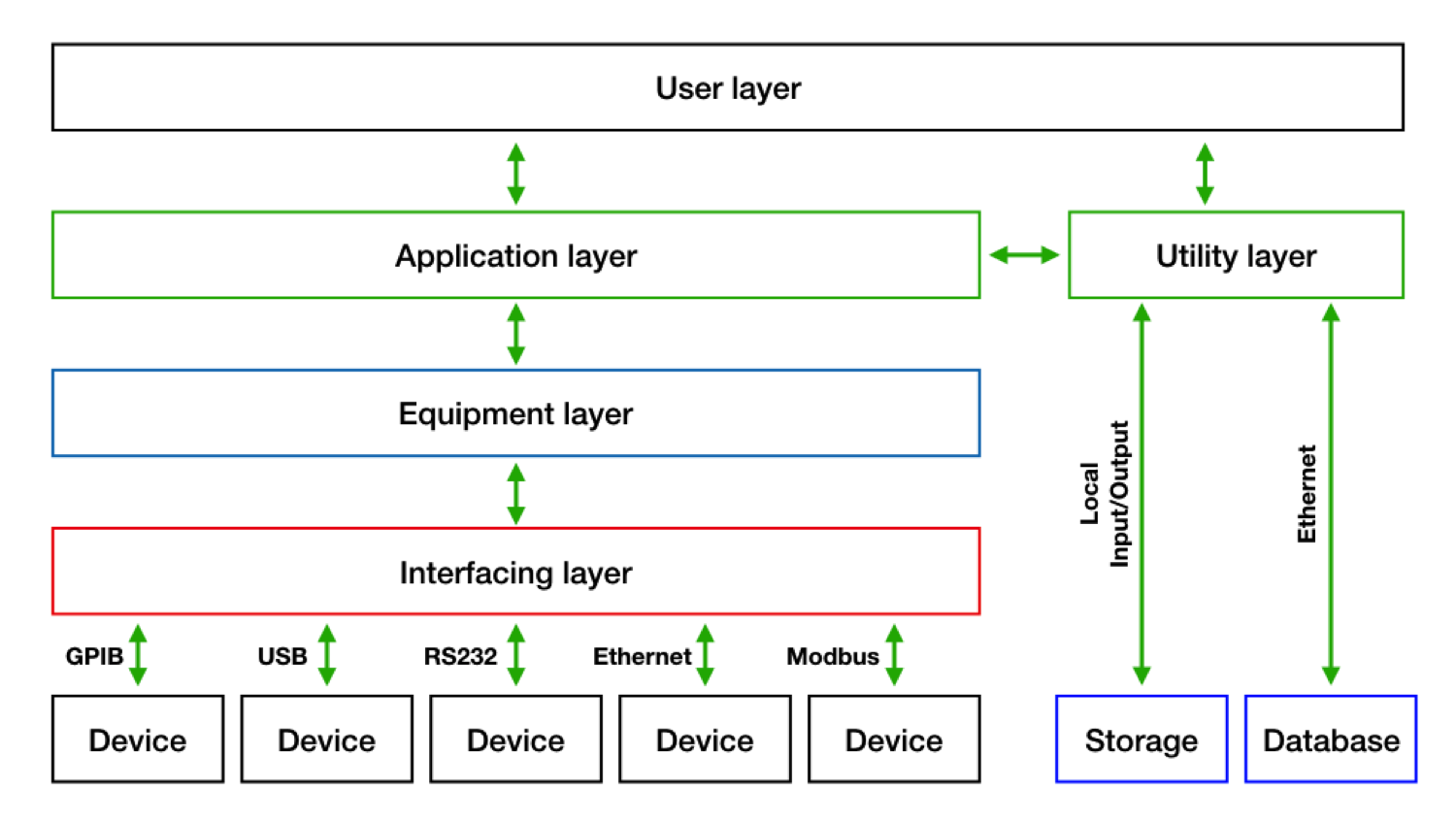}
\caption{Schematic view of layers in CULDAQ.}
\label{fig:culdaq_layers}
\end{center}
\end{figure}

CULDAQ consists of several layers that provides different roles. As described in Section \ref{section:daq_interfacing_layer}, interfacing layer is in charge of a low-level communication to devices via various protocols such as GPIB (General Purpose Interface Bus), USB (Universal Serial Bus), RS-232 (Recommended Standard 232), Modbus, and Ethernet. Equipment layer extends the interfacing layer as machine dedicated modules as described in Section \ref{section:daq_equipment_layer}. Since the equipment layer provides functions of instruction sets dedicated to devices, an application layer (Section \ref{section:daq_application_layer}) of devices provides convenient and versatile functions for practical operations of devices. The modules in the application layer can be combined to form a sequence of an experiment as a user layer. There is also a utility layer which provides convenient functions, for example, experiment management, message logging, database access, and data input/output. The utility layer is described in Section \ref{section:daq_utility_layer}. Figure \ref{fig:culdaq_layers} illustrates the layer structure of CULDAQ.

\subsection{Interfacing layer}
\label{section:daq_interfacing_layer}
Interfacing layer establishes connections to devices via various protocol such as GPIB, USB, RS-232, Modbus, and Ethernet. Since each protocol requires a different way to communicate, separate modules are made in the interfacing layer respectively.

For the communications via GPIB protocol, a hardware interface for the GPIB connection is necessary, and it is provided in a form factor of PCI-E (Peripheral Component Interconnect Express) or as a USB device. Those hardwares provides drivers to utilize as shared objects which can be linked to any C++ program. Therefore, there is a module class, \verb|AbstractGPIBPython| written in C++ to establish the GPIB communication. The \verb|AbstractGPIBPython| class also provides an interface to Python, and a wrapper class \verb|AbstractGPIB| written in Python imports it. The \verb|AbstractGPIB| class is included in equipment layer to provide the GPIB communications. 

\verb|AbstractRS232| class provides an interface via RS-232 protocol, and it is written in Python with its serial communication module. To establish the connection, the \verb|AbstractRS232| class accepts parameters related to the connection such as port name, transfer rate, number of data bits, parity, and so on. Those parameters have to be configured according to the specifications of devices. Computers do not usually have serial ports in these days, therefore, RS232 to USB converters are used for the physical connections.

USB communication is supported by \verb|AbstractUSB| class which utilizes USB modules in Python. It recognizes each USB device by its unique manufacturer and product identifiers, and those information have to be input as parameters of the class. 

Modbus is yet another serial communication protocol, and is used for several devices such as stepping motors. For the Modbus communications, \verb|AbstractModbus| class is made, and it accepts parameters of port name, transfer rate, number of data bits, parity, and so on. The \verb|AbstractModbus| class utilizes \verb|pymodbus| module \cite{web:pymodbus}.

Ethernet connections is supported by \verb|AbstractVisaEthernet| class which utilizes VISA (Virtual Instrument Software Architecture) \cite{web:visa} developed by National Instrument. Using TCP/IP protocol with VISA, the \verb|AbstractVisaEthernet| class establishes the Ethernet connections and communicates with devices.

The class modules in the interfacing layer are inherited in the modules dedicated specific devices in the equipment layer, therefore, the modules in equipment layer can use the properties and functions in the classes of interfacing layer.

\subsection{Equipment layer}
\label{section:daq_equipment_layer}
Equipment layer is a set of class modules dedicated specific devices such as a network analyzer, spectrum analyzer, temperature controller, and so on. The class modules in this layer inherit the class modules in the interfacing layer for communications via protocols accordingly. For example, a module for a spectrum analyzer which supports a GPIB connection inherits \verb|AbstractGPIB| module, and uses its functions for the communications. However, there is one exception for a controller for piezoelectric actuators. Since the controller made by Attocube uses its own set of instructions which are not open to customers but provided as C++ libraries, the class mode for the device does not need to inherit a module from the interfacing layer. Therefore, the class for the device, \verb|EquipmentPiezoControllerANC350|, is written in C++ and accesses the device by using the functions defined in the libraries. The \verb|EquipmentPiezoControllerANC350| still provides an interface to Python in the same way with modules in the interfacing layer, therefore, all the modules after the equipment layer are maintained as Python program.

Most of devices support SCPI (Standard Commands for Programmable Instruments), and those commands are hidden in the functions of the modules in this layer. Therefore, the modules provides human-readable names of the functions with parameters. It is the same for other devices which do not support SCPI, therefore, internal commands for those are hidden from outside, and functions with human-readable names are provided.

Table \ref{table:culdaq_equipment} summarizes the supported devices in the equipment layer.
\begin{sidewaystable}
\centering
\begin{tabular}{c|c|c|c|c}
\hline
Device & Class name & Manufacturer & Model & Protocol \\
\hline
Network analyzer & \verb|EquipmentNetworkAnalyzerE5063A| & KEYSIGHT & E5063A & GPIB \\
& \verb|EquipmentNetworkAnalyzerN5232A| & KEYSIGHT & N5232A & GPIB \\
& \verb|EquipmentNetworkAnalyzerZNB8| & Rohde \& Schwarz & ZNB & GPIB \\
& \verb|EquipmentNetworkAnalyzerZND| & Rohde \& Schwarz & ZNB & GPIB, Ethernet \\
Spectrum analyzer & \verb|EquipmentSpectrumAnalyzerFSV| & Rohde \& Schwarz & FSV & GPIB \\
& \verb|EquipmentSpectrumAnalyzerN90X0A| & KEYSIGHT & N9010A, N9020A, N9030A & GPIB, Ethernet \\
& \verb|EquipmentSpectrumAnalyzerModuleN9069A| & KEYSIGHT & N9069A & GPIB \\
Signal generator & \verb|EquipmentSignalGenerator81150A| & KEYSIGHT & 81150A & USB \\
& \verb|EquipmentSignalGeneratorDSG3000| & Rigol & DSG3000 & USB \\
& \verb|EquipmentSignalGeneratorE8257D| & KEYSIGHT & E8257D & GPIB \\
& \verb|EquipmentSignalGeneratorSMB100A| & Rohde \& Schwarz & SMB100A & GPIB \\
Magnet controller & \verb|EquipmentMagnetControllerModel430| & American Magnetics & Model 430 & RS-232 \\
& \verb|EquipmentMagnetControllerSMS600C| & Cryogenic Ltd. & SMS600C & RS-232 \\
Temperature controller & \verb|EquipmentTemperatureMonitorModel211| & Lakeshore & Model 211 & RS-232 \\
& \verb|EquipmentTemperatureControllerModel372| & Lakeshore & Model 372 & GPIB \\
Motor and actuator & \verb|EquipmentPiezoControllerANC350| & Attocube & ANC350 & USB \\
& \verb|EquipmentSteppingMotorControllerCRK| & Orientalmotor & CRK & Modbus \\
RF switch & \verb|EquipmentSwitchMatrixUSB2SPDTA18| & Minicircuits & USB-2SPDT-A18 & USB \\
\hline
\end{tabular}
\caption{\label{table:culdaq_equipment}Class modules in equipment layers.}
\end{sidewaystable}

\subsection{Application layer}
\label{section:daq_application_layer}
Though the equipment layer provides a full functions to access all the devices, it is necessary to provide more convenient functions to end-users. For example, if an end-user only uses a module in the equipment layer, a measurement of transmission and reflection with a network analyzer is done as
\begin{itemize}
	\item Define a window in the network analyzer,
	\item Define traces in the network anlayzer,
	\item Set data formats for traces,
	\item Retrieve frequency data,
	\item Retrieve transmission data,
	\item Retrieve reflection data,
	\item Convert binary data into a preferred format,
	\item Save the data into files and database.
\end{itemize}
This is not convenient and practical because the measurement is a common task in any experiment and end-users have to call lots of functions defined in the module to do this. The modules in the application layer provides higher level functions that combines the functions in equipment layer, and the end-users call a single function for the common task. Furthermore, the equipment layer does not provide functions to convert and store data, therefore, the application layer also includes the functions related to converting and storing from utility layer.

In addition, the application layer provides a breakthrough to communicate with components which are out of the framework. For example, a dilution refrigerator is controlled and accessed by an independent software made by the manufacturer on a different operating system, therefore, it can not be accessed by CULDAQ. Since it is necessary to record the status of the refrigerator in experiments, the status data such as temperatures and pressures have to be retrieved into the framework. For this purpose, there is a database for the data exchange between different systems. A simple Python program reads and transfers the status data from systems to the database. Modules in the application layer read the data from the database, and feed the data into the framework. 

In the user layer, end-users only need to import the modules in the application layer for the communications with devices. By combining those modules, end-users can form a sequence of an experiment. It is sometimes necessary to access lower-level functions in the equipment layer, and the application layer also provides direct accesses to the equipment layer for the purpose.

\subsection{Utility layer}
\label{section:daq_utility_layer}
The utility layer provides useful and convenient functions necessary to the operations of experiments. The class modules in this layer are combined in the user layer to complete a sequence of an experiment.

\verb|Logger| class provides a way to make logs of experiments. It displays the log messages of various categories: INFO is for general log messages; WARNING is for log messages of abnormal situations which is not critical for the operations; ERROR is for log message of critical events that may crash the operation; DEBUG is for debugging of codes in developments. The class also provides a function to save status messages into database. Those messages will be alarmed to end-users for critical events of the operations. 

\verb|RunManager| class is a module to manage the runs of experiments. It defines experiments, runs, and measurements to identify the measurements in experiments. Those experiment, run, and measurement numbers are globally maintained in the framework, therefore, a combination of those numbers is unique. The numbers are also used to associate data stored in different structures as described in Section \ref{section:daq_data_handling}.

The data retrieved from devices is stored into files in a ROOT \cite{LinuxJ_51_1998} format. \verb|OutputServer| class provides an interface to store data into ROOT files. End-users or modules in the application layer user the class to instantiate the \verb|OutputServer| object, open an output ROOT file, and write data into it by calling a function. Details of the ROOT file are described in the next section. Data from some of devices are taken in a binary format, and it requires a data conversion. For this purpose, \verb|Unpacker| class functions to unpack the binary data to readable values. 

\subsection{Data handling in CULDAQ}
\label{section:daq_data_handling}
As described in Section \ref{section:daq_utility_layer}, data taken from the experiment is stored in ROOT format. Data from devices are separately stored in \verb|TTree|, for example, measured power spectra from a spectrum analyzer is stored in \verb|culdaq_sa_data|. In \verb|TTree| as a table, relevant variables are stored in \verb|TBranch| as columns, and measurement numbers are recorded as well. One ROOT file corresponds a run in a experiment, and it contains a number of measurements. A common table, \verb|culdaq_runinfo|, contains the experiment, run, and measurement numbers with a recorded time stamp. Since the data from various devices is spread over multiple tables, a unique key is necessary to couple data. All the tables have a column named \verb|measure_no| which stands for a measurement number, and it is used as a key to associate data over tables.

There are four tables related to the data from a network analyzer; \verb|culdaq_na_sampling|, \verb|culdaq_na_data_s11|, \verb|culdaq_na_data_s22|, and \verb|culdaq_na_data_s21|. The \verb|culdaq_na_sampling| table contains measurement data of resonance such as resonant frequency, quality factor, bandwidth, loss, and coupling coefficient. The tables of \verb|culdaq_na_data_s11| and \verb|culdaq_na_data_s22| contain reflections of two ports, respectively. The columns in those tables are frequency, input power, reflected power, and impedance values in Smith chart coordinates. In \verb|culdaq_na_data_s21| table, transmission data is stored such as frequency, input power, and transmitted power.

The data structures for spectrum analyzer and signal generator are the same, and they are stored in \verb|culdaq_sa_data| and \verb|culdaq_sg_data| tables, respectively. Those contains frequency, and measured (for spectrum analyzer) or input (for signal generator) power.

Temperature data are stored in \verb|culdaq_temp_cryo| table. Temperatures at each stages in refrigerator are stored in the table, and additional temperatures of cavity and components in an experiment are also stored.

The \verb|culdaq_magnet| table contains data related to a magnet such as the present magnetic field, target current and magnetic field, current and voltage at magnet and power supply, and magnet status.

In addition to ROOT format, a snapshot data of an experiment is also stored in database. The database is configured with mariaDB \cite{web:mariadb} in a central server. Temperature data are transferred to and stored in the database from an external system as described in Section \ref{section:daq_utility_layer} for the data exchange. Measurements with a network analyzer such as resonant frequency and quality factor are also stored in the database. By keeping these snapshot data from experiments, any monitoring system which is able to access a database can be used to monitor experiments.

\subsection{Monitoring system}
\label{section:daq_monitoring}
As described in the previous section, the snapshot data of experiments are kept in a database which is accessible over internet. Therefore, any monitoring system which is capable to access a database can be used for monitoring purpose.

\begin{figure}[b]
\begin{center}
\includegraphics[width=.95\textwidth]{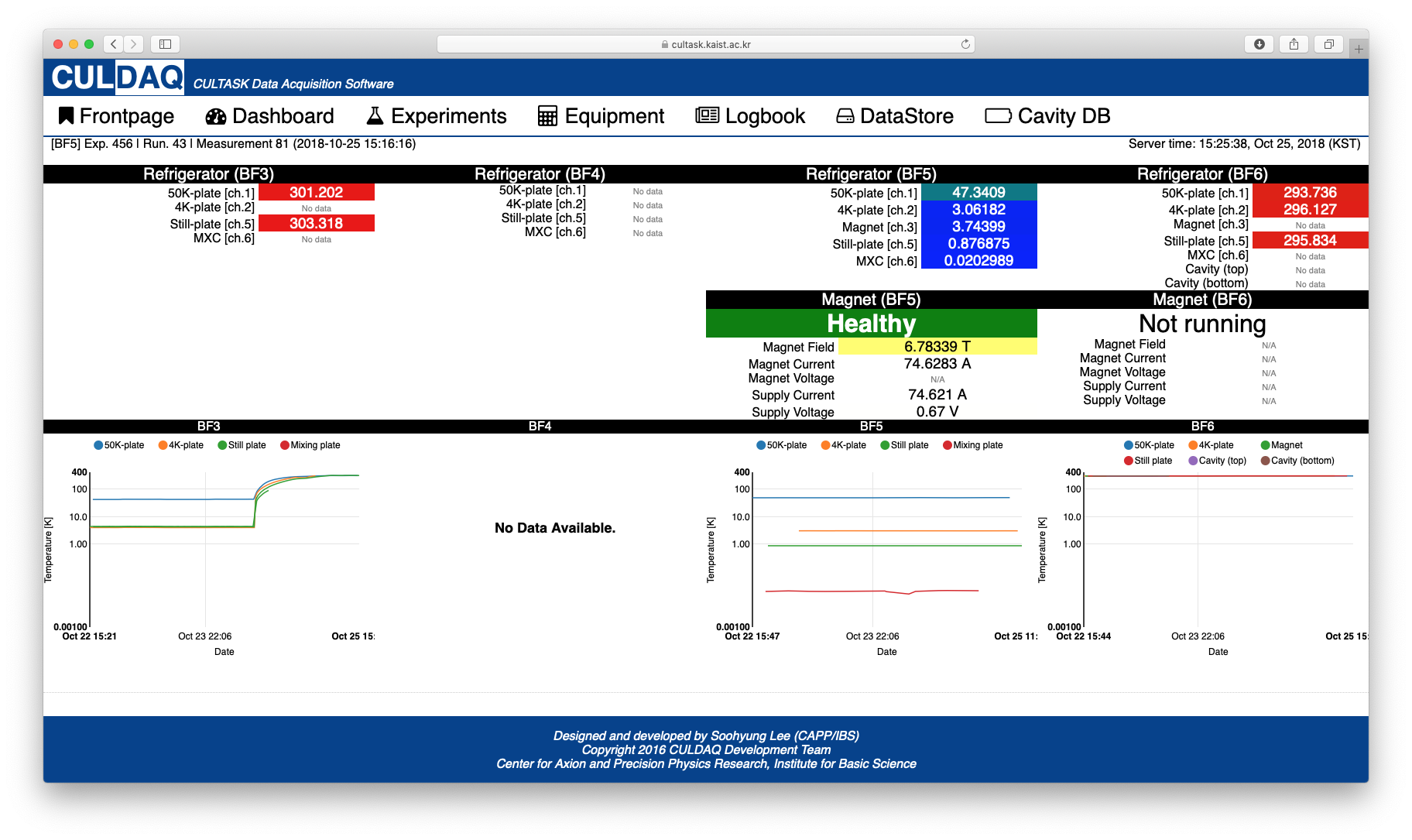}
\caption{General monitoring for refrigerator and magnets over web.}
\label{fig:culdaq_web_monitor}
\end{center}
\end{figure}

A general overview to monitor refrigerators and magnets are implemented as a web application as shown in Figure \ref{fig:culdaq_web_monitor}. The application is written in PHP with CodeIgniter \cite{web:codeigniter}. To plot temperature data, jquery \cite{web:jquery} and D3.js \cite{web:d3} are also employed. The application retrieves relevant data from the database, and displays and refreshes data and plots automatically.

\begin{figure}[h]
\begin{center}
\includegraphics[width=.95\textwidth]{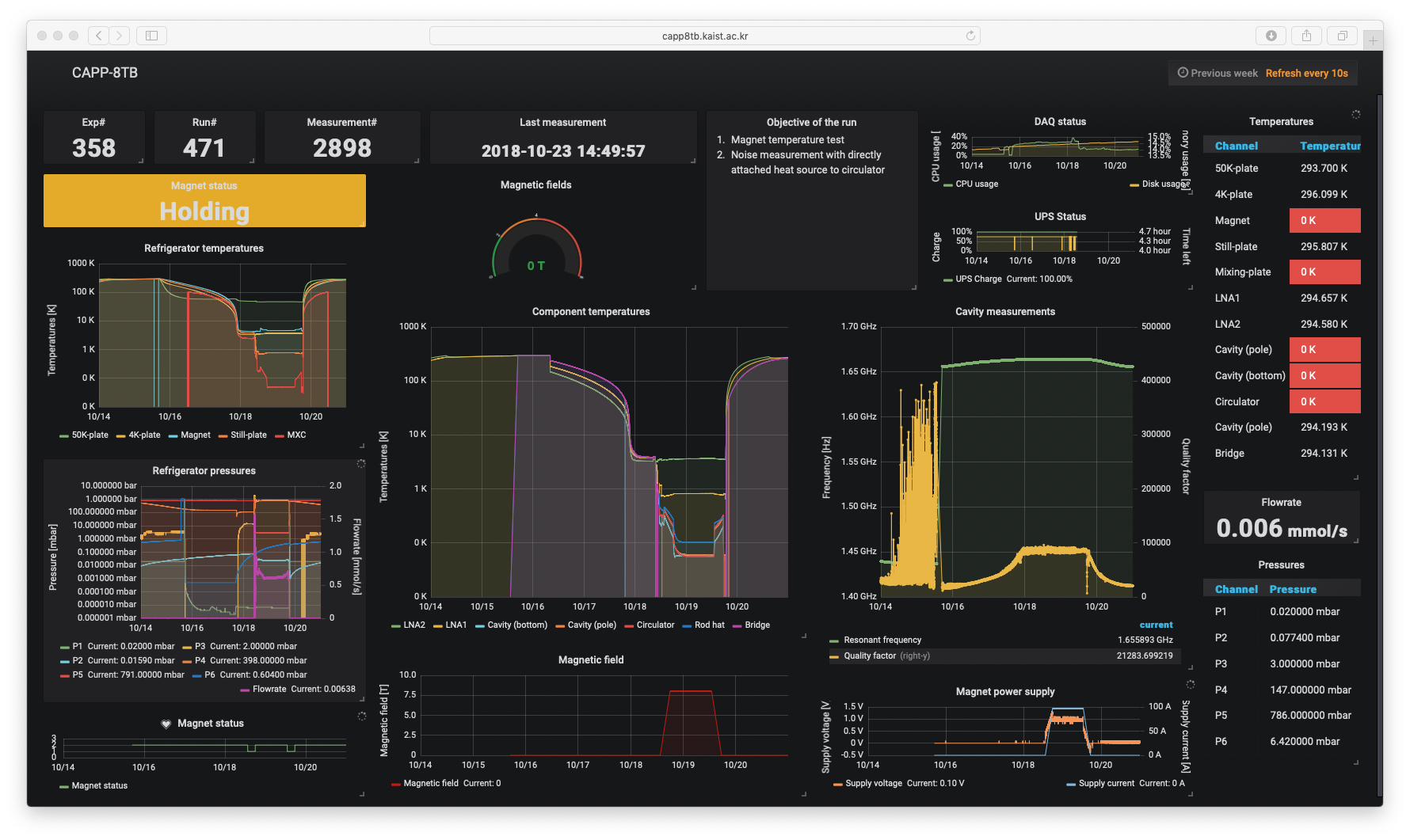}
\caption{Detailed monitoring with Granafa.}
\label{fig:culdaq_grafana}
\end{center}
\end{figure}

For a detailed view of monitoring, Grafana \cite{web:grafana} is employed as shown in Figure \ref{fig:culdaq_grafana}. Some variables monitored such as the status of DAQ computers in this system are stored in a separate database using influxDB \cite{web:influxdb}.

\subsection{Status and plans}
The core part of CULDAQ has been developed. Since the equipment and application layers depend on specific devices, further implementation for new devices may be necessary as new devices are employed and introduced. The source codes of CULDAQ is maintained with Git \cite{book:git} to store, manage, and cooperation.

Since the sequences of experiments are slightly different, the programs in the user layer need to be slightly different as well. Unified and flexible control program to run experiment needs to be developed with a GUI interface. The program will also provide functions to manage experiment shifts. Emergency alarm system and further useful functions will be also included.

%% file: 1.6.1/main.tex
\subsubsection{Executive summary of the project}
This chapter describes Global Network of Optical Magnetometers to Search for Exotic Physics (GNOME).

\paragraph{Overview}
Recently, light axions and axion-like particles have appeared as alternate dark matter candidates. According to some theories, axion field would oscillate at a specific frequency due to its non-zero vacuum energy. Such scenario might generate stable topological defects, e.g. a domain structure~\cite{1982PhRvL..48.1156S, 1982PhLB..115...21L}. When Earth crosses a domain wall (DW) separating regions with different vacuum expectation values of axion-like field, a torque can be exerted on leptonic or baryonic spins. Such a DW-crossing event could lead to a transient signal detectable with modern state-of-the-art optical magnetometers (OMAGs)~\cite{2013PhRvL.110b1803P}. New technique for detecting transient signals of exotic origin using a global network of synchronized optical magnetometers has been proposed through an international collaborative project called "Global Network of Optical Magnetometers to Search for Exotic physics" (GNOME)~\cite{2013AnP...525..659P, 2018arXiv180709391A}. The GNOME project is an international collaborative effort over more than 10 institutes in Europe, North America, and Asia. The project successfully finished a preliminary run in 2017 and is now in full scale operation.

\paragraph{Scope of the project}
The project is to develop a state-of-art optical magnetometer system and operate it as local station at CAPP for GNOME experiment. This enables the investigation of transient exotic spin coupling through the network of optical magnetometer system. The scheme is based on synchronous measurement of optical-magnetometer signals from several devices operating in magnetically shielded environments at distant locations separated by over 100\,km on Earth. Although a signature of such exotic couplings can be shown in the signal from a single magnetometer already, it would be challenging to distinguish it from the noise. By analyzing correlations between signals from multiple geographically separated magnetometers, it is not only possible to identify the exotic transient, but also to characterize its nature. The ability of the network to probe presently unconstrained physics beyond the Standard Model is examined by considering the spin coupling to stable topological defects of axion-like field.

\paragraph{Implementation of the project}
The initial focus at CAPP is to develop a high sensitivity optical magnetometer. The magnetometers for GNOME are optically pumped atomic magnetometers  that measure the spin-precession frequency of alkali atoms by observing the time-varying optical properties of the alkali vapor with a probe laser beam. The target sensitivity and bandwidth of the magnetometer are $\sim 100\,\rm{fT}/\sqrt{\rm{Hz}}$ and $\sim 100\,\rm{Hz}$. A self-oscillating magnetometer based on nonlinear magneto-optical rotation using amplitude-modulated pump light and unmodulated probe light (AM-NMOR) in $^{133}{\rm{Cs}}$  will be constructed and tested towards a goal of the GNOME network. 

\paragraph{Progress and results}
Since we joined GNOME collaboration in 2016, CAPP has been developing a high-sensitive optical magnetometer to run local station for GNOME network. We have designed Cs based optical magnetometer for this purpose.  

\begin{itemize} 
	\item Development of an optical magnetometer: A high-sensitive optical magnetometer has been developed at CAPP/IBS. This magnetometer
	\item IBS/CAPP Station operation: Local Station for GNOME has been in operation at CAPP since 2017
	\item GNOME collaboration: CAPP has been actively involved in data analysis for GNOME
\end{itemize}

\paragraph{Significance}
GNOME is the only experiment being able to investigate transient exotic spin coupling. GNOME is especially sensitive to transient events of axion or axion-like field with mass range below $\rm{neV}$. None of experiment so far has attempted to look for axion or axion-like particles in this ultralight mass range. GNOME is capable to search for other terrestrial events such as Q-ball or axion clump that has been recently postulated in Astro-particle physics. 


\subsubsection{GNOME project}

\paragraph{Introduction}
Most experimental Dark Matter (DM) searches aim at direct detection of some variety of particles that feebly interact with ordinary baryonic matter such as Weakly Interacting Massive Particles (WIMPs) or axions. However, all these dark matter searches so far have produced only upper limits on the interaction strength between DM and ordinary matter so far. If DM consists of light axions or axion-like particles, it behaves more like a coherent field than a collection of uncorrelated particles. The axion field may oscillate at a specific frequency because the vacuum energy of the axion field is non-zero and hence it would not produce static effects on matter. Such theory indicates that the non-zero vacuum energy of axion fields may generate topological defects called axion domain walls. Remarkably, axion models like, KSVZ and DFSZ, also predict the formation of such axion domain wall in early Universe.

\begin{figure}[tp]
\begin{center}
\includegraphics[width=.50\textwidth]{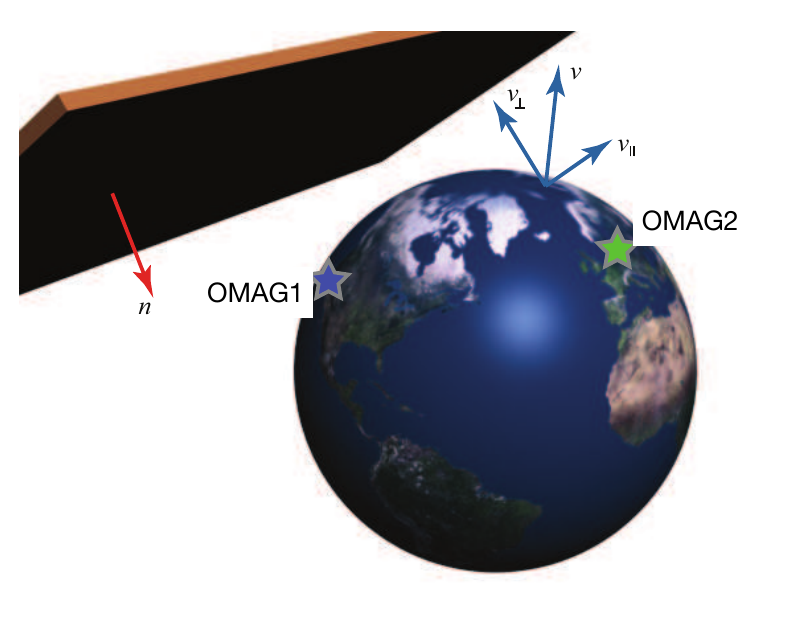}
\caption{Concept of the synchronized magnetometer arrangement. Optical magnetometers located at globally separated locations record signals with time synchronization provided by GPS. Transient events of global character can be identified by synchronous detection and correlation of magnetometer signals}
\label{fig:GNOME_FIG_1}
\end{center}
\end{figure}

The Global Network of Optical Magnetometers to search for Exotic physics (GNOME) is an experiment to search for transient events of axion domain walls based on this novel scheme. If domain wall exists in our Universe, one can image an exotic interaction between such axion domain walls and baryons rather than conventional ones. Especially when Earth crosses boundaries between axion domain walls separating with different vacuum expectation values of axion fields, a torque could be exerted on leptonic or baryonic spins. Therefore, such domain wall crossing events are possible to detect with state-of-art optical magnetometers. It is because optical magnetometry presently offers the possibility of the most sensitive magnetic-field measurements among all magnetometric techniques. The intrinsic sensitivity of optical magnetometers to spin dynamics would enable us to investigate of other types of interactions to the spin including non-magnetic ones especially the density fluctuation of vacuum expectation values for axion fields in this case.

Since such events could be very rare as well as fleeting, the major issue for detecting such events would be the isolation of signals induced from such domain wall crossing events from any other signals generated by environmental noise. A veto of other effects in optical magnetometers bring us into new approach of experiments: measurements of high precision optical magnetometer signals from multiple stations around the Earth~\cite{2014NIMPA.763..150W}.

\paragraph{Experimental approach}

The project at CAPP for GNOME is to develop a state-of-art optical magnetometer system and operate it as a local station in GNOME network. This enables the investigation of transient exotic spin coupling through the network of optical magnetometer system. The scheme is based on synchronous measurement of optical-magnetometer signals from several devices operating in magnetically shielded environments.

The GNOME consists of more than 10 dedicated atomic magnetometers located at geographically separated stations on the Earth. The target magnetometric sensitivity and bandwidth of each GNOME sensor are generally anticipated to be better than $\sim 1\rm{fT}/\sqrt{\rm{Hz}}$ over a bandwidth on the order of 100\,Hz, targets achievable with existing state-of-the-art atomic magnetometers. Each magnetometer is located within a multi-layer magnetic shield to reduce the influence of magnetic noise and perturbations. Even with shielding and co-magnetometry techniques, there will inevitably be some level of transient signals and noise associated with the local environment (and possibly with global effects like the solar wind, changes to the Earth's magnetic field, etc.). Therefore, each GNOME sensor uses auxiliary magnetometers and other sensors (such as accelerometers and gyroscopes) to measure relevant environmental conditions, allowing for exclusion/vetoing of data with known issues.

The signals from the GNOME sensors are recorded with accurate timing provided by the global positioning system (GPS) using a custom GPS-disciplined data acquisition system and will have a characteristic temporal resolution of  $\le 10\,\rm{ms}$ (determined by the magnetometer bandwidth), enabling resolution of events that propagate at the speed of light (or slower) across the Earth. Because of the broad geographical distribution of sensors, the GNOME should in principle be able to achieve good spatial resolution, acting as an exotic physics "telescope" with a baseline comparable to the diameter of the Earth.

If one assumes the axion domain walls are the predominant contribution to the cold dark matter, the domain wall may have a quasi-Maxwellian velocity distribution in the galactic reference frame with characteristic virial velocity $v \approx {10^{ - 3}}c$ where $c$ is the speed of light. We assume the rate of encounter with domain wall to be larger than at least once per year to make the experiment feasible.  In that case, the accessible parameter space is  $L \le {10^{ - 3}}{\rm{ly}}$. The thickness of the such domain wall is assumed as to be on the order of Compton wavelength as $d \approx \frac{{2\hbar }}{{{m_a}c}} \approx 400{\rm{m}} \times \frac{{1{\rm{neV}}}}{{{m_a}{c^2}}}$ . From it, one can estimate the duration of the transient signal  as $\tau  \approx \frac{d}{v} \approx 1{\rm{ms}} \times \frac{{1{\rm{neV}}}}{{{m_a}{c^2}}}$. If the bandwidth of GNOME magnetometer is set to be $\sim100{\rm{Hz}}$, it is sensitive to the axion mass ${m_a} \le 0.1{\rm{neV}}$. The coupling of the pseudoscalar field gradient to the atomic spin ${\bf{S}}$ of particle $i$ can be expressed through the interaction Hamiltonian as ${H_a} = \frac{{\hbar c}}{{{f_i}}}{\bf{S}} \cdot \nabla a({\bf{r}})$, where ${\bf{S}}$ is in the unit of $\hbar$  and the $f_{i}$ is a coupling constant for the considered particle $i$. This leads to estimates for the energy shift or torques experienced by the spins of fermion $i$ which is induced from the interaction with axion domain wall as $\Delta E(i) \approx \frac{{\hbar {c^2}}}{{4{f_i}}}\sqrt {{\rho _{{\rm{DM}}}}{m_a}L}$. 

\begin{figure}[bp]
\begin{center}
\includegraphics[width=.50\textwidth]{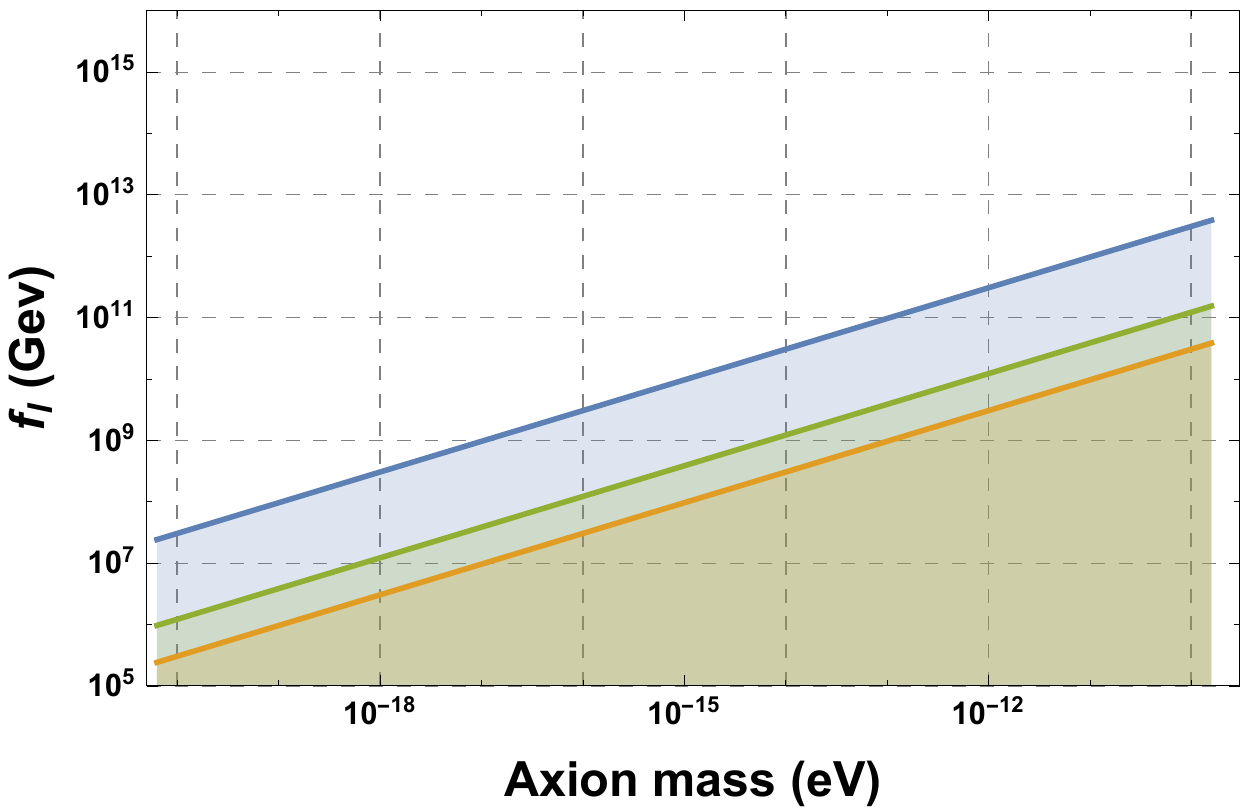}
\caption{Parameter space of the axion-like field with a domain structure that can be probed with the GNOME for various DW energy densities. The vertical line at ${10^{ - 10}}{\rm{eV}}$ is set by the bandwidth of the measurements. The horizontal line at ${10^9}{\rm{eV}}$  corresponds to the lower bound on electron, neutron, and proton decay constants. The lines/shades correspond to the demonstrated sensitivity of the devices $\delta \rm{E}$ and a measurement bandwidth of 100\,Hz. Blue $1{\rm{fT}}/\sqrt {{\rm{Hz}}}$ , Green $25{\rm{fT}}/\sqrt {{\rm{Hz}}}$:$1{\rm{fT}}/\sqrt {{\rm{Hz}}}$  , Orange: $100{\rm{fT}}/\sqrt {{\rm{Hz}}}$}
\label{fig:GNOME_FIG_2}
\end{center}
\end{figure}

The more sensitive a GNOME magnetometer is to torque or energy shift, the larger the value of ${f_i}$ that can be probed. From the basic concept of GNOME, one can estimate characteristics and potential sensitivity for the domain wall search with GNOME. First, the sensitivity of GNOME magnetometer, $\delta B$, can be interpreted as energy sensitivity via $\delta E = {g_F}{\mu _B}\delta B$, where  ${g_F}$ is ${g_F}$ the Lande factor and  ${\mu _B}$is the Bohr magneton. With ${g_F} = 1/3$, energy shift becomes $\delta E \approx {10^{ - 18}}{\rm{eV}}/\sqrt {{\rm{Hz}}}  \times \frac{{\delta E}}{{100{\rm{fT}}/\sqrt {{\rm{Hz}}} }}$. The actual energy uncertainty scales with the square root of the duration of signal, which in turn is inversely proportional to  ${m_a}$ as follow: $\Delta E \approx {10^{ - 18}}{\rm{eV}} \times \frac{{\delta B}}{{100{\rm{fT/}}\sqrt {{\rm{Hz}}} }} \times \sqrt {\frac{{{m_a}{c^2}}}{{{{10}^{ - 12}}{\rm{eV}}}}} $, where the parameterization of the mass relative to ${10^{ - 12}}{\rm{eV}}$ corresponding to the signal duration  $\sim1{\rm{s}}$. Figure~\ref{fig:GNOME_FIG_2} shows the parameter space that can be probed with various sensitivity of OMAGs that currently used in GNOME. 

\begin{figure}[tp]
\begin{center}
\includegraphics[width=0.85\textwidth]{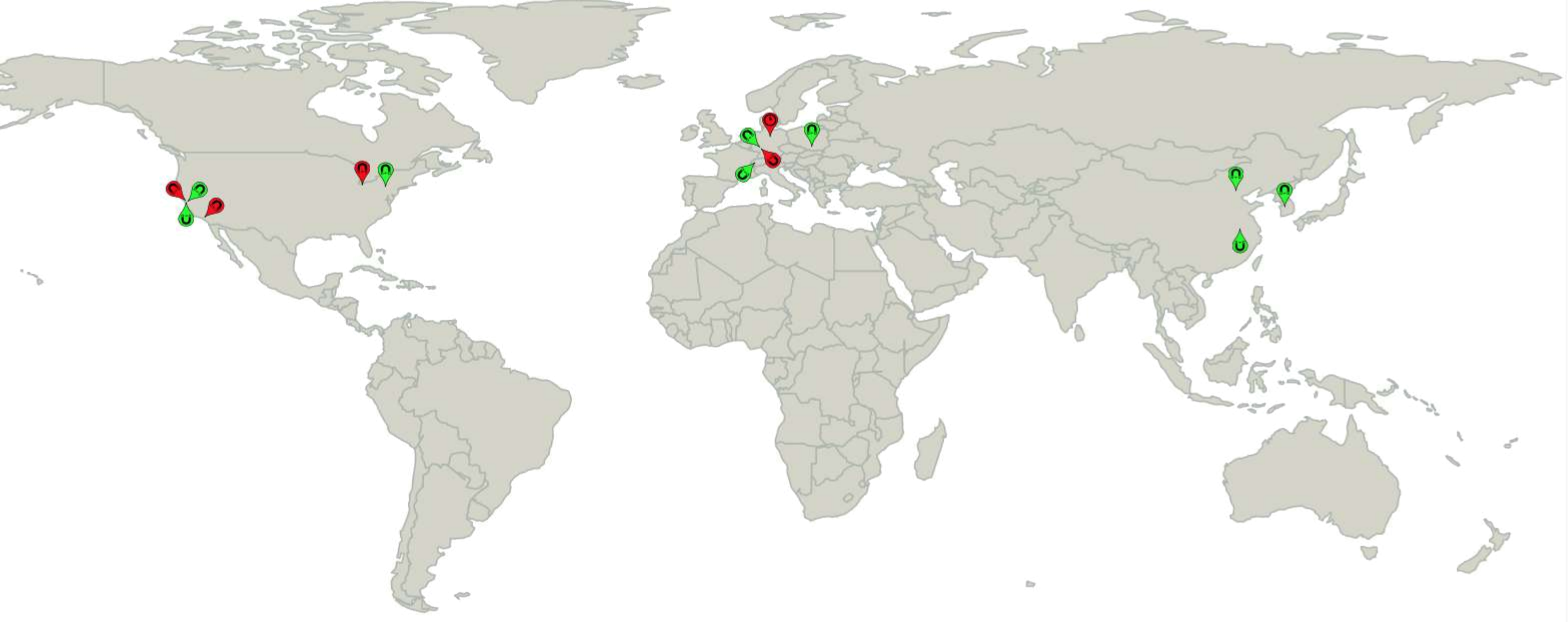}
\caption{Map of local stations in GNOME network. Green light represents on stage status and Red light represents off stage status of each station}
\label{fig:GNOME_FIG_3}
\end{center}
\end{figure}

In principle, the parameters of the model can be constrained with a single magnetometer. A particular problem for a search carried out with a single OMAG is the appearance of brief spikes in the OMAG signal related to technical noise or abrupt magnetic field changes. In a single device, rejection of these false-positive signals is difficult. At the same time, coindent measurements between two or more instruments are helpful in rejecting such signals.  They also provide consistency checks, since a signal would be expected to exist in all instruments whereas environmentally induced events are not typically correlated in the time window required for coincidence. Furthermore, information about a putative event such as its impinging direction can be determined by triangulation if several instruments (at least four) are taking data simultaneously. These features clearly show that synchronous operation of multiple synchronized, geographically separated OMAGs within the proposed global network may facilitate searches for such transient signals of astrophysical origin.

\paragraph{Project progress and results}

\begin{figure}[bp]
\begin{center}
\includegraphics[width=0.75\textwidth]{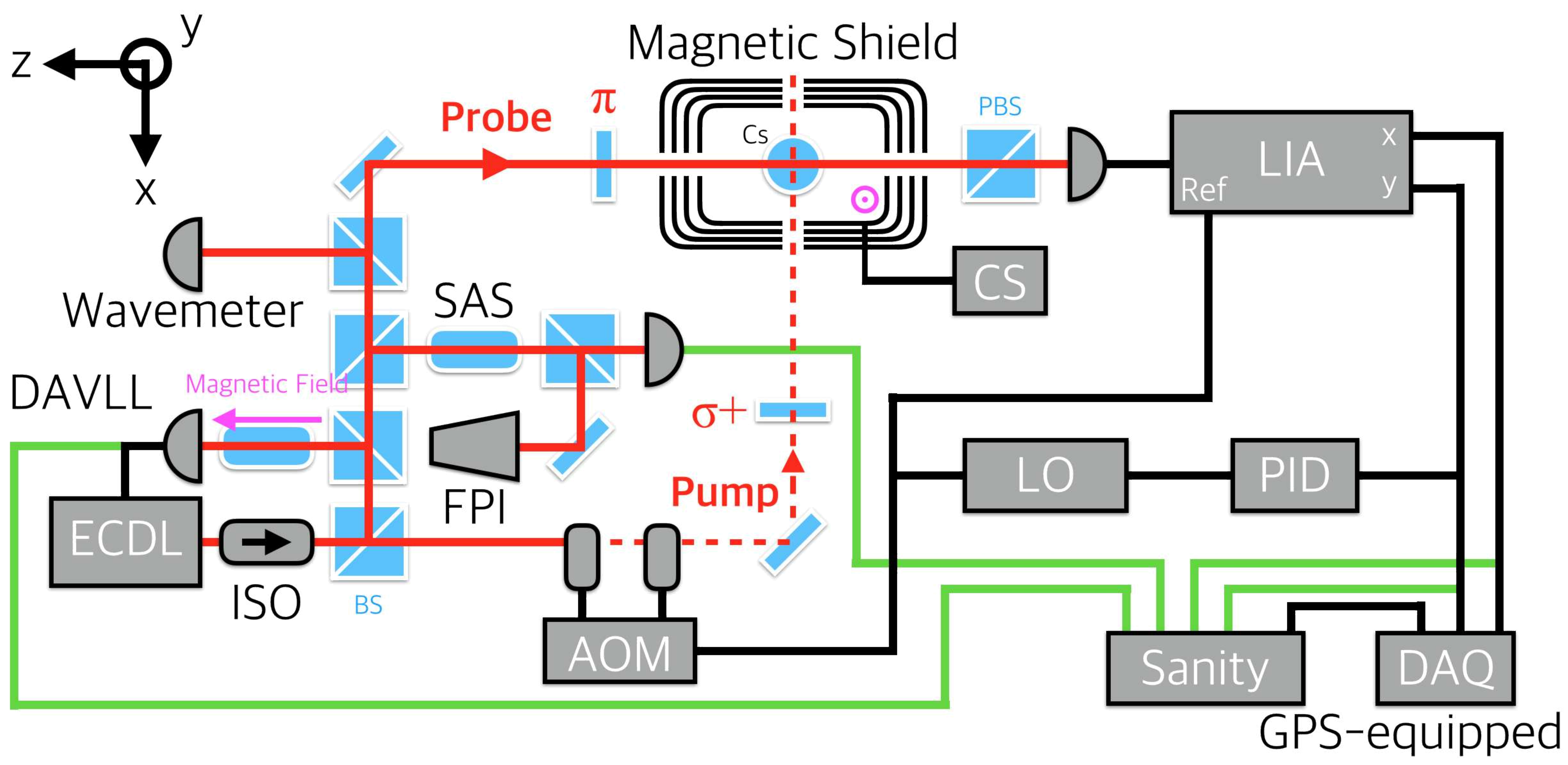}
\caption{Diagram of the optical magnetometer setup at CAPP/IBS}
\label{fig:GNOME_FIG_4}
\end{center}
\end{figure}

\begin{figure}[tp]
\begin{center}
\includegraphics[width=0.95\textwidth]{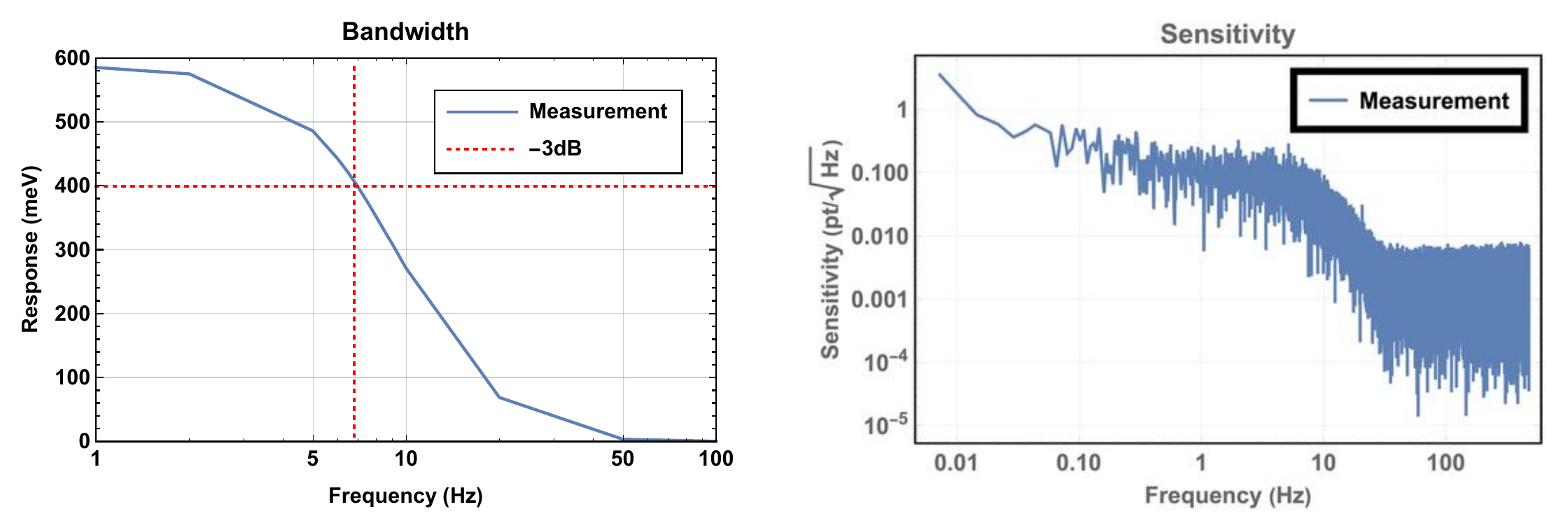}
\caption{Measured sensitivity (a) and bandwidth (b) of the optical magnetometer}
\label{fig:GNOME_FIG_5}
\end{center}
\end{figure}

\subparagraph{Development of an optical magnetometer}
The optical magnetometer in the IBS/CAPP is a cesium (Cs) atomic amplitude-modulated nonlinear magneto-optical rotation (AM NMOR) magnetometer with a single laser beam source from external cavity diode laser (ECDL). Figure~\ref{fig:GNOME_FIG_4} show the diagram of our magnetometer setup. This magnetometer is installed in the basement room C001, Creation Hall at KAIST Munji Campus, Daejeon, South Korea. Cesium vapor cells are used for the main magnetometer in this setup. Cesium is an alkali metal with a single valence electron. And the D2 transition of cesium was chosen as the main pumping line for the laser. The Cesium atomic magnetometer using single beam NMOR configuration with modulated pump beam has been installed and characterized. The sensitivity is achieved as $\sim100 \rm{fT}/ \sqrt{\rm{Hz}}$ and the bandwidth are measured as $\sim75\,\rm{Hz}$. Figure~\ref{fig:GNOME_FIG_5} shows the measured sensitivity and bandwidth of the magnetometer.

The magnetometer cell is placed in the four layers of cylindrical $\mu$-metal magnetic shield (Twinleaf MS-2). The DC shielding factor has been measured by at the center of the shield. The remaining residual field can be compensated by internal magnetic field generation coils.  An external cavity diode laser (ECDL) source (MOGLabs ECD004) has been used to pump and probe cesium atoms. The diode generates continuous wave (CW) laser, whose frequency is tuned to cesium D2 transition about 852\,nm. The laser frequency is stabilized using by DAVLL to the ${\rm{F  =  4 }} \to {\rm{ F }^{'}} {\rm{  =  5}}$ hyperfine transition. The laser generation and frequency lock are controlled by the laser diode controller (MOGLabs DLC202). 

The laser beam is split to two beams for orthogonal pump/probe beam configuration. Each beam is polarized by wave plates. The pump beam is polarized to circular by the quarter-wave plate and probe beam is polarized to linear by the half-wave plate. The pump beam intensity is modulated by the acousto-optic modulator (AOM, AAOptoelectronic MT80- A1,5-IR). The RF synthesizer (MOGLabs XRF421) provides 80 MHz RF signal to the AOM and this RF signal is modulated to the Larmor frequency with 50\% of duty cycle. The Larmor frequency is referred from a waveform generator (Keysight 33250A). A polarization rotation of the probe beam is detected by the balanced photodetector (Thorlabs PDB210A). The rotated phase signal from the balanced photodetector is oscillating around at the Larmor frequency. This signal is passing band-pass filter from 1\,kHz to 3\,kHz (SRS SR560 Low-noise pre-amplifier), then measured signal strength by the lock-in amplifier (LIA, SR865 Lock-in amplifier). 

Four RF channels are monitored to trigger environmental noise from the lock-in signals. Including two outputs of the LIA, laser locking error signal and laser intensity are connected to the specially designed unit called the sanity box. In addition, the sanity box also has temperature and magnetometer channels, so it records external environment to veto local noise fluctuation. The lock-in signals X, Y and the sanity channel are connected to the customized data acquisition hardware called the GPS box (DM Technologies GDL 110). This box is connected to a GPS antenna (Trimble Bullet III) placed at the rooftop of Creation Hall. The acquired data is digitized by 512\,Hz with GPS timestamp. These data are stored in the local personal computer (PC) and sent to the GNOME data server at Mainz, Germany.

\subparagraph{Domain wall signal parameters estimation}
CAPP team in the GNOME collaboration is developing data analysis tools for analyzing concurrent time-series data from each station. Three tools are mainly in progress in GNOME data analysis; Excess Power Search, Coincidence Search, and Coherence Analysis. The Excess Power Search is to detect burst signals of unknown waveform by dividing both time and frequency to tiny tiles. The Coincidence Search is looking for time windows such that most of magnetometers have burst signal. The Coherent Analysis is testing consistent of detected signals with single values of domain wall parameters. The Coincidence Search may lose a signal of scalar magnetometer due to its sensitive axis. This can be complemented with the Coherent Analysis. Therefore, the Coherent Analysis should be done after the Coincidence Search or Excess Power Search. Both Coincidence Search and Coherent Analysis may be tested at the same time if a reliable physics model for the interaction of axion domain wall can be built theoretically. It means that multiple time series data from magnetometer stations can be generated by arbitrary domain wall parameters through the physics model. Then the time series data from real magnetometer stations can be compared with the data generated from plausible domain wall model parameters. This method is based on a physics model of axion domain wall interaction and will be a strong method to estimate domain wall parameters from given simulated multiple time series data from the model.

\subparagraph{Effective approximation of axion electromagnetic field interaction}
CAPP team is also developing concepts for new type of experiment to search for axion and axion domain wall. The theoretical basis of this idea has been in development in our center based on effective approximation of Maxwell's equations that include axion-electromagnetic field interaction. The modified Maxwell's equations developed at CAPP has been submitted for being published. This new model provides broader scope in axion interaction and it enables us to look for axion from new kind of experiments. For example, with allowing the gradient of axion field, one can estimate a formation of charged domain-wall from the modified Maxwell's equations. It provides an idea for the network of haloscope to detect such charged domain wall structure. This idea shares many features with GNOME. We are also in design an experiment with Fabry-Perot type optical cavity that can scan broad range of axion mass by adjusting the distance between two mirrors. Unlike other haloscope searches, the electromagnetic field is sourced from high-finesse laser. Therefore, this experiment can explore for axion from a table-top setup.

\paragraph{Summary and prospect}
A new experiment for axion search enables investigations of transient exotic spin couplings. It is based on synchronous operation of globally separated optical magnetometers enclosed inside magnetic shields. Correlation of magnetometers' readouts enables filtering local signals induced by environmental and/or technical noise. Moreover, application of vetoing techniques, e.g., via correlation of optical magnetometer readouts with signals detected with non optical magnetic-field sensors, enables suppression of influence of global disturbances of magnetic origins, such as solar wind, fluctuation of the Earth's magnetic field, on the operation of the magnetometers. In such an arrangement, the network becomes primarily sensitive to spin coupling of non-magnetic origins, thus it may be used for searches of physics beyond the Standard Model. A specific example of such searches was discussed here by considering coupling of atomic spins to domain walls of axion-like fields. It was demonstrated that with modern state-of-the-art optical magnetometers probing a significant region of currently unconstrained space of parameters of the fields is feasible.

%% file: 1.6.2/main.tex
\subsubsection{Executive summary of the project}
This section describes about Axion Resonant InterAction DetectioN Experiment (ARIADNE).
\paragraph{Overview}
Axions are CP-odd scalar particles appearing in many extensions of the Standard Model \cite{2006JHEP...06..051S},\cite{2010PhRvD..81l3530A}. The well-motivated Peccei-Quinn (PQ) axion can explain the smallness of the neutron electric dipole moment, and is also a promising Dark Matter candidate \cite{1977PhRvL..38.1440P}. In addition, axions and axion-like particles could generate macroscopic P-odd and T-odd spin-dependent interactions which can be sought in sensitive laboratory experiments \cite{1977PhRvL..38.1440P}. Possible existence of such interactions has been postulated by many theories for decades.  New experiment to test axion induced spin-dependent interactions between nuclei at sub-millimeter ranges could be possible with magnetometer technique from a dense ensemble of polarized species.

\paragraph{The scope of the project}
This project is to develop a new magnetometry experiment to search for axion mediated spin-dependent interactions between nuclei at sub-millimeter ranges. Such an interaction would produce nuclear magnetic resonance (NMR) frequency as the distance between unpolarized and polarized mass is modulated. The axion field originates from the nuclear mass can act similarly to an equivalent magnetic field. This enables the investigation of exotic interaction mediated by axion though the magnetometer system. 

\paragraph{The implementation of the Project}
The technique we employed is based on NMR detection of the anomalous magnetic field sourced from matter. The experiment involves a segmented rotating cylindrical mass which sources the axion field, and laser-polarized 3He nuclei which can resonantly sense the effective axion "magnetic field" from the rotating mass. The radius of each successive segment of the cylinder is modulated in order to generate a time-varying potential at the nuclear spin precession frequency, due to the difference in the axion-mediated interaction as each section passes by the NMR sensor. Although there have been many experiments based on a similar concept, they have never reached enough sensitivity to detect axion-mediated force. Unlike in previous experiments, the key advantage is that by rotating the mass so that the segments pass by the NMR medium at the resonant frequency, the sensitivity is enhanced by the quality factor $Q = \omega \rm{T}_{2}$.

\paragraph{Progress and results}
Since CAPP joined ARIADNE collaboration in 2015, CAPP have focused to develop an infrastructure for this experiment. It includes lab space environment suitable for optical test to prepare this experiment. CAPP have also been developed 3He polarization system mainly for this experiment. The 3He polarization system have finish the first test stage with small optical test setup. Now we are developing a full scale setup for large quantity optical pumping system. A list of progress in this project are following:

\begin{itemize} 
	\item Development of an 3He optical pumping system: A full size 3He polarization system with Metastability Exchange Optical Pumping (MEOP) method is under development at CAPP/IBS
	\item High Sensitivity SQUID development: SQUID magnetometer has been developed with KRISS
	\item Superconducting shield development: Nb superconducting magnetic shield has been developed with sputtering method at CAPP
	\item ARIADNE collaboration: CAPP has been actively involved in data analysis for ARIADNE collaboration
\end{itemize}

\paragraph{Significance}
This experiment has a potential to improve on axions bounds from all other previous and current experiments and astrophysics by several orders of magnitude, and to probe deep into the theoretically interesting regime for PQ (Peccei-Quinn) axion. With a new precision magnetometry method, we will be able to reach deep sensitivity limit that can verify PQ axions. Otherwise it is possible to detect new type of interaction that can prove the existence of such particles. 

\newpage
\subsubsection{ARIADNE Project}

\paragraph{Introduction}
Axions are CP-odd scalar particles appearing in many extensions of the Standard Mode  \cite{2006JHEP...06..051S, 2010PhRvD..81l3530A}. Well-motivated Peccei-Quinn (PQ) axion can explain the smallness of the neutron electric dipole moment (nEDM), and is also a promising dark matter candidate.  Allowed axion windows can be parameterized in terms of axion decay constant and mass. For PQ axion, the mass and the decay constant are inversely related by  ${{\rm{m}}_a} \approx  6 \times {\rm{1}}{{\rm{0}}^{-3}}{\rm{ eV }}\left[ {\frac{{{{10}^9}{\rm{GeV}}}}{{{f_a}}}} \right]$ . Currently, the direct search for axion is the main objective of the current CAPP/IBS experiment, involving axion to photon conversion in a resonant cavity. Interestingly enough, axions and axion-like particles also generate macroscopic P-odd and T-odd spin-dependent interactions that can be sought in sensitive laboratory experiments\cite{1977PhRvL..38.1440P}. Axion mass determines the range of short-range interactions between nuclei, set by Compton wavelength ${\lambda _a} = \frac{\hbar }{{{m_a}c}}$. 

The tabletop experiment can measure the interaction between matter objects at short range with precision magnetometry and it is a new way to search for axion-like particles. This experiment employs high precision magnetometry in order to detect PQ axion. The experiment involves a rotating non-magnetic mass to source the axion field, and a dense ensemble of laser-polarized 3He nuclei to detect the axion field by NMR. The signal from an axion field can be resonantly enhanced by properly modulating the axion potential at the nuclear spin precession frequency. This experiment will be sensitive to axion mass ranging between  $\sim{10^{ - 2}}{\rm{eV}}$ and  $\sim{10^{ - 5}}{\rm{eV}}$ for PQ axions. This range corresponds to the interaction distance between unpolarized and polarized masses in  $30\mu m\sim30{\rm{mm}}$ range. This proposed experiment is called Axion Resonant InterAction DetectionN Experiment (ARIADNE).

In ARIADNE, pseudo magnetic field from the spin-dependent interaction will be resonantly enhanced by employing a rotating mass with grooves, effectively generating a distance difference between the source mass and 3He. We aim at investigating axion mass range of ${10^{ - 2}}$ to ${10^{ - 5}}\rm{eV}$, which is largely complementary to the reach of current CAPP/IBS axion experiment with resonant cavities. This tabletop experiment can exceed present constraints on spin-dependent short-range forces by up to 8 orders of magnitude, and can improve on the combined laboratory/astrophysical limits by a factor of  in the prime axion range of ${f_a} = {10^9}\sim{10^{12}}{\rm{GeV}}$, probing deep into the theoretically interesting regime for PQ axion \cite{2009PhRvL.103z1801V, 2013PhRvL.111j0801T, 2015PhRvD..91j2006C, 2013PhRvL.111j2001B}.

\begin{figure}[h]
\begin{center}
\includegraphics[width=.45\textheight]{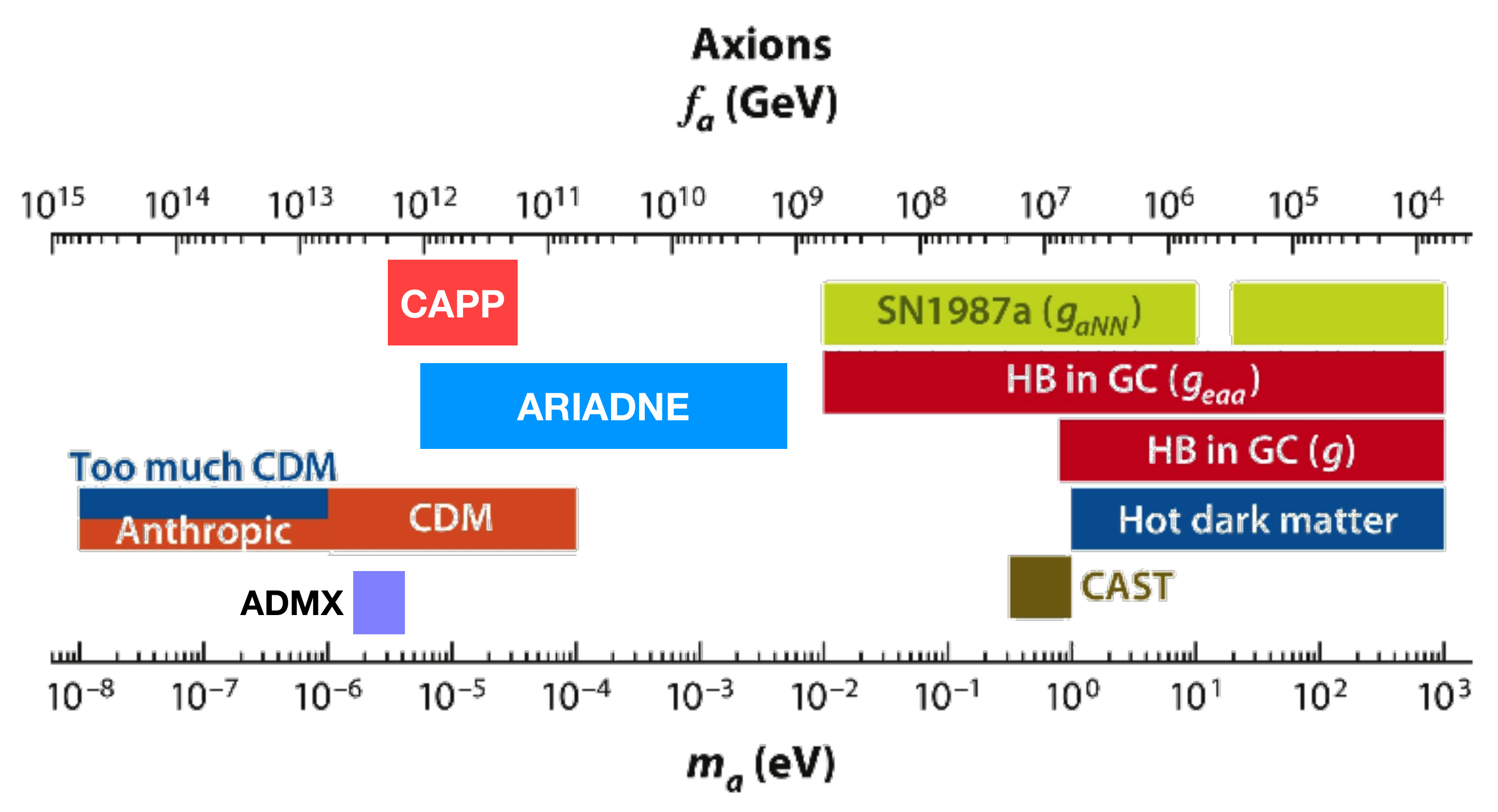}
\caption{Allowed axion window can be parameterized in terms of axion decay constant $f_{a}$ and mass  $m_{a}$. ARIADNE experiment is sensitive to ${f_a} = {10^9}\sim{10^{12}}{\rm{GeV}}$ }
\label{fig:ARIADNE_FIG_1}
\end{center}
\end{figure}

This experiment is especially relevant, as there have been no experiments that could probe the range ${f_a} = {10^9}\sim{10^{12}}{\rm{GeV}}$ (Fig.~\ref{fig:ARIADNE_FIG_1}). The main difference between ARIADNE and other direct axion searches is that it is still sensitive to axions even if they do not make up dark matter, as axion field is actually sourced by the local ensemble of nuclei. In addition, unlike any other axion haloscope experiments, this experiment does not necessarily scan over a wide axion mass range.

\begin{figure}[h]
\begin{center}
\includegraphics[width=.20\textheight]{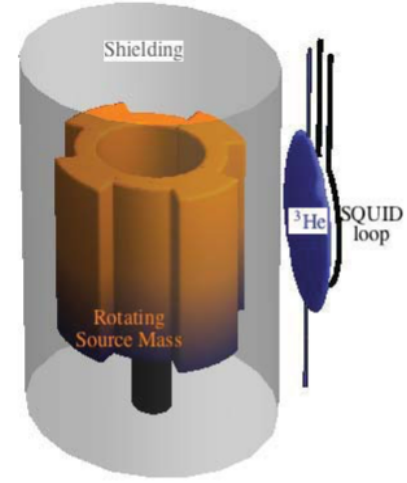}
\caption{Conceptual  design of ARIADNE experiment}
\label{fig:ARIADNE_FIG_2}
\end{center}
\end{figure}

\paragraph{Experimental approach}
ARIADNE involves a segmented rotating cylinder mass that sources axion field, and a laser polarized NMR sample that can resonantly sense the effective axion magnetic field from the rotating mass. The radius of each segment of the cylinder is modulated in order to generate a time-varying potential at the nuclear spin precession frequency, due to the difference in the axion-mediated interaction as each segment passes by NMR sensor. The conceptual design of the system is shown in Fig.~\ref{fig:ARIADNE_FIG_2}.

In presence of an axion-mediated interaction between the rotating mass and the NMR sample, nuclear spins in the hyperpolarized sample will resonantly precess off the axis of polarization. Superconducting QUantum Interference Device (SQUID) can detect any change in magnetization due to the interaction. Interaction energy between particles due to monopole-dipole axion exchange as a function of distance is \cite{1984PhRvD..30..130M}

\begin{equation}
{U_{sp}}(r) = \frac{{\hbar {g_s}{g_p}}}{{8\pi {m_f}}}\left( {\frac{1}{{r{\lambda _a}}} + \frac{1}{{{r^2}}}} \right){e^{ - \frac{r}{{{\lambda _a}}}}}\left( {\hat \sigma  \cdot \hat r} \right),
\label{eq:ARIADNE_EQ_1}
\end{equation}
where $m_{f}$ is fermion mass. For the PQ axion $g_{s}$ and $g_{p}$ are directly correlated to the axion mass as they are fixed by the axion decay constant $f_{a}$.

\begin{eqnarray}
6 \times {10^{ - 27}}\left( {\frac{{{{10}^9}{\rm{GeV}}}}{{{f_a}}}} \right) < {g_s} < {10^{ - 21}}\left( {\frac{{{{10}^9}{\rm{GeV}}}}{{{f_a}}}} \right)&,\nonumber \\
{g_p} = \frac{{{C_f}{m_f}}}{{{f_a}}} = {C_f}{10^{ - 9}}\left( {\frac{{{m_f}}}{{1{\rm{GeV}}}}} \right)\left( {\frac{{{{10}^9}{\rm{GeV}}}}{{{f_a}}}} \right)&.
\label{eq:ARIADNE_EQ_2}
\end{eqnarray}

The scalar coupling $g_{s}$ of the PQ axion is indirectly constrained from above by the EDM searches and the lower bound is set by the amount of CP violation in the Standard Model. In the PQ axion coupling to spin, $C_{f}$ is a model dependent constant. If we write this interaction in Eq. \ref{eq:ARIADNE_EQ_1} using axion potential  $V_{a_{s}}\left({r}\right)$ as

\begin{equation}
{U_{sp}}(r) =  - \vec \nabla {V_{{a_s}}}(r) \cdot {\hat \sigma _2}
\label{eq:ARIADNE_EQ_3}
\end{equation}
where ${V_{{a_s}}}(r) = \frac{{\hbar {g_s}{g_p}}}{{8\pi {m_f}}}\left( {\frac{{{e^{ - \frac{r}{\lambda }}}}}{r}} \right)$ is for monopole-dipole interaction. Axion generated potential by an unpolarized mass acts on a nearby fermion just like an effective magnetic field, and it can be expressed as

\begin{equation}
{\overrightarrow {\rm{B}} _{{\rm{eff}}}} = \frac{{2\vec \nabla {V_a}(r)}}{{\hbar {\gamma _f}}}
\label{eq:ARIADNE_EQ_4}
\end{equation}
where $\gamma_{f}$ is fermion's gyromagnetic ratio. 

However, the effective magnetic field is different from standard magnetic field. Since it couples to the spin of the particle, it is independent of the fermion's magnetic moment. It also does not couple to ordinary angular moment or charge. This magnetic field is not subject to the Maxwell equations. Therefore, magnetic shielding does not screen the effective magnetic field generated from the axion exchange. Hence, a superconducting magnetic shield, should be placed between the source mass and the detector to minimize the background electromagnetic noise.

A spin polarized nucleus near the rotating segmented cylinder with radius $R$ will feel an effective magnetic field of approximately ${B_{{\rm{eff}}}} \approx \frac{1}{{\hbar {\gamma _N}}}\nabla {V_a}(r)(1 + \cos (n{\omega _{{\rm{rot}}}}t))$, where $\gamma_{N}$ is the nuclear gyromagnetic ratio and $n$ is the number of segments, for sample thickness of order $\lambda$. ${\vec B_{{\rm{eff}}}}$ is parallel to the radius of the cylinder.

From the Bloch equation, an NMR sample with net polarization ${M_z}$ parallel to the axis of the cylinder will develop a time-varying perpendicular magnetization ${M_x}$ in response to the resonant effective axion field   ${\vec B_{{\rm{eff}}}}$ given by

\begin{equation}
{M_x}(t) \approx \frac{1}{2}{n_s}p{\mu _N}{\gamma _N}{B_{{\rm{eff}}}}{T_2}\left( {{e^{ - t/{T_1}}} - {e^{ - t/{T_2}}}} \right)\cos (\omega t)
\label{eq:ARIADNE_EQ_5}
\end{equation}
where $p$ is the polarization fraction, $n_{s}$ is the spin density in the sample, and $\mu_{N}$ is the nuclear magnetic moment. ${M_x}(t)$  grows linearly until $t\sim{T_2}$  , the transverse relaxation time, and then decay at the longer relaxation time $T_{1}$ . ${M_x}(t)$ can be detected by a SQUID with its pickup coil oriented radially. The SQUID detects the changing magnetization of the sample, not the axion field directly as it is not a real magnetic field. 

A fundamental limitation of this technique comes from the transverse noise in the sample itself as $\sqrt {{M_N}^2}  = \sqrt {\frac{{\hbar \gamma n{\mu _{{\rm{3He}}}}{T_2}}}{{2V}}} $  and the minimum transverse magnetic resonant field is sensitive to
\begin{eqnarray}
{B_{{\rm{min}}}} &\approx& {p^{ - 1}}\sqrt {\frac{{2\hbar b}}{{{n_s}{\mu _{{\rm{3He}}}}\gamma V{T_2}}}} \nonumber \\ 
&=& 3 \times {10^{ - 19}}{\rm{T}} \times \left( {\frac{1}{p}} \right)\sqrt {\left( {\frac{b}{{1{\rm{Hz}}}}} \right)\left( {\frac{{1{\rm{m}}{{\rm{m}}^3}}}{V}} \right)\left( {\frac{{{{10}^{ - 21}}{\rm{c}}{{\rm{m}}^{ - 3}}}}{{{n_s}}}} \right)\left( {\frac{{1000{\rm{s}}}}{{{T_2}}}} \right)} 
\label{eq:ARADNE_EQ_6}
\end{eqnarray}
where $V$ is the volume of the sample, $\gamma$ is the gyromagnetic ratio for $^3{\rm{He}} = \left( {2\pi } \right) \times 32.4{\rm{MHz/T}}$,  $b$ is the measurement bandwidth, and ${\mu _{{\rm{3He}}}} =  - 2.12 \times {\mu _n}$, where is the nuclear Bohr magneton.

\begin{figure}[bp]
\begin{center}
\includegraphics[width=.50\textheight]{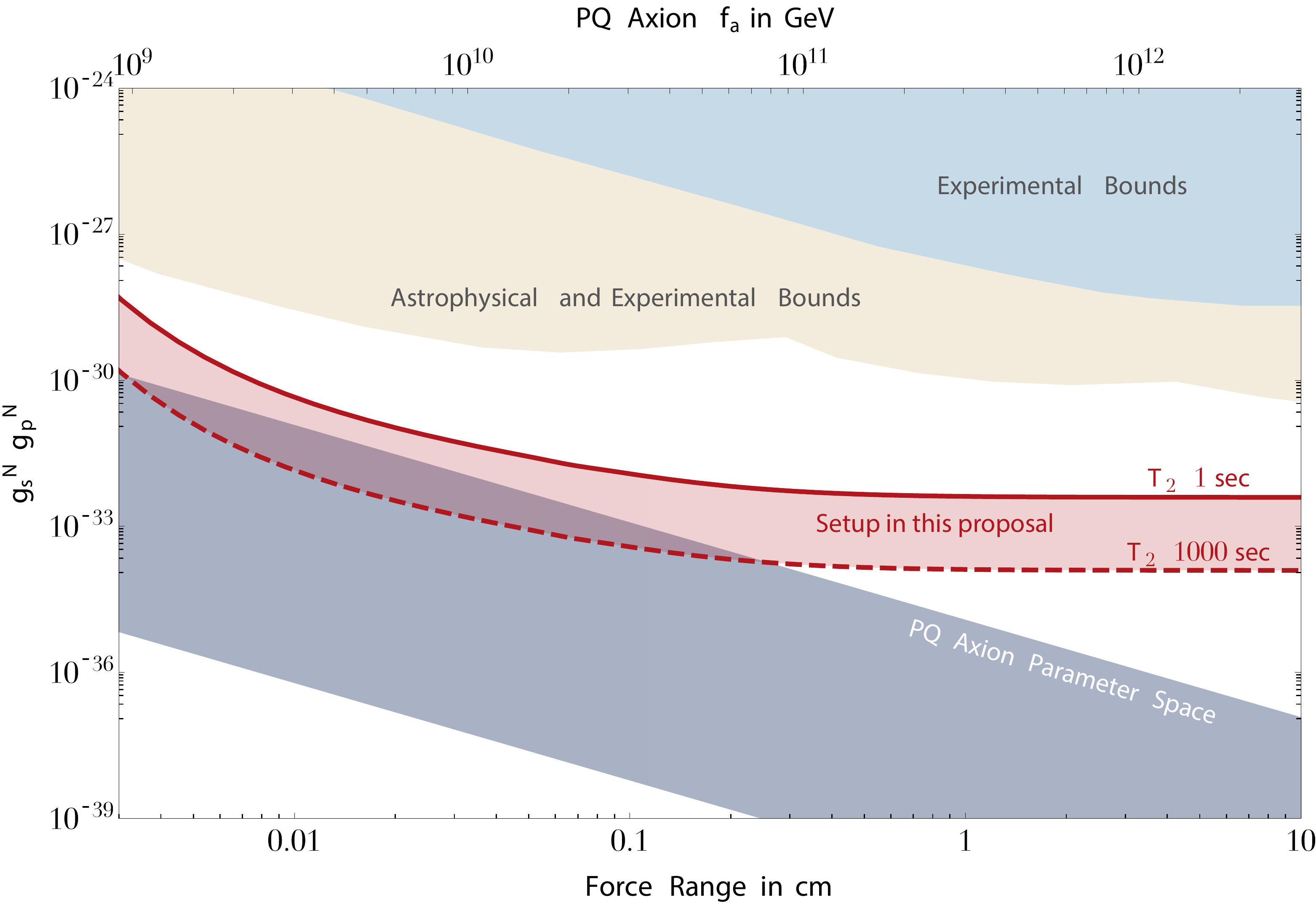}
\caption{Projected reach for monopole-dipole axion mediated interactions. The band bounded by the red (dark) solid line and dashed line denotes the limit set by transverse magnetization noise for this setup, with  $\rm{T}_{2}$ ranging from 1 to 1000\,s. The total integration time is $10^{6}\,$sec. The shaded band is the parameter space for the PQ axion. Experimental as well as combined experimental and astrophysical bounds are also presented near the sample, and the resonant frequency    of NMR sample. Superconducting shields screen the NMR sample from source mass}
\label{fig:ARIADNE_FIG_3}
\end{center}
\end{figure}

The projected reach from Eq. \ref{eq:ARADNE_EQ_6} is shown in Fig.~\ref{fig:ARIADNE_FIG_3}. This shows where the tremendous boost in the sensitivity is: (a) the resonant enhancement of the signal improves the sensitivity due to an effective quality factor $Q = \omega T_{2}$, (b) large number of nuclei ${n_s}V$, (c) long coherent time and high polarization rate of 3He.

\paragraph{Comparison with haloscope experiments}
Axion haloscope experiments at CAPP or other institutes such as CULTASK and ADMX are all based on resonant axion-to-photon conversion in a background magnetic field in a microwave cavity. The cavity is tuned so that its resonant frequency matches the axion field resonance frequency which is determined by the axion mass. The relevant coupling that is probed is ${g_{a\gamma \gamma }}$ describing the axion coupling to two photons. This experiment senses a complementary coupling, namely the axion scalar-dipole coupling between nucleons. The axion haloscope experiment may be able to detect cosmic axions with $f_{a}$ in a band that overlaps with part of the allowed dark matter window around ${f_a} = 3 \times {10^{11}}\sim3 \times {10^{12}}$, corresponding to masses of $2 \times {10^{ - 6}}\sim2 \times {10^{ - 5}}{\rm{eV}}$. 

There is also another important difference in this experiment from other haloscope experiments. This experiment does not depend on whether or not the axion is Dark Matter. In fact, there could be cosmological scenarios of late entropy production, for example where the axion is not dark matter, even in the region where haloscope experiments are sensitive. So, in reality all axion haloscope experiments can do is exclude the axion as a Dark Matter candidate and not necessarily from the PQ theory. In our case, the result does not care about the cosmological evolution of the axion field. In addition, since it is probing a different coupling than the one in CAPP cavity experiment, this proposal should be considered as a truly complementary experiment to CAPP's axion search with resonant cavities. The allowed axion window is shown in Fig.~\ref{fig:ARIADNE_FIG_3}.

\paragraph {Design concept}
\subparagraph{Experimental setup}
Three fused quartz vessels containing hyperpolarized 3He will serve as resonant magnetic field sensors. Three such sensors will be used to cancel common-mode noise and to search for appropriate correlations between their signals. A SQUID pickup coil will sense the magnetization in each of the samples. The inside of the quartz containers will be polished to a spheroidal shape, of internal dimensions $10\,{\rm{mm}} \times 3\,{\rm{mm}} \times 150\,\mu {\rm{m}}$ . The spheroidal shape will allow the magnetization to remain relatively constant throughout the sample volume. Such a shape can be fabricated by fusing together two pieces of quartz after polishing a hemi-spheroidal cavity in each piece. 

\begin{figure}[bp]
\begin{center}
\includegraphics[width=.50\textheight]{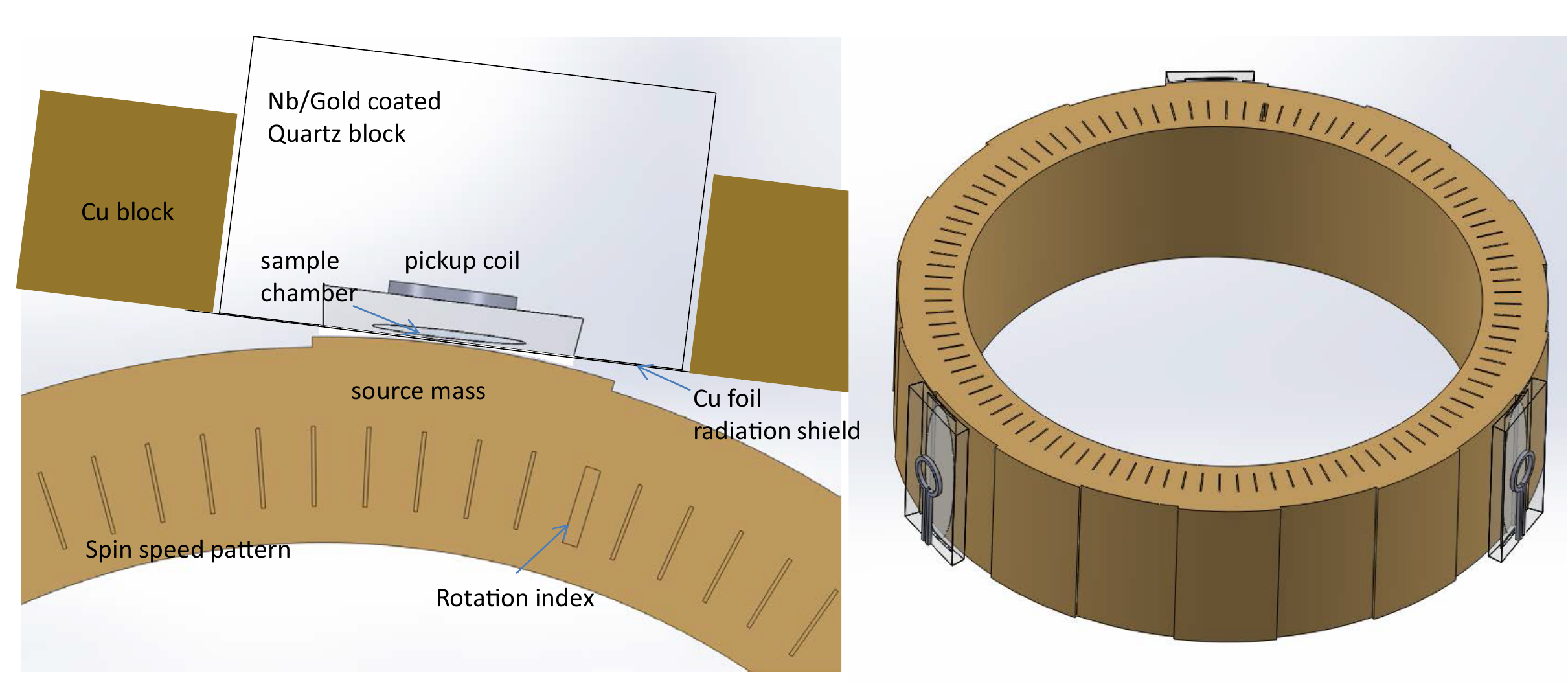}
\caption{Illustration of proposed setup. (Right) A source mass consisting of a segmented cylinder with 11 sections is rotated around its axis of symmetry at a fixed frequency ${\omega _{{\rm{rot}}}}$, which results in a resonance between the frequency  $\omega  = 10{\omega _{{\rm{rot}}}}$ at which the segments pass near the sample and the resonant frequency  $2\vec {\mu}_{\rm{N}} \cdot {\vec {\rm{B}} _{{\rm{ext}}}}/\hbar $ of the NMR sample. The NMR sample has an oblate spheroidal geometry to minimize magnetic gradients while allowing close proximity to the mass. Superconducting films shield the NMR sample from background fluctuating magnetic fields associated with the external environment and the rotating source mass. (Left) Cross sectional view from top of region near mass and detector.}
\label{fig:ARIADNE_FIG_4}
\end{center}
\end{figure}

The internal magnetic field remains constant in a spheroidal geometry regardless of the magnetization direction, so is insensitive to small misalignments of the polarization with the principal axes. The quartz sample container will be affixed to a larger quartz block, which will be sputter coated with a $10\mu {\rm{m}}$ layer of Niobium and  $200\,{\rm{nm}}$ of Ti/Cu. The Cu will be electroplated with of Gold as a blackbody reflector. The dimensions of the outer block will be chosen to minimize the effect of the magnetic gradients at the sample location due to the Meissner image currents in the superconducting walls (see Fig.~\ref{fig:ARIADNE_FIG_4}). 

The use of superconducting shielding (as opposed to e.g. $\mu$-metal shielding) is essential to mitigate magnetic field noise from thermal currents (i.e. Johnson noise) in ordinary conducting materials. A stretched copper radiation shield foil of dimensions $25\mu {\rm{m}}$ by $1{\rm{cm}} \times 1{\rm{cm}}$ will be inserted between the rotating mass and the quartz block. The foil will be supported by larger copper-coated tungsten blocks affixed to the cold plate of the cryostat. These tungsten blocks will be joined together at a central support point to minimize effects from the thermal contraction of the copper base plate. The vessel wall thickness is limited to  $75\mu {\rm{m}}$ on the side that attaches to the shield to allow close proximity to the source mass.

The rotating source mass will consist of a segmented cylindrical shell with segments of varying thickness. We will construct a cylindrical shell of length $1{\rm{cm}}$, thickness  $4{\rm{mm}}$, and outer diameter  $3.8{\rm{cm}}$ divided into 20 sections of length  $5.4{\rm{mm}}$. The radius of each section is modulated by approximately $200\mu {\rm{m}}$ in order to generate a time-varying potential at frequency $\omega  = 10{\omega _{{\rm{rot}}}}$, due to the difference in the axion-mediated interaction as each section passes by the sensor. To decouple mechanical vibration from the signal of interest, ${\omega _{{\rm{rot}}}}$ is chosen to be 10 times lower than $\omega$. The shell will be connected to spokes and the rotation stage by a tungsten rod. The wobble of the mass will be measured using fiber coupled laser interferometers and counterweights will be applied as necessary to maintain the wobble below $0.003{\rm{cm}}$ at the outer radius.

The samples will be housed in a liquid Helium cryostat. The separation between the rotating mass and quartz cell will be designed to be $75\mu {\rm{m}}$ at cryogenic temperature. Upon thermal contraction, the tungsten mass will decrease in radius by approximately  $16\mu {\rm{m}}$ and the radius of the copper base plate at the point where the quartz is contacted will contract by approximately $60\mu {\rm{m}}$. The room-temperature separation between the mass and quartz samples will be larger to account for this thermal contraction. Finite element modeling will be performed to aid in a final design which accounts for thermal contraction. It is unnecessary that the source mass itself remains cryogenic, however the superconducting shield needs to below the superconducting transition temperature. The shield can be thermally anchored to copper to allow sufficient thermal conduction. If ${\omega _{{\rm{ro}}t}}/2\pi  = 10{\rm{Hz}}$ and  $\omega /2\pi $ is $100{\rm{Hz}}$, then the net ${{\rm{B}}_{{\rm{ext}}}}$ needed at the sample is of order $30\;{\rm{mG}}$.  ${{\rm{B}}_{{\rm{ext}}}}$  is the sum of the internal magnetic field of the sample, which is roughly  $0.2\;{\rm{Gauss}}$ for  $2 \times {10^{21}}{\rm{c}}{{\rm{m}}^3}$ density of 3He, and a field generated by superconducting coils.

In such a field, SQUID can operate neat its optimal sensitivity of  $1.5\rm{fT}/\sqrt {{\rm{Hz}}} $. We expect the current in the coils needs to be maintained constant at low frequencies to within $\sim10{\rm{ppm}}$ (${\rm{1000s/}}{{\rm{T}}_2}$ ). A triple-layer of  $\mu $-metal shield will enclose the cryostat while the superconducting shields are cooled through the superconducting phase transition. The total DC magnetic field should be kept below  ${10^{ - 7}}$T during this process to avoid "freezing-in" flux.

The rotation of the cylinder can be accomplished by an in-vacuum piezoelectric transducer. Commercial solutions for ultrasonic or direct drive fast ($ \gg $3Hz) compact rotation stages that are both vacuum compatible and cryogenic are not readily available. Therefore, the rotational mechanism will need to be maintained at a higher temperature ($ - 30{\rm{C}}^\circ $) than the surrounding components. This will be achieved with heaters and heat-shielding. For a design such as this, the expected heat load can be estimated to be approximately 1 W. Direct drive stages offer faster rotation rates (up to 25\,Hz) with the caveat that local magnetic fields are larger.

These can be attenuated with additional  $\mu$-metal shielding (which lies well outside of the superconducting coating on the quartz sample vessels). Both approaches (direct and ultrasonic PZT) will be tested with regard to vibration, wobble, and stray magnetic fields, and the optimum approach will be chosen for the final cryostat design. The rotational speed must be kept constant, so that resonance between the Larmor frequency and rotational frequency can be maintained. The velocity stability for direct drive stage has been specified to better than 0.01\% for a complete rotation at frequencies between . This allows sample to remain on resonance to utilize  in excess of 100\,s. 

The top surface of the cylindrical mass shell will be patterned with a periodic reflective metallic coating deposited by electron beam evaporation through a shadow-mask. The spin speed can be determined by sensing this reflectivity modulation using a laser bounced from the top surface of the cylinder. An index mark will allow the determination of the phase of the cylinder rotation, for correlation with the spin precession in the sample. The mechanical wobble of the cylinder can be assessed by measuring light reflected from the top and bottom surfaces onto quadrant photodetectors or using fiber-coupled laser interferometers.

The hyper-polarized 3He gas can be prepared by metastability exchange optical pumping (MEOP). MEOP is especially well-suited to the needs of this experiment: it can polarize 3He at total pressures of a few mbar in an arbitrary mixture of 3He and 4He. The MEOP polarized 3He compression system can deliver polarized 3He gas at pressures from $\sim 1\,\rm{mbar}$ to $\sim 0.1\,\rm{bar}$ at room temperature and can, therefore, be used to conduct higher temperature tests in the same pressure regime that the 4K cryogenic cell will operate in. A manifold will be designed to deliver the gas from the polarization region into the sample area, and then to recirculate it for subsequent experimental cycles. The low spin relaxation valve technology may be needed both at room temperature and at low temperature.

The temperature gradient of the Pyrex tubing connecting the optical pumping region to the low temperature cell will be engineered to avoid the rapid spin relaxation which can occur on glass in the $5\sim30{\rm{K}}$ region. The system must provide a spin density of  $2 \times {10^{19}}$ up to $2 \times {10^{21}}{\rm{c}}{{\rm{m}}^{ - 3}}$ (the density of helium gas) just above 4.2\,K. SQUID measurements using 3He at densities similar to those we propose with even longer and relaxation times have been performed previously in co-magnetometer experiments searching for longer-range monopole-dipole forces and Lorentz/CPT violation. We expect a 3He polarization fraction near 0.5 based on the very extensive experience documented by the Paris group in their decades of work on NMR studies of polarized 3He and 3He/4He gas and liquid mixtures from  $\sim{\rm{mK}}$ temperatures to 4.2\,K.

The source mass assembly and rotation mechanism will be inserted into the region enclosed by the magnetic shield. The superconducting shields will be cooled in low magnetic field to avoid trapped flux. The sample will be polarized in the vertical direction, and a field will be applied using superconducting coils to set the required Larmor frequency at approximately $30\sim100{\rm{Hz}}$ . The source mass will be rotated, such that the segments of the cylinder pass by the NMR sample at the Larmor frequency to induce spin precession in each sample, which will be detected via a SQUID pickup coil. A measurement will occur for a time ${{\rm{T}}_2}$. At this point the gas will be extracted from the chamber and repolarized and the measurement will be repeated. Measurements will be averaged together and correlations between the signals in the detectors and the phase of the rotating source mass will be analyzed.

\subparagraph{Sensitivity}
In Fig.~\ref{fig:ARIADNE_FIG_3}, we present the reach of the setup assuming a total integration time of for ${10^6}$\,s a monopole-dipole axion mediated interaction for both  $T_2 = 1$\,s and 1000\,s. The limitation is due to noise indicated in Eq.~\ref{eq:ARADNE_EQ_6}, which lies significantly above the SQUID sensitivity. We estimate a minimum field sensitivity of  ${B_{{\rm{min}}}} = 3 \times {10^{-19}}\,{\rm{T}}$ for  ${T_2} = 1000$\,s in this setup. For example, with ${B_{{\rm{eff}}}} = 3 \times {10^{-19}}$\,T  at 100\,Hz, the signal at the SQUID due to ${M_x}$ will be $\sim{10^{-15}}$ to ${10^{-12}}\;{\rm{T}}$ for ${{\rm{T}}_2} = 1$ to 1000\,s , respectively. Finally, we draw curves for the PQ axion parameter space assuming  ${{\rm{C}}_f} = 1$ as well as the current astrophysical and experimental bounds or a combination thereof. Not only does the proposed setup compete with astrophysical bounds, but it probes a large part of the traditional axion window of ${10^9}\,{\rm{GeV}} < {f_a} < {10^{12}}\,{\rm{GeV}}$ .

\paragraph{Project progress and results}
\subparagraph{Development of an 3He polarization system}
The main component of 3He magnetometer system is a 3He polarization unit. It is designed to deliver at least  of spin-polarized 3He gas in measurement cell during every measurement cycle. 3He polarization unit has been specially designed to fulfill the needs of CAPP/IBS experiment. Working principle of the polarized unit is as follow. First, 3He from a high-pressure reservoir is fed into the polarizer's gas system. To avoid complications in transport of polarized 3He, magnetic field from the optical pumping unit to the measurement cell will be aligned in the same vertical direction.

The pressure inside the system is controlled by a non-magnetic pressure sensor. Gas will be purified by means of a getter-based purifier from SAES. After purification, 3He gas is released through a specific non-magnetic valve into the optical pumping cells. Optical pumping cells will consist of two quartz glass tubes, 1\,m length and 50\,mm diameter, with vacuum tight anti-reflection-coated glass tube with fused silica windows at the ends.

\begin{figure}[bp]
\begin{center}
\includegraphics[width=\textwidth]{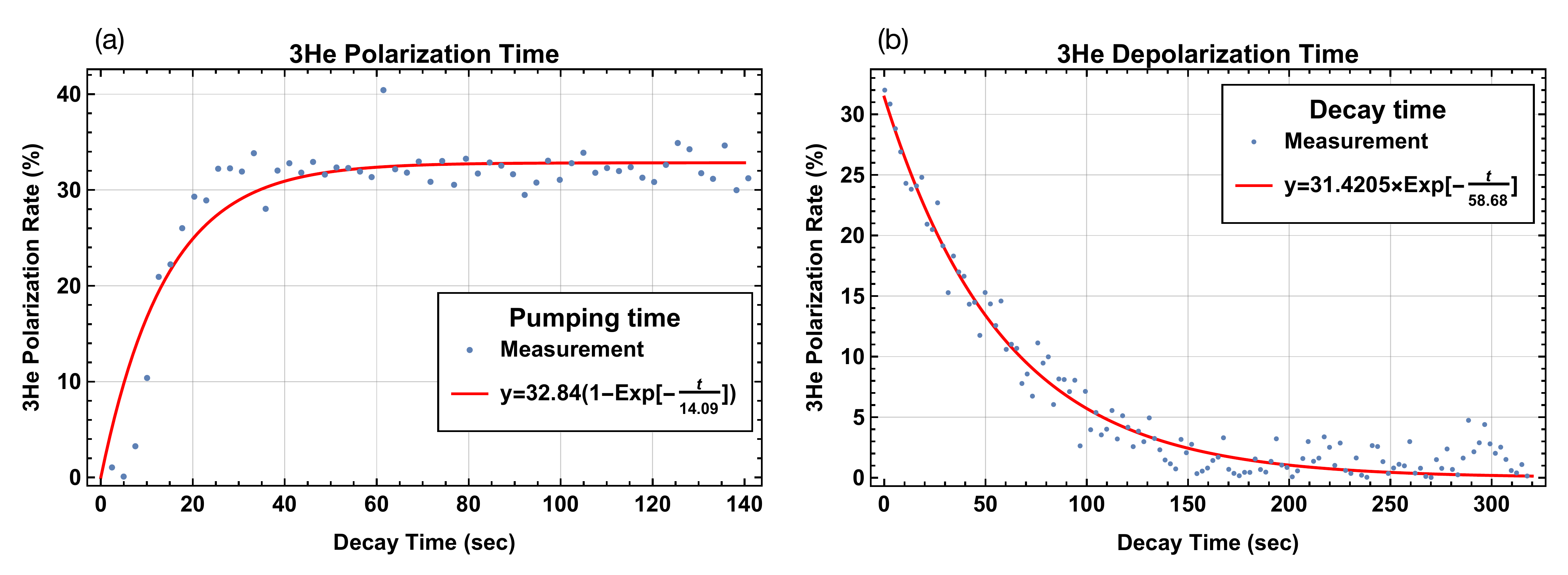}
\caption{(Left) Polarization up to 33\% has been achieved with a test cell filled with  3He gas. (Right) T1 relaxation time after 3He is polarized.}
\label{fig:ARIADNE_FIG_5}
\end{center}
\end{figure}

CAPP is now developing a polarization setup to produce high density of hyperpolarized 3He gas at room temperature. It consists of two 1m long cells made of Pyrex glass that contains 1 mbar pressure of 3He gas to be polarized. CAPP developed a prototype 3He polarization setup and a large magnetic guide coil system for a full-size polarization setup to produce high density of hyperpolarized 3He gas at room temperature. The prototype 3He polarization consists of 5cm long cell made of Pyrex glass that contains 1 mbar pressure of 3He gas to be polarized. CAPP successfully measured 3He polarization up to ~33\% and longitudinal relaxation time T1 time up to ~60 sec from the prototype setup as shown in Fig.~\ref{fig:ARIADNE_FIG_5}.

In the full-scale setup of 3He polarization, two 1m long Pyrex glass cells will be installed in the central region of the magnetic guide coil system and will polarize 3He gas up to ~40\% of polarization rate. 3\,m long magnetic guide coil system has been designed for this purpose. The guide coil system consists of total eight sets of square coils. Each coil has 1\,m$\times$ 1\,m dimensions of square shape and is designed to hold 1mm thick copper coil up to 300 turns. The number of turn and the position of each coil were main variables in the analytical study to create uniform magnetic field with strength up to 2\,mT and 99.5\% of homogeneity along 1m distance from the center. The result from analytical study to set the turn number and position of each coil was cross-checked with FEM simulation with COMSOL. The real setup consists of eight set of Al frame to hold up to ~250 turns of copper wire, frame holders to locate the Al frame in the position and 3m long skeleton to support all structures in position.

\begin{figure}[bp]
\begin{center}
\includegraphics[width=0.95\textwidth]{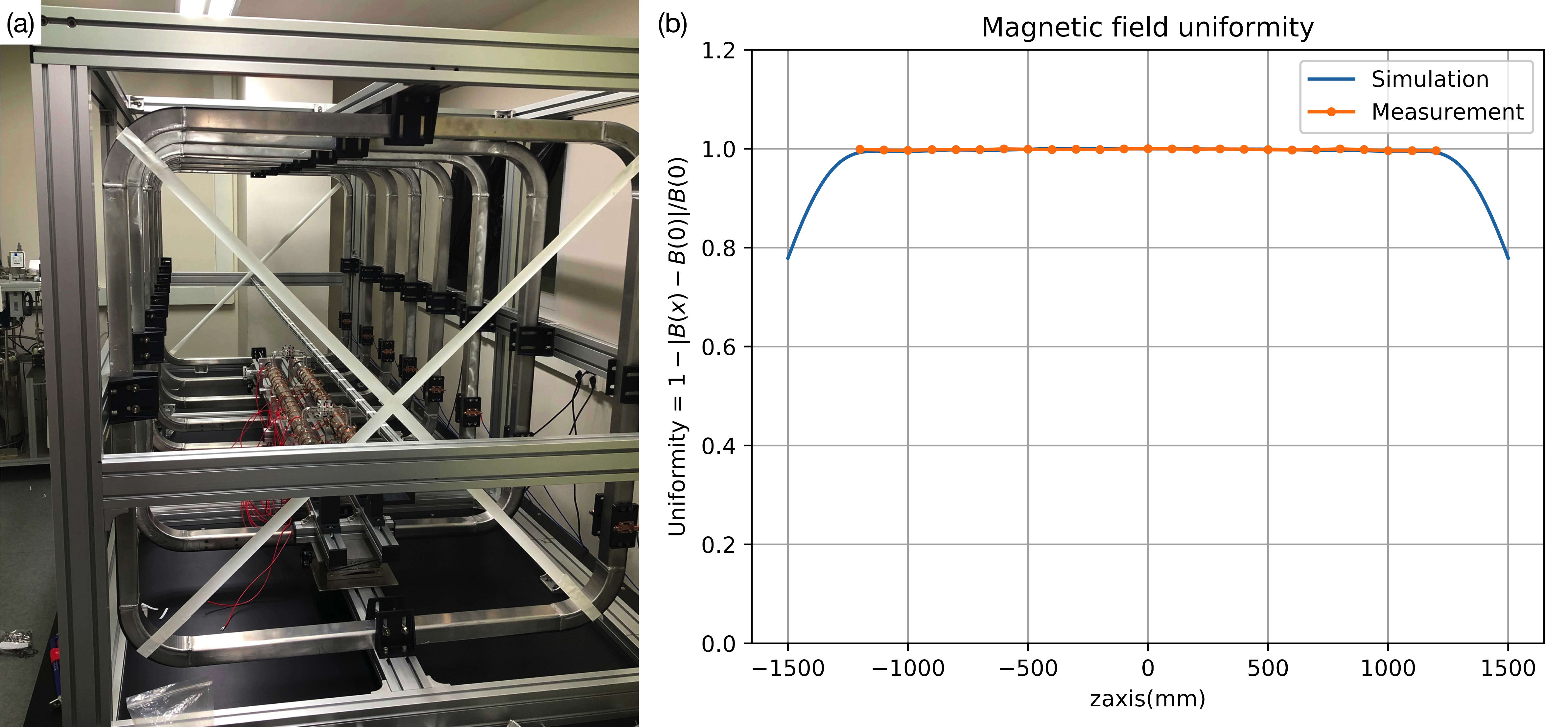}
\caption{(Left) The polarization up to 33\% has been achieved with a test cell filled with  3He gas. (Right) T1 relaxation time after 3He is polarized.}
\label{fig:ARIADNE_FIG_6}
\end{center}
\end{figure}

The magnetic field is controlled with total 4 set of power supply. Each of power supply provide up to 5\,A of current into two individual channels. Since all turn numbers are assumed to provide 1\,A of current, we simply multiplied current to 4\,A to generated ~13\,G of magnetic field. The homogeneity of magnetic field was, then, surveyed with a Gauss meter in every 10cm position from the center to each end. Figure~\ref{fig:ARIADNE_FIG_6} shows (a) the guide coil system (b) the measured magnetic field homogeneity comparing with the result from FEM simulation.

After optical pumping, polarized 3He gas will be compressed in a compression unit with a non-magnetic piston, and will be stored at $\sim0.1{\rm{atm}}$ pressure in a storage volume made with GE180 glass. All units will be installed on three different sides of a triangular skeleton post. Six sets of coils will be wound around the posts to provide a uniform magnetic field while 3He gas is polarized and transported. 

When a tungsten source mass having high nucleon density is rotating, a fictitious magnetic field   is generated from the source mass and it generate spin dependent force to 3He nucleus spins. Combined with the external field , which determines the Larmor frequency of the 3He, the effect of is measured using the NMR technique. Since the force range is short and weak. a sensitive SQUID magnetometer system is needed for detecting the NMR signals with high precision.

\subparagraph{High sensitivity SQUID development  with KRISS}
For low-noise SQUID magnetometer, CAPP has worked with KRISS for designing a magnetometer based on DC-SQUID having low intrinsic noise. For proper damping of resonances in the SQUID loop, and flux transformer, we added damping resistor in the SQUID loop, and resistor-capacitor circuit in the input coil part. To eliminate the cross talk between pickup coil and 3He cell, external feedback scheme was used, where the screening current is the flux transformer circuit is nulled constantly. To remove trapped flux in the SQUID, thin film Pd heater is placed near the SQUID. 

\begin{figure}[bp]
\begin{center}
\includegraphics[width=.95\textwidth]{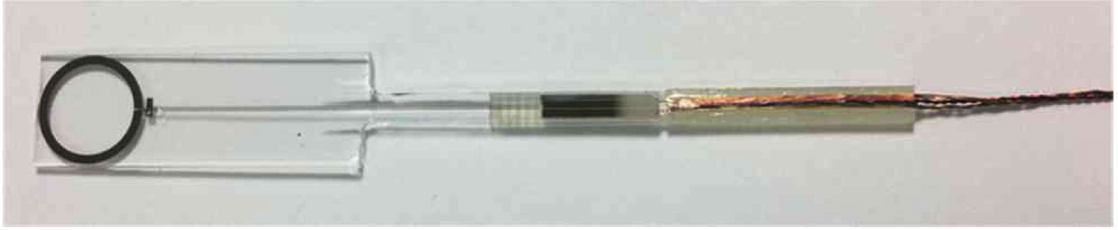}
\caption{Fabricated SQUID magnetometer. Assembled magnetometer fabricated on quartz substrate and wired}
\label{fig:ARIADNE_FIG_7}
\end{center}
\end{figure}

Since the space for 3He cell and SQUID pickup loop should be heavily shielded using Nb superconductive shielding layer, the wiring snout from the shielding enclosure also needs to be shielded using Nb tube of inner diameter. Pd thin film heater is placed near the SQUID loop, and the SQUID loop is parallel connected to reduce the SQUID inductance and to eliminate its response to uniform external field inside the Helmholtz field coil. Figure~\ref{fig:ARIADNE_FIG_7} showed the SQUID magnetometer fabricated on quartz tube.

Characterization of the SQUID magnetometers were done with the magnetometers inside a magnetically shielded room (MSR) having 2 layer mu-metal and 1 layer aluminum. The magnetometers are cooled by direct immersing into liquid He Dewar, made of non-magnetic fiberglass reinforced plastic. The flux-voltage curve shows large modulation voltage ($100\mu {\rm{V}}$) and quite steep flux-to-voltage transfer coefficient (about $1 \rm{mV}/\Phi_{0}$), which was possible with careful control of fabrication parameters to give near hysteretic region of the SQUID operation.

\begin{figure}[bp]
\begin{center}
\includegraphics[width=.7\textwidth]{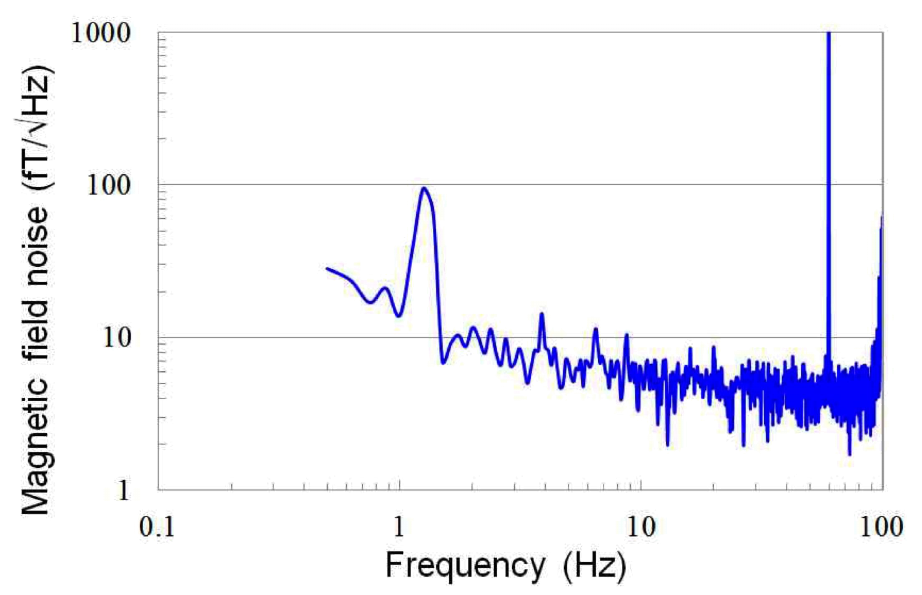}
\caption{Measured magnetic noise limit of SQUID.}
\label{fig:ARIADNE_FIG_8}
\end{center}
\end{figure}

The magnetic field noise of the magnetometer was about $4.5{\rm{f}}{{\rm{T}}_{\rm{rms}}}/\sqrt {{\rm{Hz}}}$ at $100{\rm{Hz}}$, including all the noise contributions from SQUID intrinsic noise, preamplifier input noise, thermal noise of liquid He Dewar, thermal noise of the MSR, residual environmental magnetic noise and digital (analog-to-digital circuit) noise. Figure~\ref{fig:ARIADNE_FIG_8} showed the measured magnetic flux limit of the SQUID. This prototype SQUID will be installed and tested in a prototype experiment of ARIADNE at Indiana University some time in 2019.

\subparagraph{Collaboration with US institutes}
While participating in the proposed research, a team of postdocs, graduate students, and undergraduate researchers will be broadly trained in the techniques of experimental atomic physics, optical pumping, nuclear magnetic resonance, low-temperature physics, micro-fabrication, magnetic shielding, vacuum systems, and modeling. This will be valuable preparation for work in basic or applied research, either in the U.S. or international work force or scientific community. An effort will be made to recruit members of under-represented groups as students in the laboratory, including women and minorities. Opportunities will exist for student involvement in an international collaboration, allowing travel between Korea and the United States. The CAPP periodically work with students from Indiana University on the apparatus setup through KUSP summer program.

\begin{itemize}

\item  Mofan Jang, Undergraduate, Indiana University, KUSP 2015 
\item  Inbum Lee, Graduate Student, Indiana University, KUSP 2016 
\item  Jooyun Woo, Graduate Student, Indiana University, KUSP 2017 

\end{itemize}

\paragraph{Summary and prospect}
A new experiment for axion search enables investigations of transient exotic spin couplings. It is based on synchronous operation of globally separated optical magnetometers enclosed inside magnetic shields. Correlation of magnetometers' readouts enables filtering local signals induced by environmental and/or technical noise. Moreover, application of vetoing techniques, e.g., via correlation of optical- magnetometer readouts with signals detected with non- optical magnetic-field sensors, enables suppression of influence of global disturbances of magnetic origins, such as solar wind, fluctuation of the Earth's magnetic field, on the operation of the magnetometers. In such an arrangement, the network becomes primarily sensitive to spin coupling of non-magnetic origins, thus it may be used for searches of physics beyond the Standard Model. A specific example of such searches was discussed here by considering coupling of atomic spins to domain walls of axion-like fields. It was demonstrated that with modern state-of-the-art optical magnetometers probing a significant region of currently unconstrained space of parameters of the fields is feasible.

%% file: 1.6.3/main.tex
\subsubsection{Introduction}
Large dipole magnetic field developments are fueled by the need for larger accelerator energies than currently existing at LHC of CERN, and IBS/CAPP is positioning itself to take first advantage of those magnets.  They provide a large $B^2V$ but the geometry is different, so the frequency tuning technology needs to be developed.  CAST-CAPP can be considered as a successful prototype towards that direction.

Axion dark matter search is compelling over a large mass span, arguably in the $10^{-6}\sim10^3\,\mu$eV range. Searches using microwave cavities immersed in high magnetic fields, above $20\sim30\,\mu$eV, able to reach QCD sensitivity are particularly challenging both due to the required high mechanical precision, and to their reduced volume. Alternative searches using dielectric materials present even bigger challenges. Large bore magnets with high magnetic fields in the range of 20$-$30\,T are an essential ingredient, and are within reach, but it will still take a few years to be fully implemented in solenoid configuration and made available for axion search, both in Korea and elsewhere.

Recent developments have successfully proven that multiple, tunable rectangular cavities, conceived and built at CAPP, can be inserted and operated in large dipole magnets, such as the 10\,m long, 42\,mm twin-bore, 9\,T CAST superconducting magnet at CERN. These existing resonators operate in the axion mass range 18$\sim$23\,$\mu$eV. This has been the focus of the CAST-CAPP project over the past three years.

It should be emphasized that this is the first time that tunable rectangular microwave cavities for axion searches are built and operated. And also, the first time that a system of multiple cavities is implemented, inserted, and operated in any magnet. Despite the complicated configuration of this installation, we also demonstrated that multiple rectangular cavities can be phase matched, thus optimizing sensitivity.

At CERN, dipole magnets are currently operating, with fields up to 15\,T and large bore. Dipole magnets with fields up to 20\,T are being developed. If exploited, large dipoles offer an immediate advantage compared to cylindrical geometry. The impact in the axion field can be considerable, due to the large $B^2V$ offered by these magnets. We recall that the power generated by an axion converting into a photon, inside a cavity of volume $V$, under a magnetic field $B$, is proportional to $B^2V$. The magnetic field intensity being equal, volume ratios of existing dipole to solenoid magnet bore can be a factor of 4 to 30 larger.

For the year 2019, we propose to continue the development of the CAST-CAPP project at CERN with the enhancement of the current setup, able to take data and deliver sensitivity comparable to HAYSTAC. This requires a very modest investment, in the order of a few tens of millions of KRW. We also propose to study multi-cell tunable rectangular cavities able to operate in the range of $30\sim40\,\mu$eV, also requiring a very modest investment.

A sizable advantage of rectangular geometry compared to cylindrical geometry is the lack of mode crossing. If a rectangular cavity tuning is properly chosen, higher modes will not interfere, whereas in cylindrical geometry mode crossing is unavoidable, either leaving gaps in the sensitivity, or forcing complicated tuning schemes that partially depend on simulations.

This project has leveraged international resources involving top expertise at CERN and Brookhaven National Laboratory (USA), at a fraction of the cost of other CAPP axion projects, and has brought benefits back to CAPP, for instance the idea of a split cavity, now widely used in our center.

\subsubsection{Status of CAST-CAPP}
\begin{figure}[h]
\centering
\includegraphics[width=1.0\textwidth]{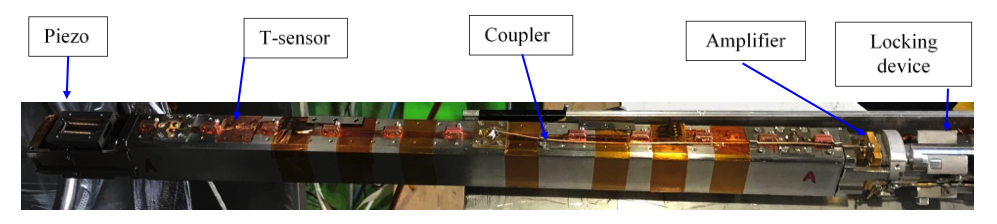}
\caption{Typical cavity assembly.}
\label{fig:cast-capp_fig_1}
\end{figure}

Four tunable, $23\times25\times390$ mm$^3$ rectangular cavities, made of stainless steel electroplated with $\sim$30\,$\mu$m of copper, have been installed inside the CAST magnet during the month of July 2018 (Fig.~\ref{fig:cast-capp_fig_1}). The full tuning range is $\sim$1\,GHz from 4.5\,GHz through 5.5\,GHz, corresponding to an axion mass range of 19$-$23\,$\mu$eV. 
The cavities are placed deep inside the magnet bore, several meters from its entrance. Currently, two cavities (cavity 3 and 4) can be immediately used for data taking and frequency scanning. The other two cavities (cavity 1 and 2) were demonstrated to work properly immediately after their installation. However, due to cabling issues developed after installation, access to the magnet is required in order to recover their full functionality. This usually needs to be scheduled well ahead of time, since the CAST magnet cooling cycle requires a few weeks involving CERN magnet experts. 

\begin{figure}[h]
\centering
\includegraphics[width=1.0\textwidth]{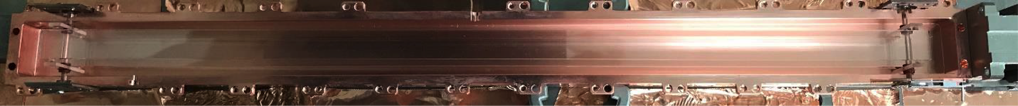}
\caption{A half-cavity hosting the assembled sapphire tuner.}
\label{fig:cast-capp_fig_2}
\end{figure}

The all sapphire tuning mechanism of each cavity, of unique design, has been entirely developed, and realized at CAPP (Fig.~\ref{fig:cast-capp_fig_2}). It is composed of two long, parallel sapphire plates activated by a single piezoelectric device through a locomotive mechanism, delivering a tuning resolution of better than 100\,Hz, when operated in stable conditions. This tuning device is mechanically robust and easily portable between cavities.

\begin{figure}[h]
\centering
\includegraphics[width=1.0\textwidth]{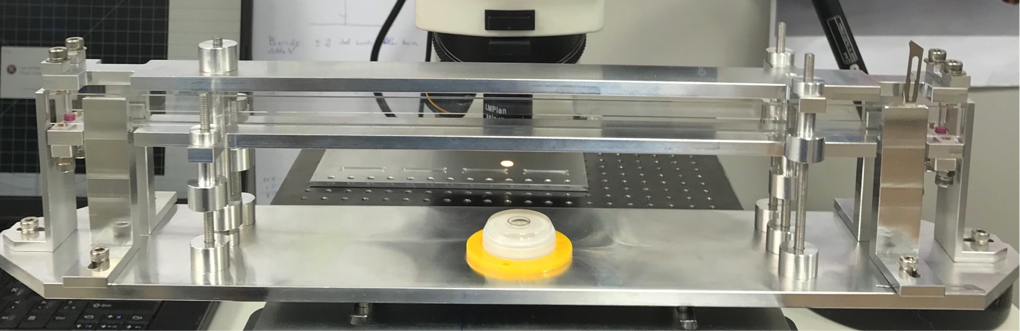}
\caption{Sapphire tuner in its assembly device before heat treatment.}
\label{fig:cast-capp_fig_3}
\end{figure}

Figure~\ref{fig:cast-capp_fig_3} shows the tuner on its final assembly device. This original device is designed to compensate and absorb all fabrication errors and insure smooth operations and robustness of the tuner.

Provisions to compensate for the differential thermal expansion between sapphire and stainless steel make these cavities able to sustain a liberal number of cooling cycles. The quality factor of each cavity is reduced by $\sim$20\% with the introduction of the tuner. The unique combination of a split cavity and the all-sapphire tuner, has made possible safe transportation, installation, and operation several meters inside the magnet bore. A delivery/locking/thermal anchoring/recovery system makes the system easily installable. This device is partially shown in the right-hand side of Fig.~\ref{fig:cast-capp_fig_2}. 

\begin{figure}[h]
	\centering
	\begin{subfigure}[]{0.49\textwidth}
		\centering
		\includegraphics[width=\linewidth]{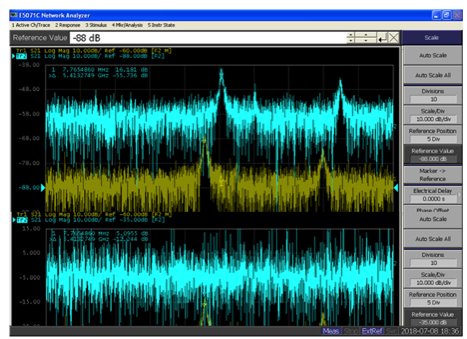}
		\subcaption{}
	\end{subfigure}
	\begin{subfigure}[]{0.48\textwidth}
		\centering
		\includegraphics[width=\linewidth]{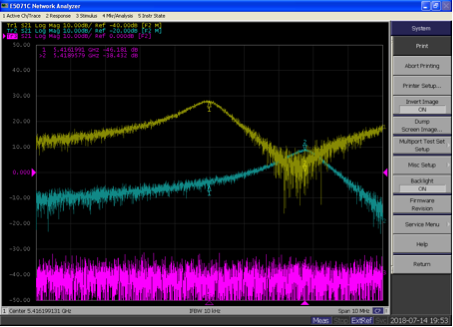}
		\subcaption{}
	\end{subfigure}
	\caption{Transmission spectra of cavity-1 (a) and cavity-2 (b) after installation. The bottom part of both figures shows the transmission measurement with the amplifier turned off. The top part shows both cavities tuned to two different frequency configurations (yellow and blue colors) by the piezo-actuator.}
	\label{fig:cast-capp_fig_4}
\end{figure}

In the following we show that all four cavities are properly working, the cabling issue that developed inside the magnet has temporarily disabled two cavities. The proper working of cavity 1 and cavity 2 is shown in Fig.~\ref{fig:cast-capp_fig_4}.

\begin{figure}[h]
\centering
\includegraphics[width=0.7\textwidth]{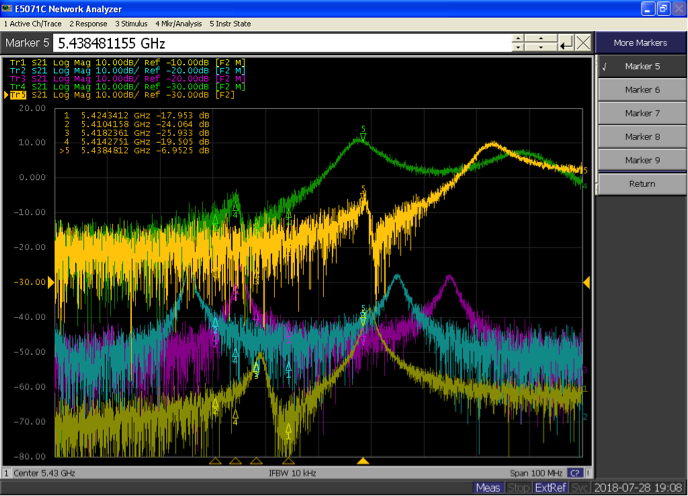}
\caption{Cavity-2, 3, and 4 transmission measurements. Yellow trace (bottom): cavity-2 at a fixed frequency. Blue and magenta traces: cavity-3 tuned to two different frequency configurations. Green and yellow traces (top): cavity-4 tuned to two different configurations.}
\label{fig:cast-capp_fig_5}
\end{figure}

Room temperature transmission measurements for cavities 2, 3, and 4 are shown in Fig.~\ref{fig:cast-capp_fig_5}. These measurements were taken on July 28, 2018.

\begin{figure}[h]
\centering
\includegraphics[width=0.8\textwidth]{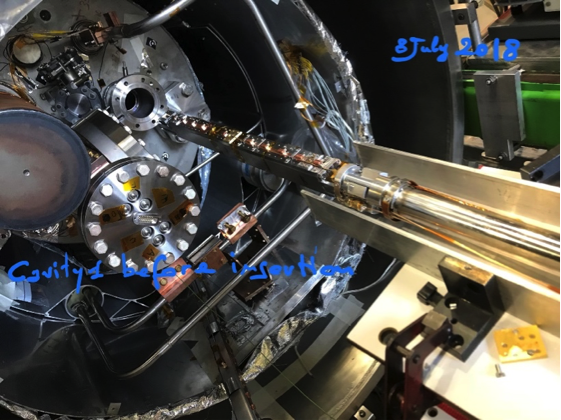}
\caption{Cavity-1 attached to the insertion tool before installation, connected to the delivery/recovery tool.}
\label{fig:cast-capp_fig_6}
\end{figure}

Finally, in Fig.~\ref{fig:cast-capp_fig_6} we show one of the four cavities just before installation inside the CAST magnet.

This has been a successful installation, despite the drawbacks of the two disconnected cables that disabled the first cavity, and prevent the tuning of the second cavity. If compared with the research proposal of 2017 at this time, the progress has been remarkable; the more so if considered that it was realized with minimal and discontinuous human resources, and a small investment, if compared to other CAPP projects. Each cavity assembly, particularly the tuner, turned out to be robust and reliable. The circumstance that all four sapphire tuners would survive intact to multiple manipulations and travel over multiple flights was not obvious. 

\begin{figure}[h]
\centering
\includegraphics[width=0.8\textwidth]{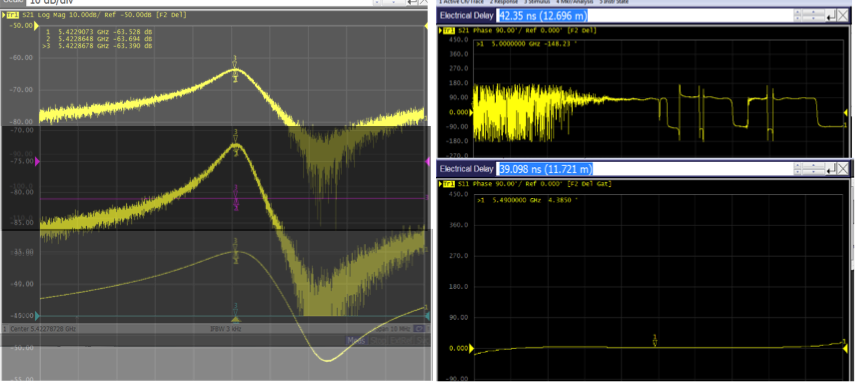}
\caption{Demonstration of phase matching of three cavities (Sep 28, 2018).}
\label{fig:cast-capp_fig_7}
\end{figure}

Phase matching among cavities is important in order to optimize the sensitivity of a multi-cavity assembly. Phase matching of a system of three cavities, inside the same CAST magnet was demonstrated at room temperature, to $\sim$10\%. The procedure is based on adding appropriate delay lines to each cavity output, deduced from a combination of broadband phase measurements of transmission signals with broadband phase/time domain measurements of properly gated input reflection losses. The right-hand side of Fig.~\ref{fig:cast-capp_fig_7} shows a sample of these two types of measurements for one of the matched cavities, after the three cavities were tuned to the same frequency, as seen on the left-hand side. This is also a first in axion search.
This installation should be regarded as a first phase whose purpose is to demonstrate that an axion search experiment is feasible with multiple, phase matched tunable cavities in a large dipole magnet. Data acquisition is now in the commissioning phase.

\subsubsection{Prospects}

\begin{figure}[h]
\centering
\includegraphics[width=0.8\textwidth]{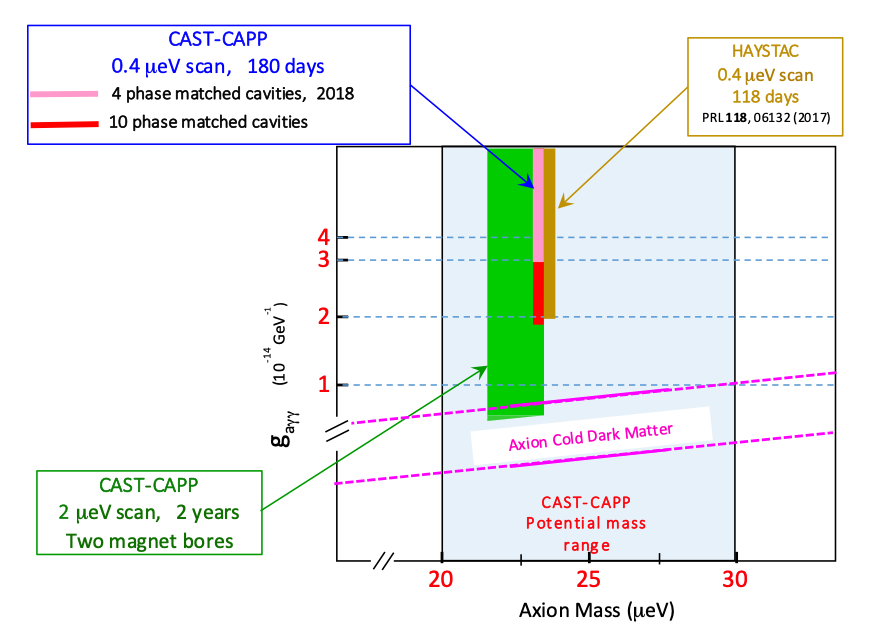}
\caption{Projected sensitivity of the CAST-CAPP project.}
\label{fig:cast-capp_fig_8}
\end{figure}

The projected sensitivity of the CAST CAPP project using four cavities is represented by the pink bar bar of Fig.~\ref{fig:cast-capp_fig_8}, compared to the recent measurement of HAYSTAC (mustard). The plot shows the potential if more cavities are installed, similar to those already installed. 

We refrain from making any statements on a possible reach beyond 2019, since this plot makes clear what the potential is, while our employment position and the available budget are both undefined at the time when this report is generated. 

\begin{figure}[h]
\centering
\includegraphics[width=1.0\textwidth]{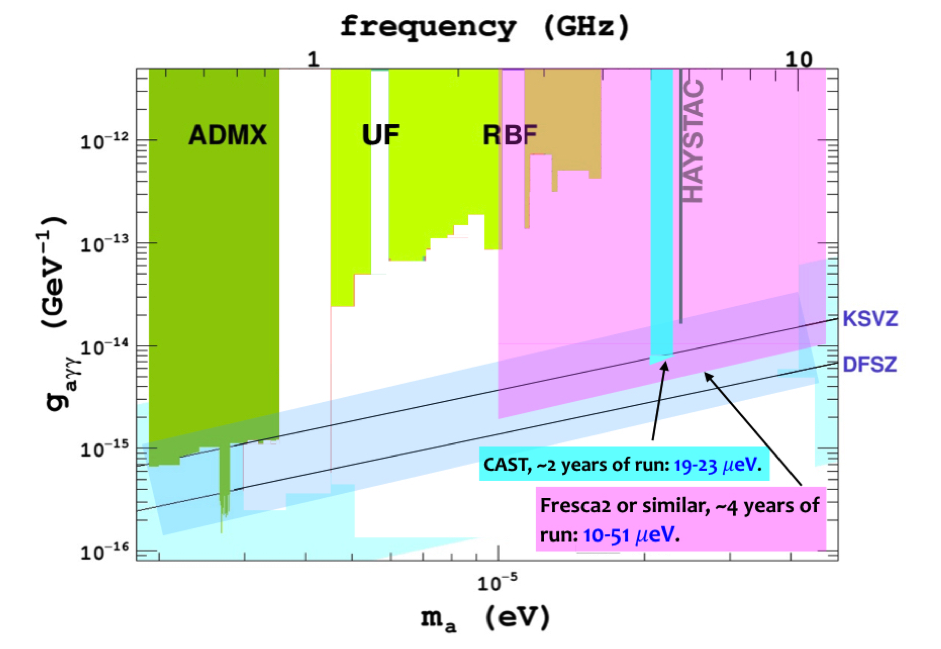}
\caption{The projected sensitivity with the existing Fresca2 dipole magnet (pink band), compared to all published data, and to the projected CAST-CAPP sensitivity (blue).}
\label{fig:cast-capp_fig_9}
\end{figure}

More striking is how the accumulated experience with the CAST-CAPP project could be applied in existing, higher field dipole magnets. By combining properly designed multiple cavities (filters) and phase matching, with new tuning concepts, remarkable sensitivities to axion masses ranging from 10\,$\mu$eV to 50\,$\mu$eV can be achieved. One such magnet is Fresca2, a 15\,T, 100\,mm bore, 1.5\,m long dipole magnet in operation at CERN since 2017. The sensitivity plot of Fig.~\ref{fig:cast-capp_fig_9} shows the potential of an axion search experiment with Fresca2, with the insertion of a cold finger at 50\,mK, using quantum-noise limited amplifiers. 

